%% file: main.tex
\documentclass[
11pt, 
english, 
singlespacing, 
headsepline, 
]{MastersDoctoralThesis} 

\usepackage[utf8]{inputenc} 
\usepackage[T1]{fontenc} 
\usepackage{wrapfig}
\usepackage{amsmath, amssymb, cancel}
\usepackage{amsfonts,amscd,yhmath,bbm}
\usepackage{amsthm}
\usepackage{mathtools}
\usepackage{amscd}
\usepackage{bbm, stmaryrd}
\usepackage{yfonts}
\usepackage{tabu}
\usepackage{multicol}
\usepackage{stix}
\usepackage{palatino} 
\usepackage{graphicx}

\theoremstyle{definition}
\newtheorem{teo}{Theorem}[section]
\newtheorem{defi}{Definition}[section]
\newtheorem{oss}{Remark}[section]
\newtheorem{Lemma}{Lemma}[section]
\newtheorem{cor}{Corollary}[section]

\newtheorem{prop}{Proposition}[section]

\usepackage{tikz}
\usetikzlibrary{patterns,matrix,through,arrows,decorations.pathreplacing,decorations.markings,snakes,shadows,shapes.geometric,positioning,calc,backgrounds,shapes.arrows,fit,backgrounds}

\newcommand{\e}{{\textbf{e}}} 
\newcommand{\n}{{\textbf{n}}}
\newcommand{\B}{{\texttt{B}}}
\newcommand{\1}{{\overrightarrow{\emph{1}}_{\infty}}}

\tikzset{anchorbase/.style={baseline={([yshift=-0.5ex]current bounding box.center)}},
int/.style={thick},
  cross line/.style={preaction={draw=white,line width=6pt,-}},
  wall/.style={thin,double,blue},
  middlearrow/.style={postaction=decorate,decoration={markings,mark=at
    position .55 with {\arrow{stealth};}}},
  middlearrowrev/.style={postaction=decorate,decoration={markings,mark=at
    position .55 with {\arrowreversed{stealth};}}},
  ev/.style={shape=rectangle, draw},
  every path/.style={line width=1pt}
}  
\input{tikzcommands.tex}

\RequirePackage{verbatim}

\makeatletter
\newwrite\bibinl@out
{\immediate\closeout\bibinl@out\@esphack}
\makeatother

\usepackage[autostyle=true]{csquotes} 

\thesistitle{Elliptic multiple zeta values, modular graph functions and genus 1 superstring scattering amplitudes} 
\supervisor{Prof. Dr. Don \textsc{Zagier}} 
\examiner{} 
\degree{} 
\author{Federico Zerbini} 
\addresses{Vivatsgasse 7, 53111, Bonn, Germany} 

\subject{Mathematics} 
\university{} 
\department{} 
\group{}{} 
\faculty{}{} 

\AtBeginDocument{
\hypersetup{pdftitle=\ttitle} 
\hypersetup{pdfauthor=\authorname} 
\hypersetup{pdfkeywords=\keywordnames} 
}

\begin{document}

\frontmatter 

\pagestyle{plain} 


\begin{titlepage}
\begin{center}

\vspace*{.06\textheight}

{\huge \bfseries \ttitle\par}\vspace{2.4cm} 
 

\Large \textbf{Dissertation}\\[0.3cm] 
\Large \textbf{zur}\\[0.3cm]
\Large \textbf{Erlangung des Doktorgrades (Dr. rer. nat.)}\\[0.3cm]
\Large \textbf{der}\\[0.3cm]
\Large \textbf{Mathematisch-Naturwissenschaftlichen Fakult\"{a}t}\\[0.3cm]
\Large \textbf{der}\\[0.3cm]
\Large \textbf{Rheinischen Friedrich-Wilhelms-Universit\"{a}t Bonn}\\[2.4cm]
\Large \textbf{vorgelegt von}\\[0.4cm]
\huge \textbf{Federico Zerbini}\\[0.3cm]
\Large \textbf{aus}\\[0.3cm]
\Large \textbf{Mailand, Italien}\\[0.8cm]

\vfill

{\large Bonn, 13/08/2017}\\[4cm] 

\vfill
\end{center}
\newpage
\begin{center}
\vfill
Angefertigt mit Genehmigung der Mathematisch-Naturwissenschaftlichen Fakultät der Rheinischen Friedrich-Wilhelms-Universität Bonn
\end{center}
\vspace{8.5cm}
1. Gutachter: Prof. Dr. Don Bernard Zagier\\[0.3cm]
2. Gutachter: Prof. Dr. Werner Ballmann\\[0.3cm]
Tag der Promotion: 18. Oktober 2017\\[0.3cm]
Erscheinungsjahr: 2018\\[0.3cm]

\end{titlepage}



 
 

\cleardoublepage






\begin{abstract}
\addchaptertocentry{\abstractname} 
We study holomorphic and non-holomorphic elliptic analogues of multiple zeta values, namely elliptic multiple zeta values and modular graph functions. Both classes of functions have been discovered very recently, and are involved in the computation of genus one superstring amplitudes. In particular, we obtain new results on the asymptotic expansion of these functions that allow us to perform explicit computations and point out analogies between genus zero and genus one amplitudes.
\end{abstract}


\begin{acknowledgements}
\addchaptertocentry{\acknowledgementname} 
First of all, I would like to thank my (first) advisor, Don Zagier, for help, encouragement, funny jokes, interesting discussions, and in particular for teaching me a lot of beautiful mathematics. Many thanks also to my second advisor, Werner Ballmann, for doing his best as a referee of my PhD thesis despite his different mathematical interests, and to the rest of my PhD committee, Catharina Stroppel and Albrecht Klemm. 

Thanks a lot to D. Zagier, N. Matthes, C. Dupont, G. Bogo and O. Schlotterer for their comments on preliminary drafts of the thesis, and to E. Garcia Failde for the beautiful dibujitos.

Thanks to all the mathematicians and physicists who have helped and guided me in these years: P. Vanhove, M.B. Green, F. Brown, Y. Manin, H. Gangl, A. Valentino, R. Rios Zertuche, J. Fres\'an, G. Borot, A. Ros Camacho, E. Panzer, O. Schlotterer, N. Matthes, J. Br\"odel, D. Radchenko, and especially to C. Dupont, whose help and encouragement have been crucial at the beginning of my PhD studies.

Thanks to E. Panzer for providing me with a version of \texttt{HyperInt} specially designed for performing efficient computations of conical sums.

Thanks to my family and to many new and old friends who have shared great moments with me during these years, and particularly to all of you who have helped me and visited me during the long months at the hospital.

Grazie mille Lorena per il supporto che non mi hai mai fatto mancare, specialmente durante queste ultime settimane.

Finally, thanks to those who taught me mathematics before my PhD, and particularly to M. Acchiappati, G. Molteni, K.R. Payne, M. Bertolini and Y. Bilu.

\end{acknowledgements}


\tableofcontents 

\dedicatory{Dedicato ai miei genitori.}


\mainmatter 

\pagestyle{thesis} 


\include{Chapter1}
\include{Chapter2} 
\include{Chapter3}
\include{Chapter4} 
\include{Chapter5}
\include{Chapter6}


\appendix 


\include{AppendixA}


\newpage

\bibliographystyle{plain}

\end{document}

%% file: tikzcommands.tex
\def \myweightstyle {}
\def \myweightx {0.09}
\def \myweighty {0.2}
\newcommand{\mydrawdown}[1]%
  {\draw [\myweightstyle] #1++(0,-\myweighty) -- ++(\myweightx,\myweighty);%
    \draw [\myweightstyle] #1++(0,-\myweighty) -- ++(-\myweightx,\myweighty);}
\newcommand{\mydrawup}[1]%
  {\draw [\myweightstyle] #1 -- ++(-\myweightx,-\myweighty);%
    \draw [\myweightstyle] #1 -- ++(\myweightx,-\myweighty);}

%% file: Chapter1.tex

\chapter{Introduction} 

\label{Introduction} 


\newcommand{\keyword}[1]{\textbf{#1}}
\newcommand{\tabhead}[1]{\textbf{#1}}
\newcommand{\code}[1]{\texttt{#1}}
\newcommand{\file}[1]{\texttt{\bfseries#1}}
\newcommand{\option}[1]{\texttt{\itshape#1}}


The aim of this thesis is to study two families of functions on the complex upper-half plane $\mathbb{H}$: elliptic multiple zeta values and modular graph functions. In particular, we focus on working out explicit formulae for their asymptotic behaviour, motivated by the fact that these two families play a key r\^{o}le in the computation of genus one scattering amplitudes of open and closed strings, respectively. In particular, we have applied our results to carry out many new explicit computations of closed string amplitudes in genus one: these computations revealed a fascinating structure, which suggests a parallel with the beautiful mathematics describing the genus zero case. We will also give further evidence of the expectation that elliptic multiple zeta values and modular graph functions should be closely related. This has to do with the expectation that open and closed superstring theories, generalizing gauge theories and gravity, respectively, should be deeply connected\footnote{This expectation is based on observations at genus zero, as well as on a recently observed connection between perturbative quantum gravity and a double-copy of gauge theory \cite{BCJ}.}.

\section{Scattering amplitudes and multiple zeta values}

One of the most important goals of modern theoretical physics is to predict probability amplitudes of scattering processes. It is a great source of experimental confirmations for quantum field theories, and string theory was born out of the study of amplitudes \cite{Veneziano}. 

In the \emph{path integral} approach to quantum physics, amplitudes are computed by averaging the exponential of the action over all possible paths, as in the equation below\footnote{Here we are greatly simplifying, for the sake of brevity. In particular, the integral (\ref{pathintegral}) computes vacuum amplitudes: to address external legs, one needs to insert other quantities in the integral.},
\begin{equation}\label{pathintegral}
\int_{\phi\in\Phi}e^{-\frac{i}{\hslash}S(\phi)}d\phi,
\end{equation}
where the space of paths $\Phi$, the measure $d\phi$, and the action $S$ depend on the theory. The classical field configurations, which are classical solutions of lagrangian equations, are stationary points of the action, and therefore by stationary phase they give the biggest contributions to the integral for $\hslash$ small. 

One of the problems of computing (\ref{pathintegral}) is that, typically, the integration over the (infinite-dimensional) space of paths is ill-defined and depends on the integration procedure. The approach proposed by Feynman goes as follows. One can approximate amplitudes by associating to certain graphs, called \emph{Feynman graphs}, which discretize the problem and schematically encode all possible trajectories, certain corresponding \emph{Feynman integrals}. The idea then is just to sum all these integrals up over all the possible Feynman graphs, organized by growing complexity (=number of cycles in the graph). This reminds of the power series expansion of a function around a point, and for this reason this approach goes under the name of \emph{perturbative quantum field theory}. The ultimate goal would be to compute the whole amplitude with a non-perturbative approach. However, already this approximation turns out to be extremely powerful, as it fournishes a great supply of predictions, that have been successfully matched to high precision with particle accelerators experiments.

The link that we want to emphasize between number theory and computations of scattering amplitudes probably goes back to the paper of Broadhurst and Kreimer \cite{BroadKreimer}, two physicists who realized that many scattering amplitudes could be computed in terms of the special holomorphic functions 
\begin{equation}\label{defMPL}
\mbox{Li}_{k_1,\ldots ,k_r}(z_1,\ldots ,z_r)=\sum_{0<v_1<\cdots <v_r}\frac{z_1^{v_1}\cdots z_k^{v_k}}{v_1^{k_1}\cdots v_r^{k_r}}
\end{equation}
of $r\geq 1$ complex variables such that $|z_1\cdots z_r|<1$, where $k_1,\ldots ,k_r\in\mathbb{N}^r$. These functions, called \emph{multiple polylogarithms}, generalize the classical polylogarithm function
\begin{equation}
\mbox{Li}_{k}(z)=\sum_{v\geq 1}\frac{z^v}{v^{k}},
\end{equation}
an important class of special functions that is related in a deep way to hyperbolic geometry, K-theory and algebraic number theory (see for instance \cite{ZagierPolylog}) and that includes for instance the complex logarithm
\[
\mbox{Li}_1(z)=\sum_{v\geq 1}\frac{z^v}{v}=-\log(1-z).
\]
If $k_r\geq 2$ (needed for absolute convergence) one can extend multiple polylogarithms to the point $z_1=\cdots =z_r=1$, and get the following real numbers:
\begin{equation}\label{defMZV}
\zeta(k_1,\ldots ,k_r)=\sum_{0<v_1<\cdots <v_r}\frac{1}{v_1^{k_1}\cdots v_r^{k_r}}.
\end{equation}
These numbers are called \emph{multiple zeta values} (or, in short, \emph{MZVs}). They generalize the special values at integers $s\geq 2$ of the Riemann zeta function
\[
\zeta(s)=\sum_{n\geq 1}\frac{1}{n^s}.
\]
They have been considered already by Euler, but they have been studied systematically only over two hundred years later by Zagier \cite{ZagierMZV}.

MZVs are precisely the numbers that Broadhurst and Kreimer found in \cite{BroadKreimer} computing amplitudes in quantum field theories. Since then, a great effort has been spent in trying to understand which kind of numbers are expected to come from physics, and why. 

It soon became clear that each amplitude computation dictates a certain type of geometry, which in turn determines which kind of numbers we should expect to find. These numbers are \emph{periods}. 

The simplest way to define periods, due to Kontsevich-Zagier \cite{KontsZagier}, is to say that they are complex numbers whose real and imaginary part can be written as absolutely convergent integrals of rational function with rational coefficients over domains in $\mathbb{R}^n$ given by polynomial inequalities with rational coefficients. This definition gives a convenient way to recognize periods, but in fact, classically, they are defined as the entries of the matrices giving the isomorphism between complex (relative) Betti and de Rham cohomologies of smooth quasi-projective varieties. As we will soon see, MZVs can be seen as periods of compactified moduli spaces $\overline{\mathcal{M}}_{0,n}$ of punctured Riemann spheres. 

In the $90$'s it was not clear whether the geometry of quantum field theory was complicated enough to produce other functions than just multiple polylogarithms. We now know (after \cite{BlochEsnaultKreimer}, \cite{BelBros}, \cite{BrownSchnetz}, \cite{BlochVanhove}, \cite{Weinzierl14} and many more)  that this is actually not the case, and that we should in general expect more complicated special functions, such as elliptic generalizations of polylogarithms. A natural playground to see how one would naturally want to go beyond multiple polylogarithms and MZVs is given by superstring theory.

\section{Superstring amplitudes}

A very rough picture that a reader who is unfamiliar with superstring theory should keep in mind is the following: in quantum field theory (massless) particle states can be thought of as points, and Feynman graphs describe the processes coming from the interaction of particles as time evolves. Strings (massless string states) should be thought of as small one-dimensional simple curves parametrized by the unit interval: in case the endpoints coincide, we talk of closed strings (think of little circles), while if they do not coincide we talk of open strings. Therefore, by analogy with quantum field theory, perturbative scattering amplitudes of $n$ strings are computed by a Feynman graph-like expansion organized by the number of holes of two-dimensional smooth connected surfaces with boundaries with $n$ string insertions, as in the pictures below\footnote{Many thanks to E. Garcia Failde for the pictures.}, where $n=4$, red boundary components represent (massless external) string states, and black lines show the evolution of strings in time.

\begin{figure}[h!]
\begin{center}
\def\svgwidth{\columnwidth}
 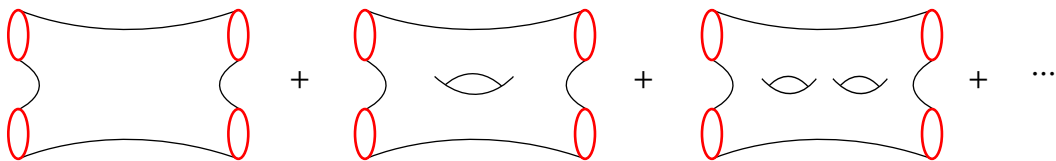
 \caption{Four closed strings}
    \end{center}
\end{figure}

\begin{figure}[h!]
\begin{center}
\def\svgwidth{\columnwidth}
 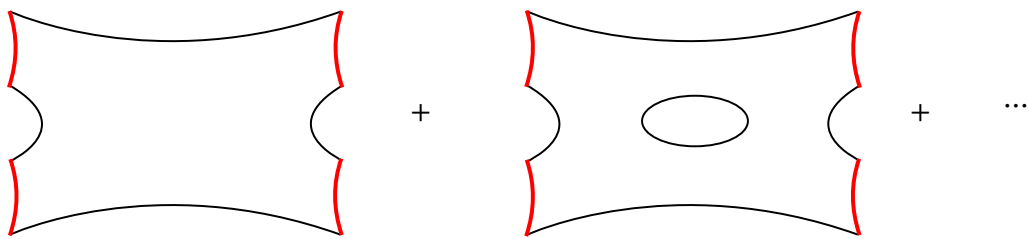
 \caption{Four open strings}
    \end{center}
\end{figure}

It is well known \cite{GSW} that actually one can shrink boundary components to points and classify these surfaces only by their conformal structure (see pictures 1.3 and 1.4 below, where in the open string case we have given only one example of the possible genus one topologies that could contribute).

\begin{figure}[h!]
\begin{center}
\def\svgwidth{\columnwidth}
 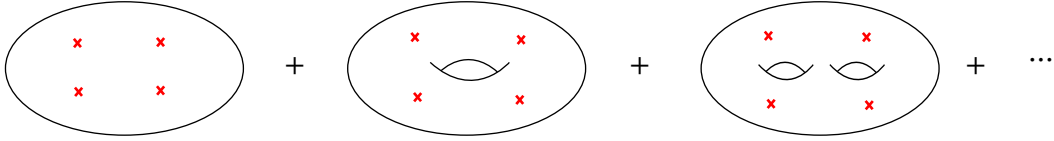
 \caption{Four closed strings}
    \end{center}
\end{figure}

\begin{figure}[h!]
\begin{center}
\def\svgwidth{\columnwidth}
 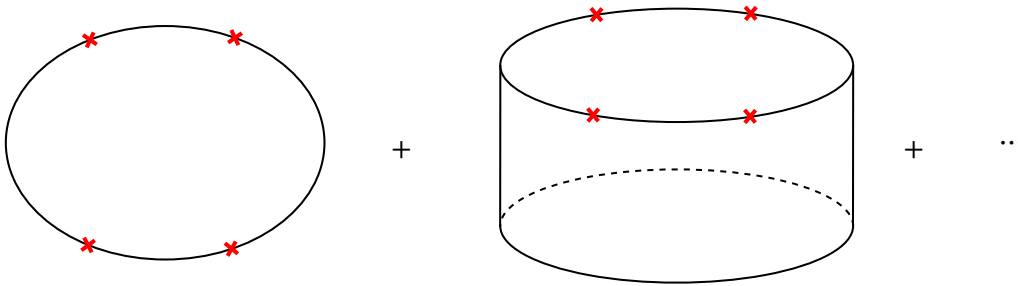
 \caption{Four open strings}
    \end{center}
\end{figure}

In other words, the analogues of Feynman graphs in superstring theory are given by (super) Riemann surfaces with $n$ marked points, and therefore we have the great advantage of having a nice mathematical description of the domain of Feynman integrals: the amplitude is, very roughly speaking, given by the sum over all genera of certain integrals of Green functions over compactified moduli spaces $\overline{\mathcal{M}}_{g,n}$ of genus~$g$ Riemann surfaces with $n$ marked points. In the case $g=0$ very much is already known or conjectured for any number of both open and closed strings, thanks to the introduction of tools coming from number theory and geometry, like motives and associators. The case $g=1$ is more challenging and not yet understood; it has inspired very active research, initiated by Green and Vanhove in 1999 \cite{GV2000}, and it is related to the development of new number theoretical objects in mathematics, as  we will see in this thesis. Almost nothing is known for higher genera.

More in details, the state of art at the moment goes as follows. For any fixed genus, each superstring amplitude can be written as a power series expansion in the \emph{Mandelstam variables}, which depend on the strings' tension and momenta. This is called the \emph{low-energy expansion}. The quantities that one wants to compute are the coefficients of this expansion. 

In the genus zero case it was observed long ago that, for the simplest physically meaningful four-string case, these coefficients are simple Riemann zeta-values $\zeta(k)$ \cite{GSW}. In particular, while for open strings one can find all simple zetas, in the closed string case one finds odd simple zetas $\zeta(2k+1)$ only. Recently, this has been generalized to the following picture \cite{BSST}, \cite{ScSt}: the coefficients of the low energy expansion are multiple zeta values, and if the number of strings is bigger than four, one finds also multiple zeta values which are not reducible to simple Riemann zeta values. In the closed case the coefficients (conjecturally) lie in a subset of the set of  multiple zeta values, whose elements are called \emph{single-valued multiple zeta values} (we will explain in the next chapter the origin of this name). This reflects what happens with four strings, because the only simple zeta values $\zeta(k)$ belonging to the set of single-valued MZVs are the odd zeta values.

The situation in genus one is more challenging, both because of the more complicated Green functions that we need to integrate and because of the less trival geometry of the moduli space. The quantities that one wants to compute now are functions of the modulus of the elliptic curve $\tau\in\mathbb{H}$ (instead of numbers). 

In the case of open strings, these functions are holomorphic but not modular, at least for the choice of topology of figure 1.4, and admit a Fourier expansion in the variable $q=\exp(2\pi i\tau)$ of the form
\begin{equation}\label{ExpOpenStrings}
\sum_{n\geq 0}a_nq^n,
\end{equation}
where all $a_n$'s are multiple zeta values\footnote{Actually one also needs to take into account $(2\pi i)^{\pm 1}$.}. It was remarked in \cite{BMMS} that they can be written in terms of \emph{elliptic multiple zeta values}, functions recently introduced by Enriquez in \cite{Enriquez} that constitute a genus one generalization of classical multiple zeta values.

In the case of closed strings, these functions are modular but not holomorphic (they are real analytic), and understanding their Fourier expansion was one of the goals of this thesis. They can be expressed\footnote{When the number of strings is $\geq 5$ this is not quite correct, as we will see in Chapter \ref{ChapterClosedStrings}. However, our focus was on computing four-point amplitudes.} in terms of \emph{modular graph functions}, a class of real analytic modular forms introduced in \cite{DGGV}, that include for instance non-holomorphic Eisenstein series. 

\section{Elliptic multiple zeta values and modular graph functions}

Let us now briefly introduce the two main mathematical objects studied in this thesis.

The first kind of functions that we want to consider, Enriquez's elliptic multiple zeta values, are divided into two families: \emph{A-elliptic MZVs} and \emph{B-elliptic MZVs}. This distinction comes from the fact that they originate from the elliptic associator, a pair $A(x_0,x_1;\tau)$ and $B(x_0,x_1;\tau)$ of formal power series\footnote{They are power series in the non-commutative variables $x_0,x_1$, whose coefficients are functions of~$\tau$.} describing the regularized monodromy of a certain differential equation (the \emph{elliptic KZB equation} \cite{EnriquezAss}) along the two canonical paths of a complex torus $\mathbb{C}/(\tau\mathbb{Z}+\mathbb{Z})$: $A(x_0,x_1;\tau)$ corresponds to $[0,1]$, $B(x_0,x_1;\tau)$ to $[0,\tau]$. Here there is a subtlety: Enriquez's A-elliptic and B-elliptic MZVs are defined as the coefficients of two generating series related to (but not equal to) $A(x_0,x_1;\tau)$ and $B(x_0,x_1;\tau)$, respectively; other authors, like Matthes, found more convenient\footnote{Matthes's definition gives functions defined as homotopy invariant iterated integrals on the two canonical paths.} to define elliptic MZVs directly as the coefficients of the associator. We will adopt Enriquez's viewpoint, because it gives access to explicit formulae. However, we want to remark that one can explicitly write Matthes's version of elliptic MZVs in terms of the ones considered in this thesis. 

The main results on elliptic MZVs are contained in \cite{Enriquez}, \cite{BMS}, \cite{MatthesDouble}, \cite{MatthesMeta}, \cite{MatthesDecomposition}, \cite{LoMaSc}. A good introduction to the subject, containing many of these results, is Matthes's recent PhD thesis \cite{MatthesThesis}. Let us briefly mention the main known properties, postponing the definition to Chapter \ref{ChapterEllMZV}. A-elliptic MZVs admit a Fourier expansion of the same form as (\ref{ExpOpenStrings}). B-elliptic MZVs are (almost) given by the image of A-elliptic MZVs under $\tau\rightarrow -1/\tau$, and admit a Fourier expansion
\begin{equation}\label{IntroBexp}
\sum_{n\geq 0}b_{n}(\tau)q^n,
\end{equation}
where $b_{n}(\tau)$ are Laurent polynomials whose coefficients can be written in terms of multiple zeta values and negative powers of $2\pi i$. Moreover, it is known that both kinds of elliptic MZVs satisfy a differential equation involving elliptic MZVs of simpler nature and (holomorphic) Eisenstein series
\[
G_k(\tau)=\sum_{(m,n)\in\mathbb{Z}^2\setminus\{(0,0)\}}\frac{1}{(m\tau+n)^k},
\] 
and that consequently they can be written as iterated integrals of Eisenstein series on the moduli space of elliptic curves. Many explicit formulae for the $q$-expansion of A-elliptic MZVs are already available in the literature. However, it seems that, in order to uncover the link between open and closed string amplitudes in genus one, it would be desirable to have an analogue of these formulae for B-elliptic MZVs \cite{Zerb17}. One of the goals of this thesis is to fill this hole in the literature.

The second kind of functions that we want to consider is indexed by undirected graphs~$\Gamma$ without self-edges, where we allow for multiple edges between two vertices. To any of these graphs we associate a modular graph function, which is a real analytic function $D_\Gamma(\tau)$ on the upper-half plane $\mathbb{H}$, invariant under the standard action of SL$_2(\mathbb{Z})$. The simplest examples of modular graph functions are given by the special values at $s\in\mathbb{Z}_{\geq 2}$ of the non-holomorphic Eisenstein series
\[
E(s,\tau)=\bigg(\frac{\Im(\tau)}{\pi}\bigg)^s\sum_{(m,n)\in\mathbb{Z}^2\setminus\{(0,0)\}}\frac{1}{|m\tau+n|^{2s}}.
\]
It is conjectured (and known for some infinite sub-families \cite{DGV2015}, \cite{AxelValentin}) that modular graph functions are solutions of certain kind of inhomogeneous Laplace equations generalizing the well known equation $(\Delta-n(n-1))E(n,\tau)=0$, where $\Delta$ is the hyperbolic Laplacian.

By modular invariance, these functions have a Fourier expansion in $\Re(\tau)$, and their \emph{zeroth Fourier mode} $\int_0^1D_\Gamma(\tau)d\Re(\tau)$ can be expanded as $d_\Gamma(\pi\Im(\tau))+O(\exp(-\pi\Im(\tau)))$ for big $\Im(\tau)$, where~$d_\Gamma$ is a Laurent polynomial. In the simplest case where~$\Gamma$ has two vertices, it is known \cite{GRV} that the coefficients of $d_\Gamma(\pi\Im(\tau))$ are MZVs, and Zagier recently proved that they can be expressed in terms of just odd zetas~$\zeta(2k+1)$ \cite{ZagierStrings}. This is related (as one can see directly from Zagier's proof) to the fact that genus zero amplitudes of four closed strings involve only odd zetas. 

Another goal of this thesis consisted in obtaining general results for these zeroth Fourier modes, because of the dominant r\^{o}le that they seem to have in understanding
\begin{itemize}
\item algebraic relations among modular graph functions \cite{DGV2015},\\
\item the connection between modular graph functions and genus zero amplitudes [Zagier, unpublished],\\
\item the connection between open and closed string amplitudes in genus one \cite{Zerb17}.
\end{itemize}

\section{Main results}

Our first goal was to investigate the Fourier expansion of modular graph functions, in order to extend the results on genus one closed string amplitudes contained in~\cite{GRV}. To do so, we were led to consider a generalization of MZVs that we have called \emph{conical sums}, which in the literature goes also under the name of conical zeta values \cite{TerasomaConical}.

Conical sums generalize MZVs in the sense that we allow for more flexibility on the choice of the domain of summation and on the denominators of equation (\ref{defMZV}).\footnote{The definition that we will give in Chapter \ref{ChapterClosedStrings} is actually even more general.} A result of Terasoma \cite{TerasomaConical} tells us that conical sums can be expressed in terms of special values of multiple polylogarithms evaluated at roots of~1. 

Our main result concerning modular graph functions is the following.
\begin{teo}\label{main}
For every graph $\Gamma$ we have the following expression for the associated modular graph function:
\begin{equation}\label{FourierModGraph}
D_{\Gamma}(\tau)=\sum_{\mu,\nu\geq 0}d_{\Gamma}^{(\mu,\nu)}(\pi\Im(\tau))\,q^\mu\,\overline{q}^\nu,
\end{equation}
where for every $\mu,\nu\geq 0$
\begin{equation*}
d_{\Gamma}^{(\mu,\nu)}(y)=\sum_{j=0}^{2l-1}a_j^{(\mu,\nu)}y^{l-j}
\end{equation*}
is a Laurent polynomial with coefficients $a_j^{(\mu,\nu)}$ lying in the $\mathbb{Q}$-algebra of conical sums~$\mathcal{C}$, and $l=l(\Gamma)$ (the weight of the graph) is the total number of edges of $\Gamma$.
\end{teo}
This means, by Terasoma's result, that the Laurent polynomials $d_{\Gamma}^{(0,0)}(\pi\Im(\tau))$ can always be written in terms of special values of multiple polylogarithms at roots of~1. This is, conjecturally, not the strongest possible result: we expect these number to be expressible just in terms of MZVs. However, the proof gives also access to explicit formulae for the coefficients of $d_{\Gamma}^{(0,0)}(\pi\Im(\tau))$ in terms of conical sums, and this has two main consequences.

First of all, finding sufficient conditions characterizing conical sums that are reducible to MZVs allows to see whether the Laurent polynomials $d_{\Gamma}^{(0,0)}(\pi\Im(\tau))$ can be a priori written in terms of MZVs. Using results of Brown \cite{BrownModuliSpace}, we found a sufficient condition (cf. Lemma \ref{consecutiveones}), and we used it to prove that we get MZVs for an infinite family of modular graph functions (cf. Theorem \ref{teoConsOnes}).

Moreover, since conical sums have a simple integral representation, we could use \texttt{HyperInt}, a Maple program developed by Panzer \cite{Panzer}, to perform explicit computations in the easiest unknown case of three vertices. This produced the first instances of coefficient of $d_{\Gamma}^{(0,0)}(\pi\Im(\tau))$ that are MZVs, but that cannot be reduced to products of simple zeta values. Moreover, we noticed that they can be written in terms of single-valued MZVs only: this reminds of what happens in the genus zero case, starting from the five-point amplitude. 

Let us give an example: suppose that $\Gamma$ is a graph with three vertices and seven edges, such that there is exactly one edge between two of the three pairs of vertices, and five edges between the last pair. Then we have obtained the Laurent polynomial\footnote{Note that the coefficient of $y^{-4}$ is slightly different from that reported in \cite{Zerb15}, where a typo had occurred.}
\begin{multline*}
d_{\Gamma}^{(0,0)}(y)=\frac{1}{4^7}\Big(\frac{62}{10945935}y^7+\frac{2}{243}\zeta(3)y^4+\frac{119}{324}\zeta(5)y^2 +\frac{11}{27}\zeta(3)^2y+\frac{21}{16}\zeta(7)\\
+\frac{46}{3}\frac{\zeta(3)\zeta(5)}{y}+\frac{7115\zeta(9)-3600\zeta(3)^3}{288y^2}+\frac{1245\zeta(3)\zeta(7)-150\zeta(5)^2}{16y^3}\\
+\frac{288\zeta(3)\zeta(3,5)-288\zeta(3,5,3)-5040\zeta(5)\zeta(3)^2-9573\zeta(11)}{128y^4}\\
+\frac{2475\zeta(5)\zeta(7)+1125\zeta(9)\zeta(3)}{32y^5}-\frac{1575}{32}\frac{\zeta(13)}{y^6}\Big),
\end{multline*}
which can be re-written (and this is highly non-trivial, because we will see that the algebra of single-valued MZVs is smaller than the algebra of MZVs) as
\begin{multline*}
d_{\Gamma}^{(0,0)}(y)=\frac{1}{4^7}\Big(\frac{62}{10945935}y^7+\frac{1}{243}\zeta_{\rm sv}(3)y^4+\frac{119}{648}\zeta_{\rm sv}(5)y^2 +\frac{11}{108}\zeta_{\rm sv}(3)^2y+\frac{21}{32}\zeta_{\rm sv}(7)\\
+\frac{23}{6}\frac{\zeta_{\rm sv}(3)\zeta_{\rm sv}(5)}{y}+\frac{7115\zeta_{\rm sv}(9)-900\zeta_{\rm sv}(3)^3}{576y^2}+\frac{1245\zeta_{\rm sv}(3)\zeta_{\rm sv}(7)-150\zeta_{\rm sv}(5)^2}{64y^3}\\
-\frac{288\zeta_{\rm sv}(3,5,3)+1620\zeta_{\rm sv}(5){\zeta_{\rm sv}(3)}^2+9573\zeta_{\rm sv}(11)}{256y^4}\\
+\frac{2475\zeta_{\rm sv}(5)\zeta_{\rm sv}(7)+1125\zeta_{\rm sv}(9)\zeta_{\rm sv}(3)}{128y^5}-\frac{1575}{64}\frac{\zeta_{\rm sv}(13)}{y^6}\Big),
\end{multline*}
where $y:=\pi\Im(\tau)$, and the notation $\zeta_{\rm sv}$ will be introduced in Section \ref{SectionSVMZVS}.

All the results mentioned so far are contained in our paper \cite{Zerb15} (formulated in a slightly less general context, because modular graph functions were introduced afterwards \cite{DGGV}), where we have also suggested the following:

\textbf{Conjecture.} The coefficients of the Laurent polynomials $d_{\Gamma}^{(\mu,\nu)}(\pi\Im(\tau))$ are given by single-valued multiple zeta values.

Evidence towards this conjecture is given in the subsequent papers \cite{DGGV} and~\cite{BrownNewClass}. 

The second goal of this thesis was to investigate the Fourier expansion of B-elliptic MZVs. This was motivated by the fact that their $q$-expansion (\ref{IntroBexp}) seemed to lend itself to a comparison with the expansion (\ref{FourierModGraph}), in contrast with the $q$-expansion of A-elliptic MZVs. This comparison is also motivated by recent indications that writing open string amplitudes in terms of B-elliptic MZVs would be helpful in order to compare them to closed string amplitudes, and therefore to modular graph functions \cite{Zerb17}.

Our approach to study elliptic MZVs is different from the approach of Enriquez. We derive their properties (known and new) from the definition in terms of iterated integrals of explicit elliptic functions, rather then using their associator origin.

The first result that we obtain is the following more precise version of equation~(\ref{IntroBexp}) (cf Theorem \ref{asymptB}):
\begin{teo}
B-elliptic multiple zeta values\footnote{Actually we will need to exclude some cases where terms like $\log(\tau)$ may appear.} of weight $w$ and length $r$ (we will define these quantities in Chapter \ref{ChapterEllMZV}) admit a Fourier expansion
\[
\sum_{i=1-r}^{w-r}\sum_{j\geq 0}b_{i,j}\tau^iq^j,
\]
where the coefficients $b_{i,j}$'s can be written in terms of MZVs and integer powers of~$2\pi i$.
\end{teo}
Note that this expansion is indeed very similar to the expansion (\ref{FourierModGraph}).

Moreover, we give explicit formulae for the Laurent polynomials
\[
\sum_{i=1-r}^{w-r}b_{i,0}\tau^i
\]
of the infinite family of so-called \emph{B-elliptic MZVs of depth one} (cf. Proposition \ref{DepthOneBinf}), as well as for their full Fourier expansion in terms of iterated integrals of Eisenstein series (cf. Theorem \ref{B=Gamma}).

In Theorem \ref{ThmFullMod} we give a formula describing the modular behaviour of the generating function of A-elliptic MZVs of depth one with respect to any element of SL$_2(\mathbb{Z})$, deduced making use of Brown's theory of multiple modular values \cite{MMV}. Moreover, we prove (cf. Theorem \ref{ThmModBehaviour}) that all these A-elliptic MZVs can be seen as (components of vector valued) modular forms of weight $1-r$.

Finally, as corollaries of our construction, we explain how to explicitly relate A-elliptic MZVs of depth one to special values of holomorphic elliptic polylogarithms\footnote{The same result was already obtained in \cite{MatthesMeta}, formulated and obtained in a different way.} and to non-holomorphic Eisenstein series (cf. Propositions \ref{eMZVellPol} and \ref{eMZVsNonHolo}).

Let us now come back to the general picture. Since non-holomorphic Eisenstein series are special values of single-valued multiple polylogarithms \cite{ZagierBWR}, and also particular cases of modular graph functions, these theorems (as well as other equivalent statements in the recent literature) give evidence towards the expected fact that elliptic MZVs and modular graph functions are elliptic analogues of MZVs and single-valued MZVs, respectively. Since they also naturally describe open and closed genus one superstring amplitudes, respectively, this ties in perfectly with the picture that we have of genus zero. 

As it is remarked in \cite{BrownNewClass}, we also expect modular graph functions to be modular invariant versions of elliptic MZVs, this relation being comparable with that between quasi-modular forms and almost holomorphic modular forms, or between mock modular forms and weak harmonic Maass forms \cite{1-2-3ModForms}, \cite{ZagierBourbaki}. 

To conclude, there are two ways to think of the (conjectural) relationship between elliptic multiple zeta values and modular graph functions. Both ways should eventually lead to a proof of the many open conjectures on modular graph functions. We will comment more on this in Chapter \ref{ChapterConclusions}.

\section{Content} 

Chapters \ref{ChapterMathBackground} and \ref{ChapterPhysicsBackground} contain respectively the mathematical and physical background needed in the rest of the thesis, and also describe the mathematical and physical frameworks (resp.) where one should place our results. The material presented is not new, the originality consists just in our presentation. 

More in details, in Chapter \ref{ChapterMathBackground} we introduce the notion of iterated integral, and its connection with the fundamental group of punctured smooth manifolds. Then we use iterated integrals to give a second definition of multiple polylogarithms and MZVs, we discuss some of their main features, and we introduce single-valued multiple polylogarithms and single-valued MZVs. We use this approach with iterated integrals to introduce in a natural way some genus one generalizations of these objects. In particular, we define (multiple) elliptic polylogarithms, following \cite{Levin} and \cite{BrownLevin}, and we conclude the chapter by mentioning partial results towards the definition of single-valued multiple elliptic polylogarithms.

Chapter \ref{ChapterPhysicsBackground} contains the background on the mathematical aspects of superstring amplitudes: after an introduction where we try to justify the great simplification of the actual physics behind our work, we give a brief review of some aspects of the actual state of art of amplitude's computations in superstring theory. We start with genus zero, working out explicitly the four-point case for both open and closed strings, and then giving some of the main results or conjectures for the~$n$-point case. After that, we focus on the genus one case: we introduce the Green's function on elliptic curves, and after an explanation of the main features of the open and closed case, we give the definition of modular graph functions, and mention the main results known before our work.

Chapter \ref{ChapterClosedStrings} is based on the author's paper \cite{Zerb15}. First of all, we introduce conical sums and we mention the main known results. Then we discuss their relation with MZVs, and we give new results and conjectures, partly based on a collaboration with Dupont. In the rest of the chapter, we present our main results on the Fourier expansion of modular graph functions, distinguishing the cases of two, three and~$n$ vertices. The new examples of Laurent polynomials and our main conjecture are contained in the section where we study the three-point case.

Chapter \ref{ChapterEllMZV} contains an original analytic presentation of the theory of elliptic multiple zeta values. This includes proofs of various known results on the Fourier expansion of elliptic MZVs and on their connection with MZVs and iterated integrals of Eisenstein series. Moreover, this chapter contains all our new contributions, from explicit asymptotic expansions of B-elliptic MZVs to the modular behaviour of A-elliptic MZVs of depth one, as well as a brief introduction to Eichler integrals and to multiple modular values. At the end of the chapter, we discuss the relation between elliptic MZVs and elliptic polylogarithms.

Finally, in Chapter \ref{ChapterConclusions} we incorporate some very recent new observations contained in Brown's paper \cite{BrownNewClass}, and we discuss future possible developments.

%% file: 1.pdf_tex
\begingroup%
  \makeatletter%
  \providecommand\color[2][]{%
    \errmessage{(Inkscape) Color is used for the text in Inkscape, but the package 'color.sty' is not loaded}%
    \renewcommand\color[2][]{}%
  }%
  \providecommand\transparent[1]{%
    \errmessage{(Inkscape) Transparency is used (non-zero) for the text in Inkscape, but the package 'transparent.sty' is not loaded}%
    \renewcommand\transparent[1]{}%
  }%
  \providecommand\rotatebox[2]{#2}%
  \ifx\svgwidth\undefined%
    \setlength{\unitlength}{2604.225878bp}%
    \ifx\svgscale\undefined%
      \relax%
    \else%
      \setlength{\unitlength}{\unitlength * \real{\svgscale}}%
    \fi%
  \else%
    \setlength{\unitlength}{\svgwidth}%
  \fi%
  \global\let\svgwidth\undefined%
  \global\let\svgscale\undefined%
  \makeatother%
  \begin{picture}(1,0.14129725)%
    \put(0,0){\includegraphics[width=\unitlength,page=1]{1.pdf}}%
    \put(0.26200688,0.08290735){\color[rgb]{0,0,0}\makebox(0,0)[lt]{\begin{minipage}{0.05833862\unitlength}\raggedright $+$\end{minipage}}}%
    \put(0.58041799,0.08290757){\color[rgb]{0,0,0}\makebox(0,0)[lt]{\begin{minipage}{0.06702736\unitlength}\raggedright $+$\end{minipage}}}%
    \put(0.89084208,0.08290768){\color[rgb]{0,0,0}\makebox(0,0)[lt]{\begin{minipage}{0.07447481\unitlength}\raggedright $+$\end{minipage}}}%
    \put(0.94793941,0.08290735){\color[rgb]{0,0,0}\makebox(0,0)[lt]{\begin{minipage}{0.09929977\unitlength}\raggedright $\cdots$\end{minipage}}}%
  \end{picture}%
\endgroup%

%% file: 2.pdf_tex
\begingroup%
  \makeatletter%
  \providecommand\color[2][]{%
    \errmessage{(Inkscape) Color is used for the text in Inkscape, but the package 'color.sty' is not loaded}%
    \renewcommand\color[2][]{}%
  }%
  \providecommand\transparent[1]{%
    \errmessage{(Inkscape) Transparency is used (non-zero) for the text in Inkscape, but the package 'transparent.sty' is not loaded}%
    \renewcommand\transparent[1]{}%
  }%
  \providecommand\rotatebox[2]{#2}%
  \ifx\svgwidth\undefined%
    \setlength{\unitlength}{1746.24170585bp}%
    \ifx\svgscale\undefined%
      \relax%
    \else%
      \setlength{\unitlength}{\unitlength * \real{\svgscale}}%
    \fi%
  \else%
    \setlength{\unitlength}{\svgwidth}%
  \fi%
  \global\let\svgwidth\undefined%
  \global\let\svgscale\undefined%
  \makeatother%
  \begin{picture}(1,0.21183758)%
    \put(0,0){\includegraphics[width=\unitlength,page=1]{2.pdf}}%
    \put(0.37426394,0.1243022){\color[rgb]{0,0,0}\makebox(0,0)[lt]{\begin{minipage}{0.09996004\unitlength}\raggedright $+$\end{minipage}}}%
    \put(0.83720932,0.12430237){\color[rgb]{0,0,0}\makebox(0,0)[lt]{\begin{minipage}{0.11106665\unitlength}\raggedright $+$\end{minipage}}}%
    \put(0.92236038,0.12430187){\color[rgb]{0,0,0}\makebox(0,0)[lt]{\begin{minipage}{0.14808891\unitlength}\raggedright $\cdots$\end{minipage}}}%
    \put(0,0){\includegraphics[width=\unitlength,page=2]{2.pdf}}%
  \end{picture}%
\endgroup%

%% file: 3.pdf_tex
\begingroup%
  \makeatletter%
  \providecommand\color[2][]{%
    \errmessage{(Inkscape) Color is used for the text in Inkscape, but the package 'color.sty' is not loaded}%
    \renewcommand\color[2][]{}%
  }%
  \providecommand\transparent[1]{%
    \errmessage{(Inkscape) Transparency is used (non-zero) for the text in Inkscape, but the package 'transparent.sty' is not loaded}%
    \renewcommand\transparent[1]{}%
  }%
  \providecommand\rotatebox[2]{#2}%
  \ifx\svgwidth\undefined%
    \setlength{\unitlength}{2595.5622035bp}%
    \ifx\svgscale\undefined%
      \relax%
    \else%
      \setlength{\unitlength}{\unitlength * \real{\svgscale}}%
    \fi%
  \else%
    \setlength{\unitlength}{\svgwidth}%
  \fi%
  \global\let\svgwidth\undefined%
  \global\let\svgscale\undefined%
  \makeatother%
  \begin{picture}(1,0.125809)%
    \put(0,0){\includegraphics[width=\unitlength,page=1]{3.pdf}}%
    \put(0.25954355,0.07332108){\color[rgb]{0,0,0}\makebox(0,0)[lt]{\begin{minipage}{0.05853335\unitlength}\raggedright $+$\end{minipage}}}%
    \put(0.57901747,0.07332131){\color[rgb]{0,0,0}\makebox(0,0)[lt]{\begin{minipage}{0.06725109\unitlength}\raggedright $+$\end{minipage}}}%
    \put(0.89047773,0.07332141){\color[rgb]{0,0,0}\makebox(0,0)[lt]{\begin{minipage}{0.07472339\unitlength}\raggedright $+$\end{minipage}}}%
    \put(0.94776564,0.07332108){\color[rgb]{0,0,0}\makebox(0,0)[lt]{\begin{minipage}{0.09963122\unitlength}\raggedright $\cdots$\end{minipage}}}%
    \put(0,0){\includegraphics[width=\unitlength,page=2]{3.pdf}}%
  \end{picture}%
\endgroup%

%% file: 4.pdf_tex
\begingroup%
  \makeatletter%
  \providecommand\color[2][]{%
    \errmessage{(Inkscape) Color is used for the text in Inkscape, but the package 'color.sty' is not loaded}%
    \renewcommand\color[2][]{}%
  }%
  \providecommand\transparent[1]{%
    \errmessage{(Inkscape) Transparency is used (non-zero) for the text in Inkscape, but the package 'transparent.sty' is not loaded}%
    \renewcommand\transparent[1]{}%
  }%
  \providecommand\rotatebox[2]{#2}%
  \ifx\svgwidth\undefined%
    \setlength{\unitlength}{1703.1608324bp}%
    \ifx\svgscale\undefined%
      \relax%
    \else%
      \setlength{\unitlength}{\unitlength * \real{\svgscale}}%
    \fi%
  \else%
    \setlength{\unitlength}{\svgwidth}%
  \fi%
  \global\let\svgwidth\undefined%
  \global\let\svgscale\undefined%
  \makeatother%
  \begin{picture}(1,0.25965948)%
    \put(0,0){\includegraphics[width=\unitlength,page=1]{4.pdf}}%
    \put(0.3584362,0.13334152){\color[rgb]{0,0,0}\makebox(0,0)[lt]{\begin{minipage}{0.1024885\unitlength}\raggedright $+$\end{minipage}}}%
    \put(0.83309159,0.13334168){\color[rgb]{0,0,0}\makebox(0,0)[lt]{\begin{minipage}{0.11387604\unitlength}\raggedright $+$\end{minipage}}}%
    \put(0.92039651,0.13334118){\color[rgb]{0,0,0}\makebox(0,0)[lt]{\begin{minipage}{0.15183476\unitlength}\raggedright $\cdots$\end{minipage}}}%
    \put(0,0){\includegraphics[width=\unitlength,page=2]{4.pdf}}%
  \end{picture}%
\endgroup%

%% file: Chapter2.tex

\chapter{Classical and elliptic multiple polylogarithms} 

\label{ChapterMathBackground} 

The goal of this chapter is to recall the construction of some special functions and periods that will appear throughout this work. Classical polylogarithms are special multi-valued functions on the punctured complex projective line $\mathbb{P}^1_{\mathbb{C}}\setminus\{0,1,\infty\}$ that generalize the complex logarithm. They are connected to a surprising number of areas in mathematics, such as algebraic number theory, hyperbolic geometry, knot theory, quantum field theory and string theory, and their special values include the special values of the Riemann $\zeta$-function at positive integers. We will see that a generalization of polylogarithms, called \emph{multiple polylogarithms}, generate all homotopy invariant iterated integrals on $\mathbb{P}^1_{\mathbb{C}}\setminus\{0,1,\infty\}$. Brown and Levin recently generalized this pictured to the genus one case, which led them to define \emph{multiple elliptic polylogarithms}. This chapter does not contain original contributions: it is intended to give the mathematical background needed for the next chapters, as well as to put our results into a context.

\section{Iterated integrals}\label{SectionItInt}

The content of this section is standard: we want to briefly recall the main features of Chen's theory of iterated integrals, and we use as references \cite{Chen}, \cite{HainItInt} and \cite{BurgosFresan}.

\subsection{Definition and first properties}
\begin{defi}
Let $M$ be a smooth manifold over $\mathbb{C}$, let $\omega_1,\ldots,\omega_r$ denote smooth complex-valued\footnote{$\mathbb{C}$ can be replaced everywhere by $\mathbb{R}$, which is more standard, but for the purpose of this work we prefer to work over the complex numbers.} 1-forms on $M$ and let $\gamma:[0,1]\rightarrow M$ be a parametrization of a piecewise smooth path. Write $\gamma^*\omega_i=f_i(t)dt$ for some piecewise smooth function $f_i:[0,1]\rightarrow\mathbb{C}$, where $1\leq i\leq r$. The \emph{iterated integral} of $\omega_1,\ldots,\omega_r$ along $\gamma$ is
\begin{equation}\label{itint}
\int_\gamma \omega_1\cdots\omega_r:=\int_{1\geq t_1\geq\cdots\geq t_r\geq 0}f_1(t_1)\cdots f_r(t_r)\,dt_1\cdots dt_r.
\end{equation}
We will call $r$ the \emph{length} of the iterated integral.
\end{defi}
\begin{oss}
More generally, we will call iterated integrals also all linear combinations of iterated integrals and~1, which will be thought of as an iterated integral of length zero. The length is then the maximum of the lengths of the summands.
\end{oss}
The first property that we want to mention is the following:
\begin{prop}[Functoriality]
Let $M$ and $N$ be smooth manifolds, $f:M\longrightarrow N$ be smooth, $\omega_1,\ldots,\omega_r$ be smooth 1-forms on $N$ and $\gamma$ be the parametrization of a piecewise smooth path on $M$. Then
\[
\int_\gamma f^*\omega_1\cdots f^*\omega_r=\int_{f\circ \gamma}\omega_1\cdots\omega_r.
\]
\end{prop}
In particular this means that iterated integrals do not depend on the parametrization of the path, and we will call $\gamma$ also the path itself. Moreover, from now on for brevity we will write just \emph{path} instead of piecewise smooth path. It is easy to verify the following
\begin{prop}[Integration by parts]
Let $\omega_1,\ldots,\omega_r$ be smooth 1-forms on $M$ and $\gamma$ be a path on $M$. If $f$ is a smooth function on $M$, then we have
\begin{itemize}
\item[(i)]
\begin{equation}\label{ItIntParts1}
\int_\gamma \omega_1\cdots\omega_r\,df=\int_\gamma \omega_1\cdots\omega_{r-1}(f\omega_r)-(f\circ\gamma)(0)\int_{\gamma}\omega_1\cdots\omega_r,
\end{equation}
\item[(ii)]
\begin{multline}\label{ItIntParts2}
\int_\gamma\omega_1\cdots\omega_i\,df\omega_{i+1}\cdots\omega_r=\\
=\int_\gamma \omega_1\cdots(f\omega_i)\omega_{i+1}\cdots\omega_r-\int_\gamma \omega_1\cdots\omega_i(f\omega_{i+1})\cdots\omega_r,
\end{multline}
\item[(iii)]
\begin{equation}\label{ItIntParts3}
\int_\gamma\,df\omega_1\cdots\omega_r=(f\circ\gamma)(1)\int_{\gamma}\omega_1\cdots\omega_r-\int_\gamma (f\omega_1)\omega_2\cdots\omega_r.
\end{equation}
\end{itemize}
\end{prop}
Finally, we want to recall the algebraic properties of iterated integrals:
\begin{prop}\label{algebprop}
Let $\omega_1,\ldots,\omega_r$ be smooth 1-forms on $M$ and $\gamma,\gamma_1,\gamma_2$ be paths on $M$ such that $\gamma_2(1)=\gamma_1(0)$.\footnote{Our convention is that we compose paths as we compose functions.} Then we have
\begin{itemize}
\item[(i)](Inversion of paths).
\begin{equation}
\int_{\gamma^{-1}} \omega_1\cdots\omega_r=(-1)^r\int_\gamma \omega_r\cdots\omega_1.
\end{equation}
\item[(ii)](Composition of paths).
\begin{equation}
\int_{\gamma_1\gamma_2} \omega_1\cdots\omega_r=\sum_{i=0}^r\int_{\gamma_1}\omega_1\cdots\omega_i\int_{\gamma_2}\omega_{i+1}\cdots\omega_r
\end{equation}
\item[(iii)](Shuffle product). Let $\shuffle(r,s)$ denote the set of permutations $\sigma$ of $\{1,\ldots ,r+s\}$ such that $\sigma(1)<\sigma(2)<\cdots <\sigma(r)$ and $\sigma(r+1)<\sigma(r+2)<\cdots <\sigma(r+s)$ (shuffles of type $(r,s)$). Then we have
\begin{equation}\label{shuffleitint}
\int_\gamma\omega_1\cdots\omega_r\int_\gamma\omega_{r+1}\cdots\omega_{r+s}=\sum_{\sigma\in\shuffle(r,s)}\int_\gamma\omega_{\sigma^{-1}(1)}\cdots\omega_{\sigma^{-1}(r+s)}.
\end{equation}
\end{itemize}
\end{prop}

\subsection{Chen's de Rham theorem} \label{sectionChen}
The reason to consider this generalization of the usual line integrals is that of constructing functions on the space of paths which depend only on the homotopy class of the path. We call these functions \emph{homotopy functionals}. By Stokes, the single $\int\omega$ is a homotopy functional if and only if $\omega$ is closed. Just by definition, one can easily see that $\int\omega$ vanishes on every $\gamma_1^{-1}\gamma_2^{-1}\gamma_1\gamma_2$, which means that this homotopy functional can see only elements of $\pi_1(M,x)$ that are visible in $H_1(M,\mathbb{C})$. The main feature of iterated integrals is that in general they can detect more. To make this more precise, let us consider the space $\pi_1(M,y,x)$ of homotopy classes of paths from\footnote{We switch the order of $x$ and $y$ in $\pi_1(M,y,x)$ to make clear what is our convention for the composition of paths.} $x\in M$ to $y\in M$, called the \emph{fundamental groupoid of M with basepoints $x,y$.} The fundamental group $\pi_1(M,x)$ acts on it on the right. Consider also the group module $\mathbb{Q}[\pi_1(M,y,x)]$, the group ring $\mathbb{Q}[\pi_1(M,x)]$, and the \emph{augmentation ideal} $J$, defined as the kernel of the map $\mathbb{Q}[\pi_1(M,x)]\rightarrow \mathbb{Q}$ sending all $\gamma\mapsto 1$. It is easy to see that, if $\int \omega_1\cdots \omega_r$ is a homotopy functional, then it gives rise to a $\mathbb{Q}$-linear function
\[
\mathbb{Q}[\pi_1(M,y,x)]/J^{r+1}\longrightarrow \mathbb{C}.
\]
The upshot is that there is a simple way to tell whether an iterated integral is a homotopy functional, and these iterated integrals suffice to describe all homotopy functionals. To make this more precise, let us consider the complex $E^*(M)$ of smooth differential $\mathbb{C}$-valued forms on $M$, equipped with the exterior product $\wedge$ and the differential $d$, which gives $E^*(M):=\bigoplus_{n\geq 0}E^n(M)$ the structure of a \emph{dg-algebra} (differential graded algebra). Consider any dg-subalgebra\footnote{One can repeat all the bar complex construction considering algebras over other fields, such as $\mathbb{Q}$. We will see that we will be interested in this later.} $A^*$ with $A^0\simeq\mathbb{C}$ such that $A^*\hookrightarrow E^*(M)$ is a quasi-isomorphism, and let $A^*=A^0\oplus A^+$. Then the \emph{reduced bar complex of A} is\footnote{The bar complex can be defined for any dg-algebra.}
\[
B^*(A^*)=B^*(A):=\mathbb{C}\oplus A^+\oplus \big(A^+\big)^{\otimes 2} \oplus \ldots,
\]
where one usually denotes $[x_1|\cdots |x_n]:=x_1\otimes\cdots\otimes x_n$, together with the grading $\deg[x_1|\cdots |x_n]=\sum_{i=1}^n\deg(x_i)-n$ and a differential
\begin{multline*}
\delta[x_1|\cdots |x_n]=-\sum_{i=1}^n(-1)^{\sum_{j=1}^{i-1}\deg[x_j]}[x_1|\cdots ,|dx_i|\cdots ,|x_n]\\
+\sum_{i=1}^{n-1}(-1)^{\sum_{j=1}^{i}[x_j]}[x_1|\cdots |x_i\wedge x_{i+1}|\cdots |x_n].
\end{multline*}
By definition, $B^0(A)$ is spanned by the empty $[]$ and all $[x_1|\cdots |x_n]$, with $x_1,\ldots ,x_n$ 1-forms. It comes with some extra structure:
\begin{itemize}
\item A \textbf{length filtration} $L_NB^0(A)$ given by the span of all $[x_1|\cdots |x_n]$ with $n\leq N$,
\item a \textbf{product} 
\[
[x_1|\cdots |x_r]\cdot[x_{r+1}|\cdots |x_{r+s}]=\sum_{\sigma\in\shuffle (r,s)}\eta(\sigma)[x_{\sigma^{-1}}(1)|\cdots |x_{\sigma^{-1}}(r+s)],
\]
where we omit the precise definition of $\eta(\sigma)=\pm 1$, and
\item a \textbf{coproduct}
\[
\Delta [x_1|\cdots |x_n]=\sum_{i=0}^n[x_1|\cdots |x_i]\otimes [x_{i+1}|\cdots |x_n].
\]
\end{itemize}
This (together with properly defined antipode and counit) makes it into a \emph{Hopf algebra}. Iterated integrals can therefore be used to define a pairing between $B^0(A)$ and (the group ring on) the space of paths on $M$, and Chen's differential $\delta$ is shaped in such a way that one has
\begin{teo}[Chen]
Let $\eta\in B^0(A)$. The iterated integral $\int\eta$ is a homotopy functional if and only if $\delta\eta=0$.
\end{teo} 
Then the result announced before can be stated in the following form:
\begin{teo}[Chen's de Rham theorem]
Integration induces an isomorphism
\begin{equation}\label{firstChen}
L_NH^0(B^*(A))\tilde{\longrightarrow} \mbox{Hom}_\mathbb{Q}(\mathbb{Q}[\pi_1(M,y,x)]/J^{N+1},\mathbb{C}).
\end{equation}
Passing to the limit, this gives an isomorphism of Hopf algebras\footnote{There is a standard way to endow any pro-unipotent completion of a group with a Hopf algebra structure. This induces a Hopf algebra structure on the right hand side.}
\begin{equation}
H^0(B^*(A))\simeq \mathcal{O}\big(\pi^{un}_1(M,y,x)\big)\otimes_\mathbb{Q}\mathbb{C}
\end{equation}
with the ring of functions on the pro-unipotent completion completion of $\pi_1(M,y,x)$.
\end{teo}
This theorem can be interpreted as a de Rham isomorphism for the fundamental group: the de Rham side is given by the bar complex, and the Betti side is given by the functions on the fundamental group. In particular, homotopy invariant iterated integrals can be seen as periods of the pro-unipotent fundamental groupoid of~$M$. Moreover, an important theorem of Beilinson, which relates $\mathbb{Q}[\pi_1(M,y,x)]/J^{N+1}$ with some relative cohomology of $M^N$, gives an interpretation of (algebraic) iterated integrals as periods in the usual sense (see \cite{DeligneGoncharov}, Proposition 3.4).

\subsection{Tangential base points}

In the rest of the chapter we will also need an important construction, due to Deligne, that allows to make sense of iterated integrals on smooth projective varieties with paths including points that are not in $M$. This is the theory of \emph{tangential base points}. We want to give a very brief sketch of this construction for $M=\mathbb{P}^1_{\mathbb{C}}\setminus\{0,1,\infty\}$, which will be relevant when talking about multiple zeta values, and we use as reference \cite{BurgosFresan}\footnote{This construction generalizes to any smooth projective curve minus a finite number of points.}. A tangential base point is the datum $\textbf{x}=(x,v)$ of a point $x\in \mathbb{P}^1_{\mathbb{C}}\setminus\{\infty\}$ and a tangent vector $v\in T_x M$.
\begin{defi}
A path from $\textbf{x}=(x,v)$ to $\textbf{y}=(y,w)$ is a piecewise smooth map $\gamma:[0,1]\rightarrow \mathbb{P}^1_{\mathbb{C}}\setminus\{\infty\}$ such that
\begin{itemize}
\item $\gamma(0)=x$, $\gamma(1)=y$.
\item $\frac{\partial\gamma}{\partial t}(0)=v$ and $\frac{\partial\gamma}{\partial t}(1)=-w$.
\item $\{t\in [0,1]:\gamma(t)\in\{0,1\}\}$ is a finite set. We call these points cusps.
\item $\frac{\partial\gamma^+}{\partial t}(t_0)=\frac{\partial\gamma^-}{\partial t}(t_0)$ at all cusps $t_0$.
\end{itemize}
\end{defi}
Then, after defining in a clever way the composition of this extended notion of paths, one can talk of the fundamental groupoid $\pi(M,\textbf{y},\textbf{x})$, and repeat all the constructions of the rest of this section in the tangential base point-case.

\section{Multiple polylogarithms and multiple zeta values}\label{sectionMPL}

We want now to specialize the previous construction to the case $M=\mathbb{P}^1_{\mathbb{C}}\setminus\{0,1,\infty\}$. Let us consider the dg-algebra $A^*=A^0\oplus A^1$, with $A^0:=\mathbb{Q}$ and $A^1:=\mathbb{Q}\omega_0\oplus\mathbb{Q}\omega_1$, where
\[
\omega_0(z)=\frac{dz}{z},
\]
\[
\omega_1(z)=\frac{dz}{1-z}.
\]
These are closed forms, and their classes are a basis of $H^1$, which is the only non-trivial cohomology group. Therefore $A^*\otimes_{\mathbb{Q}}\mathbb{C}$ is quasi-isomorphic to the de Rham complex $E^*(\mathbb{P}^1_{\mathbb{C}}\setminus\{0,1,\infty\})$ ($A^*$ is a \emph{rational model}). Note that in this case we have $H^0(B^*(A))= B^0(A)$, which is nothing but the Hopf algebra of words $\mathbb{Q}\langle \omega_0,\omega_1\rangle$ in two non-commutative letters $\omega_0, \omega_1$ with shuffle product and deconcatenation coproduct. This is sometimes called \emph{Hoffman's Hopf algebra} \cite{BurgosFresan}. Therefore, by what said in the previous section, we conclude that all homotopy invariant iterated integrals over $\mathbb{P}^1_{\mathbb{C}}\setminus\{0,1,\infty\}$ are spanned by iterated integrals of words $[\omega_{\epsilon_1}|\cdots |\omega_{\epsilon_r}]$, where $\epsilon_i\in\{0,1\}$. Moreover, one can consider the straight path $\gamma_z$ from the tangential base point $\textbf{0}=(0,1)$ to any $0<z<1$, and get
\begin{Lemma}\label{lemmaMPL}
Let $\mbox{Li}_{k_1,\ldots ,k_r}(z_1,\ldots ,z_r)$ be the multiple polylogarithm function defined in (\ref{defMPL}) for $z_1\cdots z_r<1$. Then
\begin{equation}\label{integralMPL}
\int_{\gamma_z}\underbrace{\omega_0\cdots \omega_0}_{k_r-1}\omega_1\cdots\underbrace{\omega_0\cdots \omega_0}_{k_1-1}\omega_1=\mbox{Li}_{k_1,\ldots ,k_r}(1,\ldots ,1,z).
\end{equation}
\end{Lemma}
\textbf{Proof.} Just expand $(1-t)^{-1}=\sum_{k\geq 0}t^k$.\\
$\square$

Taking the limit as $z\mapsto 1$ leads to
\begin{cor}
Let $\textbf{1}=(1,-1)$, and let ``$\rm dch$'' (from the french expression \emph{droit chemin}) be the straight path starting in $\textbf{0}$ and ending in $\textbf{1}$, which can be parametrized by $t\mapsto t$. Then, for $k_r\geq 2$,
\begin{equation}\label{integralMZV}
\int_{\rm dch}\underbrace{\omega_0\cdots \omega_0}_{k_r-1}\omega_1\cdots\underbrace{\omega_0\cdots \omega_0}_{k_1-1}\omega_1=\zeta(k_1,\ldots ,k_r),
\end{equation}
where $\zeta(k_1,\ldots ,k_r)$ are the multiple zeta values (MZVs) defined by (\ref{defMZV})
\end{cor}
This fact was first noticed by Kontsevich, and immediately implies that MZVs are periods\footnote{Here we mean periods in the Kontsevich-Zagier sense \cite{KontsZagier}.}. Note that the integrals (\ref{integralMZV}) are exactly the iterated integrals of words $[\omega_{\epsilon_1}|\cdots |\omega_{\epsilon_r}]$ which converge on $dch$. It is however possible to define regularized integrals on a path $\gamma\in\pi(M,\textbf{1},\textbf{0})$ of any $[\omega_{\epsilon_1}|\cdots |\omega_{\epsilon_r}]$ in the following way: one can show that, for each $\textbf{e}=(\epsilon_1,\ldots ,\epsilon_r)$ and any small enough $\delta$, the iterated integral $f_{\gamma,\textbf{e}}(\delta)=\int_{\gamma_\delta} [\omega_{\epsilon_1}|\cdots |\omega_{\epsilon_r}]$ on the path $\gamma_\delta$ from $\gamma(\delta)$ to $\gamma(1-\delta)$ has an asymptotic expansion
\[
f_{\gamma,\textbf{e}}(\delta)=f^0_{\gamma,\textbf{e}}(\delta)+\sum_{k\geq 0}a_{k,\gamma,\textbf{e}}\log(\delta)^k
\]
for some $a_{k,\gamma,\textbf{e}}\in\mathbb{C}$ and $f^0_{\gamma,\textbf{e}}(\delta)=O(\delta^\nu)$ as $\delta\rightarrow 0$, with $\nu> 0$. We define
\begin{equation}\label{regitint}
\int^{reg}_{\gamma}[\omega_{\epsilon_1}|\cdots |\omega_{\epsilon_r}]:=a_{0,\gamma,\textbf{e}}.
\end{equation}
One can prove that the isomorphism (\ref{firstChen}) predicted by Chen's theorem in the case of $\pi(M,\textbf{1},\textbf{0})$ is induced by the map
\[
L:\mathbb{Q}[\pi(M,\textbf{1},\textbf{0})]/J^{N+1}\longrightarrow\mbox{Hom}(L_NB^0(A),\mathbb{C})
\]
sending
\begin{equation}\label{mapL}
\gamma \longmapsto\Big([\omega_{\epsilon_1}|\cdots |\omega_{\epsilon_r}]\mapsto\int^{reg}_{\gamma}[\omega_{\epsilon_1}|\cdots |\omega_{\epsilon_r}]\Big),
\end{equation}
and it is not difficult to show that these regularized iterated integrals evaluated on $dch$ are rational linear combinations of MZVs, which implies that all homotopy invariant iterated integrals from $\textbf{0}$ to $\textbf{1}$ are given by MZVs. In other words, MZVs are the periods of the pro-unipotent fundamental group $\pi(M,\textbf{1},\textbf{0})$. More generally, all homotopy invariant iterated integrals on $\mathbb{P}^1_{\mathbb{C}}\setminus\{0,1,\infty\}$ can be expressed in terms of (regularized) multiple polylogarithms (\ref{integralMPL}), which can be seen as multi-valued functions on the whole $\mathbb{P}^1_{\mathbb{C}}\setminus\{0,1,\infty\}$ with monodromies around~$0$ and~$1$. 

Moreover, Beilinson's theorem mentioned at the end of Section \ref{sectionChen} suggests that MZVs should be periods of the moduli space $\mathcal{M}_{0,n}$ of genus zero Riemann surfaces with $n$ marked points. This was proved in \cite{GonchManin} by Goncharov and Manin. They also asked whether all periods of $\mathcal{M}_{0,n}$ can be expressed in terms $\mathbb{Q}[2\pi i]$-linear combinations of MZVs, and this was proved by Brown in \cite{BrownModuliSpace}. Even more spectacularly, Brown recently proved in \cite{BrownMTM} that all periods of all mixed Tate motives unramified over $\mathbb{Z}$ are $\mathbb{Q}[(2\pi i)^{-1}]$-linear combinations of MZVs, which make them in some sense the geometrically simplest interesting algebra of periods, but this story goes much beyond the scope of this introductory section.

Finally, it is important to mention that $L(dch)$, which can be seen as the generating series of MZVs\footnote{We have seen that $B^0(A)$ is isomorphic to Hoffman's Hopf algebra. Passing to the dual of $L_NB^0(A)$ and taking the topological limit one gets series in 2 non-commutative letters $\mathbb{C}\langle\langle e_0,e_1\rangle\rangle$ (see \cite{BurgosFresan}).}, is the so-called Drinfel'd associator, which was introduced as a special monodromy of the \emph{Knizhnik-Zamolodchikov equation} (see \cite{Drinfeld}, \cite{LeMurakami})
\begin{equation}\label{KZequation}
\frac{d}{dz}\Phi(z)=\Big(\frac{e_0}{z}+\frac{e_1}{1-z}\Big)\Phi(z).
\end{equation}

\subsection{The algebra of multiple zeta values}

By what we have said so far, it is clear that the shuffle product $\shuffle$ of iterated integrals makes rational combinations of MZVs into a $\mathbb{Q}$-algebra, that we will denote by $\mathcal{A}^\shuffle$ when we want to stress that the product is given by $\shuffle$, or otherwise just by $\mathcal{A}$.\footnote{This notation is inspired by the notation used in \cite{Zerb15}, and will be justified by the notation employed in Chapter \ref{ChapterClosedStrings}. However, we want to warn the reader that this notation is not standard, and usually, for instance in \cite{BurgosFresan}, this algebra is denoted by $\mathcal{Z}$.} Shuffle product gives rise to many algebraic relations among MZVs. For instance\footnote{When we write $\zeta(2)\shuffle\zeta(2)$ we mean, by abuse of notation, the shuffle product of their integral representations.}, 
\begin{equation}
\zeta(2)\shuffle\zeta(2)=4\zeta(1,3)+2\zeta(2,2).
\end{equation}
However, not all the possible relations between MZVs come from the shuffle product. There is another obvious way to multiply MZVs and make them into a $\mathbb{Q}$-algebra, using their series representation (\ref{defMZV}): it is called the \emph{stuffle product}\footnote{This product is sometimes called \emph{quasi-shuffle}.}. We do not want to give a formal definition, but will instead explain it through the depth one case: for each $r,s\geq 2$,
\begin{equation*}
\zeta(r)\zeta(s)=\sum_{n>0}\frac{1}{n^r}\sum_{m>0}\frac{1}{m^s}=\sum_{n,m>0}\frac{1}{n^rm^s}=\sum_{0<n<m}\frac{1}{n^rm^s}+\sum_{0<m<n}\frac{1}{n^rm^s}+\sum_{n>0}\frac{1}{n^{r+s}}.
\end{equation*}
It is straightforward to extend this to any depth. We denote this product by $*$, and we write $\mathcal{A}^*$ if we want to stress that we think of $\mathcal{A}$ as a $\mathbb{Q}$-algebra with stuffle product. Note that this gives rise to a second infinite family of algebraic relations between MZVs: for instance\footnote{This is again an abuse of notation.}
\begin{equation}
\zeta(2)*\zeta(2)=2\zeta(2,2)+\zeta(4).
\end{equation}
Moreover, comparing $\zeta(2)*\zeta(2)$ with $\zeta(2)\shuffle\zeta(2)$ we obtain the linear relation
\begin{equation}
\zeta(4)=4\zeta(1,3).
\end{equation}
We call this kind of linear relations \emph{double-shuffle relations}. Note that, in particular, the depth of an MZV does not give a grading on $\mathcal{A}$, but only a filtration. We call an MZV \emph{irreducible} if it cannot be written in terms of MZVs of smaller depth. Finally, it is possible (see \cite{IKZ} or \cite{BurgosFresan}) to define shuffle or stuffle-regularized MZVs by considering $\zeta(1)$ as a formal variable and using formally shuffle or stuffle products\footnote{We want to mention that, setting $\int_0^1\omega_0=\int_0^1\omega_1=0$, shuffle-regularized MZVs coincide with the regularized iterated integrals \ref{regitint}.}. This gives rise to new linear relations between honest MZVs, like for example
\begin{equation}\label{weight3}
\zeta(1,2)=\zeta(3).
\end{equation}
These linear relations are called \emph{extended double shuffle relations}, and are conjectured to be the only linear relations in $\mathcal{A}$ (\emph{double-shuffle conjecture}). Approximating MZVs numerically in a very clever and precise way, and using the LLL-algorithm, Zagier in the beginning of the 90's conjectured the following:

\textbf{Conjecture (Zagier).} Let $\mathcal{A}_k$ be the vector space of rational linear combinations of MZVs of weight $k$, and let $d_k$ be the dimension of this vector space. Then
\[
\mathcal{A}=\bigoplus_{k\geq 0}\mathcal{A}_k,
\]
and $d_k=d_{k-2}+d_{k-3}$.

The first part of Zagier's conjecture is implied by the double-shuffle conjecture, because extended double-shuffle relations are homogeneous. Note that, just by vector counting, $d_0=1$, $d_1=0$, and $d_2=1$. Since the only weight 3 MZVs are $\zeta(3)$ and $\zeta(1,2)$, by (\ref{weight3}) $d_3= 1$. Moreover, one can easily check that the relations above imply that $d_4=1$. These are the only $d_k$'s for which we can prove the second part of Zagier's conjecture, because to go further we would need some irrationality results that are not at our disposal at the moment (see below). A very deep result, obtained by Terasoma, and independently by Deligne and Goncharov, asserts that $d_k$ is bounded above by the numbers given by Zagier's conjecture \cite{TerasomaMZV}, \cite{DeligneGoncharov}. Here is a table listing the dimensions and a basis of irreducible MZVs of the first $\mathcal{A}_k$, assuming all the conjectures above, for $k\leq 8$:

\vspace{5mm}

\begin{tabular}{ | c | c | c | c | c | c  | c | c | c | c |}
  \hline
  $k$ & 0 & 1 & 2 & 3 & 4 & 5 & 6 & 7 & 8 \\
  \hline 
  $d_k$  & 1 & 0 & 1 & 1 & 1 & 2 & 2 & 3 & 4  \\
  \hline
  Basis  & 1 &  & $\zeta(2)$ & $\zeta(3)$ & $\zeta(4)$ & $\zeta(5)$ & $\zeta(6)$ & $\zeta(7)$ & $\zeta(8)$  \\
    & &  & & & & $\zeta(3)\zeta(2)$ & $\zeta(3)^2$ & $\zeta(3)\zeta(2)^2$ & $\zeta(3)^2\zeta(2)$ \\
     & &  & & & & & & $\zeta(5)\zeta(2)$ & $\zeta(5)\zeta(3)$ \\
   & &  & & & & & & & $\zeta(3,5)$ \\
  \hline
\end{tabular}

\vspace{5mm}

It is remarkable that Zagier's conjecture is now known (after the work of Terasoma, Deligne-Goncharov and Brown) to imply that $1,\pi,\zeta(3),\zeta(5),\cdots,\zeta(2k+1),\cdots$ are algebraically independent (whence in particular all the so-called \emph{odd} $\zeta$-\emph{values} $\zeta(2k+1)$ would be transcendental), but the known irrationality results are still very far from proving this statement. Indeed, here is a brief overview of what has been proven so far: 
\begin{itemize}
\item By the well known Euler's formula for the even $\zeta$-values
\begin{equation}
\zeta(2k)=-\frac{\B_{2k}(2\pi i)^{2k}}{2(2k)!},
\end{equation}
where $\B_{n}\in\mathbb{Q}$ are the \emph{Bernoulli numbers} defined by
\begin{equation}\label{BernoulliNumDef}
\frac{t}{e^t-1}=\sum_{n\geq 0}\B_nt^n,
\end{equation}
and therefore by the transcendentality of $\pi$ we get that all even $\zeta$-values are transcendental. 
\item Ap\'ery proved in \cite{Apery} that $\zeta(3)$ is irrational (this is the only odd $\zeta$-value which is known to be irrational).
\item Rivoal proved in \cite{Rivoal} that infinitely many odd $\zeta$-values are irrational
\item Zudilin proved that at least one among $\zeta(5),\zeta(7),\zeta(9),\zeta(11)$ is irrational \cite{Zudilin}.
\end{itemize}

\subsection{Single-valued multiple zeta values}\label{SectionSVMZVS}

\begin{defi}
Let $M=\mathbb{P}^1_{\mathbb{C}}\setminus\{0,1,\infty\}$, and let $p:\tilde{M}\rightarrow M$ be a universal cover. We call a holomorphic function on $\tilde{M}$ \emph{multi-valued} if it lifts a holomorphic function defined in a positive radius disk $D\subset M$ which admits a non-unique holomorphic continuation to the whole $M$. By abuse of terminology, we will usually just speak of multi-valued functions on $M$.
\end{defi}
For instance, we say that $\log(z)$ is a holomorphic multi-valued function on $M$, in the sense that it is defined in a neighbourhood of $z=1$ and it has a multi-valued holomorphic continuation to $\mathbb{P}^1_{\mathbb{C}}\setminus\{0,\infty\}$, and we say that its real part $\log|z|$ is a single-valued version of the logarithm. As remarked before, the multiple polylogarithms in one variable $\mbox{Li}_{k_1,\ldots ,k_r}(1,\ldots ,1,z)$ considered in eq. (\ref{integralMPL}) can be defined as series in a neighbourhood of zero, and by their integral representation they extend to a holomorphic multi-valued function on $\mathbb{P}^1_{\mathbb{C}}\setminus\{1,\infty\}$. We want to briefly present the construction, due to Brown, of their single-valued analogues, and to take a look at their special values at~1. We should mention that a single-valued version of the dilogarithm $\mbox{Li}_2(z)$, called the \emph{Bloch-Wigner dilogarithm} and given by
\[
D(z):=\Im(\mbox{Li}_2(z)+\log|z|\log(1-z)),
\]
was constructed a long time ago in \cite{BlochRegulators}, and was then related to many branches of mathematics, such as $K$-theory, hyperbolic manifolds and algebraic number theory \cite{ZagierDilog}. Moreover, at least two different generalizations for all classical polylogarithms have been constructed by Zagier \cite{ZagierBWR} and Wojtkowiak \cite{Wojtk}. Brown's construction, which appeared in \cite{BrownSVMPL}, is the most general, and includes all the others as special cases.

Let $Z(e_0,e_1)$ denote the Drinfel'd associator, defined in Section \ref{sectionMPL} as the image of $[\textbf{0},\textbf{1}]$ in $\mathbb{C}\langle\langle e_0,e_1\rangle\rangle$ via $L$, which is the map given by (\ref{mapL}). Moreover, if $U:=\mathbb{C}\setminus\{(-\infty,0]\cup[1,\infty)\}$, for any $z\in U$ we denote $L_{e_0,e_1}(z):=L([\textbf{0},z])$, where $L$ now is the obvious extension of the map (\ref{mapL}) and $[\textbf{0},z]$ is any path in $U$. This can be seen as a generating series of multiple polylogarithms in one variable. It is known (see \cite{Cartier}) that they constitute the unique family of holomorphic functions that satisfy on $U$ the recursive differential equations
\[
\frac{\partial}{\partial z}\mbox{Li}_{e_0w}(z)=\frac{\mbox{Li}_{w}(z)}{z},
\]
\[
\frac{\partial}{\partial z}\mbox{Li}_{e_1w}(z)=\frac{\mbox{Li}_{w}(z)}{1-z},
\]
where $w$ is any word in $e_0,e_1$, and such that $\mbox{Li}_e(z)=1$ ($e$ denotes the empty word), $\mbox{Li}_{e_0^n}(z)=\log^n(z)/n!$ and $\lim_{z\rightarrow 0}\mbox{Li}_w(z)=0$ if $w\neq e_0^n$. They satisfy the shuffle relations, and they are linearly independent over $\mathcal{O}:=\mathbb{C}\Big[z,\frac{1}{z},\frac{1}{1-z}\Big]$. One can also prove that, for every path $\gamma$, $L(\gamma)\in\mathbb{C}\langle\langle e_0,e_1\rangle\rangle$ is group-like, and thus it admits an inverse $L^{-1}(\gamma)\in\mathbb{C}\langle\langle e_0,e_1\rangle\rangle$.
\begin{defi}
Let $\eta\in\mathbb{R}\langle\langle e_0,e_1\rangle\rangle$ be the unique solution of the fixed point equation
\[
Z(-e_0,-\eta)\eta Z(-e_0,-\eta)^{-1}=Z(e_0,e_1)e_1Z(e_0,e_1)^{-1}.
\]
We define the generating function of single-valued multiple polylogarithms by
\begin{equation}\label{genfunsvMPL}
\mathcal{L}(z)=\widetilde{L_{e_0,\eta}(\overline{z})}L_{e_0,e_1}(z),
\end{equation}
where $\sim$ denotes reversal of words.
\end{defi}
The main result of \cite{BrownSVMPL} is the following
\begin{teo}[Brown]\label{TeoSVMPL}
The family $\mathcal{L}_w(z)$ generated by the series \ref{genfunsvMPL}, where $w$ is any non-commutative word in $e_0$ and $e_1$, is the only family of single-valued functions on $M$ which satisfy the differential equations
\[
\frac{\partial}{\partial z}\mathcal{L}_{e_0w}(z)=\frac{\mathcal{L}_{w}(z)}{z},
\]
\[
\frac{\partial}{\partial z}\mathcal{L}_{e_1w}(z)=\frac{\mathcal{L}_{w}(z)}{1-z},
\]
such that $\mathcal{L}_e(z)=1$, $\mathcal{L}_{e_0^n}(z)=\log^n|z|^2/n!$ and $\lim_{z\rightarrow 0}\mathcal{L}_w(z)=0$ if $w\neq e_0^n$. Moreover, the functions $\mathcal{L}_w(z)$ satisfy the shuffle relations, are linearly independent over $\mathcal{O}\overline{\mathcal{O}}$, and every single-valued linear combination of functions $\mbox{Li}_w(z)\overline{\mbox{Li}_{w^\prime} (z)}$ can be written as a unique linear combination of functions $\mathcal{L}_w(z)$.
\end{teo}
In particular, Zagier's and Wojtkowiak's single-valued poylogarithms can be obtained from this construction \cite{BrownSVMPL}. Let us turn now to their special values, that have been studied by Brown in \cite{BrownSVMZV}.
\begin{defi}
Let $r\geq 1$, $k_1,\ldots ,k_{r-1}\geq 1$ and $k_r\geq 2$. We denote
\begin{equation}
\zeta_{\rm sv}(k_1,\ldots ,k_r):=\mathcal{L}_{k_1,\ldots ,k_r}(1),
\end{equation}
and we call these numbers \emph{single-valued multiple zeta values}.
\end{defi}
Single-valued multiple polylogarithms are in general complex-valued, but since by construction $\mathcal{L}(1)=(Z(-e_0,-\eta))^{-1}Z(e_0,e_1)$, we deduce that $\zeta_{\rm sv}(k_1,\ldots ,k_r)\in\mathcal{A}\subset\mathbb{R}$. It is easy to see that single-valued MZVs constitute an algebra over $\mathbb{Q}$. We denote this algebra by $\mathcal{A}^{\rm sv}$, and the weight $k$ sub-vector spaces by $\mathcal{A}^{\rm sv}_k$. This algebra is actually (expected to be, assuming all the previous conjectures) much smaller than $\mathcal{A}$. For instance, it is easy to see that $D(1)=0$, which implies that $\zeta_{\rm sv}(2)=0$, and since one can show that $\rm sv:\mathcal{A}\rightarrow\mathcal{A}^{\rm sv}$ is a (graded) ring homomorphism, we deduce that $\zeta_{\rm sv}(2k)=0$ for any $k$. On the other side, one can show that $\zeta_{\rm sv}(2k+1)=2\zeta(2k+1)$. It will be very important in Chapter \ref{ChapterClosedStrings} to mention that, while in $\mathcal{A}$ the first irreducible MZV of depth two, i.e. $\zeta(3,5)$, occurs in weight eight, in the single-valued setting we have $\zeta_{\rm sv}(3,5)=-10\zeta(3)\zeta(5)$, and only in weight eleven we find the first irreducible single-valued MZV of depth greater than one (when viewed as an element of $\mathcal{A}$):
\begin{equation}
\zeta_{\rm sv}(3,5,3)=2\zeta(3,5,3)-2\zeta(3)\zeta(3,5)-10\zeta(3)^2\zeta(5).
\end{equation}
Here we give the list, taken from \cite{BrownSVMZV}, of the conjectured dimensions and generators of the vector spaces $\mathcal{A}^{\rm sv}_k$ for $2\leq k\leq 10$:

\vspace{5mm}

\begin{tabular}{ | c | c | c | c | c | c | c  | c | c | c | }
  \hline
  $k$ & 2 & 3 & 4 & 5 & 6 & 7 & 8 & 9 & 10 \\
  \hline 
  $d^{\rm sv}_k$  & 0 & 1 & 0 & 1 & 1 & 1 & 1 & 2 & 2  \\
  \hline
  Basis  & & $\zeta_{\rm sv}(3)$ &  & $\zeta_{\rm sv}(5)$ & $\zeta_{\rm sv}(3)^2$ & $\zeta_{\rm sv}(7)$ & $\zeta_{\rm sv}(3)\zeta_{\rm sv}(5)$ & $\zeta_{\rm sv}(9)$ & $\zeta_{\rm sv}(5)^2$ \\
    & & & & & & & & $\zeta_{\rm sv}(3)^3$ & $\zeta_{\rm sv}(7)\zeta_{\rm sv}(3)$ \\
  \hline
\end{tabular}

\vspace{5mm}

We conclude by mentioning that in \cite{BrownSVMZV} Brown proved also that $\mathcal{L}(1)$ coincides with the so-called Deligne's associator.

\section{Multiple elliptic polylogarithms}

\subsection{The Kronecker function}

Let us fix the notation, following \cite{BrownLevin}. From now on, for any complex number $\xi$ we define $\e(\xi):=\exp(2\pi i\xi)$. Let us denote by $\tau\in\mathbb{H}$ the modulus of the complex torus $\mathcal{E}_\tau:=\mathbb{C}/\Lambda_\tau$, where $\Lambda_\tau=\tau\mathbb{Z}+\mathbb{Z}$, let $q=\e(\tau)$, let $\xi\in\mathbb{C}$ be the complex coordinate on $\mathcal{E}_\tau$, and let $u=\e(\xi)$.

Let us introduce the odd Jacobi $\theta$-function\footnote{We refer to \cite{MumfordTataI} for a proof of all claims concerning $\theta$.}
\begin{equation}\label{Theta}
\theta(\xi,\tau)=\sum_{\nu\in\mathbb{Z}+\frac{1}{2}}(-1)^{\nu-1/2}q^{\nu^2/2}u^{\nu}.
\end{equation}
It is the unique (up to constants) entire function in $\xi$ satisfying
\begin{itemize}
\item $\theta(\xi+1,\tau)=-\theta(\xi,\tau)$
\item $\theta(\xi+\tau,\tau)=-q^{-1/2}u^{-1}\theta(\xi,\tau)$
\item Its zeros are all simple and located at the points of the lattice $\Lambda_\tau$.
\end{itemize}
Its modular transformations read 
\[
\theta(\xi,\tau+1)=\e(1/8)\theta(\xi),
\]
\[
\theta(\xi/\tau,-1/\tau)=-i\sqrt{-i\tau}\e(\xi^2/2\tau)\theta(\xi),
\]
and one can express it as an infinite product, using Jacobi's triple product formula
\begin{equation}\label{Jacobi}
\theta(\xi,\tau)=q^{1/8}(u^{1/2}-u^{-1/2})\prod_{j\geq 1}(1-q^j)(1-q^ju)(1-q^ju^{-1}).
\end{equation}
\begin{defi}
In the setting specified above, let $\alpha\in\mathbb{C}$\footnote{The variable $\alpha$ will in general take the r\^{o}le of a formal variable.}, and let $v=\e(\alpha)$. We define the \emph{Kronecker function} as
\begin{equation}\label{DefF}
F(\xi,\alpha,\tau):=\frac{\theta^{\prime}(0,\tau)\theta(\xi+\alpha,\tau)}{\theta(\xi,\tau)\theta(\alpha,\tau)},
\end{equation}
where we denote
\[
\theta^{\prime}(\xi,\tau)=\frac{\partial}{\partial\xi}\theta(\xi,\tau).
\]
\end{defi}
The main properties of this function are summarized in Zagier's \cite{ZagierPeriodsJacobi}\footnote{Note that the notation is quite different. In particular, it may be useful to remark that the two sets of variables $\{\xi,\alpha\}$ and $\{u,v\}$ are interchanged and scaled by $2\pi i$.}. We recall here some of them, and refer to Zagier's paper for the proofs. By definition we have the symmetry
\begin{equation}\label{Fsymm}
F(\xi,\alpha,\tau)=F(\alpha,\xi,\tau).
\end{equation}
Using the elliptic and modular properties of $\theta$ listed above, one gets: 
\begin{equation}\label{ellpropF}
F(\xi+1,\alpha,\tau)=F(\xi,\alpha,\tau), \,\,\,\,\,\,  F(\xi+\tau,\alpha,\tau)=\e\Big(-\frac{\Im(\xi)}{\Im(\tau)}\alpha\Big)F(\xi,\alpha,\tau),
\end{equation}
\begin{equation}\label{modpropF}
F\Big(\frac{\xi}{c\tau+d},\frac{\alpha}{c\tau+d},\frac{a\tau+b}{c\tau+d}\Big)=(c\tau+d)\e\Big(\frac{c\xi\alpha}{c\tau+d}\Big)F(\xi,\alpha,\tau).
\end{equation}
Moreover, $F$ has the $q$-expansion
\begin{equation}
F(\xi,\alpha,\tau)=-2\pi i\bigg(\frac{u}{1-u}+\frac{1}{1-v}+\sum_{m,n\geq 0}\big(u^mv^n-u^{-m}v^{-n}\big)q^{mn}\bigg).
\end{equation}
A straightforward consequence of this property is the so-called \emph{mixed heat equation}
\begin{equation}\label{mixedheat}
2\pi i\frac{\partial}{\partial \tau}F(\xi,\alpha,\tau)=\frac{\partial^2}{\partial \xi \partial \alpha}F(\xi,\alpha,\tau).
\end{equation}
In order to state the last property (\emph{logarithm}), for $k\geq 1$ and $\xi\in\mathbb{C}$ let us define\footnote{If $k=1$ and $k=2$ our convention is to take the \emph{Eisenstein summation} $\lim_{M\rightarrow\infty}\lim_{N\rightarrow\infty}\sum_{m=-M}^M\sum_{n=-N}^N$. See for instance \cite{Weil}.}
\begin{equation}\label{GenFunEisSeries}
G_k(\xi,\tau):=\sum_{(m,n)\in\mathbb{Z}^2}\frac{1}{(\xi+m\tau+n)^k}.
\end{equation}
Note that (holomorphic) Eisenstein series for $\mbox{SL}_2(\mathbb{Z})$
\begin{equation}\label{EisSeries}
G_k(\tau)=\sum_{(m,n)\neq (0,0)}\frac{1}{(m\tau+n)^k}=(1+(-1)^k)\Big(\zeta(k)+\frac{(2\pi i)^k}{(k-1)!}\sum_{n\geq 1}\frac{n^{k-1}q^n}{1-q^n}\Big)
\end{equation}
are obtained as special values: $G_k(\tau)=(G_k(\xi,\tau)-1/\xi^k)|_{\xi=0}$. These functions have the following properties (the first one follows from (\ref{Jacobi}), the second follows directly from the definition and the last is a consequence of (i) and (ii)):
\begin{itemize}
\item[(i)]
\begin{equation}\label{propG3}
\frac{\partial}{\partial\xi}\log(\theta(\xi,\tau))=G_1(\xi,\tau)=-\sum_{n\geq -1}G_{n+1}(\tau)\xi^n,
\end{equation}
where we set $G_0(\tau)\equiv -1$ and $G_{n}(\tau)\equiv 0$ for $n$ odd.
\item[(ii)]
\begin{equation}
\frac{\partial}{\partial\xi}G_j(\xi,\tau)=-jG_{j+1}(\xi,\tau).
\end{equation}
\item[(iii)]
\begin{equation}\label{propG2}
G_2(\xi,\tau)=\wp(\xi,\tau)+G_2(\tau)=\sum_{n\geq -1}nG_{n+1}(\tau)\xi^{n-1},
\end{equation}
where $\wp$ denotes the Weierstrass $\wp$-function.
\end{itemize}
It is an easy exercise to deduce from these properties that
\begin{equation}\label{logarithm}
F(\xi,\alpha,\tau)=\frac{1}{\alpha}\exp\bigg(-\sum_{j\geq 1}\frac{(-\alpha)^j}{j}\big(G_j(\xi,\tau)-G_j(\tau)\big)\bigg).
\end{equation}
If we denote the formal expansion of $F$ with respect to $2\pi i\alpha$ by\footnote{Here we deviate from the usual convention of expanding with respect to $\alpha$, in order to get cleaner statements in the rest of the paper.}
\begin{equation}\label{expansionF}
F(\xi,\alpha,\tau)=:\sum_{n\geq 0}f_n(\xi,\tau)(2\pi i\alpha)^{n-1},
\end{equation}
then one can see directly from the definition (\ref{DefF}) that $f_0(\xi,\tau)=2\pi i$ and $f_1(\xi,\tau)=\theta^\prime (\xi,\tau)/\theta(\xi,\tau)$ (=$G_1(\xi,\tau)$, by (\ref{propG3})). In general one can easily see by (\ref{logarithm}), together with the properties of $G_j(\xi,\tau)$, that
\begin{align}\label{explicitformulas}
f_n(\xi,\tau) &= \left\{ \begin{array}{cl}
2\pi i
 &: \ n=0 \\
\pi\cot(\pi \xi)-2\pi i\sum_{m\geq 1}\big(\e(m\xi)-\e(-m\xi)\big)\sum_{p\geq 1}q^{mp}
 &: \ n=1 \\
\frac{2\pi i}{(n-1)!}\big(\frac{\B_n}{n}-\sum_{m\geq 1}\big(\e(m\xi)+(-1)^n\e(-m\xi)\big)\sum_{p\geq 1} p^{n-1}q^{mp}\big)
 &: \ n\geq 2,
\end{array} \right .
\end{align}
where $\B_n$ is the $n$-th Bernoulli number. Note that every $f_n$ is holomorphic, except for $f_1$, which is meromorphic with a simple pole at every lattice point.

The function $F$ must be slightly modified in order to have nicer elliptic (with respect to the first variable) and modular behaviour, at the cost of losing holomorphicity. Let $r_{\tau}(\xi):=\Im(\xi)/\Im(\tau)$. Then we define, following \cite{BrownLevin},
\begin{equation}\label{Omega}
\Omega(\xi,\alpha,\tau):=\e(r_{\tau}(\xi)\alpha)F(\xi,\alpha,\tau)=\sum_{n\geq 0}\omega_n(\xi,\tau)(2\pi i\alpha)^{n},
\end{equation}
where the first $\omega_n$'s read $\omega_0(\xi,\tau)=2\pi i$ and $\omega_1=\theta^\prime (\xi,\tau)/\theta(\xi,\tau)+r_\tau(\xi)$ (again, this is the only $\omega_n$ which has a pole), and in general
\begin{equation}
\omega_n(\xi,\tau)=\sum_{k=0}^n\frac{r_{\tau}(\xi)^k}{k!}f_{n-k}(\xi,\tau).
\end{equation}
The fact that $\theta$ is odd implies that $\Omega(-\xi,-\alpha,\tau)=-\Omega(\xi,\alpha,\tau)$, which in turns implies that
\begin{equation}\label{omegaodd}
\omega_n(-\xi,\tau)=(-1)^n\omega_n(\xi,\tau).
\end{equation}
It is easily verified from (\ref{ellpropF}) and (\ref{modpropF}) that for all $p,q\in\mathbb{Z}$
\begin{equation}
\Omega(\xi+p\tau+q,\alpha,\tau)=\Omega(\xi,\alpha,\tau)
\end{equation}
and that, for $\left( \begin{array}{ccc}
a & b \\
c & d \end{array} \right)\in\mbox{SL}_2(\mathbb{Z})$,
\begin{equation}
\Omega\left(\frac{\xi}{c\tau+d},\frac{\alpha}{c\tau+d},\frac{a\tau+b}{c\tau+d}\right)=(c\tau+d)\Omega(\xi,\alpha,\tau),
\end{equation}
which implies that
\begin{equation}\label{elliptic}
\omega_n(\xi+p\tau+q,\tau)=\omega_n(\xi,\tau)
\end{equation}
and that
\begin{equation}\label{modular}
\omega_n\left(\frac{\xi}{c\tau+d},\frac{a\tau+b}{c\tau+d}\right)=(c\tau+d)^n\omega_n(\xi,\tau).
\end{equation}

\subsection{Homotopy invariant iterated integrals and averages of classical polylogarithms on punctured elliptic curves}

We want to briefly outline the idea of the construction of a genus one analogue of multiple polylogarithms, due to Brown and Levin \cite{BrownLevin}. We refer to their paper for all details. We have seen that multiple polylogarithms are homotopy invariant iterated integrals on the genus zero punctured Riemann surface $\mathbb{P}^1_{\mathbb{C}}\setminus\{0,1,\infty\}$. By genus one analogue we mean that we will consider homotopy invariant iterated integrals on the genus one punctured Riemann surface $\mathcal{E}_\tau^*=\mathbb{C}/\Lambda_\tau\setminus\{0\}$. Let us introduce the \emph{elliptic Knizhnik-Zamolodchikov-Bernard form}, which takes values in the graded completion of the free complex Lie algebra on 2 generators $e_0$ and $e_1$:
\begin{equation}\label{KZB-form}
\omega_{KZB}:=-\nu(\xi) e_0+\big(\mbox{ad}_{e_0}\Omega(\xi,\mbox{ad}_{e_0},\tau)d\xi\big)(e_1)=-\nu (\xi) e_0+\sum_{n\geq 0}\big(\omega_n(\xi)d\xi\big)(2\pi i)^n\mbox{ad}_{e_0}^n(e_1),
\end{equation}
where $\nu(\xi):=dr_{\tau}(\xi)$, $\Omega$ is given by (\ref{Omega}) and $\mbox{ad}_{x}(\cdot)=[x,\cdot]$. It is the elliptic analogue of the form\footnote{Recall that $\omega_0=dz/z$ and $\omega_1=dz/(1-z)$.}
\begin{equation}
\omega_{KZ}:=\omega_0 e_0+\omega_1 e_1,
\end{equation}
whose coefficients generate all homotopy invariant iterated integrals in the genus zero case: one can prove that the tensors generated by the coefficients of any word in $e_0, e_1$ belong to $H^0(B^*(\mathcal{E}_\tau^*))$, and therefore give homotopy invariant iterated integrals on $\mathcal{E}_\tau^*$. One of the main results of \cite{BrownLevin} asserts the following:
\begin{teo}[Brown-Levin]\label{ThBrLev}
Let $A^*$ be the graded $\mathbb{Q}$-algebra generated by $\nu(\xi)$ and all $\omega_n(\xi)d\xi$. It gives a rational model for the de Rham complex $E^*(\mathcal{E}_\tau^*)$ (meaning that $A^*\otimes\mathbb{C}\hookrightarrow E^*(\mathcal{E}_\tau^*)$ is a quasi-isomorphism), and therefore every homotopy invariant iterated integral on $\mathcal{E}_\tau^*$ can be written as a $\mathbb{C}$-linear combination of iterated integrals of the coefficients of (\ref{KZB-form}).
\end{teo}
The second main result of Brown and Levin was to write these iterated integrals in terms of averages of classical multiple polylogarithms. Let us denote
\[
I_{k_1,\ldots ,k_r}(u_1,\cdots ,u_r):=\mbox{Li}_{k_1,\ldots ,k_r}\Big(\frac{u_1}{u_2},\cdots ,\frac{u_{r-1}}{u_r},u_r\Big).
\]
It is possible to show, along the lines of Section \ref{sectionMPL}, that they span all iterated integrals on the moduli space of genus zero punctured Riemann surfaces
\[
\mathcal{M}_{0,r+3}(\mathbb{C})=(\mathbb{P}^1_{\mathbb{C}}\setminus\{0,1,\infty\})^r\setminus\{\Delta\}=\{(u_1,\ldots ,u_r)\in(\mathbb{C}\setminus\{0,1\})^r:u_i\neq u_j\}.
\]
\begin{prop}[Brown-Levin]
Let $r\geq 1$, $k_1,\ldots ,k_r\in\mathbb{N}$, $1<v_1,\ldots ,v_r<|q|^{-1}$, $0<|q|<|u_1|<\cdots <|u_r|<1$, $u_i\notin q^\mathbb{R}$ and $u_iu_j^{-1}\notin q^\mathbb{R}$. Then the series
\begin{equation}\label{genfunellMPL}
E_{k_1,\ldots ,k_r}(u_1,\ldots ,u_r,v_1,\ldots ,v_r,\tau)=\sum_{m_1,\ldots m_r\in\mathbb{Z}}v_1^{m_1}\cdots v_r^{m_r}I_{k_1,\ldots ,k_r}(q^{m_1}u_1,\cdots ,q^{m_r}u_r)
\end{equation}
converges absolutely and defines a generating series of functions on the configuration space $\mathcal{E}_{\tau}^{(r)}$ of $r$ distinct points on $\mathcal{E}_\tau^*$, with poles given by $v_i=|q|^{-1}$ and $\prod_{i\leq k\leq j}v_k=1$
\end{prop}
\begin{defi}[Brown-Levin]
Let $u_i=\e(\xi_i)$, $v_i=\e(\alpha_i)$, we call \emph{multiple elliptic polylogarithms} the coefficients $E_{k_1,\ldots ,k_r}(\xi_1,\ldots ,\xi_r,\tau)$ of the non-polar part of the Laurent series-expansion around $(\alpha_1,\ldots ,\alpha_r)=(0,\ldots ,0)$ of (\ref{genfunellMPL}). We will call $r$ the \emph{depth}.
\end{defi}
An informal version of the second main result of \cite{BrownLevin} is then the following:
\begin{teo}[Brown-Levin]
All homotopy invariant iterated integrals on $\mathcal{E}_\tau^*$ can be written in terms of multiple elliptic polylogarithms. More precisely, allowing some of the arguments $u_i$ in (\ref{genfunellMPL}) to degenerate to~1, one obtains multi-valued functions on $\mathcal{E}_\tau^*$ which can be written as iterated integrals of coefficients of (\ref{KZB-form}).
\end{teo}
This can be considered as the genus one analogue of Lemma \ref{lemmaMPL}. Note that the definition of multiple elliptic polylogarithms is not explicit, and it is in fact very involved to understand the behaviour of the poles in the $\alpha_i$'s. In \cite{BrownLevin} this was carried out up to depth $r=2$. We report on their result in the case of depth $r=1$. Let us consider the modified generating function of depth one classical polylogarithms
\begin{equation}
\Lambda(u,\beta)=u^{-\beta}\sum_{k\geq 1}\mbox{Li}_k(u)\beta^{k-1}.
\end{equation} 
This can also be thought of as the generating series of a version of polylogarithms called \emph{Debye polylogarithms}. Let us consider the series
\begin{equation}
\texttt{E}(u,v,\beta,\tau)=\sum_{n\in\mathbb{Z}}v^n\Lambda(q^nu,\beta).
\end{equation}
This series converges absolutely for $1<u<|q|^{-1}$, and may have poles at $u=1$, which are given by the asymptotics of $\Lambda(u,\beta)$ at $u=\infty$.
\begin{prop}[Brown-Levin, correcting a typo in the polar part]
Recall the notation $u=\e(\xi)$, $v=\e(\alpha)$. The (regularized) generating series of elliptic (Debye) polylogatihms is given by:
\begin{multline}
\texttt{E}^{reg}(\xi,\alpha,\beta,\tau)=\\
=\sum_{n\in\mathbb{Z}}\e(n\alpha)\Lambda(\e(\xi+n\tau),\beta)-\frac{\e(-\xi\beta)}{2\pi i\beta(\alpha-\tau\beta)}-\frac{1}{\alpha(1-\e(\beta))}=:\sum_{m,n\geq 0}\Lambda_{m,n}(\xi,\tau)\alpha^m\beta^n.
\end{multline}
\end{prop}
The functions $\Lambda_{m,n}(\xi,\tau)$ were already studied in the~90's by Beilinson and Levin in the papers \cite{BeilinsonLevin}, \cite{Levin}, and served as a prototype for the general definition of Brown-Levin. In particular, the definition given in \cite{Levin} was more direct, and Levin studied all the main analytic properties. Here we report, omitting all proofs, the statements of some of the main results of \cite{Levin}, translated in terms of the construction presented above (noticing a typo in the translation made in \cite{BrownLevin}).
\begin{prop}[Levin]
We have
\begin{equation}
\frac{1}{2\pi i}\frac{\partial}{\partial\xi}\texttt{E}\Big(\xi,\frac{X}{2\pi i},\frac{Y}{2\pi i},\tau\Big)=-e^{-Y\xi}F\Big(\xi,\frac{X-\tau Y}{2\pi i},\tau\Big),
\end{equation}
\begin{equation}
\frac{1}{2\pi i}\frac{\partial}{\partial\tau}\texttt{E}\Big(\xi,\frac{X}{2\pi i},\frac{Y}{2\pi i},\tau\Big)=-e^{-Y\xi}\frac{\partial}{\partial X}F\Big(\xi,\frac{X-\tau Y}{2\pi i},\tau\Big),
\end{equation}
where $F$ is the Kronecker function defined in (\ref{DefF}).
\end{prop}
\begin{teo}[Levin]
Let $\gamma=\left( \begin{array}{cc}
a & b \\
c & d \end{array} \right)\in\mbox{SL}_2(\mathbb{Z})$. Then there exists a Laurent series $c_\gamma(X,Y)$ with rational coefficients such that
\begin{equation}
\texttt{E}\Big(\frac{\xi}{c\tau+d},aX+bY,cX+dY,\frac{a\tau+b}{c\tau+d}\Big)=\texttt{E}(\xi,X,Y,\tau)+2\pi i c_\gamma\Big(\frac{X}{2\pi i},\frac{Y}{2\pi i}\Big).
\end{equation}
\end{teo}
\begin{teo}[Levin]\label{TeoLEVIN}
Let $\xi=r_{\tau}(\xi)\tau+s_{\tau}(\xi)$, for $r_{\tau}(\xi), s_{\tau}(\xi)\in [0,1)$, and let
\begin{equation}\label{DefXi}
\Xi(\xi,X,Y,\tau)=-\frac{1}{2\pi i}\exp(rX+sY)\texttt{E}\Big(\xi,\frac{X}{2\pi i},\frac{Y}{2\pi i},\tau\Big).
\end{equation}
Then for $r=r_{\tau}(\xi), s=s_{\tau}(\xi)\in\mathbb{Q}$ and $(r,s)\neq (0,0)$ we have
\begin{equation}
\Xi(\xi,X,Y,\tau)=\frac{-\tau}{X(X-\tau Y)}+\sum_{k\geq 2}(-1)^{k-1}\frac{(k-1)}{(2\pi i)^k}\mathcal{G}_k^{r,s}(\tau,X,Y)+C^{r,s}(X,Y),
\end{equation}
and if $\xi=0$ we have
\begin{multline}\label{Xi}
\Big(\Xi(\xi,X,Y,\tau)-\frac{1}{2\pi i}\log(2\pi i\xi)\Big)\Big|_{\xi=0}=\\
=\frac{-\tau}{X(X-\tau Y)}+\sum_{k\geq 2}(-1)^{k-1}\frac{(k-1)}{(2\pi i)^k}\mathcal{G}_k^{0,0}(\tau,X,Y)+C^{0,0}(X,Y),
\end{multline}
where $\mathcal{G}_k^{r,s}(\tau,X,Y)$ is defined to be the primitive of the Eisenstein series\footnote{If $r=a/p$, $s=b/q$, one can prove that they are modular forms of weight $k$ for the congruence subgroup $\Gamma(l.c.m(p,q))$.}
\[
\sum_{m,n}\frac{\e(nr-ms)}{(m\tau+n)^k}
\]
given by
\begin{equation}\label{EisCongruence}
\int_{\tau}^{i\infty}\sum_{\substack{m\neq 0\\n\in\mathbb{Z}}}\frac{\e(nr-ms)}{(mz+n)^k}(X-zY)^{k-2}dt-\int_0^{\tau}\sum_{n\neq 0}\frac{\e(nr)}{n^k}(X-zY)^{k-2}dz,
\end{equation}
and $C^{r,s}(X,Y)\in\mathbb{C}[X,Y]$ is some integration constant\footnote{This constant was not worked out by Levin, who actually just states the theorem in terms of the indefinite integrals of (\ref{EisCongruence}). We will report its value in the special case $(r,s)=(0,0)$ in Proposition \ref{eMZVellPol}.}.
\end{teo}
In Chapter \ref{ChapterEllMZV} we will see much more about this kind of primitives of modular forms, and how they are involved in Brown's definition of multiple modular values \cite{MMV}. Note that Brown-Levin did not consider special values of multiple elliptic polylogarithms at points of the lattice $\Lambda_\tau$ and limited themselves to iterated integrals with non-tangential base points. It is clear to experts that these special values are essentially given by Enriquez's elliptic multiple zeta values, but this was never worked out in details. The only case where this was made more precise is for Levin's depth one elliptic polylogarithms \cite{MatthesMeta}, as we will see in Chapter \ref{ChapterEllMZV}.

\subsection{Towards single-valued multiple elliptic polylogarithms}\label{SectionSVEllPol}

For any $\xi\in\mathbb{C}$ and $\omega\in\Lambda_\tau$ let us consider the character on the lattice $\Lambda_\tau$
\begin{equation}
\chi_\xi(\omega):=\e\Big(\frac{\bar{\omega}\xi-\omega\bar{\xi}}{\tau-\bar{\tau}}\Big).
\end{equation}
Let $a,b\geq 1$ and $r=a+b-1$, and let us denote\footnote{For $a=b=1$ these series are not absolutely convergent, and we sum using the Eisenstein convention described in \cite{Weil}.}
\begin{equation}\label{svEllPol}
e_{a,b}(\xi,\tau)=\frac{\Im(\tau)^r}{\pi}\sum_{\omega\in\Lambda_\tau\setminus\{0\}}\frac{\chi_\xi(\omega)}{\omega^a\overline{\omega}^b}.
\end{equation}
Just by definition, it is trivial to see that for all $p,q\in\mathbb{Z}$
\[
e_{a,b}(\xi+p\tau+q,\tau)=e_{a,b}(\xi,\tau),
\]
which implies that these are single-valued functions on $\mathcal{E}_\tau$.
Moreover, for $\left( \begin{array}{ccc}
a & b \\
c & d \end{array} \right)\in\mbox{SL}_2(\mathbb{Z})$ it is an easy exercise to show that
\[
e_{a,b}\Big(\frac{\xi}{c\tau+d},\frac{a\tau+b}{c\tau+d}\Big)=\frac{(c\tau+b)^a(c\overline{\tau}+d)^b}{|c\tau+d|^r}e_{a,b}(\xi,\tau)
\]
Setting $a=b$ and $\xi=0$ one gets back the non-holomorphic Eisenstein series
\begin{equation}
e_a(\tau)=\frac{\Im(\tau)^{2a-1}}{\pi}\sum_{\omega\in\Lambda_\tau\setminus\{0\}}\frac{1}{|\omega|^{2a}}.
\end{equation}
\begin{oss}\label{ossEis}
After multiplying by $(\pi\Im(\tau))^{1-a}$, these functions are usually denoted in the literature by $E(a,\tau)$, where $a$ is in general allowed to belong to $\mathbb{C}$, and are modular invariant. We will see soon how they appear in the context of closed superstring amplitudes. 
\end{oss}
The functions $e_{a,b}(\xi,\tau)$ are in some sense a single-valued analogue of elliptic polylogarithms, because of the following results of Zagier and Levin, that we present in chronological order (see \cite{ZagierBWR}, \cite{Levin}):
\begin{teo}[Zagier]\label{teoZagEis}
Let
\begin{multline}
D_{a,b}(u)=(-1)^{a-1}\sum_{k=a}^r2^{r-k}\binom{k-1}{a-1}\frac{(-\log|u|)^{r-k}}{(r-k)!}\mbox{Li}_k(u)\\
+(-1)^{b-1}\sum_{k=b}^r2^{r-k}\binom{k-1}{b-1}\frac{(-\log|u|)^{r-k}}{(r-k)!}\overline{\mbox{Li}_k(u)}.
\end{multline}
By Theorem \ref{TeoSVMPL}, these are single-valued polylogarithms. Then we have that\footnote{Recall that we denote, as always, $u=\e(\xi)$.}
\begin{equation}
e_{a,b}(\xi,\tau)=\sum_{l\geq 0}D_{a,b}(q^lu)+(-1)^{r-1}\sum_{l\geq 1}D_{a,b}(q^lu^{-1})+\frac{(-2\log|q|)^r}{(r+1)!}\B_{r+1}\Big(\frac{\log|u|}{\log|q|}\Big),
\end{equation}
where $\B_{r+1}(x)$ is the $(r+1)$st \emph{Bernoulli polynomial}, defined by the generating series 
\[
\frac{te^{xt}}{e^t-1}=\sum_{r\geq 0}\B_r(x)\frac{t^r}{r!}.
\]
\end{teo}
In other words, all $e_{a,b}(\xi,\tau)$ can be obtained as averages of single-valued polylogarithms. Moreover, if we consider the modified
\[
\tilde{e}_{a,b}(\xi,\tau)=\frac{\pi}{\Im(\tau)^r}e_{a,b}(\xi,\tau),
\]
and their generating series\footnote{This series is not absolutely convergent, and again we sum it using the Eisenstein convention.}
\[
K(\xi,\alpha,\tau)=\sum_{\omega\in\Lambda_\tau\setminus\{0\}}\frac{\chi_{\xi}(\omega)}{|w+\alpha|^2}=\frac{1}{|\alpha|^2}+\sum_{a,b\geq 1}\tilde{e}_{a,b}(\xi,\tau)(-\alpha)^{a-1}(-\overline{\alpha})^{b-1},
\]
we have
\begin{teo}[Levin]
\begin{equation}
\Xi(\xi,X,Y,\tau)-\overline{\Xi(\xi,-X,-Y,\tau)}=-\frac{\tau-\overline{\tau}}{(2\pi i)^2}K\Big(\xi,\frac{X-\tau Y}{2\pi i},\tau\Big),
\end{equation}
where $\Xi$ was defined in (\ref{Xi}).
\end{teo}
This means that all $e_{a,b}(\xi,\tau)$ can be obtained as a combination of elliptic polylogarithms (defined as avarages of classical holomorphic polylogarithms) and their complex conjugates, and gives us the right to call them \emph{single-valued elliptic polylogarithms.} At the moment, one of the next goals in this field (which is, as we will see, related to our work on superstring amplitudes) is to define and study single-valued elliptic polylogarithms of higher depth.

%% file: Chapter3.tex

\chapter{Number theoretical aspects of superstring amplitudes} 

\label{ChapterPhysicsBackground} 

\section{Superstring amplitudes in a nutshell}

The goal of this chapter is to give an overview, aimed at mathematicians, of scattering amplitudes in superstring theory. In particular, we want to highlight the aspects which are related to the mathematics discussed in the previous chapter. To do this, we will present the computation of scattering amplitudes in terms of simple mathematical problems. This involves a great simplification of the original physics issues. Explaining how to exactly relate the simplified problems considered here to actual superstring theories certainly goes beyond the scope of this work, therefore we now briefly give, once for all, an account of the physics jargon used throughout this thesis, and refer the reader to the literature for all details. 

First of all, we will just divide strings\footnote{We will not explain how supersymmetry enters into the picture, and we will always mean \emph{superstring} when we write \emph{string}.} between \emph{open} and \emph{closed}, implicitly meaning that (massless vibration modes of) open strings are \emph{gluons} in maximally supersymmetric type I superstring theory, that (massless vibration modes of) closed strings are \emph{gravitons} in maximally supersymmetric type IIB superstring theory\footnote{Type IIB and type IIA have very subtle differences, which disappears in the case of four gravitons, at least up to genus three. Therefore in this context we will speak only of \emph{type II superstring theory}.}, and that all strings are massless external states in the uncompactified ten-dimensional space-time with signature $(1,9)$, denoted $\mathbb{R}^{1,9}$. As depicted in the introduction in figures 1.3 and 1.4, open strings and closed strings give rise to very different kind of worldsheets. In particular, while in the closed string case it is easy to define what the $n$-point-amplitude is, in the open string case different topologies or different positions of the strings' insertions produce different amplitudes, as we will see later, and the $n$-point-amplitude is the average over all these possibilities. Each (massless) string carries a momentum vector $k_i\in\mathbb{R}^{1,9}$ and a polarization tensor $\zeta_i$. Momenta need to satisfy $k_1+\cdots +k_n=0$ (momentum conservation) and $k_i^2=0$, where by $k_i^2$ we mean the scalar product in  $\mathbb{R}^{1,9}$ (on-shell condition). This implies that the first physically meaningful amplitudes that we want to consider involve at least four strings. All strings depend also on a parameter $\alpha^\prime$, which is the inverse of the \emph{fundamental string tension}, and the limit $\alpha^\prime\mapsto 0$ gives back point particles and the underlying field theories. One defines the (dimensionless) \emph{Mandelstam variables} cited in the introduction as $s_{i,j}=\alpha^\prime(k_i+k_j)^2\in\mathbb{C}$. Momentum conservation and on-shell conditions give relations among these variables, as we will soon see for instance in the four-point case. Let us denote by $\textbf{s}$ the vector of all Mandelstam variables. Then for any genus $g$ the $n$-point amplitude (it will always be clear from the context whether we consider open or closed strings) $\textbf{A}_{g,n}=\textbf{A}_{g,n}(\alpha^\prime,k_1,\ldots ,k_n,\zeta_1,\ldots ,\zeta_n)$ is given by\footnote{As remarked above, in the open string case one also needs to specify the position of the insertions and the topology chosen. Therefore, for instance, speaking of open string amplitudes in genus zero we will write $A_{g,n}^\sigma$, where $\sigma\in S_n$ is a permutation of the counter-clockwise ordering of $n$ open strings from~1 to~$n$ on the boundary of a disk. However, in this attempt of giving universal statements, we prefer to keep things slightly imprecise, but simpler.}
\begin{equation}\label{genAmpl}
\textbf{A}_{g,n}=I_{g,n}(\textbf{s})\textbf{R}_{g,n}(\alpha^\prime,k_1,\ldots ,k_n,\zeta_1,\ldots ,\zeta_n),
\end{equation}
where $\textbf{R}$ is some overall kinematic factor (a column-vector or a scalar, depending on the situation, which is well understood in the cases that we will consider, but not in general), and $I$ is some Feynman integral\footnote{It would be perhaps more correct to call $I$ just a moduli-space integral, because usually the word Feynman integral raises the expectation to integrate over a momentum with the same dimension as spacetime, but this words helps mathematicians to visualize the integral as an analogue of the more familiar Feynman integrals in QFT.} (or a row-vector of Feynman integrals) depending only on the Mandelstam variables. The full $n$-point amplitude is given by $\textbf{A}_n=\sum_{g\geq 0}\textbf{A}_{g,n}$. We will not be interested in the kinematic part $\textbf{R}$ of $\textbf{A}_{g,n}$, so for now on, when we speak of amplitudes, we mean $I_{g,n}(\textbf{s})$. These Feynman integrals in general are not meromorphic functions of the Mandelstam variables in the \emph{low-energy limit} $\textbf{s}\mapsto\textbf{0}$, as they may have logarithmic singularities \cite{GRV}. We are going to be interested only in the part of $I_{g,n}(\textbf{s})$ which is meromorphic in a neighborhood of zero. Therefore, by an abuse of notation, we will call it $I_{g,n}(\textbf{s})$ too, and we will be interested in its low-energy expansion
\begin{equation}
I_{g,n}(\textbf{s})=\sum_{\textbf{m}}\alpha_{g,n,\textbf{m}}\textbf{s}^\textbf{m},
\end{equation}
where by $\textbf{s}^\textbf{m}$ we mean $s_{1,2}^{m_{1,2}}s_{1,3}^{m_{1,3}}\cdots $, and the summation runs over integer numbers $m_{i,j}$ bounded below by some $M_{g,n}\in\mathbb{Z}$. We will give the precise formula for the Feynman integrals $I_{g,n}(\textbf{s})$ only in some specific case, also because it is not clear how to define them for general $g$ and $n$ \cite{DonagiWitten}. However, we want now to sketch the idea of a general recipe to construct these integrals. The domain of integration is given by the relevant moduli space, as remarked in the introduction. For instance, in the closed string case, for genus $\leq 1$, one has to integrate over the Deligne-Mumford compactification $\overline{\mathcal{M}}_{g,n}$ of the moduli space of punctured Riemann surfaces, with $2g+n-2>0$ (if the genus is at least two, one really needs to consider super Riemann surfaces \cite{WittenSuperstringPerturbationRevisited}). The integrand is defined in terms of Green's functions, which are symmetric real analytic functions $G_\mu(z,w)$ on $C\times C\setminus \Delta$, where $C$ is a compact Riemann surface and $\Delta$ is the diagonal, associated to a metric $\mu$ compatible with the conformal structure of $C$. Green's functions are required to satisfy $G_\mu(z,w)=\log |t(z)|^2+O(1)$ as $z\rightarrow w$, where $t$ is a local coordinate near $w$ such that $t(w)=0$, and such that for all $w_1,w_2$ the function $z\rightarrow G_\mu(z,w_1)-G_\mu(z,w_2)$ is harmonic on $C\setminus\{w_1,w_2\}$. In superstring theory it turns out that for any genus $g$ there are canonical choices of Green's function $G_g(z,w)$, called \emph{propagators}. The prototype of integrand of superstring amplitudes is roughly speaking the product of
\begin{equation}\label{stringintegrand}
\prod_{i<j}\exp(s_{i,j}G_g(z_i,z_j))
\end{equation} 
with some extra term (whose complexity grows with the number of strings) defined in terms of the propagators and their derivatives\footnote{Explaining what is the precise recipe to build these integrals goes beyond the scope of this thesis. We will write down explicitly the integrals needed in the cases that we present, and refer to the literature for further details.}.

In the second section of this chapter we will introduce genus zero superstring amplitudes for open and closed strings. The structure of these amplitudes is now fairly well understood, thanks to beautiful advances accomplished during the last ten years, and we will see that this structure is related in a fascinating way to the theory of multiple zeta values developed in Chapter \ref{ChapterMathBackground}. In the third section we will discuss the less understood genus one case, which constitutes the main motivation for most of the results of this thesis. In particular, we will give an account of the state of art at the moment when we started our investigation, and we will see how the two classes of functions that we have studied in this thesis, namely elliptic MZVs and modular graph functions, are naturally related to respectively open and closed strings. Since this domain of research is very active, great progress has been made in the last three years, partly building on the results that we have obtained in Chapter~\ref{ChapterClosedStrings}. An updated account of the state of art is postponed to the last chapter. Finally, we want to mention that only very little is known for higher genera, and we refer the interested reader to \cite{DG2014}, \cite{DGPR}, \cite{GomezMafraSc} for genus two and \cite{GomezMafra} for genus three.

\section{Superstring amplitudes in genus zero}

As explained in the introduction, tree-level scattering amplitudes of superstrings are given by iterated integrals along the boundary of a disk for open strings and by integrals over the whole Riemann sphere for closed strings. The propagator in this case is given by
\begin{equation}
G_0(z,w)=\log|z-w|^2.
\end{equation}
This is actually a one-variable function, and by abuse of notation we will also write $G_0(z-w)=G_0(z,w)$. Let us see first what happens in the case of four strings, and then we will mention without giving details the main known or conjectured results for the general case.

\subsection{The four-point case}

Let us start with the open string-case. First of all, it is a simple exercise to see that there are only two independent Mandelstam variables, that we simply call~$s$ and~$t$. By SL$_2(\mathbb{C})$-invariance, one can fix the three points $0,1,\infty$ along the boundary of a disk in~$\mathbb{P}^1_{\mathbb{C}}$. Moreover, one can assume that the fourth string is inserted in~$[0,1]$ \cite{BBDamgaardVanhove}. The amplitude is given by \cite{GSW}
\begin{equation}
I(s,t)=s\int_{0}^1z^{s-1}(1-z)^{t}dz,
\end{equation}
which is indeed (almost) of the form \ref{stringintegrand}. This kind of integral, called \emph{beta function}, is known since Euler, and was related to an open bosonic string amplitude by Veneziano already at the end of the sixties \cite{Veneziano}. By a standard computation,
\begin{equation}
I(s,t)=\frac{s\Gamma(s)\Gamma(1+t)}{\Gamma(1+s+t)}=\frac{\Gamma(1+s)\Gamma(1+t)}{\Gamma(1+s+t)},
\end{equation}
where
\begin{equation}
\Gamma(z)=\int_{0}^\infty t^{z-1}e^{-t}dt
\end{equation}
is the classical $\Gamma$-function. By the well known property \cite{Lang}
\begin{equation}\label{expansionGamma}
\Gamma(1+z)=\exp\Big(-\gamma z+\sum_{n\geq 2}\zeta(n)\frac{(-z)^n}{n}\Big),
\end{equation}
where $\gamma$ is the \emph{Euler-Mascheroni constant}, we deduce immediately that
\begin{equation}\label{expansion4openstrings}
I(s,t)=\exp\Big(\sum_{n\geq 2}\frac{(-1)^n\zeta(n)}{n}(s^n+t^n-(s+t)^n)\Big).
\end{equation}
In particular, this means that the coefficients of the Taylor expansion are rational polynomials in Riemann zeta values.

Let us now consider the closed string case. Now the insertions can be everywhere on the complex projective line. Again by conformal invariance, we can fix three of them in $0, 1, \infty$. Then the amplitude is given essentially (see \cite{GSW}) by the  \emph{complex beta function}
\begin{equation}\label{4closedStrings}
B_{\mathbb{C}}(s,t):=\int_{\mathbb{C}}|z|^{2s-2}|1-z|^{2t-2}\frac{idzd\overline{z}}{2}.
\end{equation}
The computation of this integral in terms of $\Gamma$-functions is well known to physicists\footnote{It was known already by Virasoro in \cite{Virasoro}, where he related (\ref{4closedStrings}) to the amplitude of closed bosonic strings, shortly after Veneziano's pioneering computation of the open case.} but not to mathematicians. For this reason, I will go through the details, that I have learned from Zagier. Using polar coordinates, (\ref{4closedStrings}) becomes
\begin{multline*}
\int_{0}^\infty r^{2s-1}\int_0^{2\pi}(r^2-2r\cos\theta+1)^{t-1}d\theta dr \\
= \sum_{n\geq 0}\binom{t-1}{n}\int_0^\infty r^{2s+n-1}(r^2+1)^{t-n-1}dr\int_0^{2\pi} (-e^{i\theta}-e^{-i\theta})^n d\theta\\
= 2\pi \sum_{n\geq 0}\binom{t-1}{n}\binom{2n}{n}\int_0^\infty r^{2s+2n-1}(r^2+1)^{t-2n-1}dr.
\end{multline*}
The next step is to do the substitution $r^2=u/(1-u)$, which leads to
\[
\pi \sum_{n\geq 0}\binom{t-1}{n}\binom{2n}{n}\int_0^1 u^{s+n-1}(1-u)^{n-s-t}du.
\]
This is just the classical real beta integral, and therefore we get
\[
\pi \sum_{n\geq 0}\binom{t-1}{n}\binom{2n}{n}\frac{\Gamma(s+n)\Gamma(n-s-t+1)}{\Gamma(2n-t+1)}.
\]
Recall now the well known properties of the $\Gamma$-function \cite{Lang}
\begin{itemize}
\item $\Gamma(n+z)=(z+n-1)_n\Gamma(z)$, where $(x)_n:=x(x-1)\cdots (x-n+1)$ is the \emph{descending Pochhammer symbol},
\item $\Gamma(z)\Gamma(1-z)=\frac{\pi}{\sin(\pi z)}$.
\end{itemize}
Using this we get
\begin{multline*}
\pi\Gamma(s)\Gamma(t)\Gamma(1-s-t)\sum_{n\geq 0}\frac{(s+n-1)_n(n-s-t)_n}{(n!)^2\Gamma(t-2n)\Gamma(1+2n-t)}\\
=\Gamma(s)\Gamma(t)\Gamma(1-s-t)\sum_{n\geq 0}\frac{(s+n-1)_n(n-s-t)_n\sin(\pi(t-2n))}{(n!)^2}\\
=\Gamma(s)\Gamma(t)\Gamma(1-s-t)\sum_{n\geq 0}\frac{(s+n-1)_n(n-s-t)_n\sin(\pi t)}{(n!)^2}\\
=\pi\frac{\Gamma(s)\Gamma(1-s-t)}{\Gamma(1-t)}\sum_{n\geq 0}\frac{(s+n-1)_n(n-s-t)_n}{(n!)^2}.
\end{multline*}
To conclude the computation, let us recall the definition (for $|z|<1$) of the hypergeometric function
\begin{equation}\label{hypergeometric}
{}_{(k+1)}F_k(\alpha_1,\ldots ,\alpha_{k+1};\gamma_1,\ldots ,\gamma_{k};z):=\sum_{n\geq 0}\frac{(\alpha_1+n-1)_n\cdots (\alpha_{k+1}+n-1)_n}{(\gamma_1+n-1)_n\cdots (\gamma_k+n-1)_n}\frac{z^n}{n!}.
\end{equation}
A classical formula of Gauss reads
\[
{}_2F_1(\alpha,\beta;\gamma;1)=\frac{\Gamma(\gamma)\Gamma(\gamma-\alpha-\beta)}{\Gamma(\gamma-\alpha)\Gamma(\gamma-\beta)}.
\]
Using this, and again the properties of $\Gamma$, we finally get that
\begin{equation}
B_{\mathbb{C}}(s,t)=\frac{\pi(s+t)}{st}\frac{\Gamma(1+s)\Gamma(1+t)\Gamma(1-s-t)}{\Gamma(1-s)\Gamma(1-t)\Gamma(1+s+t)}.
\end{equation}
By (\ref{expansionGamma}), we have that
\[
\frac{\Gamma(1+z)}{\Gamma(1-z)}=\exp\Big(-2\gamma z-2\sum_{n\geq 1}\frac{\zeta(2n+1)}{(2n+1)}z^{2n+1}\Big),
\]
and therefore we get
\begin{equation}\label{expansioncomplexbeta}
B_{\mathbb{C}}(s,t)=\frac{\pi(s+t)}{st}\exp\Big(-2\sum_{n\geq 1}\frac{\zeta(2n+1)}{(2n+1)}(s^{2n+1}+t^{2n+1}-(s+t)^{2n+1})\Big).
\end{equation}
In particular, while in the low-energy expansion of four open strings we get all Riemann zeta-values, in the closed case we only get odd Riemann zetas. We will see that this is (conjectured to be) related with the fact that $\zeta_{\rm sv}(2k)=0$ and $\zeta_{\rm sv}(2k+1)=2\zeta(2k+1)$, or more precisely to the fact that if we apply the map $\rm sv$ defined in the previous chapter to (\ref{expansion4openstrings}) we get exactly $\frac{st}{\pi(s+t)}B_{\mathbb{C}}(s,t)$.

\subsection{The $n$-point case}

It was shown in \cite{MafraScSt1} that, for any number $n$ of open strings with associated polarization tensors $\zeta_i$ and Mandelstam variables $s_{i,j}$, and any permutation $\phi$ of the insertions' ordering, the amplitude is given by
\begin{equation}
A^\phi_{\zeta_1,\ldots ,\zeta_n}(\textbf{s})=\sum_{\sigma\in S_{n-3}}I^\phi_{\sigma}(\textbf{s})\textbf{R}^\phi_{\zeta_1,\zeta_{\sigma(2)},\ldots ,\zeta_{\sigma(n-2)},\zeta_{n-1},\zeta_n},
\end{equation}
where $S_{n-3}$ is the group of permutations on $n-3$ letters, the kinematic factors $\textbf{R}$ are partial tree amplitudes in the super Yang-Mills theory obtained in the point particle limit $\alpha^\prime\mapsto 0$, and all $I^\phi_\sigma(\textbf{s})$ look like
\begin{equation}\label{SelbergInt}
I^{id}_\sigma(\textbf{s})=\int_{0\leq z_2\leq \cdots \leq z_{n-2}\leq 1} \prod_{i<j}(z_j-z_i)^{s_{i,j}}\prod_{k=2}^{n-2}\sum_{m=1}^{k-1}\frac{s_{\sigma (m),\sigma(k)}}{z_{\sigma(m)}-z_{\sigma(k)}}dz_2\cdots dz_{n-2},
\end{equation}
which is a \emph{generalized Selberg integral}, or \emph{multi-beta function} \cite{BrownModuliSpace}. It follows directly from Corollary $8.5$ of \cite{BrownModuliSpace} that the Taylor expansion of (\ref{SelbergInt}) in the Mandelstam variables belong to the algebra $\mathcal{A}$ of multiple zeta values considered in the previous chapter. Moreover, some explicit formula for these integrals for $n\leq 6$ in terms of the $\Gamma$-function and the hypergeometric function (\ref{hypergeometric}) can be found in the physics literature: for instance, the integral corresponding to $\sigma=id$ in the five-point case is given by \cite{Kitazawa}
\begin{multline*}
I^{id}_{id}(s_{1,2},\ldots ,s_{4,5})=\frac{\Gamma(1+s_{1,2})\Gamma(1+s_{2,3})\Gamma(1+s_{3,4})\Gamma(1+s_{4,5})}{\Gamma(1+s_{1,2}+s_{2,3})\Gamma(1+s_{3,4}+s_{4,5})}\times \\
\times {}_3F_2(s_{1,2},1+s_{4,5},-s_{2,4};1+s_{1,2}+s_{2,3},1+s_{3,4}+s_{4,5};1),
\end{multline*}
and Taylor expanding this function one finds instances of higher depth MZVs, such as $\zeta(3,5)$. Before talking about the closed string-case, it is worth mentioning that one can prove that a certain vector-valued deformation of the Selberg integrals (\ref{SelbergInt}) is a solution of the KZ-equation (\ref{KZequation}), for certain matrices $e_0$ and $e_1$ whose entries are linear functions of the Mandelstam variables, and gives back the $n$-point integral (resp. $(n-1)$-point integral) when one sets $z=1$ (resp. $z=0$). In particular, since it is known that special values at $z=0$ and $z=1$ of this solution must be related by the Drinfel'd associator $Z(e_0,e_1)$, one can deduce in this way not only that the coefficients of the $n$-point amplitude must be MZVs, but also an elegant recursive structure based on the Drinfel'd associator (see \cite{TerasomaSelberg}, \cite{BSST}).

Let us now very briefly mention what is known for $n$ closed strings. A well known method used by physicists, called \emph{KLT relation} \cite{KLT}, allows to reduce the closed string amplitude, which is given just in terms of a single integral $I(\textbf{s})$ on the whole complex plane, to products of integrals on $[0,1]$ that are related to the open string amplitude. This, together with the nowadays good knowledge that we have of the open string-case, allows to do explicit computations of the first terms in the Taylor expansion for a small number of strings (we refer to the literature for all details). The upshot is that all computations performed so far strongly suggest to conjecture that closed string-amplitudes are the image under a suitable map based on Brown's single-valued map defined in the previous chapter (see \cite{ScSt}, \cite{Stieb2014} and \cite{StiebTaylor} for the heterotic string-analogue). In particular, all coefficients of the closed string amplitude are conjectured to lie in the algebra of single-valued MZVs~$\mathcal{A}^{\rm sv}$, and it was discovered in \cite{ScSt} that in the five point case Riemann zetas do not suffice anymore (the same that happened in the open case). For instance, one coefficient contained $\zeta_{\rm sv}(3,5,3)$, which, as we have seen in the previous chapter, is the first irreducible single-valued MZV of higher depth.

\section{Superstring amplitudes in genus one}\label{SectionModGraph}

In genus one, as we have remarked in the introduction, some new difficulties appear in the computation of superstring amplitudes. First of all, the Feynman integrals in general depend not only on the position of the insertions on the Riemann surface, but also on the conformal structure of the surface, which in genus zero was trivial, but starting from genus one plays a non-trivial r\^{o}le. Another difficulty is given by the fact that, in the open string case, one must consider various possible topologies (cylinders with insertion on both boundaries \cite{BMRS} and M\"{o}bius strips). However, in this work we will not focus on these points. We will instead focus on the third difficulty, which is given by the fact that the propagator on the torus is much more complicated than in genus zero. This leads to integrals over configuration spaces of points on (genus one) Riemann surfaces that are much more complicated than their genus zero analogues considered in the previous section, and are related to the genus one generalizations of polylogarithms that we have introduced in Chapter \ref{ChapterMathBackground}.

\begin{oss}
In the physics literature, the real and the imaginary part of a complex number $z$ are often denoted by $z_1$ and $z_2$. We will adopt this notation in the rest of this chapter, as well as in Chapter \ref{ChapterClosedStrings}. This may create some confusion when we want to give general formulae, because we will need to label variables: we hope that it will always be clear from the context whether $z_1$ is a labelled complex variable, or the real part of $z$.
\end{oss}

\subsection{The Green function on the torus}

For $\tau=\tau_1+i\tau_2\in\mathbb{H}$ and $\Lambda_\tau=\tau\mathbb{Z}+\mathbb{Z}$, let $\mathcal{E}_{\tau}=\mathbb{C}/\Lambda_\tau$ and $\mathcal{E}^*_{\tau}=\mathcal{E}_{\tau}\setminus\{0\}$. The genus one superstring propagator is the real analytic function on the complement of the diagonal of two copies of the complex torus $\mathcal{E}_{\tau}$ given by\footnote{We normalize it as in \cite{GRV}.}
\begin{equation}\label{PropagatorGenus1}
G_1(\xi,\nu,\tau)=-\frac{1}{4}\log\left|\frac{\theta(\xi-\nu,\tau)}{\eta(\tau)}\right|^2+\frac{\pi (\xi_2-\nu_2)^2}{2\tau_2},
\end{equation}
where $\xi=\xi_1+i\xi_2$ and $\nu=\nu_1+i\nu_2$ belong to $\mathcal{E}_{\tau}$, $\theta$ is the Jacobi $\theta$-function defined in (\ref{Theta}), and~$\eta$ is the \emph{Dedekind $\eta$-function}, a modular form of weight~$1/2$ whose infinite product representation reads 
\begin{equation}\label{eta}
\eta(\tau)=q^{1/24}\prod_{n\geq 1}(1-q^n),
\end{equation}
and that satisfies for all $\gamma=\left( \begin{array}{ccc}
a & b \\
c & d \end{array} \right)\in\mbox{SL}_2(\mathbb{Z})$
\begin{equation}
\eta(\gamma\tau)=\rho(\gamma)(c\tau+d)^{1/2}\eta(\tau),
\end{equation}
where
\[
\rho(\gamma)=\exp\Big(\frac{b\pi i}{12}\Big)
\]
for $c=0$, and otherwise
\[
\rho(\gamma)=\exp\bigg(\pi i\Big(\frac{a+d}{12|c|}-\sum_{n=1}^{|c|-1}\frac{n}{|c|}\Big(\left\{\frac{dn}{|c|}\right\}-\frac{1}{2}\Big)-\frac{1}{4}\Big)\bigg).
\] 
This modular behaviour, together with the correction term $\pi (\xi_2-\nu_2)^2/2\tau_2$, imply that $G_1(\xi,\nu,\tau)$ is $\Lambda_\tau$-periodic as a function of $\xi-\nu$ and is modular invariant. Moreover, because of \ref{Jacobi}, we have that $G_1(\xi,\nu,\tau)=-\frac{1}{2}\log|\xi-\nu| + O(1)$ as $\xi\rightarrow\nu$, and $G_1$ is harmonic, and in fact one can show that it is, up to a normalization constant, equal to the Green's function associated to the flat metric on the torus. This function, as in the genus zero case, depends only on the difference of the two variables $\xi$ and $\nu$, and again by abuse of notation we will write it as a function of one variable $\xi\in\mathcal{E}^*_{\tau}$. One can write $\xi=r_{\tau}(\xi)\tau+s_{\tau}(\xi)$, with $r_{\tau}(\xi),s_{\tau}(\xi)\in\mathbb{R}$ given by $r=\xi_2/\tau_2$ and $s=\xi_1-\tau_1\xi_2/\tau_2$. To make the notation simpler, we will just write $r$ and $s$. Obviously $G_1(\xi,\tau)$ is a 1-periodic function with respect to both $r$ and $s$, hence it has a Fourier expansion with respect to both variables, which follows from
\begin{prop}\label{propPropEis}
\begin{equation}\label{FirstFourier}
G_1(\xi,\tau)=\frac{1}{4}e_{1,1}(\xi,\tau),
\end{equation}
where $e_{1,1}$ is the single-valued elliptic polylogarithm defined in (\ref{svEllPol}).
\end{prop}
\textbf{Proof.} By definition, we can write
\[
G_1(\xi,\tau)=-\frac{1}{4}\log\Big(\frac{\theta(\xi,\tau)}{\eta(\tau)}\Big)-\frac{1}{4}\log\Big(\overline{\frac{\theta(\xi,\tau)}{\eta(\tau)}}\Big)+\frac{\pi (\xi_2-\nu_2)^2}{2\tau_2}.
\]
The claim then follows from the product expansion (obtained using (\ref{Jacobi}) and (\ref{eta}))
\[
\frac{\theta(\xi,\tau)}{\eta(\tau)}=q^{1/12}(u^{1/2}-u^{-1/2})\prod_{j\geq 1}(1-q^ju)(1-q^ju^{-1})
\]
and from Theorem \ref{teoZagEis}.\\ 
$\square$

Moreover, $G_1(\xi,\tau)$ is 1-periodic also with respect to $\tau_1$, hence one can do one more Fourier expansion and get
\begin{cor}
Denote as above $\xi=r\tau+s$. Then
\begin{equation}\label{FourierProp2}
G_1(\xi,\tau)=\frac{\pi\tau_2}{2}\overline{\B}_2(r)+\frac{1}{4}P(\xi,\tau),
\end{equation}
where $\overline{\B}_2(x)$ is the only 1-periodic continuous function coinciding with the second Bernoulli polynomial $\B_2(x)$ in the interval $[0,1]$, and
\begin{equation}\label{PPart}
P(\xi,\tau)=\sum_{\substack{m\in\mathbb{Z}\setminus\{0\}\\k\in\mathbb{Z}}}\frac{\e(m(k+r)\tau_1+ms)}{|m|}e^{-2\pi\tau_2|m||k-r|}.
\end{equation}
\end{cor}
\textbf{Proof.} We start with the expression (\ref{FirstFourier}). If $m=0$ we get
\[
\frac{\tau_2}{4\pi}\sum_{n\neq 0}\frac{e^{2\pi inr}}{|n|^2},
\]
which is equal to the first term appearing in the statement. Consider now $m\neq 0$. Then
\[
f_m(\xi,\tau):=\sum_{n\in\mathbb{Z}}\frac{\e(nr-ms)}{|m\tau_1+im\tau_2+n|^2}
\]
is $|m|^{-1}$-periodic with respect to the variable $\tau_1$, and we can consider its Fourier expansion $f_m(\xi,\tau)=\sum_{k\in\mathbb{Z}}\lambda_k \e(k|m|\tau_1)$,
where
\[
\lambda_k=|m|\int_{-1/(2|m|)}^{1/(2|m|)}\sum_{n\in\mathbb{Z}}\frac{\e\big(nr-ms+mr(t-\tau_1)\big)}{|mt+im\tau_2+n|^2}e^{-2\pi ik|m|t}dt.
\]
Then one needs to separate the cases of positive or negative $m$, and exchange series and integral. For instance, when $m>0$ one gets
\[
\lambda_k=\frac{\e(-m(s+r\tau_1))}{|m|}\int_{-\infty}^{+\infty}\frac{\e(mt(r-k))}{(t+i\tau_2)(t-i\tau_2)}dt.
\]
By a standard consequence of Cauchy's theorem \cite{Lang}, this integral can be reduced to a residue computation over the poles of the integrand, and the corollary follows from summing over all positive and negative $m$'s.\\
$\square$

\subsection{The open string case}

As we have already mentioned, there are various possible topologies contributing to the genus one open string amplitude. For simplicity, let us consider a cylinder with insertions only on one boundary component, as in figure 1.4 in the introduction. Physicists think of it as half of a torus $\mathcal{E}_\tau$ (in the genus zero case, the disk can be thought of as half of a sphere) parametrized by $\tau=i\tau_2$ with $\tau_2>1$ (which implies that $\Lambda_\tau=\overline{\Lambda_{\tau}}$), and that insertions are allowed only on the boundary $[0,1]$ of the half-torus $\{\xi\in[0,1]\times [0,\tau/2] : [0,\tau/2]\equiv [1,1+\tau/2]\}$. Thanks to this, the Green's function simply reads (the decoration \emph{op} stays for open)
\begin{equation}\label{GreenOpen}
G_1^{op}(\xi,\tau)=-\frac{1}{2}\log\left(\frac{\theta(\xi,\tau)}{\eta(\tau)}\right),
\end{equation}
where $\xi\in [0,1]$, and for instance the four-point amplitude\footnote{The general case, as in genus zero, is slightly more complicated, and involves also first and second derivatives of the propagator.} is computed by \cite{BMMS}
\begin{equation}
\int_i^{i\infty}\int_{1\geq x_1\geq x_2\geq x_3\geq 0}\prod_{1\leq i<j\leq 4}\exp(s_{i,j}G^{op}_1(x_i-x_j,\tau))\,dx_1dx_2dx_3d\tau,
\end{equation} 
where $d\tau$ is some appropriate measure on $[i,i\infty]$ and $x_4\equiv  0$ is fixed. Both in the open and in the closed string-case, we will be interested only in the integral over the positions given by
\begin{equation}\label{IntOpen4Genus1}
\int_{1\geq x_1\geq x_2\geq x_3\geq 0}\prod_{1\leq i<j\leq 4}\exp(s_{i,j}G^{op}_1(x_i-x_j,\tau))\,dx_1dx_2dx_3.
\end{equation} 
To my knowledge, so far nobody has worked out the second integration over $\tau$. The fundamental remark in order to attack (\ref{IntOpen4Genus1}) is that, if we simply think of (\ref{GreenOpen}) as a function of $\tau\in\mathbb{H}$ and $\xi\in\mathcal{E}^*_\tau$, then its derivative with respect to $\xi$ is equal to the function $f_1(\xi,\tau)$ defined by (\ref{expansionF}). This leads to a connection between genus one superstring amplitudes and elliptic polylogarithms, that was first noticed in \cite{BMMS}. In particular, one can relate (\ref{IntOpen4Genus1}) in an almost straightforward way to the \emph{elliptic multiple zeta values} defined by Enriquez in \cite{Enriquez} as the coefficients of an elliptic analogue of the Drinfel'd associator. This constitutes our main motivation for the results obtained in Chapter \ref{ChapterEllMZV}, where we will define elliptic MZVs, study their properties, and among other things we will explain how elliptic MZVs can be seen as special values of Levin's elliptic polylogarithms (this was first made clear by Matthes in \cite{MatthesMeta}). In particular, the integral (\ref{IntOpen4Genus1}) and its $n$-point generalizations inherit all properties of elliptic MZVs: they have a $q$-expansion whose coefficients are given by classical MZVs (and powers of $2\pi i$), and they can be written as iterated integrals of Eisenstein series \cite{BMMS}, \cite{BMS}. We postpone explicit computations of coefficients of the low energy expansion of (\ref{IntOpen4Genus1}) in terms of these elliptic MZVs to Chapter \ref{ChapterEllMZV}. We conclude this section by mentioning that the recent paper \cite{BMRS} considers insertions on both boundary components of the cylinder in figure 1.4 of the introduction. This involves the so-called \emph{twisted elliptic MZVs}, which can be seen as iterated integrals of Eisenstein series for congruence subgroups or as special values at torsion points of Levin's elliptic polylogs, and whose Fourier expansion involves special values of multiple polylogs at roots of unity. Nevertheless, in \cite{BMRS} it is also pointed out the remarkable fact that the final expression for the amplitude does not share these new features, and can again be written in terms of untwisted elliptic MZVs only.

\subsection{The closed string-case}\label{SectionClosed}

Studying one-loop superstring amplitudes of closed strings constitutes the main motivation for the present work, and therefore we will write down the Feynman integral very precisely. We will only focus on the first physically meaningful case of four strings, and we will try to keep our notation as close as possible to that of the foundational papers \cite{GV2000} and \cite{GRV}. The amplitude is given by
\[
\textbf{A}_{1,4}(s,t,u)=I_{1,4}(s,t,u) \textbf{R}(s,t,u),
\]
where $\textbf{R}$ encodes the kinematic part\footnote{We have suppressed from the notation the dependence of $\textbf{R}$ on the polarization tensors. See \cite{GRV} for details.}, and we want to write $\textbf{A}$ as a function of the non-independent Mandelstam variables $s,t,u$, related by the condition $s+t+u=0$. The Feynman integral is given, as in the open case (\ref{IntOpen4Genus1}), by
\begin{multline}\label{4ClosedGenus1}
I_{1,4}(s,t,u)=\int_{\mathfrak{F}}\int_{(\mathcal{E}_{\tau})^3} \exp\big(s(G_1(\xi-\nu,\tau)+G_1(\omega,\tau))+t(G_1(\nu-\omega,\tau)+G_1(\xi,\tau))\\
+u(G_1(\xi-\omega,\tau)+G_1(\nu,\tau))\big) \frac{d\xi d\nu d\omega}{\tau_2^3}\frac{d\tau_1d\tau_2}{\tau_2^2},
\end{multline}
where:
\begin{itemize}
\item $\tau=\tau_1+i\tau_2\in\mathfrak{F}$, and $\mathfrak{F}$ is the fundamental domain $\mbox{PSL}_2(\mathbb{Z})\setminus\mathbb{H}$,
\item $\xi=\xi_1+i\xi_2,\nu=\nu_1+i\nu_2,\omega=\omega_1+i\omega_2$ represent three of our four strings moving on the torus (without loss of generality, we can assume that the fourth is fixed at the origin),
\item $G_1$ is the propagator defined in (\ref{PropagatorGenus1})
\end{itemize}

To perform the integration, one can expand the exponential as a power series in $s,t,u$, and get that all coefficients are given as linear combinations of integrals over\footnote{We do not discuss this second integration here. We just want to mention that this integral leads to poles and branch cuts at $s=t=0$, and a method to extract the analytic part can be found in \cite{GRV}.}~$\mathfrak{F}$ of functions $D_{\underline{l}}(\tau)$, defined for $\underline{l}=(l_1,\ldots,l_6)\in\mathbb{Z}_{\geq 0}^6$ as
\begin{equation}\label{DefDl}
D_{\underline{l}}(\tau)=\int_{(\mathcal{E}_\tau)^3} G_1(\xi-\nu,\tau)^{l_1}G_1(\omega,\tau)^{l_2}G_1(\nu-\omega,\tau)^{l_3}G_1(\xi,\tau)^{l_4}G_1(\xi-\omega,\tau)^{l_5}G_1(\nu,\tau)^{l_6} \frac{d\xi d\nu d\omega}{\tau_2^3}.
\end{equation}
The properties of $G_1$ imply that this integral is well defined, and that~$D_{\underline{l}}(\tau)$ is a modular function. We call $l_1+\cdots+l_6$ the \emph{weight} of~$D_{\underline{l}}(\tau)$. We will see in the next chapters that these functions are conjectured to be non-trivial examples of \emph{single-valued elliptic MZVs}. It is very important to mention that, just by performing the integration with the Green function~$G_1$ in the form given by Proposition \ref{propPropEis}, one can immediately deduce the following series representation of $D_{\underline{l}}$:
\begin{equation}\label{sum}
D_{\underline{l}}(\tau)=\Big(\frac{\tau_2}{4\pi}\Big)^{l_1+\cdots +l_6}\sum \prod_{j=1}^6\prod_{i=1}^{l_i}\mid\omega^{(j)}_i\mid^{-2},
\end{equation}
where the sum runs over the lattice points $\omega^{(j)}_i:=m^{(j)}_i\tau+n^{(j)}_i\in\Lambda^*_\tau:=\Lambda_\tau\setminus\{0\}$ such that
\[
\omega^{(1)}_1+\cdots+\omega^{(1)}_{l_1}+\omega^{(4)}_1+\cdots+\omega^{(4)}_{l_4}+
\omega^{(5)}_1+\cdots+\omega^{(5)}_{l_5}=0,
\]
\[
\omega^{(3)}_1+\cdots+\omega^{(3)}_{l_3}+\omega^{(6)}_1+\cdots+\omega^{(6)}_{l_6}=
\omega^{(1)}_1+\cdots+\omega^{(1)}_{l_1},
\]
\[
\omega^{(3)}_1+\cdots+\omega^{(3)}_{l_3}+\omega^{(5)}_1+\cdots+\omega^{(5)}_{l_5}=
\omega^{(2)}_1+\cdots+\omega^{(2)}_{l_2}.
\]
Thanks to this sum representation, it is easy to see that the functions~$D_{\underline{l}}$'s which are irreducible, i.e. which cannot be written as products of other two~$D_{\underline{l}}$'s, are the ones associated with the diagrams appearing in the following figure, where every point represents a string, and every edge is labelled by the number of propagators~$l_{i,j}$ joining two strings, or in other words by the exponent of $G_1(\xi_i-\xi_j,\tau)$ in the integral (\ref{DefDl}).

\vspace{1 cm}
\begin{minipage}{0.26\columnwidth}
\raggedleft \fcolorbox{black}{white}
{\begin{tikzpicture}[scale=1]
\draw [fill] (1,0) circle [radius=0.1];
\draw [fill] (2,1) circle [radius=0.1];
\draw [fill] (2,-1) circle [radius=0.1];
\draw [fill] (3,0) circle [radius=0.1];
\draw [ultra thick][blue](1,0) --(3,0);
\node [above] at (2,0) {$l_1$};
\node [red, right] at (3.3,0) {(a)};
\draw [fill] (6,0) circle [radius=0.1];
\draw [fill] (7,1) circle [radius=0.1];
\draw [fill] (7,-1) circle [radius=0.1];
\draw [fill] (8,0) circle [radius=0.1];
\draw [ultra thick][blue](6,0) --(8,0);
\draw [ultra thick][blue](6,0) --(7,1);
\draw [ultra thick][blue](7,1) --(8,0);
\node [below] at (7,0) {$l_3$};
\node [above] at (6.4,0.5) {$l_1$};
\node [above] at (7.6,0.5) {$l_2$};
\node [red, right] at (8.3,0) {(b)};
\draw [fill] (11,0) circle [radius=0.1];
\draw [fill] (12,1) circle [radius=0.1];
\draw [fill] (12,-1) circle [radius=0.1];
\draw [fill] (13,0) circle [radius=0.1];
\draw [blue, ultra thick] (12,0) circle [radius=1];
\node at (11.1,0.9) {$l_1$};
\node at (12.9,0.9) {$l_2$};
\node at (12.9,-0.9) {$l_3$};
\node at (11.1,-0.9) {$l_4$};
\node [red, right] at (13.3,0) {(c)};
\draw [fill] (3.5,-3) circle [radius=0.1];
\draw [fill] (4.5,-2) circle [radius=0.1];
\draw [fill] (4.5,-4) circle [radius=0.1];
\draw [fill] (5.5,-3) circle [radius=0.1];
\draw [blue, ultra thick] (4.5,-3) circle [radius=1];
\draw [ultra thick][blue](3.5,-3) --(5.5,-3);
\node at (3.6,-2.1) {$l_1$};
\node at (5.4,-2.1) {$l_2$};
\node at (5.4,-3.9) {$l_3$};
\node at (3.6,-3.9) {$l_4$};
\node [above] at (4.5,-3) {$l_5$};
\node [red, right] at (5.8,-3) {(d)};
\draw [fill] (8.634,-3.5) circle [radius=0.1];
\draw [fill] (9.5,-2) circle [radius=0.1];
\draw [fill] (10.366,-3.5) circle [radius=0.1];
\draw [fill] (9.5,-3) circle [radius=0.1];
\draw [blue, ultra thick] (9.5,-3) circle [radius=1];
\draw [ultra thick][blue](9.5,-2) --(9.5,-3);
\draw [ultra thick][blue](8.634,-3.5) --(9.5,-3);
\draw [ultra thick][blue](10.366,-3.5) --(9.5,-3);
\node at (8.6,-2.1) {$l_1$};
\node at (10.4,-2.1) {$l_2$};
\node [above] at (9.5,-4) {$l_3$};
\node at (9,-3) {$l_5$};
\node [right] at (9.5,-2.5) {$l_6$};
\node at (10,-3) {$l_4$};
\node [red, right] at (10.8,-3) {(e)};
\node[align=left, below] at (3,-4.5)%
{The 5 irreducible diagrams.};
\end{tikzpicture}}
\end{minipage}
\vspace{1 cm}

No lines between pairs of points amount to say that there are zero propagators. To give an example, in the case of diagram \textcolor{red}{(a)} the associated function can be written as $D_{(l_1,0,0,0,0,0)}$, and we will speak of two-point case because only two points are connected by a propagator. Note that $D_{(l_1,0,0,0,0,0)}$ does not depend on the position of $l_1$ in the vector $\underline{l}$, and therefore we can just write~$D_{l_1}$. The same is true also for diagrams \textcolor{red}{(b)} and \textcolor{red}{(c)}, where we speak respectively of $D_{l_1,l_2,l_3}$ and $D_{l_1,l_2,l_3,l_4}$, and the order of the $l_i$'s does not play any r\^{o}le. This is not anymore true when we consider diagrams \textcolor{red}{(d)} and \textcolor{red}{(e)}, in which the order of the indeces~$l_i$'s influences the result of the integral. For this reason, in the next section we want to introduce another way of organizing the integrals (\ref{DefDl}) using graphs, following \cite{DGGV}, that will free us from any ambiguity, and leads to a more general interesting class of modular functions. We conclude this section by giving some examples. Using (\ref{sum}), we get
\begin{eqnarray*}
D_2(\tau)&=&\Big(\frac{\tau_2}{4\pi}\Big)^2\sum_{\omega_1,\omega_2\in\Lambda_\tau^*}\frac{\delta(\omega_1+\omega_2)}{|\omega_1|^2|\omega_2|^2},\\
D_{1,1,1}(\tau)&=&\Big(\frac{\tau_2}{4\pi}\Big)^3\sum_{\omega_1,\omega_2,\omega_3\in\Lambda_\tau^*}\frac{\delta(\omega_1+\omega_2)\delta(\omega_1+\omega_3)}{|\omega_1|^2|\omega_2|^2|\omega_3|^2},\\
D_{1,1,1,1}(\tau)&=&\Big(\frac{\tau_2}{4\pi}\Big)^4\sum_{\omega_1,\omega_2,\omega_3,\omega_4\in\Lambda_\tau^*}\frac{\delta(\omega_1+\omega_2)\delta(\omega_1+\omega_3)\delta(\omega_3+\omega_4)}{|\omega_1|^2|\omega_2|^2|\omega_3|^2|\omega_4|^2},\\
D_3(\tau)&=&\Big(\frac{\tau_2}{4\pi}\Big)^3\sum_{\omega_1,\omega_2,\omega_3\in\Lambda_\tau^*}\frac{\delta(\omega_1+\omega_2+\omega_3)}{|\omega_1|^2|\omega_2|^2|\omega_3|^2},
\end{eqnarray*}
where $\delta(x)=1$ if $x=0$, and $\equiv 0$ otherwise. It is trivial to relate the first three examples with the non-holomorphic Eisenstein series $E(a,\tau)$ defined in Remark \ref{ossEis}. More precisely, one gets
\begin{equation}\label{D2E}
D_2(\tau)=4^{-2}E(2,\tau),
\end{equation}
\begin{equation}\label{D111E}
D_{1,1,1}(\tau)=4^{-3}E(3,\tau),
\end{equation}
\begin{equation}\label{D1111E}
D_{1,1,1,1}(\tau)=4^{-4}E(4,\tau).
\end{equation}
It was shown by Zagier (unpublished, another proof was given later in \cite{DGV2015}) that
\begin{equation}\label{D3E}
D_3(\tau)=4^{-3}(E(3,\tau)+\zeta(3)),
\end{equation}
but the proof is not straightforward. Note that, by a standard computation, we know that the Fourier expansion with respect to the variable $\tau_1$ of $E(n,\tau)$ reads (for $n\in\mathbb{Z}_{\geq 2}$ and $y:=\pi\tau_2$)
\begin{multline}
E(n,\tau)=\Big[(-1)^{n-1}\frac{B_{2n}}{(2n)!}(4y)^n+\frac{4(2n-3)!}{(n-2)!(n-1)!}\zeta(2n-1)(4y)^{1-n}\\
+\frac{2}{(n-1)!}\sum_{N\geq 1}N^{n-1}\sigma_{1-2n}(N)(q^N+\overline{q}^N)\sum_{m=0}^{n-1}\frac{(n+m-1)!}{m!(n-m-1)!}(4Ny)^{-m}\Big].
\end{multline}
It is remarkable to point out that these examples are the only~$D_{\underline{l}}$'s for which we can compute the Fourier expansion using an explicit formula (we will see in our main result of Chapter \ref{ChapterClosedStrings} what is the general form of these Fourier expansions). For all other~$D_{\underline{l}}$'s, it is very involved even to compute the \emph{zero mode} $\int_0^1D_{\underline{l}}(\tau_1+i\tau_2)d\tau_1$, and we will see that, in order to do it, it is more convenient to consider the expansion (\ref{FourierProp2}) of the propagator.

\subsection{Modular graph functions}\label{SectionModGraph}

Let us consider an undirected graph $\Gamma=(V,E)$ with no self-edges, where we allow for multiple edges connecting the same pair of vertices. If we choose a labelling $\xi_1,\ldots \xi_N$ of the $N$ vertices, then for $i<j$ we have $l_{i,j}$ edges between $\xi_i$ and $\xi_j$, oriented (this orientation is induced by the labelling) as going from $\xi_i$ to $\xi_j$, with the total number of edges given by the \emph{weight} of the graph
\[
l:=\sum_{1\leq i<j\leq N}l_{i,j},
\]
and we construct the \emph{incidence matrix} 
\[
(\Gamma_{i,\alpha})_{\substack{1\leq i\leq N\\ 1\leq\alpha\leq l}}
\]
of $\Gamma$ by choosing any labelling $e_\alpha$ on the set of edges, and by setting $\Gamma_{i,\alpha}=0$ if $e_\alpha$ does not touch $\xi_i$, $\Gamma_{i,\alpha}=1$ if $e_\alpha$ is oriented away from $\xi_i$ and $\Gamma_{i,\alpha}=-1$ if $e_\alpha$ is oriented towards $\xi_i$.
\begin{defi}
Let $\Gamma$ be a graph as above. For $\tau\in\mathbb{H}$, we define its \emph{modular graph function} as
\begin{equation}
D_\Gamma(\tau)=\Big(\frac{\tau_2}{\pi}\Big)^l\sum_{\omega_1,\ldots ,\omega_l\in\Lambda_\tau^*}\prod_{\alpha=1}^l|\omega_\alpha|^{-2}\prod_{i=1}^N\delta\Big(\sum_{\beta=1}^l\Gamma_{i,\beta}\omega_\beta\Big).
\end{equation}
\end{defi}
One can show that this definition does not depend on the labelling. It is an easy exercise to show that
\[
D_\Gamma(\tau)=\int_{(\mathcal{E}_\tau)^{N-1}}\prod_{1\leq i<j\leq N}G_1(\xi_i-\xi_j,\tau)^{l_{i,j}}\frac{d\xi_1\cdots d\xi_{N-1}}{\tau_2^{N-1}},
\]
where we have fixed $\xi_N\equiv 0$. These functions are modular invariant, and generalize the functions $D_{\underline{l}}$'s. For instance, the function $D_3$ comes from the graph 

\begin{center}
\tikzpicture[scale=1.2]
\scope[xshift=-5cm,yshift=-0.4cm]
\draw[thick]   (0,0) node{$\bullet$} ..controls (1,0.7) .. (2,0) node{$\bullet$} ;
\draw[thick]   (0,0) node{$\bullet$} ..controls (1,-0.1) .. (2,0) node{$\bullet$} ;
\draw[thick]   (0,0) node{$\bullet$} ..controls (1,-0.7) .. (2,0) node{$\bullet$} ;
\endscope
\endtikzpicture
\end{center}

For all graphs with $N$ vertices along one cycle, as in the figure

\begin{center}
\tikzpicture[scale=0.8]
\scope[xshift=-5cm,yshift=-0.4cm]
\node at (-0.5,0.2) {$\xi_1$};
\draw[thick]   (0,0) node{$\bullet$} ..controls (0.5,0.865) .. (1,1.73) node{$\bullet$} ;
\node at (0.5,1.73) {$\xi_2$};
\draw[thick]   (1,1.73) node{$\bullet$}  ..controls (2,1.73) .. (3,1.73) node{$\bullet$} ;
\node at (3.5,1.73) {$\xi_3$};
\draw[thick] (3,1.73) node{$\bullet$}  ..controls (3.5,0.865) ..  (4,0) node{$\bullet$} ;
\node at (4.5,0.2) {$\xi_4$};
\draw[thick, dashed]   (4,0) node{$\bullet$} ..controls (3,-1.73) .. (1,-1.73) node{$\bullet$} ;
\node at (0.5,-1.73) {$\xi_N$};
\draw[thick]   (1,-1.73) node{$\bullet$} ..controls (0.5,-0.865) .. (0,0) node{$\bullet$} ;
\endscope
\endtikzpicture
\end{center}

we have by definition that
\[
D_\Gamma(\tau)=\Big(\frac{\tau_2}{\pi}\Big)^N \sum_{\omega\in\Lambda_\tau^*}\frac{1}{|\omega_\alpha|^{2N}},
\]
and therefore we deduce that $D_\Gamma(\tau)=E(N,\tau)$. This generalizes equations (\ref{D2E}), (\ref{D111E}) and (\ref{D1111E}). 
\begin{oss}
It is easy to see that, if one vertex of a graph $\Gamma$ is reached by only one edge, then $D_\Gamma(\tau)=0$. Moreover, more generally $D_\Gamma(\tau)=0$ whenever there exists any edge whose removal would disconnect the graph. Here we have pictures of these situations:

\begin{center}
\tikzpicture[scale=1]
\scope[xshift=-5cm,yshift=-0.4cm]
\draw[thick]   (0,0) node{$\bullet$} ..controls (1,0.7) .. (2,0) node{$\bullet$} ;
\draw[thick]   (0,0) node{$\bullet$} ..controls (1,-0.1) .. (2,0) node{$\bullet$} ;
\draw[thick]   (0,0) node{$\bullet$} ..controls (1,-0.7) .. (2,0) node{$\bullet$} ;
\draw[thick]   (2,0) node{$\bullet$} ..controls (3,-0.1) .. (4,0) node{$\bullet$} ;
\draw[thick]   (4,0) node{$\bullet$} ..controls (5,0.7) .. (6,0) node{$\bullet$} ;
\draw[thick]   (4,0) node{$\bullet$} ..controls (5,-0.1) .. (6,0) node{$\bullet$} ;
\draw[thick]   (4,0) node{$\bullet$} ..controls (5,-0.7) .. (6,0) node{$\bullet$} ;
\draw[thick]   (8,0) node{$\bullet$} ..controls (9,0.7) .. (10,0) node{$\bullet$} ;
\draw[thick]   (8,0) node{$\bullet$} ..controls (9,-0.1) .. (10,0) node{$\bullet$} ;
\draw[thick]   (8,0) node{$\bullet$} ..controls (9,-0.7) .. (10,0) node{$\bullet$} ;
\draw[thick]   (10,0) node{$\bullet$} ..controls (11,-0.1) .. (12,0) node{$\bullet$} ;
\endscope
\endtikzpicture
\end{center}

Therefore the relevant graphs are the ones where all edges are part of a cycle, and we define the \emph{depth} of a graph $\Gamma$ as the number of its cycles. We will see later that this seems to be related with the depth of multiple elliptic polylogarithms and elliptic MZVs. Moreover, we call \emph{reducible} a graph $\Gamma$ such that the removal of a vertex would disconnect the graph, as in the figure below.

\begin{center}
\tikzpicture[scale=1]
\scope[xshift=-5cm,yshift=-0.4cm]
\draw[thick]   (0,0) node{$\bullet$} ..controls (1,0.7) .. (2,0) node{$\bullet$} ;
\draw[thick]   (0,0) node{$\bullet$} ..controls (1,-0.1) .. (2,0) node{$\bullet$} ;
\draw[thick]   (0,0) node{$\bullet$} ..controls (1,-0.7) .. (2,0) node{$\bullet$} ;
\draw[thick]   (2,0) node{$\bullet$} ..controls (3,0.7) .. (4,0) node{$\bullet$} ;
\draw[thick]   (2,0) node{$\bullet$} ..controls (3,-0.1) .. (4,0) node{$\bullet$} ;
\draw[thick]   (2,0) node{$\bullet$} ..controls (3,-0.7) .. (4,0) node{$\bullet$} ;
\endscope
\endtikzpicture
\end{center}

When a graph is reducible, it is easy to see that the associated modular graph functions factors into the product of the irreducible components. For instance, in the case of the figure above the modular graph function associated is $D_3^2$. This is consistent with the fact that we called irreducible the diagrams \textcolor{red}{(a)}-\textcolor{red}{(e)}, and in fact those diagrams encode all the irreducible graphs with four vertices.
\end{oss}
We conclude this chapter by mentioning the main results on modular graph functions known before our work. First of all, it was proven in the paper \cite{GRV} that, for all graphs $\Gamma$ with only two vertices and $l$ edges between them, the function $D_\Gamma(\tau)$, which in the notation of the previous section (which is taken from \cite{GRV}) corresponds to $D_l(\tau)$, is given for $y:=\pi\tau_2$ by
\begin{prop}\label{TeoGRV}
\begin{equation}
D_l(\tau)=d_{l}(y)+O(e^{-y})
\end{equation}
for $y\rightarrow\infty$, with 
\begin{eqnarray*}
d_{l}(y)&=&\Big(\frac{y}{12}\Big)^{l} {}_2F_1(1,-l,3/2;3/2)\\
&+&\frac{2}{4^l}\sum_{\substack{a+b+c+m=l\\m\geq 2}}\frac{l!(2a+b)!}{a!b!c!m!}\frac{(-1)^b}{6^c}S(m,2a+b+1)(2y)^{c-a-1},
\end{eqnarray*}
where ${}_2F_1$ is the hypergeometric function defined in (\ref{hypergeometric}) and for $m\geq 2$, $n\geq 0$
\begin{equation}
S(m,n):=\sum_{k_1,\ldots ,k_m\in\mathbb{Z}^*}\frac{\delta(k_1+\cdots +k_m)}{|k_1\cdots k_m|(|k_1|+\cdots +|k_m|)^n}.
\end{equation}
\end{prop}
We will see the proof of this proposition in the next chapter. Moreover, in the appendix \cite{ZagierApp} of \cite{GRV}, Zagier proved (an explicit version, that we will see in the next chapter, of) the following:
\begin{prop}
For all $m\geq 2$ and $n\geq 0$ $S(m,n)\in\mathcal{A}$, where we recall that $\mathcal{A}$ denotes the algebra of MZVs.
\end{prop}
Thanks to this result, we have a formula to compute $d_{l}(y)$, that is the zero mode of $D_l(\tau)$, and will be sometimes called the \emph{Laurent polynomial part} of $D_l(\tau)$, in terms of MZVs. For instance, we have (noting a few typos in the data given in \cite{GRV})
\begin{equation}
4^2d_2(y)=\frac{1}{45}y^2+\frac{\zeta(3)}{y},
\end{equation}
\begin{equation}
4^3d_3(y)=\frac{2}{945}y^3+\zeta(3)+\frac{3}{4}\frac{\zeta(5)}{y^2},
\end{equation}
\begin{equation}
4^4d_4(y)=\frac{1}{945}y^4+\frac{2}{3}\zeta(3)y+\frac{10\zeta(5)}{y}-\frac{3\zeta(3)^2}{y^2}+\frac{9}{4}\frac{\zeta(7)}{y^3},
\end{equation}
\begin{multline}
4^5d_5(y)=\frac{4}{18711}y^5+\frac{10}{27}\zeta(3)y^2+\frac{95}{6}\zeta(5)+\frac{10\zeta(3)^2}{y}+\frac{105}{4}\frac{\zeta(7)}{y^2}\\
-\frac{45}{2}\frac{\zeta(3)\zeta(5)}{y^3}+\frac{225}{16}\frac{\zeta(9)}{y^4},
\end{multline}
\begin{multline}
4^6d_6(y)=\frac{53}{729729}y^6+\frac{5}{27}\zeta(3)y^3+\frac{140}{9}\zeta(5)y+25\zeta(3)^2+\frac{1005}{4}\frac{\zeta(7)}{y}-\frac{135\zeta(3)\zeta(5)}{y^2}\\
+\frac{90\zeta(3)^3+405\zeta(9)}{2y^3}-\frac{675\zeta(5)^2+1350\zeta(3)\zeta(7)}{8y^4}+\frac{4725}{32}\frac{\zeta(11)}{y^5}.
\end{multline}
As we can see from the first examples, it seems that actually the coefficients could be expressed in terms of odd Riemann zetas only, and in fact Zagier recently managed, finding a way to write explicitly the coefficients of $d_{l}(y)$ in terms of the coefficients of the four-point amplitude in genus zero, to prove the following \cite{ZagierStrings}:
\begin{teo}[Zagier]\label{zagier}
For all integers $l$ the Laurent polynomial $d_l(y)$ has coefficients belonging to the polynomial ring generated over $\mathbb{Q}$ by the odd zeta values~$\zeta(2n+1)$.
\end{teo}
Unfortunately, things get much more cumbersome beyond the two-point case, and therefore in \cite{GRV} it was possible to give only very few Laurent polynomials appearing as zero modes of modular graph functions\footnote{We will see in the next chapter a proof that the zero modes are always Laurent polynomials in $\pi\tau_2$.}, because their computation could be done only by hand. In particular, beyond the two-point case the computation of these Laurent polynomials in weight $>5$ seemed to be hopeless. One of the main results of Chapter~\ref{ChapterClosedStrings} is the description of an algorithmic procedure that gives access to many new highly non-trivial zero modes. A great interest in computing zero modes is given by the fact that D'Hoker, Green and Vanhove noticed that it was possible to predict differential equations satisfied by modular graph functions, and algebraic relations among them, just by looking at these Laurent polynomials. Here we report a general result that they obtained for modular graph functions of depth two:
\begin{teo}[D'Hoker, Green, Vanhove]\label{theorem2cycles}
Let $\Gamma$ be an (irreducible) graph of depth two and weight~$l=a+b+c$, as depicted below.

\begin{center}
\tikzpicture[scale=0.8]
\scope[xshift=-5cm,yshift=-0.4cm]
\node at (2,1.3) {$\underbrace{}_{a-2 \mbox{ edges}}$};
\node at (2,-0.43) {$\underbrace{}_{b-2 \mbox{ edges}}$};
\node at (2,-2.16) {$\underbrace{}_{c-2 \mbox{ edges}}$};
\draw[thick]   (0,0) node{$\bullet$} ..controls (0.5,0.865) .. (1,1.73) node{$\bullet$} ;
\draw[thick, dashed]   (1,1.73) node{$\bullet$}  ..controls (2,1.73) .. (3,1.73) node{$\bullet$} ;
\draw[thick] (3,1.73) node{$\bullet$}  ..controls (3.5,0.865) ..  (4,0) node{$\bullet$} ;
\draw[thick]   (4,0) node{$\bullet$}  ..controls (3.5,-0.865) .. (3,-1.73) node{$\bullet$} ;
\draw[thick, dashed]   (3,-1.73) node{$\bullet$} ..controls (2,-1.73) .. (1,-1.73) node{$\bullet$} ;
\draw[thick]   (1,-1.73) node{$\bullet$} ..controls (0.5,-0.865) .. (0,0) node{$\bullet$} ;
\draw[thick]   (0,0) node{$\bullet$} ..controls (0.5,0) .. (1,0) node{$\bullet$} ;
\draw[thick,dashed]   (1,0) node{$\bullet$} ..controls (2,0) .. (3,0) node{$\bullet$} ;
\draw[thick]   (3,0) node{$\bullet$} ..controls (3.5,0) .. (4,0) node{$\bullet$} ;
\endscope
\endtikzpicture
\end{center}

In this case
\[
D_{\Gamma}(\tau)=\Big(\frac{\tau_2}{\pi}\Big)^l\sum_{\omega_1,\omega_2,\omega_3\in\Lambda_\tau^*}\frac{\delta(\omega_1+\omega_2+\omega_3)}{|\omega_1|^{2a}|\omega_2|^{2b}|\omega_3|^{2c}}.
\]
Then there exist a basis of elements $\mathfrak{C}_{w,s,p}$ of the rational vector space generated by these modular graph functions, where $s=w,w-1,\ldots , w-\lfloor (w-1)/2\rfloor$ and $p=0,1,\ldots ,\lfloor (s-1)/3\rfloor$, such that
\begin{equation}
(\Delta-s(s-1))\mathfrak{C}_{w,s,p}=P_{w,s,p}(E(s^\prime,\tau),\zeta(s^\prime)),
\end{equation}
where $\Delta=4\tau_2\partial_\tau\partial_{\overline{\tau}}$ is the Laplace-Beltrami operator on $\mathbb{H}$,~$P_{w,s,p}$ is a polynomial of degree $\leq 2$ and weight\footnote{By this we mean that each monomial has weight $w$ after summing the weight of the modular graph functions and zeta values involved.}~$w$,~$2\leq s^\prime \leq w$ in non holomorphic Eisenstein series $E(s^\prime,\tau)$ and odd zeta values $\zeta^{\rm sv}(s^\prime)$, with $2\leq s^\prime \leq w$.
\end{teo}

The simplest instances of this theorem are given by \cite{DGV2015}
\begin{equation*}
\Delta D_3(\tau)=\frac{6}{4^3}E(3,\tau)
\end{equation*}
and
\begin{equation*}
(\Delta -2) D_{2,1,1}(\tau)=\frac{9E(4,\tau)-E(2,\tau)^2}{4^4}.
\end{equation*}
Moreover, further conjectured differential equations, such as
\begin{equation*}
(\Delta -2)\Big(5D_4-\frac{15E(2,\tau)^2-18E(4,\tau)}{4^4}\Big)=-\frac{120}{4^4}E(2,\tau)^2,
\end{equation*}
have been proven in \cite{Basu1} (see also \cite{AxelValentin}), and some algebraic relations conjectured using Theorem \ref{theorem2cycles} and the Laurent polynomial behaviour at the cusp, such as
\begin{equation*}
D_{4}=24D_{2,1,1}+\frac{3E(2,\tau)^2-18E(4,\tau)}{4^4},
\end{equation*}
have been proven recently in \cite{DGV2} or \cite{DHokerKaidi}.

Finally, we want to remark that modular graph functions do not suffice to describe the analogues of the functions $D_{\underline{l}}$'s when the number of strings is $\geq 5$. As we have already mentioned, in that case one may have to take into account in the integrand some factor depending on the first and second derivatives of the propagator, and this led to the definition in \cite{DG} of modular graph forms. We do not want to discuss their definition here; they are functions of mixed (holomorphic and anti-holomorphic) modular weight (see Chapter \ref{ChapterConclusions}), generalizing the functions $e_{a,b}(0,\tau)$ defined in (\ref{svEllPol}) when $a$ and $b$ need not to coincide, and they should allow to describe all functions involved in genus one closed superstring amplitudes.

%% file: Chapter4.tex

\chapter{Modular graph functions and single-valued multiple zeta values} 

\label{ChapterClosedStrings} 

This chapter is based on the author's paper \cite{Zerb15}. We want to mention that throughout this chapter we will write $\underline{z}$ to denote tuples $z_1,\ldots ,z_n$, deviating from the convention of writing $\textbf{z}$ that we have adopted for most of the present work, in order to improve the readability and to follow faithfully the notation employed in \cite{Zerb15}. Moreover, since we will be interested in the genus one case only, we will drop the subscript~1 from the propagator $G_1(\xi,\tau)$ defined in the previous chapter and write just $G(\xi,\tau)$. Besides this, the notation for modular graph functions is the same as in Section \ref{SectionModGraph}. 

After a brief introduction of conical sums, we will see a detailed proof of Theorem \ref{main} for modular graph functions with two and three vertices, in order to skip some details in the similar but more cumbersome proof of the general case, that will be given in the last section. Most of the consequences, including explicit computations of Laurent polynomials involving non-trivial single-valued MZVs, are contained in the three-point section. 

\section{Conical sums}

Conical sums, also called conical zeta-values, constitute a natural generalization of multiple zeta values that has been scarcely considered in the mathematics literature. Special cases with trivial cones have been considered already in \cite{ZagierMZV}. The main references are the paper of Terasoma \cite{TerasomaConical}, and a recent paper by Guo, Paycha and Zhang \cite{Paycha}. This is one of the first times\footnote{See also \cite{AnzaiSumino} for similar computations in quantum field theory.} that conical sums have been studied systematically in connection with physics problems.

\subsection{Definition and first properties} 

A first useful remark, in order to introduce conical sums as a generalization of multiple zeta values, is that
\[
\zeta(k_1,\ldots,k_r)=\sum_{0<v_1<\cdots<v_r}\frac{1}{v_1^{k_1}\cdots v_r^{k_r}}=\sum_{\underline{x}\in\mathbb{N}^r}\frac{1}{x_1^{k_1}(x_1+x_2)^{k_2}\cdots(x_1+\cdots+ x_r)^{k_r}},
\]
where $r,k_1,\ldots,k_r\in\mathbb{N}$ and $k_r\geq 2$.
\begin{defi}
Let $v_1,\ldots,v_m\in\mathbb{Q}^n$, and let $\mathbb{R}^{+}$ denote the non-negative real numbers. Then we say that $C:=\mathbb{R}^{+}v_1+\cdots+\mathbb{R}^{+}v_m$ is a \emph{rational cone}, and we denote by $C^{0}$ its interior.
\end{defi}
\begin{defi}
Let $C$ be a rational cone in $\mathbb{R}^n$, let $l_1,\ldots,l_r$ be (possibly not distinct) linear forms\footnote{One can easily extend all what follows to the case of affine forms, in order to cover a broader range of applications \cite{AnzaiSumino}.} with integer coefficients, and suppose that $l_i(\underline{x})>0$ for all $\underline{x}\in C^{0}$ and all $i$. If $l_i(\underline{x})=\sum_{j=1}^n a_{i,j}x_j$, consider the matrix $A:=(a_{i,j})$. Then for $\chi$ a finite order character of~$\mathbb{Z}^n$ we define the following series:
\begin{equation}\label{eq}
\zeta(C,A,\chi):=\sum_{\underline{x}\in C^{0}\cap\mathbb{Z}^n}\frac{\chi(\underline{x})}{l_1(\underline{x})\cdots l_r(\underline{x})}.
\end{equation}
If this series converges to a number, we call it \emph{conical sum}, and we define $\mathcal{C}$ to be the vector space spanned over the union of all cyclotomic fields $\mathbb{Q}[\zeta_N]$ by all conical sums, where $\zeta_N=\exp(2\pi i/N)$. Note that $\zeta(C,A,\chi)$ does not depend on the order of rows and columns of $A$.
\end{defi}
One can immediately see that $\mathcal{C}$ is an algebra. Setting the cone $C$ equal to the first quadrant of~$\mathbb{R}^2$, $l_1(\underline{x})=x_1$, $l_2(\underline{x})=l_3(\underline{x})=x_1+x_2$, and $\chi$ identically equal to 1, we get $\zeta(C,A,\chi)=\zeta(1,2)$. In the same way one gets all MZVs.

Terasoma proved in \cite{TerasomaConical} that any conical sum can be reduced to a linear combination (over a cyclotomic number field $\mathbb{Q}[\zeta_N]$) of sums of the canonical form 
\begin{equation}\label{standardcone}
\zeta(A,\chi):=\sum_{\underline{x}\in\mathbb{N}^n}\frac{\chi(\underline{x})}{l_1(\underline{x})\cdots l_r(\underline{x})},
\end{equation}
where the coefficients $a_{i,j}\in\mathbb{Z}_{\geq 0}$. Sums of this form admit the integral representation
\begin{equation}\label{int}
\zeta(A,\chi)=\int_{[0,1]^r}\frac{\chi(\underline{u})\,y_1^{l_1-1}\cdots y_r^{l_r-1}dy_1\cdots dy_r}{\prod_{j=1}^n(1-\chi(\underline{e}_j)\,y_1^{a_{1,j}}\cdots y_r^{a_{r,j}})},
\end{equation}
where $\underline{e}_j$ is the canonical $j$-th element of the basis of $\mathbb{Z}^n$ and $\underline{u}:=\sum_{j=1}^n \underline{e}_j=(1,\ldots,1)^T$. To see this, first write
\begin{equation}
\frac{\chi(\underline{x})}{l_1(\underline{x})\cdots l_r(\underline{x})}=\int_{[0,1]^r}\chi(\underline{x})y_1^{l_1(\underline{x})}\cdots y_r^{l_r(\underline{x})}\frac{dy_1\cdots dy_r}{y_1\cdots y_r}.
\end{equation}
Then one can re-arrange the exponentials in the integrand as
\[
\prod_{j=1}^n\Big(y_1^{a_{1,j}}\cdots y_r^{a_{r,j}}\Big)^{x_j}.
\]
Exchanging integration and summation leads to the expression (\ref{int}).

This means, first of all, that all conical sums are periods. In order to describe these periods, we need the following
\begin{defi}
Let $N\in\mathbb{N}$, and let $\zeta_N=\exp(2\pi i/N)$. We call \emph{N-th cyclotomic multiple zeta values} the special values of the multiple polylogarithms $\mbox{Li}_{\textbf{k}}(\zeta_N^{a_1},\ldots ,\zeta_N^{a_r})$, for $a_1,\ldots ,a_r\in\mathbb{Z}$, and denote by $\mathcal{A}[N]$ the algebra that they generate over $\mathbb{Q}[\zeta_N]$.
\end{defi}
In particular, the algebra $\mathcal{A}$ of MZVs coincides with $\mathcal{A}[1]$, and the simplest instance of a cyclotomic MZV which is (conjecturally) not contained in $\mathcal{A}$ is given by
\[
\sum_{v\in\mathbb{N}}\frac{(-1)^v}{v}=-\log(2).
\]
Obviously all cyclotomic multiple zeta values can be written as conical sums, because we are allowing for a character $\chi$. The main result of \cite{TerasomaConical} is that the converse holds.
\begin{teo}[Terasoma]\label{ThTera}
\[
\mathcal{C}=\bigcup_{N\in\mathbb{N}}\mathcal{A}[N].
\]
\end{teo}
The proof of this theorem is long and complicated, and we refer the reader to the original paper. Panzer recently sketched (private communication) a shorter alternative proof, based on the machinery developed by Brown to prove that all periods of $\overline{\mathcal{M}}_{0,n}$ belong to $\mathcal{A}[2\pi i]$ \cite{BrownModuliSpace}.

\subsection{Conical sums and multiple zeta values}

Characterizing conical sums that belong to $\mathcal{A}$ is an interesting problem that we were led to consider because the coefficients of modular graph functions, which we will relate to conical sums, are widely expected to be MZVs. The answer turned out not to be so simple. Once we work with conical sums in the form (\ref{standardcone}), it is clear that we should consider only sums of the kind $\zeta(A):=\zeta(A,\chi_0)$, where $\chi_0$ is the trivial character sending everything to~1.

If the non-negative matrix $A$ contains at least one entry $\geq 2$, one cannot hope in general to get MZVs, because this entry introduces a congruence condition on the sum. For example, it is an instructive exercise to show that
\[
\sum_{x,y\geq 1}\frac{1}{x\,(x+2y)^2}=\frac{\pi^2\log(2)}{8}-\frac{5\zeta(3)}{16}.
\]
A natural subset that we may want to define (and that is the good set to consider for the three-point modular graph functions, as we will see later) is then the following:
\begin{defi}
We call $(0,1)$-matrix any matrix whose entries are only zeros and ones. Then we define $\mathcal{B}\subset\mathcal{C}$ as the vector space spanned over $\mathbb{Q}$ by the conical sums $\zeta(A)$ such that $A$ is a $(0,1)$-matrix.
\end{defi}
Examples of conical sums belonging to this set are given by the \emph{Mordell-Tornheim sums} \cite{BradleyZhou}
\begin{equation}\label{MordellTornheim}
\sum_{x_1,\ldots ,x_n\geq 1}\frac{1}{x_1^{k_1}\cdots x_n^{k_n}(x_1,\ldots ,x_n)^m},
\end{equation}
where $k_i,m\geq 1$. It is trivial to see that $\mathcal{B}$ is an algebra, and that $\mathcal{A}\subseteq \mathcal{B}$. At first glance, one could be tempted to believe that $\mathcal{A}=\mathcal{B}$. For instance, it is a long but easy exercise in partial fraction decomposition to show that all Tornheim sums belong to~$\mathcal{A}$. Moreover, one can show with a little effort that up to dimension $n=3$ these two spaces indeed coincide, the most complicated case to consider being the family
\begin{equation}\label{sumDanylo}
\sum_{x_1,x_2,x_3\geq 1}\frac{1}{(x_1+x_2)^a(x_1+x_3)^b(x_1+x_2+x_3)^c},
\end{equation}
which can be reduced to a linear combination of MZVs by writing it as
\[
\sum_{1<\max\{u,v\}<w<u+v}\frac{1}{u^av^bw^c},
\]
and using a combination of operations on the cone (\emph{cone decomposition} \cite{Paycha}) and on the summand (partial fraction decomposition) \footnote{These two kinds of manipulations are conjectured \cite{Paycha} to give all the possible relations in $\mathcal{B}$, generalizing the double-shuffle conjecture for multiple zeta values: stuffle relations can obviously be considered as some kind of cone decomposition, and it is easy to see that shuffle relations, i.e. some kind of decomposition on the domain of the integral representation of MZVs, can be written in terms of partial fraction decompositions in the sum representation.}. However, we have noticed that, starting from cones of dimension $>3$, $\mathcal{B}$ is actually bigger than $\mathcal{A}$ (assuming the transcendence conjectures on MZVs), because for instance one finds, using the method explained below, that 
\[
\sum_{x,y,z,w\geq 1}\frac{1}{(x+y)(x+y+z)(y+z+w)(x+y+w)^2}=\frac{15}{32}\zeta(5)-\frac{9}{4}\zeta(2)\zeta(3)+\frac{9}{4}\log(2)\zeta(2)^2.
\]
Using the integral representation of conical sums of the form (\ref{standardcone}) (so in particular of the numbers in $\mathcal{B}$) one can try to employ the Maple program \texttt{HyperInt} developed recently by E. Panzer\cite{Panzer}, which is based on ideas of Brown contained in \cite{BrownModuliSpace} and \cite{BrownFeynmanInt}. In some cases, \texttt{HyperInt} answers rewriting $\zeta(A)$ as a linear combination of cyclotomic MZVs, and that is how we obtained the counterexample showing that $\mathcal{B}$ is strictly bigger than $\mathcal{A}$. Unfortunately, \texttt{HyperInt} up to now is not always able to give an answer, even for numbers belonging to $\mathcal{B}$, for reasons that have to do with the fact that we do not have at our disposal a data-base with all relations between $N$-th cyclotomic MZVs beyond $N=1$ and $N=2$, except for the first trivial weights.

Nevertheless, one can characterize\footnote{I am grateful to C. Dupont for remarking this to me, as well as for suggesting the idea of using \texttt{HyperInt} to compute conical sums.} a (not optimal) subset of numbers in $\mathcal{B}$ that will always belong to the ring $\mathcal{A}$ of MZVs (and can be computed algorithmically by \texttt{HyperInt}):
\begin{Lemma}\label{consecutiveones}
Let $\mathcal{S}$ be the set  of $(0,1)$-matrices such that, up to permutations of the rows, the ones are consecutive in every column. From now on we will call this property the \emph{C1s-property}. If $A\in\mathcal{S}$ and $\zeta(A)$ converges, then $\zeta(A)\in\mathcal{A}$.
\end{Lemma}
\textbf{Proof.} We can write the matrix with consecutive ones in every column, because interchanging rows does not change the sum. Recall the integral representation (\ref{int}). In our case it reduces to
\[
\zeta(A)=\int_{[0,1]^r}\frac{x_1^{l_1-1}\cdots x_r^{l_r-1}dx_1\cdots dx_r}{\prod_{j=1}^n(1-x_1^{a_{1,j}}\cdots x_r^{a_{r,j}})},
\]
where any factor in the denominator will actually be of the form $1-\prod_{k=a}^bx_k$ for $1\leq a\leq b \leq r$.
The result follows from Theorem 8.2 of \cite{Brown2006}, where it is proven that integrals of this kind always belong to $\mathcal{A}$.\\
$\square$

This for instance immediately implies that the sums (\ref{sumDanylo}) belong to $\mathcal{A}$. Nevertheless, as announced above, this condition is not optimal. Of course there are sums which we do not expect in general to be multiple zeta values, for example involving coefficients bigger than~1, that reduce to MZVs by accident (using double subdivision relations introduced in \cite{Paycha}):
\[
\sum_{x,y\geq 1} \frac{1}{(x+y)^2(2x+y)}=\frac{\zeta(3)}{4}.
\]
However, experiments suggest that there is a set of $(0,1)$-matrices strictly bigger than $\mathcal{S}$ which gives only MZVs; an example of this is given by all Mordell-Tornheim sums (\ref{MordellTornheim}) with at least three variables.

A conjecturally optimal characterisation of the conical sums belonging to $\mathcal{A}$ was given by Dupont (private communication):

\textbf{Conjecture 1.} Let $N$ be the least common multiple of the minors of $A$. Then $\zeta(A)$ is the period of a mixed Tate motive unramified over $\mathbb{Z}[\zeta_N,1/N]$.

This conjecture is based on considerations about the geometry of the hyperplanes defined by the factors in the denominator of the integral representation of $\zeta(A)$. Moreover, recently Panzer suggested to me the following (stronger) conjecture:

\textbf{Conjecture 2.} Let $N$ be the least common multiple of the minors of $A$. Then $\zeta(A)$ can be written as a linear combination of MPLs evaluated at $N$-th roots of unity.

This conjecture is stronger (again we assume all standard transcendentality conjectures) because the algebra of periods of mixed Tate motives over $\mathbb{Z}[\zeta_N,1/N]$ is in general bigger than the algebra of special values of $N$-th cyclotomic MZVs, for instance when $N$ is a prime $\geq 5$.\footnote{See \cite{DeligneGoncharov} and \cite{Zhao}.}

If we restrict our attention to matrices $A\in\mathcal{B}$, both conjectures\footnote{Conjecture 1 is equivalent to Conjecture 2 in this case because Brown proved \cite{BrownMTM} that periods of mixed Tate motives over $\mathbb{Z}$ are (up to inverting $\pi$) MZVs.} imply the following special case:

\textbf{Conjecture 3.} Let $A\in\mathcal{B}$ be \emph{totally unimodular}, which means that every minor is equal to 0 or 1 in absolute value. Then $\zeta(A)\in\mathcal{A}$.

Note that this condition is implied by $A\in\mathcal{S}$ (see Lemma \ref{consecutiveones}):
\begin{Lemma}
Let $A\in\mathcal{S}$. Then $A$ is totally unimodular.
\end{Lemma}
\textbf{Proof.} We proceed by induction on the number $n$ of columns of $A$. If $n=1$ the statement is trivial. Let $n>1$. We just need to compute the determinant of the $n\times n$ sub-matrices, because for any smaller sub-matrix we can apply the inductive hypothesis. Let $B$ be any $n\times n$ sub-matrix. If there is any row or column having only a 1 and the rest are 0's, we are done, because the determinat is equal to the determinant of a strictly smaller matrix with C1s-property. Moreover, we can assume that there is at least a 1 in any row, otherwise the determinant is 0 and we are fine. We can re-arrange the columns of $B$ in such a way that the first row reads $1,\ldots,1,0,\ldots,0$, where the last 1 corresponds to the $i$-th column, $i\geq 2$, and the length of the strings of ones in the first $i$ $b_1,\ldots,b_i$ increases. Let us consider the matrix with columns $b_1,b_2-b_1,\ldots,b_i-b_1,b_{i+1},\ldots,b_n$. The absolute value of the determinant of this matrix is the same of the absolute value of $\det(B)$, and the first row reads $1,0,\ldots,0$; therefore this determinant can be computed in terms of a smaller sub-matrix with consecutive ones, and we get our claim.\\
$\square$

The converse is obviously not true: as we have already mentioned, just consider any Mordell-Tornheim sum involving at least three variables.

\section{The two-point case}

When only two gravitons are involved, with $l$ propagators between them (diagram~\textcolor{red}{(a)} in Section~\ref{SectionClosed}), the associated modular graph functions read:
\[
D_{l}(\tau):=\int_{\mathcal{E}_\tau} G_1(\xi,\tau)^{l} \frac{d\xi}{\tau_2}.
\]
The behaviour of this integral as $\tau_2$ tends to infinity was already studied in \cite{GRV}. Using the same ideas we generalize that result giving the following expansion of the functions $D_l$'s, which is of course a special case of Theorem \ref{main}:
\begin{teo}\label{2points}
For every $l\geq 2$
\begin{equation}\label{qqbar2}
D_l(\tau)=\sum_{\mu,\nu\geq 0}d_{l}^{(\mu,\nu)}(\pi\tau_2)q^\mu\overline{q}^\nu,
\end{equation}
where for every $\mu,\nu\geq 0$ 
\[
d_l^{(\mu,\nu)}(\pi\tau_2)=\sum_{j=0}^{2l-1}a_j^{(\mu,\nu)}(\pi\tau_2)^{l-j}
\]
is a Laurent polynomial with coefficients $a_j^{(\mu,\nu)}$ belonging to the algebra~$\mathcal{C}$ of conical sums.
\end{teo}
\textbf{Proof}. Using (\ref{FourierProp2}) and doing the change of variables $x=\xi_2/\tau_2$, we want to compute
\begin{equation*}
D_{l}(\tau)=\frac{1}{2^l}\sum_{l=r+m}\frac{l!}{r!m!}\frac{(\pi\tau_2)^{r}}{2^{m}}
\int_{\mathcal{E}_\tau} \overline{\B}_2(x)^{r}P(\xi,\tau)^{m} d\xi_1dx,
\end{equation*}
where we recall that $\mathcal{E}_\tau$ denotes the complex torus associated with the lattice $\tau\mathbb{Z}+\mathbb{Z}$. Since, by the definition (\ref{PPart}),
\begin{equation}\label{P^m}
P(z,\tau)^{m}=\sum_{\substack{\underline{n}\in\mathbb{Z}^m\\ \underline{k}\in\mathbb{Z}_0^m}}\frac{1}{|\underline{k}|}e^{2\pi i\sum_i k_i(n_i\tau_1+\xi_1)}e^{-2\pi\tau_2\sum_i |k_i||n_i-x|},
\end{equation}
where $\mathbb{Z}_{0}:=\mathbb{Z}\setminus\{0\}$ and $|\underline{k}|:=|k_1|\cdots|k_m|$, we have
\begin{equation*}
\int_{\mathbb{R}/\mathbb{Z}} P(\xi,\tau)^{m} d\xi_1
=\sum_{\substack{\underline{n}\in\mathbb{Z}^m\\ \underline{k}\in\mathbb{Z}_0^m}}\frac{\delta_0(\underline{k})}{|\underline{k}|}e^{2\pi i(\underline{k}\cdot\underline{n})\tau_1} e^{-2\pi\tau_2(\sum|k_i||n_i-x|)},
\end{equation*}
where, for any $\underline{v}\in\mathbb{Z}^m$ and for any $a\in\mathbb{Z}$, $\delta_a(v)=1$ if the sum of the coordinates is $a$, and is zero otherwise. Moreover, $\underline{x}\cdot\underline{y}$ denotes the standard inner product of $\mathbb{R}^m$.\\
In the interval $[0,1]$,
\[
\overline{\B}_2(x)^{r}=\sum_{a+b+c=r}\frac{r!}{a!\,b!\,c!}\frac{(-1)^b}{6^c}x^{2a+b}.
\]
Therefore we have
\begin{multline}\label{chaopescao}
D_l(\tau)=\\
\frac{1}{4^l}\sum_{a+b+c+m=l}\frac{l!(2\pi\tau_2)^{l-m}}{a!\,b!\,c!\,m!}\frac{(-1)^b}{6^c}\sum_{\substack{\underline{n}\in\mathbb{Z}^m\\
\underline{k}\in\mathbb{Z}_0^m}}\frac{\delta_0(\underline{k})}{|\underline{k}|}e^{2\pi i(\underline{k}\cdot\underline{n})\tau_1}\int_0^1 x^{2a+b}e^{-2\pi\tau_2(\sum|k_i||n_i-x|)} dx.
\end{multline}
To compute the last integral, let us fix $n_1,\ldots,n_m$. Since $x\in [0,1]$ we have
\[
|n_i-x|=\left\{\begin{array}{ll}
n_i-x & \mbox{if } n_i>0 \\
x-n_i & \mbox{if } n_i\leq 0,
\end{array}\right.
\]
so $\sum|k_i||n_i-x|=\sum|k_i||n_i|+x\sum\mbox{sgn}(-n_i)|k_i|$.

By repeated integration by parts one easily finds that for any $c\in\mathbb{R}_{>0}$, $\beta\in\mathbb{R}\setminus\{0\}$ and~$M\in\mathbb{N}$
\begin{equation}\label{parts}
\int_0^c x^Me^{-\beta x} dx=\frac{M!}{\beta^{M+1}}-\sum_{j=0}^M\frac{(M)_j}{\beta^{j+1}}c^{M-j}e^{-\beta c},
\end{equation}
where $(M)_j=M(M-1)\cdots (M-j+1)$ is the descending Pochhammer symbol.
In our case $\beta=2\pi\tau_2\sum\mbox{sgn}(-n_i)|k_i|$, $c=1$ and $M=2a+b$. Note that $\sum\mbox{sgn}(-n_i)|k_i|$ can be equal to zero. If it is not zero, since $|n_i|+\mbox{sgn}(-n_i)=|n_i-1|$, by (\ref{parts}) we get, as a result of the integral,
\[
\frac{(2a+b)!\;e^{-2\pi\tau_2\sum|k_i||n_i|}}{(2\pi\tau_2\sum\mbox{sgn}(-n_i)|k_i|)^{2a+b+1}}-\sum_{j=0}^{2a+b}\frac{(2a+b)!\;e^{-2\pi\tau_2\sum|k_i||n_i-1|}}{(2a+b-j)!(2\pi\tau_2\sum\mbox{sgn}(-n_i)|k_i|)^{j+1}},
\]
while if $\sum\mbox{sgn}(-n_i)|k_i|=0$ then we just get
\[
\frac{e^{-2\pi\tau_2\sum|k_i||n_i|}}{2a+b+1}.
\]
In both cases, once we fix $n_1,\ldots,n_m$ and $k_1,\ldots,k_m$, putting everything together we are left with an expression of the kind
\[
c^{(p,q)}(\pi\tau_2)e^{2\pi ip\tau_1}e^{-2\pi q\tau_2},
\]
where $p=\sum k_in_i \in\mathbb{Z}$, $q$ is a non-negative integer\footnote{This notation may generate some confusion, because $q$ also denotes $\e(\tau)$. What we mean will always be clear from the context.} equal to either $\sum|k_i||n_i-1|$ or $\sum|k_i||n_i|$, and $c^{(p,q)}(\pi\tau_2)$ is a Laurent polynomial with rational (explicitly determined) coefficients whose maximum power is $l$ and minimum power is $1-l$.\footnote{When $m=0$ and $m=1$ one can see that there is no contribution to the negative powers, and therefore one can assume that $m\geq 2$ to get this lower bound.}

Note that $e^{2\pi ip\tau_1}e^{-2\pi q\tau_2}=q^\mu\overline{q}^\nu$, with $\mu=(q+p)/2$ and $\nu=(q-p)/2$. Therefore we would like to show that $q\geq |p|$ and that $p\equiv_2 q$ (this notation is a shorthand for the standard $p\equiv q$ mod 2), in order to have that $\mu$ and $\nu$ are non-negative integers. If $q=\sum|k_i||n_i|$ the claim is trivial, so we have to take care only of the case $q=\sum|k_i||n_i-1|$.

Note that $k_in_i \equiv_2 |k_i||n_i| \equiv_2 |k_i|(1+|n_i-1|)$ and that $\sum |k_i| \equiv_2 \sum k_i=0$, so
\[
p=\sum k_in_i \equiv_2 \sum|k_i| + \sum|k_i||n_i-1| \equiv_2 q.
\]
Moreover $|p|=|\sum k_in_i|+|\sum k_i|\leq |\sum k_i(n_i-1)|\leq \sum|k_i||n_i-1|=q$. 

To conclude our proof we have to analyse more carefully the rational coefficients of $\tilde{c}^{(\mu,\nu)}(\pi\tau_2)$, which are obtained by the $c^{(p,q)}(\pi\tau_2)$'s.

Any fixed $q=\sum|k_i||n_i-\varepsilon|$ ($\varepsilon=0,1$) can be obtained with just finitely many $m$-tuples $(n_1,\ldots,n_m)$, because $|n_i-\varepsilon|\leq q$ for any $i$: otherwise, since for every $i$ $|k_i|\geq 1$, we would have $\sum |k_i||n_i-\varepsilon|\geq \sum |n_i-\varepsilon|\geq q$.
This means that for any $(\mu,\nu)$, which is uniquely determined by a couple $(p,q)$, one has to consider a finite rational linear combination of sums of the kind
\[
T(m,\underline{n},\alpha):=\sum_{\underline{k}}\frac{\delta_0(\underline{k})}{|\underline{k}|(\sum\mbox{sgn}(-n_i)|k_i|)^\alpha},
\]
where $\underline{k}\in\mathbb{Z}_0^m$ are such that $\sum\mbox{sgn}(-n_i)|k_i|\neq 0$ and $\alpha$ is a positive integer. Note that the function $T$ only depends on $n_i/|n_i|$.

Since we can split the sum defining $T$ as a sum over cones such that the factors in the denominator are either bigger or smaller than zero, it follows that our coefficients are linear combinations of conical sums.\\
$\square$

Specializing carefully this computation to the $(p,q)=(0,0)$ case, one gets the proof, first obtained in \cite{GRV}, of Proposition \ref{TeoGRV}:

\textbf{Proof of prop. \ref{TeoGRV}.} Note that, since $q=\sum|k_i||n_i-\varepsilon|$ with $\varepsilon=0$ or~$=1$, the only possible $m$-tuples $(n_1,\ldots,n_m)$ which can give $q=0$ are $(0,\ldots,0)$ and $(1,\ldots 1)$, and with them also $p=0$, because $\sum k_i=0$. Using this and looking carefully\footnote{As explained in \cite{GRV}, one exploits the fact that $\overline{B}_2(1-x)=\overline{B}_2(x)$ in order to get the nice looking formula in the corollary instead of the more complicated one that we would naively get just by performing the same steps as in the theorem's proof.} at the proof of the previous theorem one is then lead to the formula given in the statement, except for the hypergeometric coefficient of the leading term, which is more elegant than the term obtained with the integration process described in the proof, and can be deduced by making use of the identity
\[
\frac{1}{x(x+1)\cdots (x+n)}=\sum_{j=0}^n \frac{(-1)^j}{j!(n-j)!}\frac{1}{x+j},
\]
easily obtained by partial fraction decomposition.\\
$\square$

As remarked in the previous chapter, the function $S(m,\alpha)$ was proven by Zagier in \cite{ZagierApp} to be equal to an explicit linear combination of MZVs, allowing to algorithmically compute in terms of MZVs the non-exponentially small part of $D_l$:
\begin{prop}[Zagier]
\begin{equation}
S(m,\alpha)=m!\sum_{r\geq 0}\sum_{\substack{a_1,\ldots ,a_r\in\{1,2\}\\a_1+\cdots +a_r=m+2}}2^{2(r+1)-m-\alpha}\zeta(a_1,\ldots ,a_r,\alpha+2).
\end{equation}
\end{prop}
We will not repeat the proof here, because it is essentially contained in the more complicated proof of Theorem \ref{th1}, that can be found in Appendix \ref{AppendixCalcolo}. 

Unfortunately, knowing the Laurent polynomial $d_{l}=d_l^{(0,0)}$ is not enough to perform the integration over the moduli space of complex tori, so one would like to understand better the behaviour of the functions $D_l$. This can be achieved by looking at the more general Theorem \ref{2points}, because it allows us to predict other coefficients of the expansion (\ref{qqbar2}).

To make an example, let us recall that the non-holomorphic Eisenstein series are defined, for $s\in\mathbb{C}$ with $\Re(s)>1$ and $\tau\in\mathbb{H}$, by
\begin{equation}\label{eisenstein}
E(s,\tau)=\bigg(\frac{\tau_2}{\pi}\bigg)^s{\sum_{(m,n)\in\mathbb{Z}^2\setminus\{(0,0)\}}} \frac{1}{|m\tau +n|^{2s}}.
\end{equation}
Its expansion at the cusp is given, setting $y=\pi\tau_2$, by
\begin{multline}\label{eisensteincusp}
E(n,\tau)=(-1)^{n-1}\frac{\B_{2n}}{(2n)!}(4y)^n+\frac{4(2n-3)!}{(n-2)!(n-1)!}\zeta(2n-1)(4y)^{1-n}\\
+\frac{2}{(n-1)!}\sum_{N\geq 1}N^{n-1}\sigma_{1-2n}(N)(q^N+\overline{q}^N)\sum_{m=0}^{n-1}\frac{(n+m-1)!}{m!(n-m-1)!}(4Ny)^{-m},
\end{multline}
where $\B_n$ is the $n$-th Bernoulli number, and $\sigma_k(N)=\sum_{d|N}d^k$ is a finite power sum running over the positive divisors of $N$.

As we have mentioned in the last chapter, using the series representations (\ref{sum}) for $D_l(\tau)$ and (\ref{eisenstein}) for the Eisenstein series one can immediately see that $D_2(\tau)=E(2,\tau)/16$.

Let us briefly see how can we get the same result by comparing the expansion given by Theorem \ref{2points} and the expansion (\ref{eisensteincusp}), which becomes in this case
\[
E(2,\tau)=\frac{y^2}{45}+\frac{\zeta(3)}{y}+2\sum_{N\geq 1}(2N+y^{-1})\sigma_{-3}(N)(q^N+\overline{q}^N).
\]
Using the intermediate step (\ref{chaopescao}) in the proof of the theorem, for $m=0$ we get $(1/16)(y^2/45)$, which is the leading term of the non-exponential part, for $m=1$ we get zero, and for $m=2$ (so $a=b=c=0$) we get
\[
\frac{1}{16}\sum_{\substack{k\in\mathbb{Z}\setminus\{0\}\\ n\in\mathbb{Z}}}\frac{e^{2\pi ip\tau_1}e^{-2\pi\tau_2(|k|(|n_1|+|n_2|))}}{|k|^2}\int_0^1 e^{-2\pi\tau_2x|k|(\text{sgn}(-n_1)+\text{sgn}(-n_2))} dx,
\]
where we denote $p=k(n_1-n_2)$.
When $p=0$, i.e. when $q$ and $\overline{q}$  have the same power in the expansion given by the theorem, then $n_1=n_2$, which implies that the argument of the exponential in the integral is never zero. Therefore, splitting the sum into the $n\geq 1$ part and the $n\leq 0$ part, one gets two telescoping sums, both giving as a result
\[
\sum_{k\in\mathbb{Z}\setminus\{0\}}\frac{1}{64|k|^3y}.
\]
We conclude that the variable $\tau_1$ does not appear only in the non-exponentially small part of $D_2(\tau)$, which is the Laurent polynomial $d_2(y)=(y^2/45+\zeta(3)/y)/16$. This fits with the expansion of $E(2,\tau)$.

Moreover, if $p>0$ one gets again telescoping sums in $n$, but now we sum only over finitely many $k$, which are the divisors of $p$. We leave as an exercise to the reader to verify that one gets exactly the same expansion as we get for the non-holomorphic Eisenstein series.

In general it is too messy to repeat the same game as above and explicitly get the full expansion for other $D_l$'s, except maybe for $D_3(\tau)$, which is already known, as we have mentioned in eq. (\ref{D3E}), to be equal to $(E(3,\tau)+\zeta(3))/64$. However, it is in principle possible to algorithmically get the coefficient of $q^\mu\overline{q}^\nu$ for any fixed $(\mu,\nu)$, which would allow to check conjectures on the full expansion or to numerically approximate the functions very precisely (the sum (\ref{qqbar2}) converges much faster than the sum~(\ref{sum})).

\section{The three-point case}

When three particles are involved the only new irreducible diagram that we have to consider is diagram \textcolor{red}{(b)} of last chapter, with all $l_i$'s strictly positive, whose associated modular graph function $D_{\underline{l}}$ reads:
\[
D_{\underline{l}}(\tau):=\int_{(\mathcal{E}_\tau)^2} G(\xi,\tau)^{l_1}G(\nu,\tau)^{l_2}G(\xi-\nu,\tau)^{l_3} \frac{d\xi d\nu}{\tau_2^2}.
\]
We now give the proof of Theorem \ref{main} also for this case, because it helps to understand how to explicitly get the coefficients. Since the ideas used here are exactly the same as in the previous section, and the notation gets much heavier, we will give less details. It is however important to understand how the generalization to this case exploits the same ideas used for two particles, because in the next section we will give only a sketch of the proof in the general case, assuming that one has already understood how to take care of the missing details.
\begin{teo}\label{trepunti}
For every $\underline{l}=(l_1,l_2,l_3)$ we have
\[
D_{\underline{l}}(\tau)=\sum_{\mu,\nu\geq 0}d_{\underline{l}}^{(\mu,\nu)}(\pi\tau_2)q^\mu\overline{q}^\nu,
\]
where for every $\mu,\nu\geq 0$ 
\[
d_{\underline{l}}^{(\mu,\nu)}(x)=\sum_{j=0}^{2(l_1+l_2+l_3)-1}a_j^{(\mu,\nu)}x^{l_1+l_2+l_3-j}
\]
is a Laurent polynomial with coefficients $a_j^{(\mu,\nu)}\in\mathcal{C}$.
\end{teo}
\textbf{Proof.} Let us introduce the following notations: $\underline{l}!:=l_1!\,l_2!\,l_3!$, and $c\,^{\underline{l}}:=c\,^{l_1+l_2+l_3}$. Moreover, for $\underline{k}=(k_1,...,k_m)$ we write $|\underline{k}|:=|k_1|\cdots |k_m|$, and $\Vert\underline{k}\Vert:=|k_1|+\cdots +|k_m|$.\\
With the substitutions $x=\xi_2/\tau_2$ and $y=\nu_2/\tau_2$ we get
\begin{multline*}
D_{\underline{l}}(\tau)=\frac{1}{2^{\underline{l}}}\sum_{\underline{r}+\underline{m}=\underline{l}}\frac{\underline{l}!}{\underline{r}!\,\underline{m}!}\frac{(\pi\tau_2)^{\underline{r}}}{2^{\underline{m}}}\times \\
\times \int_{(\mathcal{E}_{\tau})^2} \overline{\B}_2(x)^{r_1}\overline{\B}_2(y)^{r_2}\overline{\B}_2(x-y)^{r_3}P(\xi,\tau)^{m_1}P(\nu,\tau)^{m_2}P(\xi-\nu,\tau)^{m_3} d\xi_1d\nu_1dxdy.
\end{multline*}

Using (\ref{P^m}) we have
\begin{multline*}
\int_{(\mathbb{R}/\mathbb{Z})^{2}} P(\xi,\tau)^{m_1}P(\nu,\tau)^{m_2}P(\xi-\nu,\tau)^{m_3} d\xi_1d\nu_1 = \\
=\sum_{\substack{(\underline{k},\underline{h},\underline{t})\\(\underline{n},\underline{p},\underline{q})}}\frac{\delta_0((\underline{k},\underline{h}))\delta_0((\underline{k},\underline{t}))}{|\underline{k}|\,|\underline{h}|\,|\underline{t}|}e^{2\pi i(\underline{k}\cdot\underline{n}+\underline{h}\cdot\underline{p}+\underline{t}\cdot\underline{q})\tau_1} e^{-2\pi\tau_2(\sum|k_i||n_i-x|+\sum|h_i||p_i-y|+\sum|t_i||q_i-(x-y)|)},
\end{multline*}
where the sum runs over $(\underline{k},\underline{h},\underline{t})\in\mathbb{Z}_0^{m_1}\times\mathbb{Z}_0^{m_2}\times\mathbb{Z}_0^{m_3}$ and $(\underline{n},\underline{p},\underline{q})\in\mathbb{Z}^{m_1}\times\mathbb{Z}^{m_2}\times\mathbb{Z}^{m_3}$.

Then we need to calculate, for any fixed $\underline{r}$, $\underline{m}$, $(\underline{k},\underline{h},\underline{t})$ and $(\underline{n},\underline{p},\underline{q})$,
\begin{equation*}
\int_{(\mathbb{R}/\mathbb{Z})^2}\overline{\B}_2(x)^{r_1}\overline{\B}_2(y)^{r_2}\overline{\B}_2(x-y)^{r_3}e^{-2\pi\tau_2(\sum|k_i||n_i-x|+\sum|h_i||p_i-y|+\sum|t_i||q_i-(x-y)|)}dxdy.
\end{equation*}
Since $\sum|k_i||n_i-x|=\sum|k_i||n_i|+x\sum\mbox{sgn}(-n_i)|k_i|$, this is equal to
\begin{equation*}
e^{-2\pi\tau_2(\sum |k_i||n_i|+\sum |h_i||p_i|+\sum |t_i||q_i|)}\int_{[\mathbb{R}/\mathbb{Z}]^2}\overline{\B}_2(x)^{r_1}\overline{\B}_2(y)^{r_2}\overline{\B}_2(x-y)^{r_3}e^{-\gamma x}e^{-\delta y}dxdy,
\end{equation*}
where $\gamma:=2\pi\tau_2(\sum\mbox{sgn}(-n_i)|k_i|+\sum\mbox{sgn}(-q_i)|t_i|)$ and $\delta:=2\pi\tau_2(\sum\mbox{sgn}(-p_i)|h_i|+\sum\mbox{sgn}(q_i)|t_i|)$.

This is equal to
\begin{multline}\label{dueint}
\sum_{\underline{a}+\underline{b}+\underline{c}=\underline{r}}\frac{\underline{r}!}{\underline{a}!\,\underline{b}!\,\underline{c}!}\frac{(-1)^{\underline{b}}}{6^{\underline{c}}}e^{-2\pi\tau_2(\sum |k_i||n_i|+\sum |h_i||p_i|+\sum |t_i||q_i|)}\times \\
\times \Big(\int_0^1 x^{2a_1+b_1}e^{-\gamma x} dx\int_0^x y^{2a_2+b_2}(x-y)^{2a_3+b_3}e^{-\delta y} dy \\
+\int_0^1 y^{2a_2+b_2}e^{-\delta y} dy\int_0^y x^{2a_1+b_1}(y-x)^{2a_3+b_3}e^{-\gamma x} dx\Big).
\end{multline}
Since the integral that are left to compute are completely similar, let us describe just the result of the first one. After using the binomial theorem on $(x-y)^{2a_3+b_3}$, we get a $\mathbb{Q}$-linear combination of integrals of the kind
\[
\int_0^1 x^{M}e^{-\gamma x} dx\int_0^1 y^{N}e^{-\delta y} dy.
\]
Now, as in the two-point case, we use integration by parts and finally get a linear combination of $1$, $e^{-(\gamma+\delta)}$ and $e^{-\gamma}$, with coefficients that are products of polynomials in $\delta^{-1}$  and $(\gamma~+~\delta)^{-1}$ with rational coefficients (with some obvious modifications in case $\gamma$ and/or $\delta$ are zero). We do not give the exact formula here, for reasons of space, except for the special case of the non-exponentially small term, which we describe in the next corollary.
However, it is easy to see, going through the computation, that for any fixed $(\underline{k},\underline{h},\underline{t})$ and $(\underline{n},\underline{p},\underline{q})$ we get a term of the kind $c^{(p,q)}(\pi\tau_2)e^{2\pi ip\tau_1}e^{-2\pi q\tau_2}$, where $p=\sum k_in_i+\sum h_ip_i+\sum t_iq_i \in\mathbb{Z}$, $q$ is a non-negative integer equal to $\sum|k_i||n_i-1|+\sum|h_i||p_i-1|+\sum|t_i||q_i-1|+\vartheta$, with $\vartheta=0$ or $\sum\mbox{sgn}(-n_i)|k_i|+\sum\mbox{sgn}(-q_i)|t_i|$ or $\sum\mbox{sgn}(-n_i)|k_i|+\sum\mbox{sgn}(-h_i)|q_i|$ or $\sum\mbox{sgn}(-p_i)|h_i|+\sum\mbox{sgn}(q_i)|t_i|$, and $c^{(p,q)}(\pi\tau_2)$ is a Laurent polynomial with rational (explicitly determined) coefficients, whose maximum power is $l_1+l_2+l_3$ and minimum power is $1-(l_1+l_2+l_3)$.

For all the possible $\vartheta$ we can apply the method described in the two-point case to prove that $q\geq |p|$ and that $p\equiv_2 q$. Moreover, again one can prove that for any $(\mu,\nu)$ only finitely many $(\underline{n},\underline{p},\underline{q})$ are allowed, and the coefficients of the Laurent polynomials belong to $\mathcal{C}$ (and are very explicitly determined).\\
$\square$

In particular one can deduce the following (already found in \cite{GRV}):
\begin{cor}
\[
d_{\underline{l}}(x):=4^{\underline{l}}\;d^{(0,0)}_{\underline{l}}(x)=d_{\underline{l}}^{A}(x)+d_{\underline{l}}^{B}(x)+d_{\underline{l}}^{C}(x)
\]
and the three contributions are defined as follows:
\begin{equation*}
d_{\underline{l}}^{A}(x)=2(2x)^{\underline{l}}\sum_{\underline{a}+\underline{b}+\underline{c}=\underline{l}}\frac{\underline{l}!}{\underline{a}!\underline{b}!\underline{c}!}\frac{(-1)^{\underline{b}}}{6^{\underline{c}}}\frac{(2a_2+b_2)!(2a_3+b_3)!}{(2(a_2+a_3)+b_2+b_3+1)!}\frac{1}{\lambda+1}
\end{equation*}
is the contribution for $m_1=m_2=m_3=0$, where $\lambda:=2(a_1+a_2+a_3)+b_1+b_2+b_3+1$;
\[
d_{\underline{l}}^B(x)=d^B_{l_1,l_2,l_3}(x)+d^B_{l_2,l_1,l_3}(x)+d^B_{l_3,l_2,l_1}(x),
\]
with
\begin{multline*}
d^B_{l_1,l_2,l_3}(x)=\sum_{\substack{\underline{a}+\underline{b}+\underline{c}+\underline{m}=\underline{l}\\u+v=2a_3+b_3\\e+f=2a_1+b_1+u}} 2\frac{\underline{l}!}{\underline{a}!\underline{b}!\underline{c}!\underline{m}!}\frac{(-1)^{\underline{b}}}{6^{\underline{c}}}(2x)^{\underline{c}-\underline{a}-2}\times \\
\times \frac{(-1)^{v}(2a_3+b_3)!(2a_1+b_1+u)!(2a_2+b_2+v+f)!}{u!v!f!}R(m_1,m_2,m_3;2a_2+b_2+v+f+1,e+1),
\end{multline*}
is the contribution when at least two of the $m_i$'s are $>0$, with
\[
R(m_1,m_2,m_3;\alpha ,\beta ):=\sum_{(\underline{k},\underline{h},\underline{t})} \frac{\delta_0((\underline{k},\underline{h}))\delta_0((\underline{k},\underline{t}))}{|\underline{k}||\underline{h}||\underline{t}|(\Vert\underline{k}\Vert + \Vert\underline{h}\Vert)^\alpha(\Vert\underline{k}\Vert + \Vert\underline{t}\Vert)^\beta};
\]
\[
d_{\underline{l}}^C(x)=d^C_{l_1,l_2,l_3}(x)+d^C_{l_2,l_1,l_3}(x)+d^C_{l_3,l_1,l_2}(x),
\]
with
\begin{multline*}
d^C_{l_1,l_2,l_3}(x)=\sum_{\substack{\underline{a}+\underline{b}+\underline{c}+\underline{m}=\underline{l}\\u+v=2a_3+b_3}} 2\frac{\underline{l}!}{\underline{a}!\,\underline{b}!\,\underline{c}!\,\underline{m}!}\frac{(-1)^{\underline{b}}}{6^{\underline{c}}}(2x)^{\underline{c}-\underline{a}-2}\times \\
\times \frac{(2a_3+b_3)!}{u!\,v!}\frac{(-1)^{v}}{2a_2+b_2+v+1}(\lambda)!\times\\
\times \Big[S(m_1,\lambda+1)+ \sum_{j=0}^{\lambda}\frac{(-1)^j S(m_1,j+1)}{(\lambda-j)!}(2\pi\tau_2)^{(\lambda-j)}\Big],
\end{multline*}
is the contribution when two of the $m_i$'s are zero ($m_2$ and $m_3$ in $d^C_{l_1,l_2,l_3}$), where again $\lambda:=2(a_1+a_2+a_3)+b_1+b_2+b_3+1$ and 
$S(m,\alpha)=R(m,0,0,\alpha,0)$ is the sum already introduced in the previous section. In every case where at least one of the $m_i$'s in $d^B$ and $d^C$ is zero, we assume that the other $m_i$'s are $\geq 2$, otherwise it is easy to see that the contribution given is zero.
\end{cor}
\begin{oss}
Note that we are computing the non-exponentially small term without dividing it by $4^{\underline{l}}$, in order to have neater results afterwards. This does not change the proof, but may generate some confusion concerning the resulting expression.
\end{oss}
\textbf{Sketch of the proof.} It is convenient to consider as separate cases: the case where all the $m_i$'s are $=0$; the case where two of the~$m_i$'s are~$=0$ and one is~$>1$ (it cannot be~$=1$); the case where one of the $m_i$'s is~$=0$ and two are~$>1$; the case where all of them are~$>0$. The last case is the most complicated (and it gives the same result as the case where only one is~$=0$); we briefly describe how to treat it, and the same argument can be applied to the other cases.

Following the proof of the theorem above one arrives at the point (\ref{dueint}), and then should take into account only the $(\underline{n},\underline{p},\underline{q})$ leading to non-exponentially small terms. These are $(\underline{0},\underline{0},\underline{0})$, $(\underline{1},\underline{0},\underline{1})$ and $(\underline{1},\underline{1},\underline{0})$ for the first integral, $(\underline{1},\underline{1},\underline{0})$, $(\underline{0},\underline{1},\underline{-1})$ and $(\underline{0},\underline{0},\underline{0})$ for the second integral. By substituting $x=1-x$ and $y=1-y$ in the second integral one gets just two copies of the three possible cases for the first integral. We call the first one $d^B_{l_1,l_2,l_3}(x)$, and then one can easily notice that the other two are given by $d^B_{l_2,l_1,l_3}(x)$ and $d^B_{l_3,l_2,l_1}(x)$.\\
$\square$

Let us remark that, since by definition $D_{\underline{l}}$ does not depend on the order of the $l_i$'s, also $d_{\underline{l}}$ does not, even though from this formula it is not clear at first sight. So, for instance, speaking of $d_{1,2,3}$ is the same as speaking of $d_{3,1,2}$.

Evaluating by hand the functions $R(m_1,m_2,m_3;\alpha ,\beta )$ in terms of MZVs, one is able to find for the lower weights:
\begin{eqnarray*}
d_{1,1,1}(y)&=&\frac{2}{945}y^3+\frac{3}{4}\frac{\zeta(5)}{y^2},\\
d_{1,1,2}(y)&=&\frac{2}{14175}y^4+\frac{\zeta(3)}{45}y +\frac{5}{12}\frac{\zeta(5)}{y}- \frac{1}{4}\frac{\zeta(3)^2}{y^2}+\frac{9}{16}\frac{\zeta(7)}{y^3},\\
d_{1,1,3}(y)&=&\frac{2}{22275}y^5+\frac{\zeta(3)}{45}y^2+\frac{11}{60}\zeta(5)+\frac{105}{32}\frac{\zeta(7)}{y^2}-\frac{3}{2}\frac{\zeta(3)\zeta(5)}{y^3}+\frac{81}{64}\frac{\zeta(9)}{y^4},\\
d_{1,2,2}(y)&=&\frac{8}{467775}y^5+\frac{4\zeta(3)}{945}y^2+\frac{13}{45}\zeta(5)+\frac{7}{8}\frac{\zeta(7)}{y^2}-\frac{\zeta(3)\zeta(5)}{y^3}+\frac{9}{8}\frac{\zeta(9)}{y^4}.
\end{eqnarray*}

These are all the possible cases up to weight five. The same result was found in \cite{DGV2015}, where the authors corrected some mistakes made in \cite{GRV}. 

Note that up to this weight the coefficients of the Laurent polynomials are MZVs, but in the literature it is not proven yet that this will happen in any weight. Moreover, let us remark that they are MZVs of a very particular kind: they are always polynomials in simple odd zeta values, just as it happens (and is proven to be so in any weight by Theorem \ref{zagier}) in the two-point case. What we are now able to say is that, using Theorem \ref{trepunti} and Terasoma's Theorem \ref{ThTera}, they have to be cyclotomic MZVs.

For higher weights the sums $R(m_1,m_2,m_3;\alpha ,\beta )$ look impossible to be evaluated by hand; this is why no other~$d_{\underline{l}}$ was known before our work.

Since we have seen that the coefficients are conical sums, one can use \texttt{HyperInt} if the cones and the matrices involved are simple enough. This turns out to often be the case with three strings, because of the following theorem:
\begin{teo}\label{th1}
For all $l_1,l_2,l_3\in\mathbb{N}$ the coefficients of $d_{l_1,l_2,l_3}(y)$ belong to the algebra~$\mathcal{B}$ of conical sums associated to $(0,1)$-matrices.
\end{teo}
The proof of this theorem is constructive, and gives an actual formula to compute the coefficients, but the formula itself is very long and complicated. It can be found in Appendix \ref{AppendixCalcolo} (see equations \ref{zero},...,\ref{otto}), together with the proof.

Thus one gets an algorithm which will certainly compute the coefficients of~$d_{\underline{l}}$ in terms of MZVs for any $\underline{l}$ such that all matrices involved belong to~$\mathcal{S}$ (matrices with consecutive ones on the columns). For example, all the~$d_{\underline{l}}$ of weight six satisfy this condition, after some partial fraction decomposition on the conical sums, but not all the weight-seven ones: we will come back to this later. Moreover, as remarked before, the algorithm will produce an answer, either in terms of MZVs or alternating sums, for many more $(0,1)$-matrices than just the ones in~$\mathcal{S}$.

We list here the new data obtained so far with this method (we set again $y:=\pi\tau_2$):
\begin{eqnarray*}
d_{1,1,4}(y)&=&\frac{284}{18243225}y^6+\frac{2}{135}\zeta(3)y^3+\frac{5\zeta(5)}{18}y +\frac{1}{10}\zeta(3)^2+\frac{51}{20}\frac{\zeta(7)}{y}+\frac{11}{2}\frac{\zeta(5)\zeta(3)}{y^2}\\&+&\frac{79\zeta(9)-36\zeta(3)^3}{24y^3}-\frac{9}{4}\frac{\zeta(3)\zeta(7)}{y^4}+\frac{45}{16}\frac{\zeta(11)}{y^5},\\
d_{2,2,2}(y)&=&\frac{193}{11609325}y^6+\frac{1}{315}\zeta(3)y^3+\frac{59}{315}\zeta(5)y+\frac{23}{20}\frac{\zeta(7)}{y}+\frac{5}{2}\frac{\zeta(3)\zeta(5)}{y^2}-\frac{65}{48}\frac{\zeta(9)}{y^3}\\&+&\frac{21\zeta(5)^2-18\zeta(3)\zeta(7)}{16y^4}+\frac{99}{64}\frac{\zeta(11)}{y^5},\\
d_{1,2,3}(y)&=&\frac{298}{42567525}y^6+\frac{1}{315}\zeta(3)y^3+\frac{173}{1260}\zeta(5)y +\frac{3}{20}\zeta(3)^2+\frac{53}{20}\frac{\zeta(7)}{y}-\frac{5}{2}\frac{\zeta(3)\zeta(5)}{y^2}\\&+&\frac{223\zeta(9)+96\zeta(3)^3}{32y^3}-\frac{99\zeta(5)^2+162\zeta(3)\zeta(7)}{32y^4}+\frac{729}{128}\frac{\zeta(11)}{y^5},
\end{eqnarray*}
\begin{multline*}
d_{1,1,5}(y)=\frac{62}{10945935}y^7+\frac{2}{243}\zeta(3)y^4+\frac{119}{324}\zeta(5)y^2 +\frac{11}{27}\zeta(3)^2y+\frac{21}{16}\zeta(7)\\
+\frac{46}{3}\frac{\zeta(3)\zeta(5)}{y}+\frac{7115\zeta(9)-3600\zeta(3)^3}{288y^2}+\frac{1245\zeta(3)\zeta(7)-150\zeta(5)^2}{16y^3}\\
+\frac{288\zeta(3)\zeta(3,5)-288\zeta(3,5,3)-5040\zeta(5)\zeta(3)^2-9573\zeta(11)}{128y^4}\\
+\frac{2475\zeta(5)\zeta(7)+1125\zeta(9)\zeta(3)}{32y^5}-\frac{1575}{32}\frac{\zeta(13)}{y^6},
\end{multline*}
\begin{multline*}
d_{1,3,3}(y)=\frac{34}{8513505}y^7+\frac{2}{945}\zeta(3)y^4+\frac{17}{252}\zeta(5) y^2+\frac{23}{105}\zeta(3)^2y+\frac{1391}{560}\zeta(7)\\
-\frac{3\zeta(3)\zeta(5)}{y} + \frac{953\zeta(9)+144\zeta(3)^3}{32y^2}-\frac{1701\zeta(3)\zeta(7)+120\zeta(5)^2}{32y^3}\\
+\frac{324\zeta(3,5,3)-324\zeta(3)\zeta(3,5)+22299\zeta(11)+8460\zeta(5)\zeta(3)^2}{320y^4}\\
-\frac{891\zeta(5)\zeta(7)+702\zeta(9)\zeta(3)}{16y^5}+\frac{7209}{128}\frac{\zeta(13)}{y^6},  
\end{multline*}
\begin{multline*}
d_{1,2,4}(y)=\frac{592}{383107725}y^7+\frac{152}{93555}\zeta(3)y^4+\frac{44}{567}\zeta(5)y^2+\frac{148}{945}\zeta(3)^2y+\frac{277}{105}\zeta(7)\\
+\frac{62}{15}\frac{\zeta(3)\zeta(5)}{y}+\frac{191\zeta(9)-36\zeta(3)^3}{18y^2}-\frac{69\zeta(3)\zeta(7)+180\zeta(5)^2}{16y^3}\\
-\frac{72\zeta(3,5,3)-72\zeta(3)\zeta(3,5)-11893\zeta(11)-2520\zeta(5)\zeta(3)^2}{320y^4}\\
-\frac{477\zeta(5)\zeta(7)+441\zeta(9)\zeta(3)}{16y^5}+\frac{4905}{128}\frac{\zeta(13)}{y^6},
\end{multline*}
\begin{multline*}
d_{1,1,6}(y)=\frac{262}{186080895}y^8+\frac{1}{243}\zeta(3)y^5+\frac{113}{324}\zeta(5)y^3+\frac{25}{36}\zeta(3)^2y^2+\frac{749}{144}\zeta(7)y
+\frac{331}{18}\zeta(3)\zeta(5)\\
+\frac{56\zeta(9)-207\zeta(3)^3}{18y}+\frac{705\zeta(3)\zeta(7)+375\zeta(5)^2}{2y^2}\\
+\frac{2304\zeta(3,5,3)-2304\zeta(3)\zeta(3,5)-38541\zeta(11)-32400\zeta(5)\zeta(3)^2}{64y^3}+\frac{a}{y^4}\\
+\frac{b}{y^5}+\frac{179550\zeta(11)\zeta(3)+274050\zeta(9)\zeta(5)+155925\zeta(7)^2}{128y^6}-\frac{1233225}{512}\frac{\zeta(15)}{y^7},
\end{multline*}
where $a$ cannot be determined because of current limits\footnote{These limits have to do with the database of relations between MZVs. Panzer informed me that this database will soon be significantly enlarged.} of \texttt{HyperInt}, and
\begin{multline*}
b=\frac{837}{14}\zeta(5)\zeta(5,3)-\frac{3375}{4}\zeta(3){\zeta(5)}^2-\frac{6075}{8}\zeta(7){\zeta(3)}^2-\frac{675}{56}\zeta(3,7,3)\\+\frac{675}{56}\zeta(3)\zeta(7,3)+\frac{54}{7}\zeta(5,3,5)+\frac{135}{4}\zeta(5)\zeta(8)-\frac{134257}{896}\zeta(13).
\end{multline*}
Starting from weight 7, we can see something new and very interesting happening to the coefficients:
not only polynomials in odd simple zeta values are involved. For example, the coefficient\footnote{Here a sign mistake is noticed w.r.t. the result reported in \cite{Zerb15}. Note that also eq. (\ref{y^(-4)d511}) has been corrected.}  of $y^{-4}$ in $d_{1,1,5}(y)$ contains $\zeta(3,5,3)$ and $\zeta(3)\zeta(3,5)$, which are not reducible to polynomials in odd zetas. The fundamental remark is that they are still very special, because that coefficient can be written as the following linear combination of single-valued multiple zeta values:
\begin{equation}\label{y^(-4)d511}
-\frac{9}{8}\zeta_{\rm sv}(3,5,3)-\frac{405}{64}\zeta_{\rm sv}(5){\zeta_{\rm sv}(3)}^2-\frac{9573}{256}\zeta_{\rm sv}(11).
\end{equation}
This actually happens to all of the coefficients in the polynomials above (products of odd zeta values are already single-valued MZVs), the most astonishing case being the coefficient of $y^{-5}$ in $d_{1,1,6}(y)$, that we called $b$. Indeed, one can check that, in terms of the basis for single valued MZVs in weight 13 in \cite{BrownSVMZV},
\[
b=\frac{27}{7}\zeta_{\rm sv}(5,3,5)-\frac{675}{112}\zeta_{\rm sv}(3,7,3)-\frac{4995}{4}\zeta_{\rm sv}(3){\zeta_{\rm sv}(5)}^2-\frac{7425}{8}\zeta_{\rm sv}(7){\zeta_{\rm sv}(3)}^2-\frac{134257}{1792}\zeta_{\rm sv}(13).
\]
This means that a multiple zeta value belonging a priori to a vector space of dimension 16 actually belongs to the subspace of dimension\footnote{The conjectured dimensions of spaces of single-valued MZVs are given in \cite{BrownSVMZV}.} 5 of single-valued MZVs, which certainly looks more than just a coincidence. Let us now draw a parallel between our setting and the genus zero case for closed strings. We have seen that in genus zero the picture goes as follows: in the most trivial case (four particles) only odd zetas appear, while starting from the next case (five particles) one finds also MZVs of higher depth, which are conjectured to always belong to the algebra of single valued MZVs. Therefore we conclude that it is not too optimistic to conjecture that our coefficients are given by single valued MZVs only, even with so little evidence. Arguments supporting this conjecture, based on the structure of the one-loop string amplitude, have been given afterwards in the paper~\cite{DGGV}.

We conclude this section by observing that, unfortunately, the matrices appearing do not always belong to $\mathcal{S}$, even after performing standard manipulations like partial fraction decomposition, and sometimes they produce special values of polylogarithms at higher roots of unity. This happens, for instance, in the computation of $R(3,3,1;1,1)$, which is one of the sums involved in the computation of $d_{1,3,3}$. However, in this case we get only alternating sums, which is good enough to let them be computed by \texttt{HyperInt}, and in the end all the non-MZV part of $R(3,3,1;1,1)$ cancels out.

It is actually very tempting to conjecture that the numbers $R(m_1,m_2,m_3;\alpha,\beta)$ themselves always lie in $\mathcal{A}$ (but they are not single-valued), because this is what we have found so far. However, this time we do not have any other argument to support this evidence.

A very partial result in the direction of proving the conjectures above is the following:
\begin{teo}\label{teoConsOnes}
For any $n\in\mathbb{N}$ the coefficients of $d_{1,1,n}(\tau)$ are linear combinations of conical sums whose matrices belong to $\mathcal{S}$, so in particular they are (algorithmically) $\mathbb{Q}$-linear combinations of multiple zeta values.
\end{teo}
\textbf{Proof.} The proof uses the explicit formula given in Appendix \ref{AppendixCalcolo} for the numbers
$R$ in terms of elements of $\mathcal{B}$. The only $R$'s involved are of the kind $R(1,1,j;\alpha,\beta)$, $R(1,j,1;\alpha,\beta)$ and $R(j,1,1;\alpha,\beta)$, with $j\leq n$ and some $\alpha$, $\beta$.
Note that $R(1,j,1;\alpha,\beta)=R(1,1,j;\beta,\alpha)$ and $R(j,1,1;\alpha,\beta)=R(1,1,j;0,\alpha+\beta)$, so it is enough to study $R(1,1,j;\alpha,\beta)$. Only the sums (\ref{uno}) and (\ref{quattro}) are contributing to this $R$, but the sum (\ref{uno}) is easily seen to be contained in $\mathcal{A}$, so we have to study (\ref{quattro}) only, which in our case is particularly simple (assume $j\geq 2$, otherwise (\ref{uno}) suffices):
\[
\sum_{\substack{Q,F\geq 0\\Q+F=j-2}}\sum_{\substack{l_3+a>q_1>\cdots >q_Q\\l_3>f_1>\cdots >f_F}}\frac{1}{{(l_3+a)}^{\beta+1}a^{\alpha+2}l_3q_1\cdots q_Qf_1\cdots f_F}
\]
Following the stuffle procedures described in Appendix \ref{AppendixCalcolo} one is left with a linear combination of sums in $\mathcal{B}$ with associated matrices of the kind
\[ 
\begin{pmatrix}
  1 &  &  &  &  &  &  \\
  \vdots & \ddots &  &  &  &  &  \\
  1  & \cdots  & 1 &  &  &  &  \\
    &  &  & 1 &  &  &  \\
    & A &  & \vdots & \ddots &  &  \\
    &  &  & 1 & \cdots & 1 &  \\
  1 & \cdots & 1 & 0 & \cdots & 0 & 1 \\
  1 & \cdots & 1 & 1 & \cdots & 1 & 1 
 \end{pmatrix}
\]
where $A$ is a matrix with rows given either by consecutive ones or by consecutive zeros (this comes from the stuffle).
Since interchanging the rows does not change the conical sum, we can rewrite the matrix as
\[ 
\begin{pmatrix}
  1 &  &  &  &  &  &  \\
  \vdots & \ddots &  &  &  &  &  \\
  1  & \cdots  & 1 &  &  &  &  \\
  1 & \cdots & 1 &  &  &  & 0 \\
  \vdots &  & \vdots &  & B &  & \vdots \\
  1 & \cdots & 1 &  &  &  & 0 \\
  1 & \cdots & 1 & 1 & \cdots & 1 & 1 \\
  1 & \cdots & 1 & 0 & \cdots & 0 & 1 \\
  0 & \cdots & 0 &  &  &  & 0 \\
  \vdots &  & \vdots &  & C &  & \vdots \\
  0 & \cdots & 0 &  &  &  & 0 
 \end{pmatrix}
\]
where $B$ and $C$ are matrices with, from left to right, a string of ones followed by a string of zeros in every row, such that the length of the string of ones increases in $B$ with the increase of the row's index and decreases in $C$. At this point we almost have a matrix belonging to $\mathcal{S}$, the only problem being the row in the middle of the form $r=1,\ldots,1,0,\ldots,0,1$.

Note now that a partial fraction operation on the sum of the kind
\[
\frac{1}{l_i(\underline{x})\,l_j(\underline{x})}=\frac{1}{l_i(\underline{x})\,(l_i+l_j)(\underline{x})}+\frac{1}{l_j(\underline{x})\,(l_i+l_j)(\underline{x})}
\]
is reflected on the matrix just by substituting the $i$-th or the $j$-th row by the sum of the 2. Hence if we do this sum operation on $r$ together with the row immediately below we get the sum of 2 matrices, one belonging to $\mathcal{S}$ (when $r$ is deleted) and one such that the sub-matrix below $r$ is strictly smaller (after interchanging $r$ with the new row obtained as a sum). Iterating this process one finally gets that $r$ is the last row in the matrix, and in this case the matrix belongs to $\mathcal{S}$ and we are done.\\
$\square$

\section{The general case}

In this section we will briefly explain why the techniques seen in details in the previous sections allow us to prove Theorem \ref{main} in its more general statement, for a graph with $n$ vertices. It is important to have read and understood all steps of the proof in the two-point and three-point case, in order to follow the proof given below.

\textbf{Proof of Theorem \ref{main} (sketch).} The most difficult part of proving this result, after having proved the two-point case and three-point case, is probably to find an acceptable notation. First of all, let $\xi_1,\ldots ,\xi_{N}$ denote $N$ points on the torus $\mathcal{E}_\tau$. We fix $\xi_N=0$, but nonetheless we will keep writing $\xi_N$ in the integrals. Moreover, we define $x_i:=\Im(\xi_i)/\tau_2$. Suppose that our graph $\Gamma$ has $N$ vertices and $l=\sum l_{i,j}$ edges, as in the notation of Section \ref{SectionModGraph}. This gives rise to the integral
\[
D_{\Gamma}(\tau)=\int_{(\mathcal{E}_\tau)^{N-1}} \prod_{1\leq i<j\leq N}G(\xi_i-\xi_j,\tau)^{l_{i,j}}\frac{d\xi_1\cdots d\xi_{N-1}}{\tau_2^{N-1}}
\]
\begin{equation*}
=\frac{1}{2^{\underline{l}}}\sum_{\underline{r}+\underline{m}=\underline{l}}\frac{\underline{l}!}{\underline{r}!\,\underline{m}!}\frac{(\pi\tau_2)^{\underline{r}}}{2^{\underline{m}}}\int\prod_{1\leq i<j\leq N} \overline{\B}_2(x_i-x_j)^{r_{i,j}}
P(\xi_i-\xi_j,\tau)^{m_{i,j}}d\Re(\xi_1)\cdots d\Re(\xi_{N-1})dx_1\cdots dx_{N-1}
\end{equation*}
where the notation employed for the exponential or the factorial of a vector was introduced in the proof of Theorem \ref{trepunti}. The integration over the real part of the $\xi_i$'s gives (recall the definition of the incidence matrix $(\Gamma_{i,\beta})$ from Section \ref{SectionModGraph})
\begin{multline*}
\sum_{\substack{k_1,\ldots ,k_l\in\mathbb{Z}\setminus\{0\}\\ n_1,\ldots ,n_l\in\mathbb{Z}}}\prod_{i=1}^N\delta_0\Big(\sum_{\beta=1}^l\Gamma_{i,\beta}k_\beta\Big)\prod_{\alpha=1}^l \frac{\exp(2\pi i\tau_1k_\alpha n_\alpha)}{|k_\alpha|}\times\\
\times \int_{(\mathbb{R}/\mathbb{Z})^{N-1}} \prod_{1\leq i<j\leq N}\overline{\B}_2(x_i-x_j)^{r_{i,j}}\exp\bigg(-2\pi\tau_2\Big(\sum_{\alpha=t_{i,j}}^{t_{i,j}+m_{i,j}}|k_\alpha||n_\alpha-x_i+x_j|\Big)\bigg)dx_1\cdots dx_{N-1},
\end{multline*}
where $t_{i,j}$ denote certain integers in $\{1,\ldots ,l\}$, explicitly determined by an explicit choice of labelling variables when we sum all $l_{i,j}$'s propagators to obtain $l$ propagators. Let us call $I$ the integral appearing in the last step. We have
\begin{multline}\label{peacenlove}
I=\sum_{\underline{a}+\underline{b}+\underline{c}=\underline{r}}\frac{\underline{r}!}{\underline{a}!\underline{b}!\underline{c}!}\frac{(-1)^{\underline{b}}}{6^{\underline{c}}}\prod_{1\leq i<j\leq N}\exp\bigg(-2\pi\tau_2\Big(\sum_{\alpha=t_{i,j}}^{t_{i,j}+m_{i,j}}|k_\alpha||n_\alpha-x_i+x_j|\Big)\bigg)\times\\
\times \Big(\sum_{1\leq i<j\leq N}\int_{P_{i,j}}\Big)\prod_{1\leq i<j\leq N}|x_i-x_j|^{M_{i,j}}e^{-\gamma_1 x_1}\cdots e^{-\gamma_{N-1} x_{N-1}} dx_1\cdots dx_{N-1},
\end{multline}
where $P_{i,j}$ is the path $0\leq\sigma_{i,j}(x_1)\leq\cdots \leq\sigma_{i,j}(x_{N-1})\leq 1$, for $\sigma_{i,j}$ a permutation of the $N-1$ variables $x_1,\ldots ,x_{N-1}$, $M_{i,j}:=2a_{i,j}+b_{i,j}$ and the $\gamma_i$'s will depend on the path chosen.

Let us consider the path associated with the identity of the symmetric group $S_{N-1}$, i.e $0\leq x_1\leq \cdots\leq x_{N-1}$. In this case, for every $i$, we have
\[
\gamma_i=2\pi\tau_2 \delta\Big(\sum_{\beta=1}^l\Gamma_{i,\beta}\mbox{sgn}(-n_{\beta})|k_{\beta}|\Big).
\]
The integral on this path reduces to a linear combination of iterated integrals of the kind
\[
\int_{[0,1]} x_{N-1}^{Q_{N-1}}e^{-\gamma_{N-1} x_{N-1}} dx_{N-1} \cdots x_{1}^{Q_{1}}e^{-\gamma_{1} x_{1}} dx_{1},
\]
where the $Q_i$'s are non negative integers. One can solve the integral by repeatedly using integration by parts, and all the possible exponentials involved in the result are $e^{-\gamma_{N-1}},e^{-(\gamma_{N-2}+\gamma_{N-1})},\ldots ,e^{-(\gamma_1+\cdots+\gamma_{N-1})}$. Multiplying them by the exponential in front of the integral in formula (\ref{peacenlove}) tells us what are all the possible integers $q$ in the terms of the kind $e^{-2\pi q\tau_2}$.

It is not difficult to see that the argument used in the two-point case works for all of these $q$'s, and that nothing new happens to the Laurent polynomials involved and to their coefficients, which are therefore expressible as conical sums.\\
$\square$

We do not write down here an explicit formula for the Laurent polynomial part of the functions $D_{\Gamma}(\tau)$, because it gets really complicated already in the four-point case and does not really allow one to work with it. Indeed, the same method explained in the previous section to explicitly write down the conical sums as integrals produces, already in the four-point case, matrices with coefficients strictly bigger than~1, whose computation in terms of special values of polylogarithms goes beyond the current limits of \texttt{HyperInt}. 

%% file: Chapter5.tex
\newtheorem{thm}{Theorem}[section]
\newtheorem{conj}[thm]{Conjecture}
\newtheorem{defn}[thm]{Definition}
\newtheorem{ex}[thm]{Example}
\newtheorem{hyp}[thm]{Hypotheses}
\newtheorem{lemma}[thm]{Lemma}
\newtheorem{prob}[thm]{Problem}
\newtheorem{quest}[thm]{Question}
\newtheorem{remark}[thm]{Remark}

\chapter{Elliptic multiple zeta values} 

\label{ChapterEllMZV} 

Let us fix the notation for this chapter, which is the same as that of Chapter~\ref{ChapterMathBackground}. First of all, we will come back to denoting tuples by $\textbf{x}=(x_1,\ldots ,x_n)$. We will consider $\tau\in\mathbb{H}$, $q=\e(\tau)$\footnote{Recall that $\e(x):=\exp(2\pi ix)$.} and $\mathcal{E}_{\tau}=\mathbb{C}/(\tau\mathbb{Z}+\mathbb{Z})$. Moreover, for $\xi\in\mathcal{E}_{\tau}$, we will write $r_{\tau}(\xi):=\Im(\xi)/\Im(\tau)$. Recall also from Chapter~\ref{ChapterMathBackground} the Kronecker function
\begin{equation}
F(\xi,\alpha,\tau):=\frac{\theta^{\prime}(0,\tau)\theta(\xi+\alpha,\tau)}{\theta(\xi,\tau)\theta(\alpha,\tau)}=\sum_{n\geq 0}f_n(\xi,\tau)(2\pi i\alpha)^{n-1}
\end{equation}
where $\alpha$ is a formal variable, and the modified real analytic
\begin{equation}
\Omega(\xi,\alpha,\tau):=\e(r_{\tau}(\xi)\alpha)F(\xi,\alpha,\tau)=\sum_{n\geq 0}\omega_n(\xi,\tau)(2\pi i\alpha)^{n}.
\end{equation}
We have already seen in Chapter~\ref{ChapterMathBackground} that
\begin{align}\label{explicitformulas2}
f_n(\xi,\tau) &= \left\{ \begin{array}{cl}
2\pi i
 &: \ n=0 \\
\pi\cot(\pi \xi)-2\pi i\sum_{m\geq 1}\big(\e(m\xi)-\e(-m\xi)\big)\sum_{p\geq 1}q^{mp}
 &: \ n=1 \\
\frac{2\pi i}{(n-1)!}\big(\frac{\B_n}{n}-\sum_{m\geq 1}\big(\e(m\xi)+(-1)^n\e(-m\xi)\big)\sum_{p\geq 1} p^{n-1}q^{mp}\big)
 &: \ n\geq 2,
\end{array} \right .
\end{align}
and by definition $\omega_n(\xi,\tau)=\sum_{k=0}^n\frac{r_{\tau}(\xi)^k}{k!}f_{n-k}(\xi,\tau)$.
Finally, recall the KZB form (\ref{KZB-form}):
\begin{eqnarray}\label{KZB-form2}
\omega_{KZB}(x_0,x_1;\xi,\tau)&=&-\nu (\xi) x_0+\big(\mbox{ad}_{x_0}\Omega(\xi,\mbox{ad}_{x_0},\tau)d\xi\big)(x_1)\\
&=&-\nu (\xi)x_0+\sum_{n\geq 0}\big(\omega_n(\xi)d\xi\big)(2\pi i)^n\mbox{ad}_{x_0}^n(x_1),
\end{eqnarray}
where $\nu(\xi):=dr_{\tau}(\xi)$, $\mbox{ad}_{x_0}(\cdot)=[x_0,\cdot]$ and
\[
\mbox{ad}^n_{x_0}(x_1)=[\underbrace{x_0,[x_0,[\cdots [x_0}_n,x_1]]\cdots ].
\]

Elliptic multiple zeta values were defined about five years ago by Enriquez in the context of his work on elliptic associators (see \cite{CalEtiEnr}, \cite{EnriquezAss}, \cite{Enriquez}). Immediately afterwards, some physicists realized that these functions naturally appear as coefficients of genus 1 open superstring amplitudes \cite{BMMS}. 

We have mentioned in Chapter \ref{ChapterMathBackground} that multiple zeta values can be seen as the coefficients of the Drinfel'd associator, a power series in two non-commutative variables which describes the regularized monodromies of the KZ differential equation (\ref{KZequation}). Enriquez's elliptic analogue of the Drinfel'd associator is given by a pair of power series $(A(x_0,x_1;\tau),B(x_0,x_1;\tau))$ in two non-commutative variables $x_0,x_1$ describing the regularized monodromies of an elliptic analogue of the KZ equation related to the KZB form (\ref{KZB-form2}), called the \emph{KZB differential equation}\footnote{KZB is the acronym of Knizhnik-Zamolodchikov-Bernard.}, that we will not discuss here. The upshot is that (a slightly modified version\footnote{See \cite{MatthesThesis}.} of) the elliptic associator can be defined as
\begin{equation}\label{Aass}
A(x_0,x_1,\tau)=\lim_{\epsilon\rightarrow 0}(-2\pi i\epsilon)^{\mbox{ad}_{x_0}(x_1)}\exp\bigg[\int_{\epsilon}^{1-\epsilon}\omega_{KZB}(x_0,x_1;\xi,\tau)\bigg](-2\pi i\epsilon)^{-\mbox{ad}_{x_0}(x_1)},
\end{equation}
\begin{equation}\label{Bass}
B(x_0,x_1,\tau)=\lim_{\epsilon\rightarrow 0}(-2\pi i\epsilon\tau)^{\mbox{ad}_{x_0}(x_1)}\exp\bigg[\int_{\epsilon\tau}^{(1-\epsilon)\tau}\omega_{KZB}(x_0,x_1;\xi,\tau)\bigg](-2\pi i\epsilon\tau)^{-\mbox{ad}_{x_0}(x_1)}.
\end{equation}
One can show (see \cite{MatthesThesis}, \cite{HainItInt}) that these two limits exist and are finite. In order to define elliptic MZVs, Enriquez\footnote{Enriquez presented these generating series in a different way \cite{Enriquez}. Here we prefer to follow \cite{MatthesThesis}, Definition 3.4.1, but we warn the reader of a typo in formula (3.48) therein, which is corrected in our formula (\ref{BassEnr}).} considered the following modified formal series:
\begin{equation}\label{AassEnr}
A^{Enr}(x_0,x_1,\tau)=\lim_{\epsilon\rightarrow 0}(-2\pi i\epsilon)^{\mbox{ad}_{x_0}(x_1)}\exp\bigg[\int_{\epsilon}^{1-\epsilon}\mbox{ad}_{x_0}\Omega (\xi,\mbox{ad}_{x_0},\tau)(x_1)d\xi\bigg](-2\pi i\epsilon)^{-\mbox{ad}_{x_0}(x_1)},
\end{equation}
\begin{multline}\label{BassEnr}
B^{Enr}(x_0,x_1,\tau)=\\
=\lim_{\epsilon\rightarrow 0}(-2\pi i\epsilon)^{\mbox{ad}_{x_0}(x_1)}\exp\bigg[\int_{\epsilon\tau}^{(1-\epsilon)\tau}\mbox{ad}_{x_0}\Omega (\xi,\mbox{ad}_{x_0},\tau)(x_1)d\xi\bigg](-2\pi i\epsilon)^{-\mbox{ad}_{x_0}(x_1)}.
\end{multline}
Once again, these two limits exist and are finite. In analogy with the genus zero case, $A^{Enr}(x_0,x_1,\tau)$ and $B^{Enr}(x_0,x_1,\tau)$ can be considered as the generating series of two families of functions on the complex upper-half plane: Enriquez called them \emph{elliptic analogues of multiple zeta values}. 

It is important to remark, as we already mentioned in the introduction, that Enriquez defined them as the coefficients (with respect to $\mbox{ad}^{n_1}_{x_0}(x_1)\cdots \mbox{ad}^{n_r}_{x_0}(x_1)$) of the modified pair $(A^{Enr}(x_0,x_1;\tau),B^{Enr}(x_0,x_1;\tau))$, while for instance Matthes considered the coefficients (with respect to monomials in the non-commutative variables $x_0$ and~$x_1$) of the elliptic associator $(A(x_0,x_1;\tau),B(x_0,x_1;\tau))$ \cite{MatthesThesis}. This second choice is in some sense more natural, because by Theorem \ref{ThBrLev} (extended to tangential base points) it gives \emph{homotopy invariant} iterated integrals on the paths $[0,1]$ and $[0,\tau]$. This happens to be the case also for the coefficients of (\ref{AassEnr}), because $\nu(\xi)\equiv 0$ on the straight path $[0,1]$, but it is not the case for the coefficients of (\ref{BassEnr}). However, since on the straight path $[0,\tau]$ we have $\nu(\xi)=d\xi/\tau$, Matthes's elliptic MZVs can be expressed as certain (homotopy invariant) $\mathbb{Q}[\tau^{\pm 1}]$-linear combinations of Enriquez's elliptic MZVs.

In this chapter we will not use most of what we have just said, and try to give a self-contained analytic theory of elliptic MZVs, without referring to the elliptic associator. In particular, we will deduce new explicit results on the asymptotic expansion and the modular behaviour of B-elliptic MZVs. At the end of the chapter, we will compare elliptic MZVs with special values of multi-valued and single-valued elliptic polylogarithms. 

\section{Definition and first properties}

\begin{defn}
Let $r\geq 1$. For $n_1,\ldots,n_r\in\mathbb{Z}_{\geq 0}$ with $n_1\neq 1$, $n_r\neq 1$\footnote{We have imposed this condition because $\omega_1$ has a pole at $\xi\in\tau\mathbb{Z}+\mathbb{Z}$, and therefore the integral would be divergent. It is explained below how to define them for $n_1= 1$ or $n_r= 1$.}, we call \emph{A-elliptic multiple zeta values}, or in short A-elliptic MZVs, the iterated integrals over the straight path $[0,1]$
\begin{equation}\label{Aell}
A(n_1,\ldots,n_r;\tau)=\int_{[0,1]}\omega_{n_1}(t_1,\tau)dt_1\cdots \omega_{n_r}(t_r,\tau)dt_r
\end{equation}
and \emph{B-elliptic multiple zeta values}, or B-elliptic MZVs,
\begin{equation}\label{BellDef}
B(n_1,\ldots,n_r;\tau)=\tau^{n_1+\cdots +n_r}\int_{[0,1]}\omega_{n_1}(\tau t_1,\tau)dt_1\cdots \omega_{n_r}(\tau t_r,\tau)dt_r.
\end{equation}
Moreover, one can see\footnote{This was noticed in \cite{BMMS}, and it is a particular case of a general result for iterated integrals with simple poles (see Lemma 3.175 in \cite{BurgosFresan}).} that for all $\textbf{n}=(n_1,\ldots ,n_r)\in (\mathbb{Z}_{\geq 0})^r$ and for small~$\epsilon$ there exist holomorphic functions $A_{k}(\n,\tau)$ and $B_{k}(\n,\tau)$ such that
\begin{equation}\label{RegA}
\int_{[\epsilon,1-\epsilon]}\omega_{n_1}(t_1,\tau)dt_1\cdots \omega_{n_r}(t_r,\tau)dt_r=I_{\epsilon}(\n;\tau)+\sum_{k=0}^rA_{k}(\n;\tau)\log(-2\pi i\epsilon)^k
\end{equation}
and
\begin{equation}\label{RegB}
\tau^{n_1+\cdots +n_r}\int_{[\epsilon,1-\epsilon]}\omega_{n_1}(\tau t_1,\tau)dt_1\cdots \omega_{n_r}(\tau t_r,\tau)dt_r=J_{\epsilon}(\n;\tau)+\sum_{k=0}^rB_{k}(\n;\tau)\log(-2\pi i\epsilon)^k,
\end{equation}
where $I_{\epsilon}(\n,\tau)$ and $J_{\epsilon}(\n,\tau)$ are $O(\epsilon^\nu)$ for some $\nu>0$ as $\epsilon\rightarrow 0$, and we choose the branch of the logarithm such that $2\log(\pm i)=\pm \pi i$. As we did in equation (\ref{regitint}), to deal with the divergent case with $n_1=1$ or $n_r=1$ we define (regularized) A-elliptic and B-elliptic MZVs by
\begin{equation}\label{AellDef2}
A(n_1,\ldots ,n_r;\tau):=A_{0}(n_1,\ldots ,n_r;\tau),
\end{equation}
\begin{equation}\label{BellDef2}
B(n_1,\ldots ,n_r;\tau):=B_{0}(n_1,\ldots ,n_r;\tau).
\end{equation}
When $n_1,n_r\neq 1$ equations (\ref{AellDef2}) and (\ref{BellDef2}) coincide with the definitions (\ref{Aell}) and (\ref{BellDef}); we have given the latter separately, because they are easier to handle. We will call $r$ the \emph{length} and $n_1+\cdots +n_r+r$ the \emph{weight} of the elliptic MZVs. We also want to define the \emph{depth} as the number of non-zero entries of the tuple $\n=(n_1,\ldots ,n_r)$.
\end{defn}
\begin{remark}
As in the case of classical MZVs, these regularized iterated integrals satisfy the shuffle relations \cite{BMS}
\begin{equation}\label{ShuffleEllMZV}
A(n_1,\ldots ,n_r;\tau)A(n_{r+1},\ldots ,n_{r+s};\tau)=\sum_{\sigma\in\Sigma (r,s)}A(n_{\sigma^{-1}(1)}\ldots ,n_{\sigma^{-1}(r+s)};\tau),
\end{equation}
\[
B(n_1,\ldots ,n_r;\tau)B(n_{r+1},\ldots ,n_{r+s};\tau)=\sum_{\sigma\in\Sigma (r,s)}B(n_{\sigma^{-1}(1)}\ldots ,n_{\sigma^{-1}(r+s)};\tau),
\]
and therefore one can equivalently define regularized elliptic MZVs by giving initial conditions (for instance, knowing $A(1,0;\tau)$ and $B(1,0;\tau)$ is enough for the depth one case, as we will see later) using this regularization recipy, and then recursively using the shuffle product, i.e. they can be seen as \emph{shuffle-regularized elliptic MZVs}.
\end{remark}
\begin{remark}
Let us mention how our elliptic multiple zeta values are related to the coefficients of (\ref{AassEnr}) and (\ref{BassEnr}) considered by Enriquez. If we write (following \cite{MatthesThesis})
\[
A^{Enr}(x_0,x_1,\tau)=\sum_{r\geq 0}\sum_{n_1,\ldots ,n_r\geq 0}I^A(n_1,\ldots ,n_r;\tau)\,\mbox{ad}^{n_r}_{x_0}(x_1)\cdots \mbox{ad}^{n_1}_{x_0}(x_1),
\]
\[
B^{Enr}(x_0,x_1,\tau)=\sum_{r\geq 0}\sum_{n_1,\ldots ,n_r\geq 0}I^B(n_1,\ldots ,n_r;\tau)\,\mbox{ad}^{n_r}_{x_0}(x_1)\cdots \mbox{ad}^{n_1}_{x_0}(x_1),
\]
then one can show\footnote{Recall that we have different conventions both for iterated integrals and for powers of $2\pi i$: taking care of this, for $n_1,n_r\neq 1$ the proof is straightforward. The cases where $n_1=0$ or $n_r=0$ can be deduced as follows. First one must show the equality in length two: we will soon see how to explicitly work this case out from our definition, and an explicit expression for Enriquez's length two $I^A$ and $I^B$ is already available in the literature \cite{BMS}. Using that both regularizations satisfy the shuffle-product gives us the equality for any length.} that
\[
A(n_1,\cdots ,n_r;\tau)=(2\pi i)^{r-(n_1+\cdots +n_r)}I^A(n_r,\cdots ,n_1;\tau),
\]
\[
B(n_1,\cdots ,n_r;\tau)=\bigg(\frac{2\pi i}{\tau}\bigg)^{r-(n_1+\cdots +n_r)}I^B(n_r,\cdots ,n_1;\tau),
\]
This dictionary may be useful for a comparison of our results with those obtained in \cite{MatthesThesis}, \cite{BMMS}, \cite{BMS}.
\end{remark}
One of the main reasons for our normalizations of B-elliptic MZVs is the following:
\begin{lemma}\label{lemmaTS}
\begin{eqnarray}
A(n_1,\ldots,n_r;\tau+1)&=&A(n_1,\ldots,n_r;\tau)\\
A(n_1,\ldots,n_r;-1/\tau)&=&B(n_1,\ldots,n_r;\tau),\label{modS}
\end{eqnarray}
\end{lemma}
\textbf{Proof.} To prove the first assertion, just note that the functions $\omega_n$'s are replaced on $[\epsilon,1-\epsilon]$ by the $f_n$'s, because if $\xi=t\in\mathbb{R}$, then $r_{\tau}(\xi)=\Im(\xi)/\Im(\tau)=0$. Since all $f_n$'s are invariant under $T:\tau\rightarrow\tau+1$, so are their iterated integrals over $[\epsilon,1-\epsilon]$. The second assertion is also easy, and follows from the modular behaviour (\ref{modular}).\\
$\square$

This means that B-elliptic MZVs (with our normalization) are nothing but the image of A-elliptic MZVs under the involution $S:\tau\mapsto -\tau^{-1}$. Of course the picture given by Lemma \ref{lemmaTS} is not complete: we do not know what are the modular properties of B-elliptic MZVs. We will come back to this in Section \ref{SectionDepth1Mod}.

Let us now see a consequence\footnote{Enriquez's result connects explicitly the elliptic associator with the Drinfel'd associator. Since we prefer to keep things more elementary, we prefer to state and prove this weaker result.} of one of the main results deduced in Enriquez's paper~\cite{Enriquez}:
\begin{prop}[Enriquez]\label{asymptA}
Let $\textbf{n}=(n_1,\ldots ,n_r)$. Then
\begin{equation}\label{A-expansionCh5}
A(\n;\tau)=\sum_{j\geq 0}a_j(\n)q^j,
\end{equation}
where $a_j(\n)\in\mathcal{A}[(2\pi i)^{\pm 1}]$. Moreover, if\footnote{We will see that B-elliptic MZVs with $n_1=1$  or $n_r=1$ have an expansion involving also $\log(\tau)$. We will mention in Section \ref{SectionDepth1Mod} an explicit way to modify depth one B-elliptic MZVs in order to get rid of $\log(\tau)$. A systematic way in any depth consists in expanding eq. (\ref{RegB}) with respect to $\log(-2\pi i\epsilon\tau)$ instead of $\log(-2\pi i\epsilon)$: this different regularization was already considered in a first version of \cite{Enriquez}, but has the disadvantage of breaking down the modular behaviour (\ref{modS}) when $n_1=1$ or $n_r=1$.} $n_1, n_r \neq 1$ we have
\begin{equation}\label{B-expansionEnr}
B(\n;\tau)=\sum_{i\in\mathbb{Z}}\sum_{j\geq 0}b_{i,j}(\n)\tau^iq^j,
\end{equation}
where  $b_{i,j}(\n)\in\mathcal{A}[(2\pi i)^{\pm 1}]$ and for every fixed $j$ all but finitely many $b_{i,j}$ are zero.
\end{prop}
\textbf{Sketch of the proof.} The proof is a direct consequence of the definition. We have already mentioned that for A-elliptic MZVs the functions $\omega_n$'s are replaced by the $f_n$'s, while for B-elliptic MZVs, again because $t\in\mathbb{R}$, we have $r_{\tau}(\tau t)=\Im(\tau t)/\Im(\tau)=t$. One can check\footnote{It is easier to check this computation after looking at the proofs of the results obtained in the next section.} that these facts, together with the explicit developments (\ref{explicitformulas2}), imply the statement for all cases of convergent elliptic MZVs (i.e. $n_1\neq 1$ and $n_r\neq 1$). We will see in Section \ref{SectionDepthOne} that also the expansion of $A(1;\tau)$, $A(1,n;\tau)$ and $A(n,1;\tau)$ is of the form (\ref{A-expansionCh5}). Using this and the shuffle product (\ref{ShuffleEllMZV}) we get the full statement.\\
$\square$

We will see in Section \ref{SectionAsym} that for B-elliptic MZVs such that $n_i\neq 1$ for all~$i$ we can say something more precise than (\ref{B-expansionEnr}).

We conclude this section by mentioning that there are many linear and algebraic relations among elliptic MZVs. In particular, the following result was proven in~\cite{BMS}:
\begin{lemma}\label{LemmaInterestingBoring}
Suppose that the weight $n_1+\cdots +n_r+r$ is even. Then 
$A(\n;\tau)$ can be written as linear combination of products of A-elliptic MZVs of shorter length $r$. The same hold in the B-elliptic case.
\end{lemma}
\textbf{Idea of the proof.} The proof of this lemma is a simple consequence of the shuffle product of elliptic multiple zeta values and the inversion formula (\ref{inversion}) given in the next section.\\
$\square$

\section{Constant terms of elliptic MZVs}\label{sectionConstant}

Let us denote $A(\n;\tau)=A^{\infty}(\n)+A^0(\n;\tau)$ and, for $n_1,n_r\neq 1$, $B(\n;\tau)=B^{\infty}(\n;\tau)+B^0(\n;\tau)$, where, using the expansions (\ref{A-expansionCh5}) and (\ref{B-expansionEnr}), we define
\[
A^{\infty}(n):=a_0(\n),
\]
\[
B^{\infty}(n;\tau):=\sum_{i\in\mathbb{Z}}b_{i,0}(\n)\tau^i,
\] 
and the latter sum is finite, by Proposition \ref{asymptA}. It follows that $A^0(\n;\tau)=O(q)$ and $B^0(\n;\tau)=O(q)$ (for $q\mapsto 0$). We will call $B^{\infty}(n;\tau)$ the \emph{Laurent polynomial part} of $B^{\infty}(n;\tau)$. As we have mentioned in the introduction of this chapter, Enriquez found a way to relate constant terms of A-elliptic and B-elliptic MZVs to the Drinfel'd associator, which we defined in Chapter \ref{ChapterMathBackground}. This result in principle leads to the possibility of computing all $A^{\infty}(n;\tau)$ and $B^{\infty}(n;\tau)$ in terms of multiple zeta values. However, formulae are not explicit, and an implementation on the computer has a very bad running time, in particular for B-elliptic MZVs: I was recently informed by the authors of \cite{BMS} that it may take more than one hour to compute $B^{\infty}(n_1,\ldots ,n_r;\tau)$ already when $n_1+\cdots n_r+r= 10$. The goal of this section is to work out explicit formulae allowing for a very quick implementation, without referring to associators.

\subsection{Length one}

It is instructive to see first what happens in the almost trivial case of length one. Equation (\ref{omegaodd}), together with the change of variable $s=1-t$ and equation (\ref{elliptic}), gives
\begin{equation}\label{a1}
A(n;\tau)=\int_{[0,1]}\omega_n(t,\tau)dt=\int_{[0,1]}\omega_n(1-s,\tau)ds=\int_{[0,1]}\omega_n(-s,\tau)ds=(-1)^nA(n;\tau),
\end{equation}
which implies that $A(n;\tau)=0$ for every odd $n>1$. This is a special instance of the more general fact, proven exactly in the same way, that 
\begin{equation}\label{inversion}
A(n_1,\ldots ,n_r;\tau)=(-1)^{n_1+\cdots +n_r}A(n_r,\ldots ,n_1;\tau).
\end{equation}
Moreover, this symmetry is easily seen to imply that also $A(1;\tau)=0$. We can then deduce, using (\ref{modS}), that also $B(n;\tau)=0$ for $n$ odd. One can deduce this also by noting that
\begin{eqnarray*}
B(n;\tau)&=&\tau^n\int_{[0,1]}\omega_n(\tau t,\tau)dt=\tau^n\int_{[0,1]}\omega_n(\tau-\tau s,\tau)ds\\
&=&\tau^n\int_{[0,1]}\omega_n(-\tau s,\tau)=(-1)^nB(n;\tau)
\end{eqnarray*}
It is an easy exercise to check directly using formula (\ref{explicitformulas2}) that for every $n\neq 1$
\begin{equation}
A(n;\tau)=\frac{2\pi i\B_n}{n!},
\end{equation}
where $\B_n$ is the $n$-th Bernoulli number. Moreover, we can use again (\ref{modS}) to get that also
\begin{equation}
B(n;\tau)=\frac{2\pi i\B_n}{n!}.
\end{equation}
It is however a really instructive exercise to compute the Laurent polynomial part $B^{\infty}(n;\tau)$ of $B(n;\tau)$, which in this case coincides with $B(n;\tau)$ itself, just by performing the integral defining $B$. First of all it is useful to recall that, for $m\in\mathbb{Z}_{\geq 0}$ and $\alpha\in\mathbb{C}$, integrating by parts leads to the formula
\begin{equation}\label{intbyparts}
\int_{[0,1]}t^m e^{-\alpha t}=\frac{m!}{\alpha^{m+1}}-\sum_{j=0}^m\frac{(m)_j}{\alpha^{j+1}}e^{-\alpha},
\end{equation}
where $(m)_j:=m!/(m-j)!$ is the \emph{descending Pochhammer symbol}. We need now to understand from which terms in the $q$-expansion of $\omega_n(\xi,\tau)=\sum_{k=0}^n\frac{r_{\tau}(\xi)^k}{k!}f_{n-k}(\xi,\tau)$ we can extract contributions to $B^{\infty}(n;\tau)$. There are two distinct kinds of contributions (this is a general principle that applies to every length $r$ B-elliptic MZVs). The obvious one, that we will call $B^{\infty,1}(n;\tau)$, is given by considering the left hand side part of the developments (\ref{explicitformulas2}):
\begin{multline}
\tau^n\int_{[0,1]}\bigg((2\pi i)\frac{t^n}{n!}+\frac{t^{n-1}}{(n-1)!}\pi\cot (\pi\tau t)+\sum_{k=2}^n(2\pi i)\frac{\B_{k}}{k!}\frac{t^{n-k}}{(n-k)!}\bigg)dt\\
=\tau^n\sum_{k=0}^n(2\pi i)\frac{\B_{k}}{k!(n-k+1)!}-(-1)^n\frac{\zeta(n)}{(2\pi i)^{n-1}}+O(q)=:B^{\infty,1}(n;\tau)+O(q).
\end{multline}
In the computation above we have used that
\begin{equation}\label{cotangent}
\pi\cot (\pi\tau t)=-\pi i -2\pi i\sum_{h\geq 1}\e(h\tau t),
\end{equation}
and we have performed the integration using formula (\ref{intbyparts}). The second contribution is given by the $p=1$-part of the right hand side terms of the $f_n$'s. In fact, when $p=1$, one is left with $q$-expansions (we do not take into account the part already considered before) of the kind
\begin{equation}\label{inininini}
-\frac{2\pi i}{k!(n-k-1)!}\sum_{m\geq 1}\Big(\int_{[0,1]}t^k(\e(m\tau t)+(-1)^{n-k}\e(-m\tau t))dt\Big)q^m.
\end{equation}
Each of them will then contribute to a non-exponentially small term $B^{\infty,2}(n;\tau)$, because all the integrals in (\ref{inininini}) can be computed using formula (\ref{intbyparts}) and evaluate to
\begin{equation*}
-(-1)^{n-k}q^{-m}\sum_{j=0}^k\frac{(k)_j}{(2\pi im\tau)^{j+1}}+O(q^{-(m-1)}),
\end{equation*}
which gives as final answer
\begin{equation*}
B^{\infty,2}(n;\tau)=2\pi i\tau^n\sum_{k=0}^{n-1}\frac{(-1)^{n-k}}{k!(n-k-1)!}\sum_{j=0}^k\frac{(k)_j}{(2\pi im\tau)^{j+1}}.
\end{equation*}
Therefore to show that $B^{\infty}(n;\tau)=2\pi i\B_n/n!$, it is enough to prove the following two identities:
\begin{equation}\label{id1}
\sum_{k=0}^n\frac{\B_{k}}{k!(n-k+1)!}=0
\end{equation}
and
\begin{equation}\label{id2}
\sum_{k=0}^{n-1}\frac{(-1)^{n-k}}{k!(n-k-1)!}\sum_{j=0}^k\frac{k!}{(k-j)!}X^{j+1}=-X^n.
\end{equation}
Formula (\ref{id1}), which is sometimes used as the definition of Bernoulli numbers, is obtained by multiplying their generating series (\ref{BernoulliNumDef}) by $e^t-1$. To prove (\ref{id2}), one should note that for every $1\leq i\leq n$, the coefficient of $X^i$ on the left hand side is always equal to (setting $l=k-i$)
\begin{equation*}
-\sum_{l=0}^{n-i}\frac{(-1)^{n-i-l}}{l!(n-i-l)!}=-\frac{(1-1)^{n-i}}{(n-i)!}=\delta_{n,i}.
\end{equation*} 
This concludes our explicit computation of $B(n;\tau)$. 

\subsection{Length two}

As a concrete example of the proof of Lemma \ref{LemmaInterestingBoring}, let us remark that using shuffle product and (\ref{inversion}) we get
\begin{equation}
A(n_1,n_2;\tau)+(-1)^{n_1+n_2}A(n_1,n_2;\tau)=A(n_1;\tau)A(n_2;\tau)
\end{equation}
and 
\begin{equation}
B(n_1,n_2;\tau)+(-1)^{n_1+n_2}B(n_1,n_2;\tau)=B(n_1;\tau)B(n_2;\tau).
\end{equation}
This implies that the only new interesting length two elliptic MZVs need $n_1$ and $n_2$ to be of opposite parity, which is obviously a special case of Lemma \ref{LemmaInterestingBoring}. We start with the trivial computation of $A^{\infty}(n_1,n_2)$, that has been already given for instance in  \cite{BMS}. Let $n_1, n_2\neq 1$, then
\begin{equation}\label{Ainf2}
A^{\infty}(n_1,n_2)=\int_0^1\frac{2\pi i\B_{n_1}}{n_1!}\int_0^t\frac{2\pi i\B_{n_2}}{n_2!}dsdt=\frac{-2\pi^2\B_{n_1}\B_{n_2}}{n_1!n_2!}
\end{equation}
Moreover, in order to compute $A^{\infty}(1,n)$ for even $n\geq 0$, we need to compute the constant term (with respect to $\epsilon$ and $q$) of
\begin{equation*}
\frac{2\pi i\B_{n}}{n!}\int_{\epsilon}^{1-\epsilon}tf_1(t,\tau)dt,
\end{equation*}
and a simple computation using the expansion of the cotangent (\ref{cotangent}) shows that the only constant term of the $q$-expansion not depending on $\epsilon$ is  given by
\begin{equation}\label{AInf1n}
A^{\infty}(1,n)=\frac{2\pi i\B_{n}}{n!}\frac{\pi i}{2}.
\end{equation}
Let us turn now our attention to the more interesting $B^{\infty}(n_1,n_2;\tau)$.
\begin{prop}
Let $n\geq 2$. Then we have
\begin{equation}
B^{\infty}(n,0;\tau)=\frac{\B_{n+1}}{(n+1)!}(2\pi i)^2\tau^n-\frac{\zeta(n)}{(2\pi i)^{n-2}}+\frac{n\B_{n+1}}{(n+1)!}\frac{(2\pi i)^2}{\tau}.
\end{equation}
\end{prop}
\textbf{Proof.} One just has to repeat for this case the same steps already seen in the computations of $B^{\infty}(n;\tau)$. The result for the first contribution is
\begin{eqnarray*}
B^{\infty,1}(n,0;\tau)&=&(2\pi i)\tau^n\sum_{k=0}^n\frac{n-k+1}{(n-k+2)!}\frac{\B_{k}}{k!}+(-1)^n\frac{n\zeta(n+1)}{(2\pi i)^n}\tau^{-1}\\
&=&\frac{\B_{n+1}}{(n+1)!}(2\pi i)\tau^n+(-1)^n\frac{n\zeta(n+1)}{(2\pi i)^n}\tau^{-1},
\end{eqnarray*}
where in the second equality we have used the formula
\begin{equation}\label{id3}
\sum_{k=0}^n\frac{\B_k(n-k+1)}{k!(n-k+2)!}=\frac{\B_{n+1}}{(n+1)!},
\end{equation}
which is obtained by rewriting the left hand side as
\begin{equation*}
\sum_{k=0}^n\frac{\B_k}{k!(n-k+1)!}-\sum_{k=0}^n\frac{\B_k}{k!(n-k+2)!}
\end{equation*}
and then applying (\ref{id1}) to both terms. The second contribution, coming from the $p=1$ terms in (\ref{explicitformulas2}), is given by
\begin{eqnarray*}
B^{\infty,2}(n,0;\tau)&=&(2\pi i)\tau^n\sum_{m\geq 1}\sum_{k=0}^{n-1}\frac{(-1)^{n-k}(k+1)}{k!(n-k-1)!}\sum_{j=0}^{k+1}\frac{(2\pi im\tau)^{-j-1}}{(k+1-j)!}\\
&=&-\frac{\zeta(n)}{(2\pi i)^{n-1}}-\frac{n\zeta(n+1)}{(2\pi i)^n}\tau^{-1},
\end{eqnarray*}
where the last equality comes from the identity
\begin{equation*}
\sum_{k=0}^{n-1}\frac{(-1)^{n-k}(k+1)}{k!(n-k-1)!}\sum_{j=0}^{k+1}\frac{X^{j+1}}{(k+1-j)!}=-X^n-nX^{n+1},
\end{equation*}
which can be proven by noting that for every $1\leq i\leq n+1$ the coefficient of $X^i$ on the left hand side is
\[
-(i-1)\sum_{l=0}^{n-i+1}\frac{(-1)^{n-i+1-l}}{l!(n-i+1-l)!}-\sum_{l=0}^{n-i}\frac{(-1)^{n-i-l}}{l!(n-i-l)!},
\]
and therefore gives the same coefficient of $X^i$ on the right hand side. Putting everything together we get the Laurent polynomial predicted in the statement.\\
$\square$

It is interesting to remark that, in the B-elliptic case, one can obtain odd zeta values already at length $r=2$. Later we will see that this is related with the appearence of odd zeta values in the period polynomials of Eisenstein series for SL$_2(\mathbb{Z})$. In the next section we will see how one obtains odd zeta values in the A-elliptic case, as well as higher length multiple zeta values for both the~A and~B case. We conclude this section here, since Corollary \ref{CorDouble} will show that having a formula for $B^{\infty}(n,0;\tau)$ is enough to be able to compute all $B^{\infty}(n_1,n_2;\tau)$.

\subsection{The general case}
We want now to take a look to the asymptotic expansion of less trivial families of elliptic multiple zeta values.
Looking at the formulae (\ref{explicitformulas2}), it is clear that every $f_n$ with $n\neq 1$ does not give any interesting contribution to the constant term of A-elliptic MZVs (just rational numbers and powers of $2\pi i$). Just for completeness, we report here the general formula when every $n_i\neq 1$, already given in \cite{BMS}:
\begin{equation}\label{GeneralAinf}
A^{\infty}(n_1,\ldots ,n_r)=\frac{(2\pi i)^r}{r!}\prod_{i=1}^r\frac{\B_{n_i}}{n_i!}.
\end{equation}
Let us now focus on some cases where $n_i=1$ for some $i$.
\begin{prop}\label{Ainf010}
The constant term of $A(\underbrace{0,\ldots,0}_{n-1},1,0;\tau)$ is given by
\begin{equation}
A^{\infty}(\underbrace{0,\ldots,0}_{n-1},1,0)=-2\pi in\zeta^{\rm odd}(n)-\sum_{j=2}^{n-2}\frac{n-j}{j!}(2\pi i)^{j}\zeta^{\rm odd}(n+1-j),
\end{equation}
where $\zeta^{\rm odd}(n)=\zeta(n)$ for $n$ odd and $=0$ otherwise.
\end{prop}
\textbf{Proof.}
$A^{\infty}(\underbrace{0,\ldots,0}_{n-1},1,0)$ is given by
\[
(2\pi i)^n\int_{[0,1]}dt_1\cdots dt_{n-1}\pi t_n\cot (\pi t_n)dt_n.
\]
Using the expression (\ref{cotangent}) for the cotangent as well as formula (\ref{intbyparts}) one eventually gets that
\[
A^{\infty}(\underbrace{0,\ldots,0}_{n-1},1,0)=-\frac{(2\pi i)^{n+1}}{2(n+1)!}-2\pi in\zeta(n)-\sum_{j=2}^{n-1}\frac{n-j}{j!}(2\pi i)^j\zeta(n-j+1).
\]
The statement of the proposition can be deduced using the following identities:
\[
\sum_{k=0}^n\frac{(k-1)\B_k}{k!(n+1-k)!}=-\frac{\B_n}{n!}
\]
for $n\geq 2$ and
\[
\sum_{k=0}^n\frac{(k-1)\B_k}{k!(n+2-k)!}=-\frac{\B_{n+1}}{n!}
\]
for $n\geq 1$. Both identities rely on the same kind of trick already used to get (\ref{id3}).\\
$\square$

The following corollary is equivalent to a proposition already given (without proof) in \cite{BMS}
\begin{cor}\label{CorA100}
\begin{equation}
A^{\infty}(\underbrace{0,\ldots ,0}_n,1;\tau)=(-2\pi i)^{n}\Big(\frac{i\pi}{2\,n!}-\sum_{k=1}^{\lfloor \frac{(n+1)}{2}\rfloor -1}\frac{\zeta(2k+1)}{(n-2k)!(2\pi i)^{2k}}\Big)
\end{equation}
\end{cor}
\textbf{Proof.} By (\ref{inversion}) we have that $A^{\infty}(1,0;\tau)=-A^{\infty}(0,1;\tau)$. The result follows by induction on $n$, using (\ref{AInf1n}), shuffle product and Proposition \ref{Ainf010}.\\
$\square$

In general, one can prove using associators (see \cite{MatthesThesis}, Th. 5.4.2) that, up to powers of $2\pi i$, all multiple zeta values arise as constant terms of A-elliptic MZVs. One can see this considering all A-elliptic MZVs of the kind
\[
A(1,\underbrace{0,\ldots ,0}_{n_1},\ldots ,1,\underbrace{0,\ldots ,0}_{n_r};\tau).
\]
An example, borrowed from \cite{MatthesThesis}, is the following:
\begin{multline}
A(1,\underbrace{0,\ldots ,0}_3,1,\underbrace{0,\ldots ,0}_5;\tau)=2\pi i\Big(\zeta(5,3)+2\zeta(2)\zeta(3)^2+6\pi i\zeta(3)\zeta(4)\\
-12\pi i\zeta(2)\zeta(5)-\zeta(3)\zeta(5)+\frac{21\pi i}{2}\zeta(7)+10\zeta(8)\Big).
\end{multline}
Let us now turn our attention to the B-elliptic case.
\begin{prop}\label{DepthOneBinf}
Let $n\geq 2$. Then we have
\begin{multline}
B^{\infty}(n,\underbrace{0,\ldots,0}_r;\tau)=\frac{(2\pi i)^{r+1}\tau^n}{r!}\sum_{k=0}^{n}\frac{\B_k}{k!(n-k)!(n-k+r+1)}\\
-\sum_{p=0}^{r-1}\frac{1}{(r-p)!}\binom{n+p-1}{p}\frac{\zeta(n+p)}{(2\pi i)^{n+p-r-1}}\tau^{-p}
-(1+(-1)^{n+r})\binom{n+r-1}{r}\frac{\zeta(n+r)}{(2\pi i)^{n-1}}\tau^{-r}.
\end{multline}
\end{prop}
\textbf{Proof.} Since $\omega_0(\tau t,\tau)=2\pi i$, what we need to compute is the non-exponentially small part of the single integral
\begin{equation}
\frac{(2\pi i)^r}{r!}\tau^n\int_{0}^1t^r\omega_{n}(\tau t,\tau)dt.
\end{equation}
As we have seen before, we need to compute two different kinds of contributions $B^{\infty,1}$ and $B^{\infty,2}$. It is now an easy exercise to see, repeating the computations already done in length one and two, that
\begin{multline}
B^{\infty,1}(n,\underbrace{0,\ldots,0}_r;\tau)=\frac{(2\pi i)^{r+1}}{r!}\sum_{k=0}^{n}\frac{\B_k}{k!(n-k)!(n-k+r+1)}\tau^n\\
-(-1)^{n+r}\binom{n+r-1}{r}\frac{\zeta(n+r)}{(2\pi i)^{n-1}}\tau^{-r}
\end{multline}
and that
\begin{equation}\label{Binf2}
B^{\infty,2}(n,\underbrace{0,\ldots,0}_r;\tau)=\frac{(2\pi i)^{r+1}}{r!}\tau^n\sum_{m\geq 1}\sum_{k=0}^{n-1}\frac{(-1)^{n-k}}{k!(n-k-1)!}\sum_{j=0}^{k+r}\frac{(k+r)_j}{(2\pi im\tau)^{j+1}}.
\end{equation}
To conclude the proof we need to rewrite the term $B^{\infty,2}$ using the following
\begin{lemma}
\begin{equation}\label{eqlem}
\sum_{k=0}^{n-1}\frac{(-1)^{n-k}}{k!(n-k-1)!}\sum_{j=0}^{k+r}\frac{(k+r)!}{(k+r-j)!}X^{j+1}=-\sum_{p=0}^r\binom{r}{p}\frac{(n+p-1)!}{(n-1)!}X^{n+p}
\end{equation}
\end{lemma}
To prove this identity, let us first recall the \emph{Chu-Vandermonde identity}, valid for $r,k,m\in\mathbb{Z}_{\geq 0}$:
\begin{equation}\label{vandermonde}
\sum_{p=0}^m\binom{r}{p}\binom{k}{m-p}=\binom{k+r}{m}.
\end{equation}
This is proven by comparing the coefficients of $(1+x)^{k+r}=(1+x)^k(1+x)^r$. Now note that for any fixed $1\leq i\leq n+r$ the coefficient of $X^i$ on the left hand side of (\ref{eqlem}) is given by
\begin{equation}\label{coeffXi}
\sum_{k=i-1-r}^{n-1}\frac{(-1)^{n-k}(k+r)_r}{(n-k-1)!(k+1+r-i)!}.
\end{equation}
Identity (\ref{vandermonde}) implies that for any $i\geq r+1$
\begin{equation}
(k+r)_r=\sum_{p=0}^r\binom{r}{p}\frac{(i-1)!}{(i-1-p)!}\frac{(k+r-m)!}{(k+p-m)!},
\end{equation}
which in turns implies that for any $i\geq r+1$ we can rewrite (\ref{coeffXi}) as
\begin{multline}\label{finaleq}
\sum_{k=i-1-r}^{n-1}\frac{(-1)^{n-k}(k+r)_r}{(n-k-1)!(k+1+r-i)!}=\\
=\sum_{p=0}^{r}\binom{r}{p}\frac{(i-1)!}{(i-1-p)!}\sum_{k=i-1-p}^{n-1}\frac{(-1)^{n-k}}{(n-k-1)!(k+1+p-i)!}.
\end{multline}
This equality actually holds also for $1\leq i\leq r$, because $(i-1)_p$ is identically zero whenever $p>i-1$, and therefore one can just use (\ref{vandermonde}) with $m=i-1$ on the left hand side to get the terms of the right hand side with $p\leq i-1$. But then we are almost done, because we can rewrite the right hand side of (\ref{finaleq}) as
\begin{equation}
-\sum_{p=0}^{r}\binom{r}{p}\frac{(i-1)!}{(i-1-p)!}\sum_{k=0}^{n+p-i}\frac{(-1)^{n+p-i-k}}{k!(n+p-i-k)!}=-\sum_{p=0}^{r}\binom{r}{p}\frac{(i-1)!}{(i-1-p)!}\frac{(1-1)^{n+p-i}}{(n+p-i)!},
\end{equation}
and this proves identity (\ref{eqlem}), because $(1-1)^{n+p-i}\equiv 0$ unless $i=n+p$, in which case we get~1. The final step to prove the proposition simply consists in applying the lemma on (\ref{Binf2}) with $X=(2\pi im\tau)^{-1}$, and summing over $m\geq 1$.\\
$\square$

Note that, thanks to this result, we are now able to compute in less than a second terms like
\begin{equation*}
B^{\infty}(15,0)=-\frac{3617}{10670622842880000}(2\pi i)^2\tau^{15}-\frac{\zeta(15)}{(2\pi i)^{13}}-\frac{30\,\zeta(16)}{(2\pi i)^{14}\tau}
\end{equation*}
or 
\begin{multline*}
B^{\infty}(6,\underbrace{0,\ldots ,0}_8) =\frac{779}{7846046208000}(2\pi i)^6\tau^9-\frac{1}{120}\frac{\zeta(9)}{(2\pi i)^3}-\frac{3}{8}\frac{\zeta(10)}{(2\pi i)^4\tau}\\
-\frac{15}{2}\frac{\zeta(11)}{(2\pi i)^5\tau^2}-\frac{165}{2}\frac{\zeta(12)}{(2\pi i)^6\tau^3}-\frac{495\,\zeta(13)}{(2\pi i)^7\tau^4}-\frac{2574\,\zeta(14)}{(2\pi i)^8\tau^5}.
\end{multline*}
These computations were previously out of reach using methods based on Enriquez's results involving associators.

\section{Differential behaviour}

\subsection{Enriquez's differential equation}

In this section we want to recall from \cite{Enriquez} the behaviour of elliptic MZVs with respect to differentiation. To do this, for any fixed $r\geq 1$ we introduce the generating series of (regularized) elliptic MZVs
\begin{equation}
\mathcal{A}(X_1,\ldots,X_r;\tau):=\sum_{n_1,\ldots,n_r\in\mathbb{Z}_{\geq 0}}A(n_1,\ldots, n_r;\tau)X_1^{n_1-1}\cdots X_r^{n_r-1},
\end{equation}
\begin{equation}
\mathcal{B}(Y_1,\ldots,Y_r;\tau):=\sum_{n_1,\ldots,n_r\in\mathbb{Z}_{\geq 0}}B(n_1,\ldots, n_r;\tau)Y_1^{n_1-1}\cdots Y_r^{n_r-1}.
\end{equation}
We want now to introduce the following normalization of the Eisenstein series (\ref{EisSeries}):
\begin{equation}\label{modifiedEis}
\mathbb{G}_k(\tau):=\frac{1}{(2\pi i)^{k-1}}G_k(\tau)
\end{equation}
Moreover, let us consider the generating function
\[
\mathcal{G}(X,\tau):=\sum_{n\geq -1}n\mathbb{G}_{n+1}(\tau)X^{n-1}.
\]
Note that $\mathcal{G}(X,\tau):=\frac{1}{2\pi i}G_2(\frac{X}{2\pi i},\tau)$, where $G_i(\xi,\tau)$ are the functions define by (\ref{GenFunEisSeries}). 

One of the main results of Enriquez's paper \cite{Enriquez} is the following:
\begin{prop}[Enriquez]\label{PropDiffA}
\begin{multline}\label{diffeqA}
\frac{\partial}{\partial \tau}\mathcal{A}(X_1,\ldots,X_r;\tau)=\mathcal{G}(X_r,\tau)\mathcal{A}(X_1,\ldots,X_{r-1};\tau)-\mathcal{G}(X_1,\tau)\mathcal{A}(X_2,\ldots,X_r;\tau)\\
+\sum_{i=1}^{r-1} \left(\mathcal{G}(X_i,\tau)-\mathcal{G}(X_{i+1},\tau)\right)\mathcal{A}(X_1,\ldots,X_{i-1},X_i+X_{i+1},X_{i+2},\ldots,X_r;\tau).
\end{multline}
\end{prop}
\textbf{Sketch of the proof.} The main ingredient will be the mixed-heat equation (\ref{mixedheat}). First of all, one must prove the following:
\begin{lemma}\label{lemmaEnr}
We have
\begin{itemize}
\item[(i)] $\frac{\partial}{\partial X_i}F(0,X_i,\tau)=\frac{\partial}{\partial X_i}F(1,X_i,\tau)=-G_2(X_i,\tau)$.
\item[(ii)] $\Big(\frac{\partial}{\partial X_i}-\frac{\partial}{\partial X_j}\Big)F(\xi,X_i,\tau)F(\xi,X_j,\tau)=F(\xi,X_i+X_j,\tau)\Big(G_2(X_j,\tau)-G_2(X_i,\tau)\Big).$
\end{itemize}
\end{lemma}
\textbf{Proof of the Lemma.} By (\ref{Fsymm}) and (\ref{expansionF}) we have
\begin{equation}
\frac{\partial}{\partial X_i}F(0,X_i,\tau)=\sum_{n\geq 1}\frac{\partial}{\partial X_i}f_n(X_1,\tau)(2\pi i\xi)^{n-1}.
\end{equation}
Therefore $\frac{\partial}{\partial X_i}F(0,X_i,\tau)=\frac{\partial}{\partial X_i}f_1(X_i,\tau)$. We know that $f_1=\theta^\prime/\theta$, so using (\ref{propG3}), (\ref{propG2}) plus the transformation $F(\xi+1,\alpha,\tau)=F(\xi,\alpha,\tau)$ we get (i). We want now to prove (ii). Again by the elliptic behaviour of $F$ one can easily see that the left hand side transforms as the right hand side under $\xi\mapsto m\xi+n$. We want now to prove that their quotient is holomorphic, which thus by Liouville's theorem must be constant \cite{Lang}, and then we will prove that this constant is~1. To see that the quotient is holomorphic, we just need to show that the poles of the left hand side cancel with the poles of the right hand side. On the right hand side the polar part is given by
\begin{equation*}
\frac{1}{\xi}\Big(G_2(X_j,\tau)-G_2(X_i,\tau)\Big).
\end{equation*}
The left hand side can be written as
\begin{multline}
\mbox{l.h.s.}=F(\xi,X_i,\tau)F(\xi,X_j,\tau)\Bigg(\frac{\frac{\partial}{\partial X_i}F(\xi,X_i,\tau)}{F(\xi,X_i,\tau)}-\frac{\frac{\partial}{\partial X_j}F(\xi,X_j,\tau)}{F(\xi,X_j,\tau)}\Bigg)\\
=F(\xi,X_i,\tau)F(\xi,X_j,\tau)\bigg(\frac{\theta^\prime (\xi+X_i,\tau)}{\theta (\xi+X_i,\tau)}-\frac{\theta^\prime (X_i,\tau)}{\theta (X_i,\tau)}+\frac{\theta^\prime (\xi+X_j,\tau)}{\theta (\xi+X_j,\tau)}-\frac{\theta^\prime (X_j,\tau)}{\theta (X_j,\tau)}\bigg),
\end{multline}
because
\begin{multline*}
\frac{\theta (\xi,\tau)\theta (X_i,\tau)}{\theta^\prime (0,\tau)\theta (\xi+X_i,\tau)}\frac{\partial}{\partial X_i}\bigg(\frac{\theta^\prime (0,\tau)\theta (\xi+X_i,\tau)}{\theta (\xi,\tau)\theta (X_i,\tau)}\bigg)=\\
=\bigg(\frac{\theta^\prime (\xi+X_i,\tau)}{\theta (X_i,\tau)}-\frac{\theta^\prime (X_i,\tau)\theta (\xi+X_i,\tau)}{\theta (X_i,\tau)^2}\bigg)\frac{\theta (X_i,\tau)}{\theta (\xi+X_i,\tau)}.
\end{multline*}
Therefore the polar part of the left hand side is equal to
\begin{equation}
\frac{1}{\xi^2}\Bigg(\bigg(\frac{\theta^\prime (X_i,\tau)}{\theta (X_i,\tau)}\bigg)^\prime -\bigg(\frac{\theta^\prime (X_j,\tau)}{\theta (X_j,\tau)}\bigg)^\prime \Bigg)\xi=\frac{1}{\xi}(G_2(X_j,\tau)-G_2(X_i,\tau)),
\end{equation}
which coincides with the polar part of the right hand side. In the last equality we have used again (\ref{propG3}) and (\ref{propG2}). We just need to prove that the proportionality constant is~1, but this is just a consequence of the computation above.\\
$\square$

Now we want to apply the lemma and prove the statement of the proposition. Let us write
\begin{equation}
\mathcal{A}(X_1,\ldots,X_r;\tau)=\int^{reg}_{[0,1]}F\Big(\xi_1,\frac{X_1}{2\pi i},\tau\Big)d\xi_1\ldots F\Big(\xi_r,\frac{X_r}{2\pi i},\tau\Big)d\xi_r.
\end{equation}
By this we mean that we are considering the regularization explained in the definition of elliptic MZVs. We will skip the long and tedious proof that these regularized iterated integrals still satisfy all the properties of iterated integrals stated in Section~\ref{SectionItInt} (see Proposition 3.1 of \cite{Enriquez}). Let us write $\tilde{X}_i:=X_i/2\pi i$. By the mixed heat equation we have
\begin{multline}
\frac{\partial}{\partial \tau}\mathcal{A}(X_1,\ldots,X_r;\tau)=\sum_{i=1}^r\int^{reg}_{[0,1]}F(\xi_1,\tilde{X}_1,\tau)d\xi_1\ldots \frac{\partial}{\partial \tau}F(\xi_i,\tilde{X}_i,\tau)d\xi_i\ldots F(\xi_r,\tilde{X}_r,\tau)d\xi_r\\
=\frac{1}{2\pi i}\sum_{i=1}^r\int^{reg}_{[0,1]}F(\xi_1,\tilde{X}_1,\tau)d\xi_1\ldots \frac{\partial^2}{\partial \xi\partial\tilde{X}_i}F(\xi_i,\tilde{X}_i,\tau)d\xi_i\ldots F(\xi_r,\tilde{X}_r,\tau)d\xi_r.
\end{multline}
Making use of the properties of iterated integrals (\ref{ItIntParts1}), (\ref{ItIntParts2}), (\ref{ItIntParts3}), we write
\begin{multline}
2\pi i\frac{\partial}{\partial \tau}\mathcal{A}(X_1,\ldots,X_r;\tau)=\\
\frac{\partial}{\partial \tilde{X}_1}F(0,\tilde{X}_1,\tau)\mathcal{A}(X_2,\ldots,X_r;\tau)-\frac{\partial}{\partial \tilde{X}_r}F(0,\tilde{X}_r,\tau)\mathcal{A}(X_2,\ldots,X_r;\tau)\\
+\sum_{i=1}^{r-1}\Bigg(\int^{reg}_{[0,1]}F(\xi_1,\tilde{X}_1,\tau)d\xi_1\ldots F(\xi_i,\tilde{X}_{i},\tau)\frac{\partial}{\partial \tilde{X}_{i+1}}F(\xi_i,\tilde{X}_{i+1},\tau)d\xi_i\ldots F(\xi_{r-1},\tilde{X}_r,\tau)d\xi_{r-1}\\
-\int^{reg}_{[0,1]}F(\xi_1,\tilde{X}_1,\tau)d\xi_1\ldots \Big(F(\xi_i,\tilde{X}_{i+1},\tau)\frac{\partial}{\partial \tilde{X}_{i}}F(\xi_i,\tilde{X}_{i},\tau)\Big)d\xi_i\ldots F(\xi_{r-1},\tilde{X}_r,\tau)d\xi_{r-1}\Bigg),
\end{multline}
and using the two identities of the lemma this is equal to
\begin{multline*}
2\pi i\Big(\mathcal{G}(X_r,\tau)\mathcal{A}(X_1,\ldots,X_{r-1};\tau)-\mathcal{G}(X_1,\tau)\mathcal{A}(X_2,\ldots,X_r;\tau)\\
+\sum_{i=1}^{r-1} \left(\mathcal{G}(X_i,\tau)-\mathcal{G}(X_{i+1},\tau)\right)\mathcal{A}(X_1,,\ldots,X_{i-1},X_i+X_{i+1},X_{i+2},\ldots,X_r;\tau)\Big),
\end{multline*}
which concludes the proof.\\
$\square$

\begin{remark}
On the right hand side of (\ref{diffeqA}) we have poles of order higher than on the right hand side, but one can check that their contribution vanishes.
\end{remark}
Using (\ref{modS}), (\ref{diffeqA}) and taking into account the quasi-modular behaviour 
\begin{equation}
G_2\Big(\frac{a\tau+b}{c\tau+d}\Big)=(c\tau+d)^2G_2(\tau)-2\pi ic(c\tau+d),
\end{equation} 
one immediately gets
\begin{prop}\label{PropDiffB}
\begin{multline}\label{diffeqB}
\frac{\partial}{\partial \tau}\mathcal{B}(Y_1,\ldots,Y_r;\tau)=\\
\left(\mathcal{G}(\tau Y_r,\tau)-\frac{2\pi i}{\tau}\right)\mathcal{B}(Y_1,\ldots,Y_{r-1};\tau)-\left(\mathcal{G}(\tau Y_1,\tau)-\frac{2\pi i}{\tau}\right)\mathcal{B}(Y_2,\ldots,Y_r;\tau)\\
+\sum_{i=1}^{r-1} \left(\mathcal{G}(\tau Y_i,\tau)-\mathcal{G}(\tau Y_{i+1},\tau)\right)\mathcal{B}(Y_1,,\ldots,Y_{i-1},Y_i+Y_{i+1},Y_{i+2},\ldots,Y_r;\tau).
\end{multline}
\end{prop}
Comparing term by term the coefficients of all monomials in (\ref{diffeqA}) and in (\ref{diffeqB}), one gets
\begin{cor}\label{propExpl}
Every A-elliptic multiple zeta value satisfies the differential equation\footnote{Recall that, for $n\in\mathbb{Z}$ and $j\in\mathbb{Z}_{\geq 0}$, the definition of the binomial coefficient $\binom{n}{j}$ is $(n)_j/j!$, where $(n)_j$ is the descending Pochhammer symbol.} 
\begin{multline}\label{explDiffA}
\frac{\partial}{\partial\tau} A(n_1,\ldots,n_r;\tau) = n_r \mathbb{G}_{n_r+1} (\tau)A(n_1,\ldots,n_{r-1};\tau)-n_1 \mathbb{G}_{n_1+1}(\tau) A(n_2,\ldots,n_r;\tau) \\
+ \sum_{i=1}^{r-1} \Big( (-1)^{n_i} (n_{i}+n_{i+1}) \mathbb{G}_{n_{i}+n_{i+1}+1}(\tau) A(n_1,\ldots,n_{i-1},0,n_{i+2},\ldots,n_r;\tau) \\
+  \sum_{j=0}^{n_{i}+1} (n_{i}-j) { n_{i+1}+j-1 \choose j } \mathbb{G}_{n_{i}-j+1}(\tau) A(n_1,\ldots,n_{i-1},j+n_{i+1},n_{i+2},\ldots,n_r;\tau)\\
- \sum_{j=0}^{n_{i+1}+1} (n_{i+1}-j) { n_{i}+j-1 \choose j } \mathbb{G}_{n_{i+1}-j+1}(\tau) A(n_1,\ldots,n_{i-1},j+n_{i},n_{i+2},\ldots,n_r;\tau) \Big),
\end{multline}
and every B-elliptic multiple zeta value satisfies the differential equation
\begin{multline}\label{explDiffB}
\frac{\partial}{\partial\tau} B(n_1,\ldots,n_r;\tau) \\
= n_r\Big(\tau^{n_r-1}\mathbb{G}_{n_r+1}(\tau)-\frac{\delta_{n_r,1}}{\tau}\Big)B(n_1,\ldots,n_{r-1};\tau)-n_1\Big(\tau^{n_1-1}\mathbb{G}_{n_1+1}(\tau)-\frac{\delta_{n_1,1}}{\tau}\Big)B(n_2,\ldots,n_r;\tau) \\
+ \sum_{i=1}^{r-1} \Big( (-1)^{n_i} (n_{i}+n_{i+1})\tau^{n_i+n_{i+1}-1}\mathbb{G}_{n_{i}+n_{i+1}+1}(\tau) B(n_1,\ldots,n_{i-1},0,n_{i+2},\ldots,n_r;\tau) \\
+  \sum_{j=0}^{n_{i}+1} (n_{i}-j) { n_{i+1}+j-1 \choose j } \tau^{n_i-j-1}\mathbb{G}_{n_{i}-j+1}(\tau) B(n_1,\ldots,n_{i-1},j+n_{i+1},n_{i+2},\ldots,n_r;\tau)\\
- \sum_{j=0}^{n_{i+1}+1} (n_{i+1}-j) { n_{i}+j-1 \choose j } \tau^{n_{i+1}-j-1}\mathbb{G}_{n_{i+1}-j+1}(\tau) B(n_1,\ldots,n_{i-1},j+n_{i},n_{i+2},\ldots,n_r;\tau) \Big),
\end{multline}
where $\delta_{a,b}=1$ for $a=b$ and 0 otherwise.
\end{cor}
\textbf{Proof.} The equation (\ref{explDiffB}) for B-elliptic MZVs is easily deduced by (\ref{explDiffA}) using (\ref{modS}) and the modularity of the Eisenstein series, or else by comparing the monomials of (\ref{diffeqB}) in the same way as explained below for the A-elliptic case. Therefore we just need to get equation (\ref{explDiffA}) by comparing the coefficients of the monomials in (\ref{diffeqA}). This equation has been already presented in \cite{BMS}, but we could not find the details of the proof anywhere in the literature. In order to justify the algebraic manipulations needed when dealing with the poles $(X_i+X_{i+1})^{-1}$, we want to think of our variables $X_i$'s as complex variables lying in the region $|X_1|<\cdots <|X_r|$. One can show that the result is independent of this choice. We need to compute the coefficient of $X_1^{n_1-1}\cdots X_r^{n_r-1}$ on the right hand side of (\ref{diffeqA}). It is easy to see that the first two terms give
\begin{equation*}
n_r \mathbb{G}_{n_r+1} (\tau)A(n_1,\ldots,n_{r-1};\tau)-n_1 \mathbb{G}_{n_1+1}(\tau) A(n_2,\ldots,n_r;\tau).
\end{equation*}
Let us turn to the sum over the index $i$. For each fixed $1\leq i\leq r-1$ we have
\begin{multline*}
\mathcal{A}(X_1,\ldots,X_{i-1},X_i+X_{i+1},X_{i+2},\ldots,X_r;\tau)\\
=\sum_{(\mathbb{Z}_{\geq 0})^{r-1}}A(n_1,\ldots,n_{i-1},k,n_{i+2},\ldots,n_r;\tau)X_1^{n_1-1}\cdots (X_i+X_{i+1})^{k-1}\cdots X_r^{n_r-1}.
\end{multline*}
For every $k\geq 0$ we can write
\begin{equation}\label{pole}
(X_i+X_{i+1})^{k-1}=\sum_{j=0}^\infty \binom{k-1}{j}X_i^{j}X_{i+1}^{k-1-j},
\end{equation}
because $\binom{n}{j}=0$ whenever $0\leq n<j$, and for $|X_i|\leq |X_{i+1}|$ one has
\begin{equation*}
(X_i+X_{i+1})^{-1}=\frac{1}{X_{i+1}}\Big(1+\frac{X_i}{X_{i+1}}\Big)^{-1}=\sum_{j\geq 0}(-1)^j\frac{X_i^j}{X_{i+1}^{j+1}}=\sum_{j\geq 0}\frac{(-1)_j}{j!}X_i^jX_{i+1}^{-1-j}.
\end{equation*}
Therefore for every $i$ we get contributions
\begin{align}
&\sum_{j=0}^{n_{i}+1} (n_{i}-j) { n_{i+1}+j-1 \choose j } \mathbb{G}_{n_{i}-j+1}(\tau) A(n_1,\ldots,n_{i-1},j+n_{i+1},n_{i+2},\ldots,n_r;\tau)\notag \\
&- \sum_{j=0}^{n_{i+1}+1} (n_{i+1}-j) { n_{i}+j-1 \choose j } \mathbb{G}_{n_{i+1}-j+1}(\tau) A(n_1,\ldots,n_{i-1},j+n_{i},n_{i+2},\ldots,n_r;\tau).\notag
\end{align}
Finally, there is another source of contributions: multiplying the expansion (\ref{pole}) by $\mathcal{G}(X_{i+1},\tau)$ we find that for every $i$ we must take into account also the term
\begin{equation*}
(-1)^{n_i} (n_{i}+n_{i+1}) \mathbb{G}_{n_{i}+n_{i+1}+1}(\tau) A(n_1,\ldots,n_{i-1},0,n_{i+2},\ldots,n_r;\tau).
\end{equation*}
This concludes the proof.\\
$\square$

This proposition suggests the fact, already noticed in \cite{BMMS}, that one could write elliptic MZVs in terms of iterated integrals of Eisenstein series. We will come back to this in Section \ref{sectionItIntEis}.

\begin{cor}\label{CorDouble}
For $n_1+n_2$ even we have $A(n_1,n_2;\tau)=A^{\infty}(n_1,n_2)$, while for $n_1+n_2$ odd we have
\begin{align}
A(n_1,n_2;\tau)&=-(-1)^{n_1}A(n_1+n_2,0;\tau)\notag \\
&+2\sum_{p=1}^{\left \lceil{\frac{n_1-3}{2}}\right \rceil }\binom{n_1+n_2-2p-2}{n_2-1}\zeta(n_1+n_2-2p-1)A(2p+1,0;\tau)\notag \\
&-2\sum_{p=1}^{\left \lceil{\frac{n_2-3}{2}}\right \rceil}\binom{n_1+n_2-2p-2}{n_1-1}\zeta(n_1+n_2-2p-1)A(2p+1,0;\tau).
\end{align}
\end{cor}
The proof of this corollary, based on the previous result and on the triviality of length one elliptic MZVs, does not present any difficulty. This corollary explains why we have claimed that computing $B^\infty(n,0;\tau)$ was enough to get all length two $B^\infty(n_1,n_2;\tau)$. It is worth mentioning that, building on this result, Matthes proved in \cite{MatthesDouble} the following
\begin{teo}[Matthes]
The dimension of the $\mathbb{Q}$-vector space spanned by A-elliptic MZVs of length two and weight\footnote{Note that the formula given in \cite{MatthesDouble} looks different, because the weight is not defined in the same way. We will see later that our notion of weight is more natural for our purposes.} $N$ is one if $N$ is even and
\[
\left \lfloor{\frac{N}{3}}\right  \rfloor -1
\]
if $N$ is odd.
\end{teo}
Beyond length two, we do not even have conjectural dimensions for spaces of elliptic MZVs of fixed length and weight (see \cite{BMS}, \cite{MatthesDouble}).

\subsection{Asymptotic behaviour of B-elliptic MZVs}\label{SectionAsym}

For what concerns B-elliptic MZVs, Proposition \ref{asymptA} for the asymptotic expansion was not the best possible result. We prove here, as a consequence of \ref{PropDiffB}, a stronger statement:
\begin{thm}\label{asymptB}
Let $n_1,n_r\neq 1$. Then
\begin{equation}
B(\n;\tau)=\sum_{i=1-r}^{n_1+\cdots n_r}\sum_{j\geq 0}b_{i,j}(\n)\tau^iq^j,
\end{equation}
where $b_{i,j}(\n)\in\mathcal{A}[(2\pi i)^{\pm 1}]$.
\end{thm}
\textbf{Proof.} Let us first give the details of this proof in the simpler case where $n_i\neq 1$ for all $i$'s.
Let 
\[
\mathcal{N}_{1,r}:=\{(n_1,\ldots n_r)\in(\mathbb{Z}_{\geq 0})^r:n_i\neq 1\,\,\forall\,\, i\},
\]
and let $\n=(n_1,\ldots ,n_r)\in\mathcal{N}_{1,r}$. By Proposition \ref{asymptA} we can write
\begin{equation}\label{IntermediateStep}
B(\n;\tau)=\sum_{j\geq 0}\sum_{i=N_j}^{M_j}b_{i,j}(\n)\tau^iq^j,
\end{equation}
where $N_j,M_j\in\mathbb{Z}$ and $b_{i,j}(\n)\in\mathcal{A}[(2\pi i)^{\pm 1}]$. We need to prove that all $N_j$'s are bounded below by $1-r$ and all $M_j$'s are bounded above by $n_1+\cdots +n_r$. To do it we will make use of Proposition \ref{PropDiffB}. Let us denote 
\begin{equation}
\beta_j(\n;\tau):=\sum_{i=N_j}^{M_j}b_{i,j}(\n)\tau^i.
\end{equation}
Moreover, let us denote by
\begin{equation}
\mathbb{G}(\tau)=\sum_{j\geq 0}g_{j,k}q^j
\end{equation} 
the $q$-expansion of the Eisenstein series (\ref{modifiedEis}) (so in particular $g_{j,k}=0$ for every $j$ whenever $k$ is odd). Since we know that (\ref{IntermediateStep}) holds, formula (\ref{explDiffB}) leads to the following formula, holding for every $\n\in\mathcal{N}_{1,r}$ and every $j\geq 0$:
\begin{multline}\label{derivative}
\frac{\partial}{\partial\tau}\beta_j(n_1,\ldots ,n_r;\tau)+(2\pi i)j\beta_j(n_1,\ldots,n_r;\tau)\\
=n_rg_{j,n_r+1}\tau^{n_r-1}\beta_j(n_1,\ldots ,n_{r-1};\tau)-n_1g_{j,n_1+1}\tau^{n_1-1}\beta_j(n_2,\ldots ,n_{r};\tau)\\
+ \sum_{i=1}^{r-1} \Big( (-1)^{n_i} (n_{i}+n_{i+1})g_{j,n_{i}+n_{i+1}+1}\tau^{n_i+n_{i+1}-1}\beta_j(n_1,\ldots,n_{i-1},0,n_{i+2},\ldots,n_r;\tau) \\
+  \sum_{k=0}^{n_{i}+1} (n_{i}-k) { n_{i+1}+k-1 \choose k } g_{j,n_{i}-k+1}\tau^{n_i-k-1}\beta_j(n_1,\ldots,n_{i-1},k+n_{i+1},n_{i+2},\ldots,n_r;\tau)\\
- \sum_{k=0}^{n_{i+1}+1} (n_{i+1}-k) { n_{i}+k-1 \choose k } g_{j,n_{i+1}-k+1}\tau^{n_{i+1}-k-1}\beta_j(n_1,\ldots,n_{i-1},k+n_{i},n_{i+2},\ldots,n_r;\tau) \Big).
\end{multline}
We have already seen that $B(n;\tau)=2\pi i\B_n/n!$. This means in particular that $\beta_j(n;\tau)=(2\pi i\B_n/n!)\delta_{j,0}$, where $\delta_{j,0}=1$ for $j=0$ and $\delta_{j,0}=0$ otherwise, and it obviously implies the much weaker statement $\beta_j(n;\tau)=\sum_{i=0}^{n}b_{i,j}(n)\tau^i$. This will constitute the first step of the induction on $r\in\mathbb{N}$. Let us now assume by  inductive hypothesis that 
\begin{equation}
\beta_j(\n;\tau)=\sum_{i=2-r}^{n_1+\cdots n_{r-1}}b_{i,j}(\n)\tau^i
\end{equation} 
for every $\n\in\mathcal{N}_{1,r-1}$. Then the highest and lowest exponents of $\tau$ appearing in (\ref{derivative}) in the right hand side's Laurent polynomial must be $\leq n_1+\cdots +n_r-1$ and $\geq -r$, respectively. Looking at the left hand side, this means that, for every $j$, $N_j\geq 1-r$ and $M_j\leq n_1+\cdots +n_r$.\footnote{Actually, to be more precise, $M_j\leq n_1+\cdots +n_r-1$ unless $j=0$.} The missing cases ($n_1=1$ or $n_r=1$) can be proven in a completely similar way, using a refined version of Proposition \ref{asymptA}: it is long but not difficult to prove that when $n_1=1$ or $n_r=1$ B-elliptic MZVs admit an expansion of the kind
\[
\sum_{k=0}^r\sum_{j\geq 0}\beta_{j,k}(n_1,\ldots n_r;\tau)q^j\log^k(\tau),
\]
where $\beta_{j,k}(\n;\tau)$ are Laurent polynomials in $\tau$. The statement of our theorem again follows by induction on $r$ from the differential equation applied term by term to this expansion, using also that we know by Proposition \ref{asymptA} that we do not expect any $\log(\tau)$ in the final result. Since this computation is essentially similar to that with $\n\in\mathcal{N}_{1,r}$, but more cumbersome, we prefer to skip the details.\\
$\square$

\begin{remark}
As we have already mentioned, when we have $n_1=1$ or $n_r=1$ the asymptotic expansion of B-elliptic MZVs could in principle include some $\log(\tau)$. For instance, since the derivative of $A(1,0;\tau)$ is, by formula (\ref{explDiffA}), essentially equal to $\mathbb{G}_2(\tau)$, we conclude that $A(1,0;\tau)$ is essentially (up to a multiplicative constant and a degree one polynomial in $\tau$) given by $\log(\eta(\tau))$, by the well known fact \cite{1-2-3ModForms} that 
\[
\frac{\partial}{\partial\tau}\log(\eta(\tau))=-\frac{1}{2}\mathbb{G}_2(\tau).
\]
This term, using the modular transformation (\ref{eta}), is the reason for the appearence of $\log(\tau)$ in the B-elliptic counterpart. We will see in Section \ref{SectionDepthOne} that all $B(1,0,\ldots ,0;\tau)$ contain some $\log(\tau)$.
\end{remark}
\begin{remark}\label{Slash}
Let us define the \emph{weight $k$ slash operator} $|_k\,\gamma$ as the action of SL$_2(\mathbb{Z})$ on the space of functions $f:\mathbb{H}\rightarrow\mathbb{C}$ defined for $k\in\mathbb{Z}$ and $\gamma =\left( \begin{array}{ccc}
a & b \\
c & d \end{array} \right)$ by
\begin{equation}\label{ActionSL2Z}
f(\tau)|_{k}\gamma=\frac{1}{(c\tau+d)^k}f\bigg(\frac{a\tau+b}{c\tau+d}\bigg).
\end{equation}
We call a \emph{vector valued modular form of weight $k$} for SL$_2(\mathbb{Z})$ any function $F:\mathbb{H}\rightarrow\mathbb{C}^n$ which satisfies (besides appropriate analytic conditions, that in our case are holomorphicity in the upper-half plane and sub-exponential growth at the cusp $i\infty$) $F(\tau)|_{k}\gamma=\rho(\gamma)F(\tau)$ for all $\gamma\in\mbox{SL}_2(\mathbb{Z})$, where $\rho$ is a (finite dimensional) representation of SL$_2(\mathbb{Z})$.

Let us also recall that the action of the full modular group SL$_2(\mathbb{Z})$ is determined by its generators \[
S=\left( \begin{array}{ccc}
0 & -1 \\
1 & 0 \end{array} \right),
\]
and
\[
T=\left( \begin{array}{ccc}
1 & 1 \\
0 & 1 \end{array} \right).
\]
A simple consequence of Theorem \ref{asymptB} is the fact that, for any fixed $\n=n_1,\ldots ,n_r$ such that $b_{1-r,0}(\n)\neq 0$, the space of functions 
\[
\big\langle A(\n;\tau)|_{k}\gamma:\gamma\in\mbox{SL}_2(\mathbb{Z}) \big\rangle_\mathbb{C}
\]
is infinite-dimensional for all $k\geq 2-r$. This is true because, from the fact that $A(\n;\tau)|_{0}S=B(\n;\tau)$, the function $A(\n;\tau)|_{k}S$ will have a pole at $\tau=0$, and therefore for all $N\in\mathbb{N}$
\[
\dim \big\langle A(\n;\tau)|_{k}T^nS:0\leq n\leq N \big\rangle _{\mathbb{C}}= N+1
\]
In other words, if $A(\n;\tau)$ is a (component of a vector valued) modular form, we expect the weight to be $\leq 1-r$. We will prove that this is in fact the case for all depth one elliptic MZVs.
\end{remark}

\section{Elliptic MZVs as iterated integrals of Eisenstein series}\label{sectionItIntEis}

\subsection{Iterated integrals of Eisenstein series}

In this section we want to briefly recall some highlights of the construction of iterated integrals of modular forms, initiated by Manin in \cite{ManinItInt} for cusp forms and extended to Eisenstein series by Brown in \cite{MMV}. We will follow \cite{MMV}.

Let us consider the complex vector space $\mathcal{M}_k$ of all weight $k$ classical modular forms for the full modular group, i.e. holomorphic functions $f:\mathbb{H}\rightarrow \mathbb{C}$ with an expansion at the cusp
\[
f(\tau)=\sum_{m\geq 0}a_mq^m
\]
such that $f(\tau)|_{k}\,\gamma=f(\tau)$ for each $\gamma\in\mbox{SL}_2(\mathbb{Z})$. In order to follow closely \cite{MMV}, let us introduce yet another normalization of the Eisenstein series: for all $k\geq 4$ we consider
\begin{equation}\label{DefEisBern}
E_k(\tau)=\frac{-\B_k}{2k!}+\frac{1+(-1)^k}{2}\sum_{m\geq 1}\sigma_{k-1}(m)q^m,
\end{equation}
where
\begin{equation}\label{sigma}
\sigma_{n}(m)=\sum_{d|m}d^n
\end{equation}
is the sum of the $n$-th powers of all positive divisors of $m$. From (\ref{EisSeries}) and (\ref{modifiedEis}) we get
\begin{equation}\label{ConversionEis2}
E_k(\tau)=\frac{(k-1)!}{2(2\pi i)^k}G_k(\tau)=\frac{(k-1)!}{4\pi i}\mathbb{G}_k(\tau).
\end{equation}
Let $\mathcal{B}_k$ be a basis of $\mathcal{M}_k$ given by modular forms with rational coefficients, containing the Eisenstein series $E_k$'s, let $\mathcal{B}=\bigoplus_k\mathcal{B}_k$, let 
\[
M_k=\big\langle a_f:f\in\mathcal{B}_k \big\rangle_{\mathbb{Q}}
\]
be the rational vector space spanned by symbols $a_f$ indexed by $\mathcal{B}_k$ and let 
\[
M_k^*=\big\langle A_f:f\in\mathcal{B}_k \big\rangle_{\mathbb{Q}}
\]
be its dual vector space. Moreover, let us consider the rational vector spaces $V_k=\mathbb{Q}[X,Y]_k$ of homogeneous polynomials of degree $k$. The modular group SL$_2(\mathbb{Z})$ acts on them by
\[
P(X,Y)|_\gamma=P(aX+bY,cX+dY).
\]
For each modular form $f$ of weight $k$ we shall write
\begin{equation}
\underline{f}(X,Y,\tau)=(2\pi i)^{k-1}f(\tau)(X-\tau Y)^{k-2}d\tau,
\end{equation}
and by the modularity of $f$ we have
\begin{equation}\label{modularityforitint}
\underline{f}(X,Y,\tau)|_\gamma:=(2\pi i)^{k-1}\big(f(\tau)|_0\,\gamma\big)(X-\tau Y)^{k-2}|_\gamma \,d(\gamma\tau)=\underline{f}(X,Y,\tau),
\end{equation}
where 
\[
\gamma\tau:=\frac{a\tau+b}{c\tau+d},
\]
and $|_k\,\gamma$ is the action of SL$_2(\mathbb{Z})$ defined by (\ref{ActionSL2Z}). We define
\[
\Theta(X,Y,\tau)=\sum_{\mathcal{B}}A_f\underline{f}(X,Y,\tau).
\]
The iterated integrals defined for a path $\eta\subset\mathbb{H}$ by the generating function in non-commutative variables $A_f$
\begin{equation}\label{pgfrons}
1+\int_\eta \Theta(X,Y,z)+\int_\eta \Theta(X_1,Y_1,z_1)\,\Theta(X_2,Y_2,z_2)+\ldots
\end{equation}
are homotopy invariant \cite{ManinItInt} (in fact, all iterated integrals of holomorphic 1-forms on a simply connected one-dimensional smooth manifold are homotopy invariant), and therefore if $\eta(0)=\tau_0$ and $\eta(1)=\tau_1$ we denote (\ref{pgfrons}) by $I(\tau_0,\tau_1)$ (omitting the dependence on all formal variables). One can prove, using (\ref{modularityforitint}), that \cite{ManinItInt}
\[
I(\gamma\tau_0,\gamma\tau_1)|_\gamma=I(\tau_0,\tau_1).
\]
Brown extended this construction to the cusp $i\infty$ \cite{MMV}. Note that the naive integral on $[\tau,i\infty]$ of the Eisenstein series $E_k(\tau)$ is divergent, because of the constant term $E_k^{\infty}=-\B_k/2k$. Therefore, Brown defined regularized (iterated) integrals of modular forms on $[\tau,\1]$, where $\1$ denotes a certain tangential base point at infinity. If we employ the usual notation $f(\tau)=f^{\infty}+f^0(\tau)$ to denote the constant term and the exponentially small part of a modular form $f$, we define
\[
\int_{\tau}^\1 f(z)dz=\int_\tau^{i\infty}f^0(z)dz-\int_0^\tau f^{\infty}dz.
\]
Using this, one can inductively define the generating series of regularized iterated integrals 
\[
I(\tau,\infty)=1+\int_{[\tau,\1]} \Theta(X,Y,z)+\int_{[\tau,\1]} \Theta(X_1,Y_1,z_1)\,\Theta(X_2,Y_2,z_2)+\ldots.
\]
\begin{lemma}[Brown]\label{LemmaBrown}
For every $\gamma\in\mbox{SL}_2(\mathbb{Z})$ there exists a series\footnote{It is invertible and group-like, i.e. the morphism sending a non-commutative word $w$ in the formal variables $a_f$'s to the coefficient $\mathcal{C}_\gamma(w)$ respects the shuffle product.} $\mathcal{C}_\gamma$ in infinitely many non-commutative variables $A_f$ and infinitely many commutative pairs of variables $(X_i,Y_i)$ such that\footnote{Equation (\ref{defcocycle}) looks different from Brown's original result, where $\mathcal{C}_\gamma$ multiplies from the right. This is due to the fact that Brown's notation for iterated integrals in \cite{MMV} is opposite to ours.}
\begin{equation}\label{defcocycle}
I(\tau,\infty)=\mathcal{C}_\gamma I(\gamma\tau,\infty)|_\gamma.
\end{equation}
This series does not depend on $\tau$, and for all $\gamma_1, \gamma_2\in\mbox{SL}_2(\mathbb{Z})$ it satisfies the cocycle relation
\[
\mathcal{C}_{\gamma_1\gamma_2}=\mathcal{C}_{\gamma_1}\big|_{\gamma_2}\mathcal{C}_{\gamma_2}.
\]
\end{lemma}
The coefficient of the generating series $\mathcal{C}$ are called \emph{multiple modular values}. Note that, just by equation (\ref{defcocycle}) and by the fact that $-1/i=i$, we get the formula
\begin{equation}
\mathcal{C}_S=I(i,\infty)I(i,\infty)|_S^{-1}.
\end{equation}
In the case of a single integration, one gets abelian cocycles $\mathcal{C}_\gamma(a_f)$, very well known after the work of Eichler, Shimura and Manin in the case of cusp forms, and worked out for Eisenstein series by Zagier in \cite{ZagierPeriodsJacobi}. In particular, Zagier proved that
\begin{equation}\label{CocycleEis}
\mathcal{C}_S(a_{E_{2k}})=\frac{(2k-2)!}{2}\Big(\zeta(2k-1)(Y^{2k-2}-X^{2k-2})-(2\pi i)^{2k-1}\sum_{i=1}^{k-1}\frac{\B_{2i}\B_{2k-2i}}{(2i)!(2k-2i)!}X^{2i-1}Y^{2k-2i-1}\Big).
\end{equation}

\subsection{Elliptic MZVs as iterated integrals on $\overline{\mathcal{M}}_{1,1}$}

Brown's regularized single integration of a modular form on a path $[\tau,\1]$ contained in $\overline{\mathcal{M}}_{1,1}$ is nothing but (minus) the primitive of the modular form with the integration constant set to zero. This is actually something that we can do for any function $f\in\mathbb{C}[[q]][\tau]$ \cite{MatthesDecomposition}. Regularized iterated integrals are then defined inductively. Thus we can write
\begin{equation}\label{propEisInt}
A(\n;\tau)=A^{\infty}(\n) -  \int_\tau^{\1}\frac{\partial}{\partial z}A(\n;z)dz.
\end{equation}
In order to write down explicitly the connection between elliptic MZVs and Brown's iterated integrals of modular forms, we need to introduce a new class of functions, first considered in \cite{BMS}, and very closely related to Brown's iterated integrals of Eisenstein series.
\begin{defn}
For $\n=n_1,\cdots , n_r$, we define
\begin{equation}
\Gamma (\n;\tau):=\int_{[\tau,\1]}\mathbb{G}_{n_r}(z_r)dz_r\cdots \mathbb{G}_{n_1}(z_1)dz_1,
\end{equation}
where the integration over $[\tau,\1]$ is to be intended, as explained above, as an iterated primitive (with sign reversed). For instance,
\[
\Gamma (0;\tau)=\int_{\tau}^{\1}\mathbb{G}_0(z)dz=-2\pi i\int_{\tau}^{\1}dz=2\pi i\tau.
\]
As always, we will write $\Gamma (\n;\tau)=\Gamma^{\infty} (\n;\tau)+\Gamma^0 (\n;\tau)$, with $\Gamma^0 (\n;\tau)=O(q)$.
\end{defn}
It is easy to see that
\[
\Gamma^{\infty} (\n;\tau)=\frac{(2\pi i\tau)^{k_1+\cdots +k_r}}{(k_1+\cdots +k_r)!}\prod_{i=1}^r \frac{\B_{n_i}}{(n_i)!}.
\]
In order to write down explicitly the Fourier expansion of $\Gamma^0 (n_1,\ldots ,n_r;\tau)$, one must noticed, as already observed in \cite{BMS}, that\footnote{Here we are going to be a bit sloppy: writing $\int \mathbb{G}_k$ we mean $\int \mathbb{G}_k(z)dz$.}
\begin{multline}\label{q-expGamma}
\int_{[\tau,\1]}\mathbb{G}^0_{n_1}\underbrace{\mathbb{G}_{0}\cdots \mathbb{G}_{0}}_{k_1-1}\mathbb{G}^0_{n_2}\cdots \mathbb{G}^0_{n_r}\underbrace{\mathbb{G}_{0}\cdots \mathbb{G}_{0}}_{k_r-1}=\\
=\prod_{i=1}^r\frac{-2}{(n_i-1)!}
\sum_{0<m_1<\cdots <m_r}\frac{\sigma_{n_1-1}(m_1)\sigma_{n_2-1}(m_2-m_1)\cdots \sigma_{n_r-1}(m_r-m_{r-1})q^{m_r}}{m_1^{k_1}\cdots m_r^{k_r}},
\end{multline}
where $\sigma$ was defined in (\ref{sigma}). Then, by the shuffle product of iterated integrals (or otherwise by integration by parts) one can always write $\Gamma^0(\n;\tau)$'s in terms of expansions of the kind (\ref{q-expGamma}). We will work this out in details later, when we will focus on the \emph{depth one case}, where all but one entries of $\Gamma$ are equal to zero.

We have already seen how to determine the integration constant $A^{\infty}(\n;\tau)$. Using that the length of the A-elliptic MZVs appearing in the formula (\ref{explDiffA}) is strictly shorter, one can iterate the procedure and get the announced fact that A-elliptic MZVs can be written in terms of the iterated integrals $\Gamma (\n,\tau)$, and therefore B-elliptic MZVs can be computed using the modular transformations of these functions, that we can infer by comparing them with Brown's iterated integrals of Eisenstein series. Let us see this on one non-trivial example: by the constant term formula (\ref{Ainf2}) we know that $A^{\infty}(3,0)=0$, and therefore using the differential equation (\ref{diffeqA}) we conclude that
\[
A(3,0;\tau)=3\int_\tau^{\1}A(0;z)\mathbb{G}_4(z)dz-3\int_\tau^\1A(4;z)\mathbb{G}_0(z)dz.
\]
We know that $A(n;\tau)=2\pi i\B_n/n!$ for all $n\neq 1$, and $\mathbb{G}_0=-2\pi i$. Hence we get
\[
A(3,0;\tau)=6\pi i\int_\tau^{\1}\mathbb{G}_4(z)dz-\frac{\pi^2}{60}\tau.
\]
Now we want to compute $B(3,0;\tau)=A(3,0;-1/\tau)$ using Brown's theory. From now on, following \cite{MMV}, we write $[\underline{f_1},\ldots ,\underline{f_r}]$ to denote the iterated integral
\[
\int_{[\tau,\1]}\underline{f_1}(X_1,Y_1,z_1)\cdots \underline{f_r}(X_r,Y_r,z_r).
\]
In particular, we have
\begin{equation}\label{SingleEisIntBrown}
[\underline{E}_k]=(2\pi i)^{k-1}\int_\tau^\1 (X-zY)^2E_k(z)dz=\sum_{i=1}^{k-1}h_{k,i}(\tau)(2\pi iX)^{k-i-1}(2\pi iY)^{i-1},
\end{equation}
and 
\[
h_{4,1}(\tau)=2\pi i\int_\tau^\1 E_4(z)dz=3\int_\tau^\1 \mathbb{G}_4(z)dz.
\]
We can easily get, by Lemma \ref{LemmaBrown} and equation (\ref{CocycleEis}),  that
\[
h_{4,1}(-1/\tau)=h_{4,3}(\tau)-\frac{\zeta(3)}{(2\pi i)^2},
\]
and therefore we obtain
\[
B(3,0;\tau)=-\frac{\zeta(3)}{(2\pi i)^2}+\frac{\pi^2}{60}\frac{1}{\tau}+6\pi i\int_\tau^\1 z^2\mathbb{G}_4(z)dz.
\]
Working out the asymptotic expansion of the last term (this is an easy exercise, using integration by parts and (\ref{q-expGamma})) one gets back the Laurent polynomial $B^\infty(3,0;\tau)$ predicted by (\ref{Binf2}).
In Section \ref{SectionDepth1Mod} we will exploit this method systematically and work out the full modular behaviour for all depth one elliptic MZVs.

\section{Depth one}\label{SectionDepthOne}

In this section we will present the simplest example of systematic explicit computations of elliptic multiple zeta values in terms of iterated integrals of Eisenstein series, and we will present various consequences of these explicit formulae. Recall that the depth of elliptic MZVs is defined as the number of entries $n_i$ which are strictly positive. It is important to mention that this set of elliptic MZVs has been identified with the coefficients of the \emph{meta-abelian quotient} of Enriquez's associator by Matthes in \cite{MatthesMeta}, where he also connects depth one elliptic MZVs with special values of Levin's elliptic polylogarithms. In particular, some of our results are equivalent to results contained in \cite{MatthesMeta} (we will make clear which ones). As always, our approach will consist in deriving all the results without referring to associators, and we will obtain the first explicit examples of complete Fourier expansion of non-trivial B-elliptic MZVs, as well as precise results about the modular nature of A-elliptic (and therefore B-elliptic) MZVs. At the end of this section we will see how this story is related to special values of the elliptic polylogarithms introduced in Chapter \ref{ChapterMathBackground}. A generalization of the results of this section to higher depths is postponed to future investigations.

\subsection{Explicit formulae}

It is convenient to introduce some special notation: we will write
\[
A_{n,r}(\tau):=A(n,\underbrace{0,\ldots ,0}_{r-1};\tau)=\int_0^1\frac{(2\pi it)^{r-1}}{(r-1)!}f_n(t,\tau)dt,
\]
and
\begin{eqnarray*}
\mathcal{A}(X,Y;\tau):=\sum_{\substack{n\geq 0\\r\geq 1}}\frac{A_{n,r}(\tau)}{(2\pi i)^{r-1}}X^{n-1}Y^{r-1}&=&\sum_{r\geq 1}\int_0^1\frac{(tY)^{r-1}}{(r-1)!}F\Big(t,\frac{X}{2\pi i};\tau\Big)dt\\
&=&\int_0^1e^{tY}F\Big(t,\frac{X}{2\pi i};\tau\Big)dt,
\end{eqnarray*}
where we mention, once for all, that by $\int_0^1$ we mean the \emph{regularized integral} described in the definition of elliptic MZVs.

First of all, note that one can reduce all depth one elliptic MZVs to the $A_{n,r}$'s:
\begin{multline}\label{shuffleAD1}
A(\underbrace{0,\ldots ,0}_s,n,\underbrace{0,\ldots ,0}_r;\tau)=\\
=(2\pi i)^{r+s}\int_0^1\frac{t^r(1-t)^s}{r!s!}f_n(t,\tau)=\sum_{i=0}^s \frac{(2\pi i)^{s-i}(-1)^i}{(s-i)!}\binom{r+i}{r}A_{n,r+i+1}(\tau).
\end{multline}
Let us now see what is the analogue of Proposition \ref{PropDiffA} for the depth 1 case.
\begin{lemma}\label{lemmaDiffDepth1}
\begin{equation}
\frac{\partial}{\partial \tau}\mathcal{A}(X,Y;\tau)=(1-e^Y)\mathcal{G}(X,\tau)-Y\frac{\partial}{\partial X}\mathcal{A}(X,Y;\tau)
\end{equation}
\end{lemma}
\textbf{Proof.} Using the mixed heat equation, the substitution $\hat{X}=X/2\pi i$ and Lemma \ref{lemmaEnr} we get\footnote{As in the proof of Enriquez's Proposition \ref{PropDiffA}, we omit the proof that the basic properties of iterated integrals are still valid for the regularized ones. All details can be found in \cite{Enriquez}.}
\begin{multline}
\frac{\partial}{\partial \tau}\int_0^1e^{tY}F\Big(t,\frac{X}{2\pi i};\tau\Big)dt=\frac{1}{2\pi i}\int_0^1e^{tY}\frac{\partial}{\partial t \partial \hat{X}}F(t,\hat{X};\tau)dt\\
=\frac{1}{2\pi i}\Big(e^Y\frac{\partial}{\partial \hat{X}}F(1,\hat{X};\tau)-\frac{\partial}{\partial \hat{X}}F(0,\hat{X};\tau)\Big)-\frac{Y}{2\pi i}\int_0^1e^{tY}\frac{\partial}{\partial \hat{X}}F(t,\hat{X};\tau)dt\\
=(1-e^Y)\mathcal{G}(X,\tau)-Y\frac{\partial}{\partial X}\mathcal{A}(X,Y;\tau).
\end{multline}
$\square$

We want to make more precise in this setting the remark that elliptic MZVs can be written as iterated integrals of Eisenstein series. As always, we write $A_{n,r}(\tau)=A^\infty_{n,r}+A^0_{n,r}(\tau)$, as well as $\mathcal{A}(X,Y;\tau)=\mathcal{A}^\infty(X,Y)+\mathcal{A}^0(X,Y;\tau)$ at the level of generating functions. Note that, by (\ref{GeneralAinf}), we have that for all $n\neq 1$
\begin{equation}
A_{n,r}^{\infty}=\frac{(2\pi i)^r\B_n}{r!n!},
\end{equation} 
and that, by Corollary \ref{CorA100} and (\ref{inversion}),
\begin{equation}\label{FormulaAinf1r}
A_{1,r}^\infty=(2\pi i)^{r-1}\Big(\frac{i\pi}{2(r-1)!}-\sum_{k=1}^{\lfloor \frac{r}{2}\rfloor -1}\frac{\zeta(2k+1)}{(r-2k-1)!(2\pi i)^{2k}}\Big).
\end{equation}
We will denote the generating series of the exponentially small part $\mathbb{G}^0_k$ of the Eisenstein series $\mathbb{G}_k$ by
\begin{equation}
\mathcal{G}^0(X,\tau):=\sum_{n\geq 1}n\mathbb{G}^0_{n+1}X^{n-1}.
\end{equation} 
Moreover, we denote\footnote{As we have already done in the previous section, here by $\int \mathbb{G}_k$ we mean $\int \mathbb{G}_k(z)dz$.}
\[
\Gamma^L_{n,k}(\tau):=\Gamma(\underbrace{0,\ldots ,0}_{k-1},n;\tau)=\int_{[\tau,\1]}\mathbb{G}_n\underbrace{\mathbb{G}_0\cdots \mathbb{G}_0}_{k-1}
\]
and
\[
\Gamma^R_{n,k}(\tau):=\Gamma(n,\underbrace{0,\ldots ,0}_{k-1};\tau)=\int_{[\tau,\1]}\underbrace{\mathbb{G}_0\cdots \mathbb{G}_0}_{k-1}\mathbb{G}_n
\]
Again, we write
\[
\Gamma^L_{n,k}(\tau)=\Gamma^{L,\infty}_{n,k}(\tau)+\Gamma^{L,0}_{n,k}(\tau),
\]
\[
\Gamma^R_{n,k}(\tau)=\Gamma^{R,\infty}_{n,k}(\tau)+\Gamma^{R,0}_{n,k}(\tau).
\]
It is obvious that
\begin{equation}
\Gamma^{L,0}_{n,k}(\tau)=\int_{[\tau,i\infty]}\mathbb{G}^0_n\underbrace{\mathbb{G}_0\cdots \mathbb{G}_0}_{k-1}
\end{equation}
and that
\begin{equation}
\Gamma^{L,\infty}_{n,k}(\tau)=\frac{\B_n(2\pi i\tau)^k}{n!k!}.
\end{equation}
Moreover, one can easily check that
\begin{equation}\label{gammaR}
\Gamma^{R,0}_{n,k}(\tau)=\frac{(2\pi i)^{k-1}}{(k-1)!}\int_\tau^{i\infty}z^{k-1}\mathbb{G}^0_n(z)dz
\end{equation}
and that
\begin{equation}
\Gamma^{R,\infty}_{n,k}(\tau)=\Gamma^{L,\infty}_{n,k}(\tau)=\frac{\B_n(2\pi i\tau)^k}{n!k!}.
\end{equation}
It is important to keep in mind that $\Gamma^L_{n,k}(\tau)$ and $\Gamma^R_{n,k}(\tau)$ vanish identically whenever $n$ is odd. Moreover, we want to stress the fact that one can compute the $q$-expansion of any $\Gamma^0(n_1,\ldots ,n_r;\tau)$, and this is the simplest way to get the $q$-expansion of A-elliptic MZVs, as already noticed in \cite{BMS}. In depth one the $q$-expansion (\ref{q-expGamma}) just reads 
\begin{equation}\label{ExplGammaD1}
\Gamma^{L,0}_{n,k}(\tau)=-\frac{2}{(n-1)!}\sum_{m,p\geq 1}\frac{m^{n-k-1}}{p^k}q^{mp}.
\end{equation}
Let us now introduce the generating functions
\begin{equation}
\mathcal{E}(X,Y;\tau):=\sum_{n,k\geq 1}\frac{(-1)^{n-1}}{(2\pi i)^{k-1}}\frac{(n+k-1)!}{(n-1)!}\Gamma^L_{n+k,k}(\tau)X^{n-1}Y^{k-1},
\end{equation}
\begin{equation}
\mathcal{F}(X,Y;\tau):=\sum_{n,k\geq 1}\frac{1}{(2\pi i)^{k-1}}\frac{(n+k-1)!}{(n-1)!}\Gamma^R_{n+k,k}(\tau)X^{n-1}Y^{k-1}.
\end{equation}
As before, we write
\[
\mathcal{E}^0(X,Y;\tau)=\sum_{n,k\geq 1}\frac{(-1)^{n-1}}{(2\pi i)^{k-1}}\frac{(n+k-1)!}{(n-1)!}\Gamma^{L,0}_{n+k,k}(\tau)X^{n-1}Y^{k-1}
\]
and
\[
\mathcal{F}^0(X,Y;\tau)=\sum_{n,k\geq 1}\frac{1}{(2\pi i)^{k-1}}\frac{(n+k-1)!}{(n-1)!}\Gamma^{R,0}_{n+k,k}(\tau)X^{n-1}Y^{k-1}.
\]
The main consequence of the lemma above (this consequence, stated in a different way, is already contained in \cite{MatthesMeta}) is the following:
\begin{prop}\label{PropA=E}
\begin{equation}\label{A=E}
\mathcal{A}^0(X,Y;\tau)=(e^Y-1)\mathcal{E}^0(X,Y;\tau)
\end{equation}
\end{prop}
\textbf{Proof.} Iterating the statement of the proposition, and noting that, for every $k\geq 1$,
\[
\int_{[\tau,\1]}A^{\infty}_{n,r}dz_1\,dz_2\cdots dz_{k}=O(q),
\]
we deduce that
\begin{equation}
\mathcal{A}^0(X,Y;\tau)=(e^Y-1)\sum_{k\geq 1}\frac{Y^{k-1}}{(2\pi i)^{k-1}}\frac{\partial^{k-1}}{\partial X^{k-1}}\int_{[\tau,i\infty]}\mathcal{G}^0(X,z_{1})dz_1\mathbb{G}_0dz_{2}\cdots \mathbb{G}_0dz_{k}.
\end{equation}
Expanding the generating series $\mathcal{G}^0$, we get
\begin{eqnarray*}
\mathcal{A}^0(X,Y;\tau)&=&(e^Y-1)\sum_{k,m\geq 1}\frac{(-1)^{k-1}m!}{(m-k)!(2\pi i)^{k-1}}\Gamma^{L,0}_{m+1,k}(\tau)X^{m-k}Y^{k-1}\\
&=&(e^Y-1)\sum_{k,n\geq 1}\frac{(-1)^{n-1}(n+k-1)!}{(n-1)!(2\pi i)^{k-1}}\Gamma^{L,0}_{n+k,k}(\tau)X^{n-1}Y^{k-1},
\end{eqnarray*}
where to get $(-1)^{n-1}$ in the second equality we have used that $n+k$ must be even. This concludes the proof.\\
$\square$
\begin{cor}
For any $n\geq 1$ and any $r\geq 1$ we have
\begin{equation}\label{A(n,0,...,0)}
A_{n,r}(\tau)=A^{\infty}_{n,r}+ \frac{(-1)^{n-1}}{(n-1)!}\sum_{j=1}^{r-1}\frac{(2\pi i)^{r-j}(n+j-1)!}{(r-j)!}\Gamma^{L,0}_{n+j,j}(\tau).
\end{equation}
\end{cor}
\textbf{Proof.} It is a straightforward comparison term by term of the left hand side of the proposition with the right hand side.\\
$\square$

\begin{remark}
This corollary, together with (\ref{ExplGammaD1}), the formulae for the constant term $A^{\infty}_{n,r}$ and formula (\ref{shuffleAD1}), lead to completely explicit formulae for the $q$-expansion of all A-elliptic MZVs of depth one, which allow to approximate them numerically to high precision\footnote{One can easily get $500$ digits in less than a second with PARI GP.}. We have used this to check numerically all results given in the rest of this section.
\end{remark}
\begin{remark}\label{Rem2}
By formula (\ref{A(n,0,...,0)}) we deduce that all A-elliptic MZVs of the form $A_{1,r}(\tau)$ include $\Gamma(2,\tau)=-2\log(\eta(\tau))$ as part of their Fourier expansion, and therefore all $B_{1,r}(\tau):=B(1,\underbrace{0,\ldots ,0}_{r-1};\tau)$ will include $\log(\tau)$ in the asymptotic expansion. Note that for the modified
\begin{equation}\label{HatA}
\hat{A}_{1,r}(\tau):=A_{1,r}(\tau)-\frac{(2\pi i)^{r-2}}{(r-1)!}A_{1,2}(\tau)
\end{equation}
the term $\Gamma(2,\tau)$ disappears.
\end{remark}
Another consequence of the proposition above is that we can easily invert the r\^{o}le of A-elliptic MZVs and iterated integrals of Eisenstein series:
\begin{cor}\label{Gamma=A}
For any $n\geq 1$ and any $1\leq k\leq n-1$ we have
\begin{equation}
\Gamma^{L,0}_{n,k}(\tau)=\frac{(-1)^{n-k-1}(n-k-1)!}{(n-1)!}\sum_{i=2}^{k+1}\frac{\B_{k+1-i}(2\pi i)^{n-k-i}}{(k+1-i)!}A^0_{n-k,i}(\tau).
\end{equation}
\end{cor}
\textbf{Proof.} Note that $e^Y-1=Y+Y^2/2+\ldots $, and this corresponds to the fact that length one elliptic MZVs are constant. This means that, if we want to invert equation (\ref{A=E}), we need to rescale both sides, i.e. to multiply both sides by $Y$. This yields
\begin{equation}
\frac{Y}{e^Y-1}\mathcal{A}^0(X,Y;\tau)=\mathcal{E}^0(X,Y;\tau),
\end{equation} 
and comparing term by term this equation we are done.\\
$\square$

Now we want to deduce explicit formulae for B-elliptic MZVs of depth one. To do this, we need to know the behaviour of $\Gamma^{L}_{n,k}(\tau)$ under $\tau\mapsto -1/\tau$. Let us introduce another piece of notation. For any operator $\Lambda$ acting on a space of formal power series, we set $\Lambda^{(k)}:=\underbrace{\Lambda\circ\cdots\circ\Lambda}_{k}$, and $\exp(\Lambda):=\sum_{k\geq 0}\Lambda^{(k)}/k!$. Then we have the following two lemmas.
\begin{lemma}\label{E=F}
\begin{equation}
\mathcal{F}^0(X,Y;\tau)=\mathcal{E}^0(X+\tau Y,Y;\tau)
\end{equation}
\end{lemma}
\textbf{Proof.} By equation (\ref{gammaR}), we can write
\begin{equation}
\mathcal{F}^0(X,Y;\tau)=\sum_{k\geq 1}\frac{Y^{k-1}}{(k-1)!}\frac{\partial^{k-1}}{\partial X^{k-1}}\int_{\tau}^{i\infty}z^{k-1}\mathcal{G}^0(X,z)dz
\end{equation}
Since $d\int_{[\tau,i\infty]}\mathbb{G}^0_k(z)dz=-\mathbb{G}^0_k(\tau)$, repeatedly integrating by parts gives
\begin{multline}
\sum_{k\geq 1}\frac{Y^{k-1}}{(k-1)!}\frac{\partial^{k-1}}{\partial X^{k-1}}\int_{\tau}^{i\infty}z^{k-1}\mathcal{G}^0(X,z)dz=\\
=\exp\Big(\tau Y\frac{\partial}{\partial X}\Big)\sum_{k\geq 1}\frac{Y^{k-1}}{(2\pi i)^{k-1}}\frac{\partial^{k-1}}{\partial X^{k-1}}\int_{[\tau,i\infty]}\mathcal{G}^0(X,z_{1})dz_1\mathbb{G}_0dz_{2}\cdots \mathbb{G}_0dz_{k},
\end{multline}
and this, as we have already seen in the proof of Proposition \ref{PropA=E}, leads to the identity
\begin{equation*}
\mathcal{F}^0(X,Y;\tau)=\exp\Big(\tau Y\frac{\partial}{\partial X}\Big)\mathcal{E}^0(X,Y;\tau).
\end{equation*}
By Taylor's theorem $\exp\big(a\frac{\rm d}{\rm{d}x}\big)f(x)=f(x+a)$. This concludes the proof.\\
$\square$

By an abuse of notation, we denote by $\mathcal{F}^0(X,Y;-1/\tau)$ the exponentially small (with respect to $\tau\rightarrow i\infty$) part of the function $\mathcal{F}^0(X,Y;-1/\tau)$.\footnote{In general the latter yields some non-exponentially small term, as we will see later.} This will not create any confusion in what follows.
\begin{lemma}\label{F=F}
\begin{equation}
\mathcal{F}^0(X,Y;-1/\tau)=\mathcal{F}^0(Y,-X;\tau)
\end{equation}
\end{lemma}
\textbf{Proof.} We have already seen that we can write
\[
\mathcal{F}^0(X,Y;\tau)=\sum_{k\geq 1}\frac{Y^{k-1}}{(k-1)!}\frac{\partial^{k-1}}{\partial X^{k-1}}\int_{\tau}^{i\infty}z^{k-1}\mathcal{G}^0(X,z)dz.
\]
Therefore, using the transformation properties of $\mathcal{G}$, we find that the exponentially small part of $\mathcal{F}^0(X,Y;-1/\tau)$ is equal to
\[
\sum_{k\geq 1}\frac{(-1)^{k-1}Y^{k-1}}{(k-1)!}\frac{\partial^{k-1}}{\partial X^{k-1}}\int_{\tau}^{i\infty}z^{1-k}\mathcal{G}^0(zX,z)dz.
\]
Expanding this expression with respect to $X$, and using that $n+k$ must be even, we find
\[
\sum_{k,n\geq 1}\frac{(-1)^{n-1}(n+k-1)!}{(n-1)!(k-1)!}\int_{\tau}^{i\infty}z^{n-1}\mathbb{G}^0_{n+k}(z)dz\,Y^{k-1}X^{n-1}.
\]
Comparing this expression with the definition of $\mathcal{F}$ concludes the proof.\\
$\square$

Comparing the coefficients of the identity in Lemma \ref{E=F} gives
\begin{cor}
For each $n,j\geq 1$ we have
\begin{equation}
\Gamma^{R,0}_{n+j,j}(\tau)=\sum_{i=0}^{j-1}\frac{(-1)^{n+i-1}(2\pi i\tau)^i}{i!}\Gamma^{L,0}_{n+j,j-i}(\tau),
\end{equation}
\begin{equation}\label{Left=Right}
\Gamma^{L,0}_{n+j,j}(\tau)=\sum_{i=0}^{j-1}\frac{(-1)^{n+i-1}(2\pi i\tau)^i}{i!}\Gamma^{R,0}_{n+j,j-i}(\tau).
\end{equation}
\end{cor}
Moreover, Lemma \ref{F=F} implies
\begin{cor}
For $n, j\geq 1$,
\begin{equation}
\Gamma^{R,0}_{n+j,j}(-1/\tau)=(-1)^{n-1}\frac{(2\pi i)^{j-n}(n-1)!}{(j-1)!}\Gamma^{R,0}_{n+j,n}(\tau)
\end{equation}
\end{cor}
These two facts together lead to
\begin{cor}
Let $n\geq 1$ and $1\leq j\leq n-1$. Then
\begin{equation}
\Gamma^{R,0}_{n,j}(-1/\tau)=\frac{(2\pi i)^{j-1}(n-1-j)!}{(j-1)!}\sum_{i=1}^{n-j}\frac{(-1)^{i+j}\tau^{n-j-i}}{(n-j-i)!(2\pi i)^{i-1}}\Gamma^{L,0}_{n,i}(\tau)
\end{equation}
and
\begin{equation}
\Gamma^{L,0}_{n,j}(-1/\tau)=(2\pi i)^{j-1}\sum_{i=0}^{j-1}\frac{(n-j+i-1)!}{i!(j-i-1)!}\sum_{k=1}^{n-j+i}\frac{(-1)^{k-1}\tau^{n-j-k}}{(n-j+i-k)!(2\pi i)^{k-1}}\Gamma^{L,0}_{n,k}(\tau)
\end{equation}
\end{cor}
We are now ready to put all the pieces together.
Let us consider
\[
B_{n,r}(\tau)=B(n,\underbrace{0,\ldots,0}_{r-1};\tau).
\]
We write
\[
B_{n,r}(\tau)=B^{\infty}_{n,r}(\tau)+B^0_{n,r}(\tau),
\]
meaning as always that $B^{\infty}_{n,r}(\tau)$ denotes the first Laurent polynomials in the asymptotic expansion of $B_{n,r}(\tau)$ (plus a logarithmic term, in case $n=1$, cf. Remark \ref{Rem2}) and $B^0_{n,r}(\tau)$ denotes the exponentially small part. Let us consider the generating function
\[
\mathcal{B}^0(X,Y;\tau):=\sum_{n,r\geq 1}\frac{B^0_{n,r}(\tau)}{(2\pi i)^{r-1}}X^{n-1}Y^{r-1}.
\]
Then we have
\begin{thm}\label{B=Gamma}
\begin{equation}
\mathcal{B}^0(X,Y;\tau)=(e^Y-1)\mathcal{E}^0\Big(-\tau X,-X-\frac{Y}{\tau};\tau\Big).
\end{equation}
\end{thm}
\textbf{Proof.} By Proposition \ref{PropA=E} and Lemma \ref{E=F} we have
\[
\mathcal{A}^0(X,Y;-1/\tau)=(e^Y-1)\mathcal{E}^0(X,Y;-1/\tau)=(e^Y-1)\mathcal{F}^0\Big(X+\frac{Y}{\tau},Y;-1/\tau\Big),
\]
where in this case $\mathcal{F}^0(X,Y;-1/\tau)$ really means that we consider $\mathcal{F}^0(X,Y;\tau)$ and then we invert $\tau$. Using Lemma \ref{F=F} and again Lemma \ref{E=F} we conclude that
\begin{equation*}
\mathcal{B}^0(X,Y;\tau)=(e^Y-1)\mathcal{F}^0\Big(Y,-X-\frac{Y}{\tau};\tau\Big)=(e^Y-1)\mathcal{E}^0\Big(-\tau X,-X-\frac{Y}{\tau};\tau\Big),
\end{equation*}
where in this case $\mathcal{F}^0(X,Y;-1/\tau)$ was meant to be the exponentially small part of $\mathcal{F}^0(X,Y;-1/\tau)$.
\\
$\square$

The consequence of this theorem is that we are now able to express B-elliptic MZVs of depth one in terms of iterated integrals of Eisenstein series, and therefore in terms of A-elliptic MZVs:
\begin{cor}
For $n,r\geq 1$ we have
\begin{multline}
B_{n,r}(\tau)=B^{\infty}_{n,r}(\tau)+\\
\frac{(2\pi i)^{r-1}}{(n-1)!}\sum_{j=1}^{r-1}\frac{(n+j-1)!}{(r-j)!}\sum_{i=0}^{j-1}\frac{(-1)^{j-i-1}(n+i-1)!}{i!(j-i-1)!}\sum_{k=1}^{n+i}\frac{(-1)^{k-1}\tau^{n-k}}{(2\pi i)^{k-1}(n+i-k)!}\Gamma^{L,0}_{n+j,k}(\tau),
\end{multline}
and thus
\begin{multline}
B_{n,r}(\tau)=B^{\infty}_{n,r}(\tau)+\frac{(-1)^{n-1}(2\pi i)^{r-1}}{(n-1)!}\sum_{j=1}^{r-1}\frac{1}{(r-j)!}\sum_{i=0}^{j-1}\frac{(-1)^{i}(n+i-1)!}{i!(j-i-1)!}\times\\
\times\sum_{k=1}^{n+i}\frac{(n+j-k-1)!\tau^{n-k}}{(n+i-k)!}\sum_{l=2}^{k+1}\frac{\B_{k+1-l}}{(k+1-l)!(2\pi i)^{l-1}}A^0_{n+j-k,l}(\tau).
\end{multline}
\end{cor}
\textbf{Proof.} One can prove the first statement either by comparing term by term the equation of Theorem \ref{B=Gamma}, or by using the corollaries of Proposition \ref{PropA=E}, Lemma~\ref{E=F} and Lemma~\ref{F=F}. The second statement follows from the first plus Corollary~\ref{Gamma=A}.\\
$\square$

In particular, one can see that the range of powers of $\tau$ agrees with the range predicted by Theorem \ref{asymptB}. It is useful to write down the formulae given in the corollary, together with the explicit formula for the constant term, for the simplest case of length 2:
\begin{eqnarray*}
B_{n,2}(\tau)&=&B^\infty_{n,2}(\tau)+\sum_{k=1}^n\frac{(-1)^{k-1}n!\tau^{n-k}}{(n-k)!(2\pi i)^{k-2}}\Gamma^{L,0}_{n+1,k}(\tau)\\
&=&B^\infty_{n,2}(\tau)+(-1)^{n-1}n!\sum_{k=1}^n\tau^{n-k}\sum_{l=2}^{k+1}\frac{\B_{k+1-l}}{(k+1-l)!(2\pi i)^{l-2}}A^0_{n+1-k,l}(\tau)
\end{eqnarray*}

The first non-trivial examples of B-elliptic MZVs (recall that in order to have interesting examples we need $n+r$ odd, and we prefer to avoid B-elliptic MZVs with $\log(\tau)$ in the expansion) given by our construction are
\begin{equation*}
B_{3,2}(\tau)=-\frac{(2\pi i)^2}{720}\tau^3-\frac{\zeta(3)}{2\pi i}-\frac{6\zeta(4)}{(2\pi i)^2}\frac{1}{\tau}+6\pi i\tau^2\Gamma^{L,0}_{4,1}(\tau)-6\tau\Gamma^{L,0}_{4,2}(\tau)+\frac{3}{\pi i}\Gamma^{L,0}_{4,3}(\tau),
\end{equation*}
\begin{equation*}
B_{2,3}(\tau)=\frac{(2\pi i)^3}{720}\tau^2-\frac{\pi i}{2}\zeta(2)-\frac{2\zeta(3)}{\tau}-\frac{3\zeta(4)}{\pi i}\frac{1}{\tau^2}-48\pi^2\Gamma^{L,0}_{4,1}(\tau)-36\pi i\Gamma^{L,0}_{4,2}(\tau)+\frac{12}{\tau}\Gamma^{L,0}_{4,3}(\tau),
\end{equation*}
\begin{multline*}
B_{5,2}(\tau)=\frac{(2\pi i)^2}{30240}\tau^5-\frac{\zeta(5)}{(2\pi i)^3}-\frac{10\,\zeta(6)}{(2\pi i)^4\tau}+10\pi i\tau^4\Gamma^{L,0}_{6,1}(\tau)\\
-20\tau^3\Gamma^{L,0}_{6,2}(\tau)+\frac{30}{\pi i}\tau^2\Gamma^{L,0}_{6,3}(\tau)-\frac{120\,\tau}{(2\pi i)^2}\Gamma^{L,0}_{6,4}(\tau)+\frac{120}{(2\pi i)^3}\Gamma^{L,0}_{6,5}(\tau),
\end{multline*}
\begin{multline*}
B_{4,3}(\tau)=-\frac{(2\pi i)^3}{30240}\tau^4-\frac{1}{2}\frac{\zeta(4)}{2\pi i}-\frac{4\,\zeta(5)}{(2\pi i)^2\tau}-\frac{20\,\zeta(6)}{(2\pi i)^3\tau^2}\\
-40\pi i\tau^2\Gamma^{L,0}_{6,2}(\tau)+120\tau\Gamma^{L,0}_{6,3}(\tau)-\frac{180}{\pi i}\Gamma^{L,0}_{6,4}(\tau)+\frac{480}{(2\pi i)^2\tau}\Gamma^{L,0}_{6,5}(\tau),
\end{multline*}
\begin{multline*}
B_{3,4}(\tau)=-\frac{(2\pi i)^4}{5040}\tau^3-\frac{\pi i\,\zeta(3)}{6}-\frac{3\,\zeta(4)}{2\tau}-\frac{3\,\zeta(5)}{\pi i\tau^2}-\frac{20\,\zeta(6)}{(2\pi i)^2\tau^3}
+\frac{(2\pi i)^3\tau^2}{2}\Gamma^{L,0}_{4,1}(\tau)\\
-(2\pi i)^2\tau\Gamma^{L,0}_{4,2}(\tau)+2\pi i\,\Gamma^{L,0}_{4,3}(\tau)+120\pi i\Gamma^{L,0}_{6,3}(\tau)-\frac{360}{\tau}\Gamma^{L,0}_{6,4}(\tau)+\frac{360}{\pi i\tau^2}\Gamma^{L,0}_{6,5}(\tau),
\end{multline*}
\begin{multline*}
B_{2,5}(\tau)=\frac{(2\pi i)^5}{5040}\tau^2-\frac{(2\pi i)^3\zeta(2)}{24}-\frac{(2\pi i)^2\zeta(3)}{3\tau}-\frac{3\pi i\,\zeta(4)}{\tau^2}-\frac{4\,\zeta(5)}{\tau^3}-\frac{5\,\zeta(6)}{\pi i\tau^4}\\
-(2\pi i)^3\Gamma^{L,0}_{4,2}(\tau)+\frac{2(2\pi i)^2}{\tau}\Gamma^{L,0}_{4,3}(\tau)-\frac{240\pi i}{\tau^2}\Gamma^{L,0}_{6,4}(\tau)+\frac{480}{\tau^3}\Gamma^{L,0}_{6,5}(\tau).
\end{multline*}

\subsection{Modular behaviour}\label{SectionDepth1Mod}

We want now to repeat the same steps of the previous subsection, this time keeping track not only of the generating functions of the exponentially small terms, but of the full generating functions. To do this, we need to exploit Brown's theory of iterated integrals of Eisenstein series. This will make the construction more technical, but has two advantages: it gives for free the exponentially small term of B-elliptic MZVs with a different method (it will come from the period polynomial of Eisenstein series), and most importantly it leads to a nice description not only of B-elliptic MZVs, but of the transformation of depth one A-elliptic MZVs with respect to any $\gamma\in\mbox{SL}_2(\mathbb{Z})$. In what follows, for any generating function
\[
K(X_1,\ldots , X_r)=\sum_{n_1,\ldots ,n_r\geq 0}k_{n_1,\ldots ,n_r}X_1^{n_1}\cdots X_r^{n_r},
\]
we denote
\[
K^*(X_1,\ldots , X_r)=K(X_1,\ldots , X_r)-k_{0,\ldots ,0}.
\]
In particular, we have
\[
\mathcal{E}^*(X,Y;\tau)=\mathcal{E}(X,Y;\tau)-\Gamma(2;\tau)
\]
and
\[
\mathcal{F}^*(X,Y;\tau)=\mathcal{F}(X,Y;\tau)-\Gamma(2;\tau).
\]
Note that
\begin{equation}
\Gamma(2;\tau)=\int_{\tau}^{\1}\mathbb{G}_2(z)dz=-2\log(\eta(\tau)).
\end{equation}
Moreover, for a formal power series
\[
K(X_1,\ldots,X_p)=\sum_{n_1,\ldots ,n_p\geq 0}k_{n_1,\ldots ,n_p}X^{n_1}\cdots X^{n_p}
\]
we define\footnote{$\mathcal{L}$ stays for Laplace, because this is what happens to the solution of the Laplace transform of a differential equation.}
\[
\mathcal{L}_{X_{i_1},\ldots X_{i_q}}K(X_1,\ldots,X_p):=\sum_{n_1,\ldots ,n_p\geq 0}\frac{k_{n_1,\ldots ,n_p}}{(n_{i_1}+\cdots +n_{i_q})!}X^{n_1}\cdots X^{n_p}
\]
Recall now the definition (\ref{SingleEisIntBrown}) of $[\underline{E}_m]$, which gives
\begin{equation}
[\underline{E}_m]=(2\pi i)^{m-1}\sum_{j,n\geq 1}\sum_{j+n=m}\frac{(j+n-2)!}{(j-1)!(n-1)!}\int_{\tau}^{\1}z^{j-1}E_{m}(z)dz\,X^{n-1}(-Y)^{j-1}.
\end{equation}
Let us also consider the generating series of single integrals of holomorphic Eisenstein series
\begin{equation}
\mathcal{H}(X,Y;\tau):=\sum_{m\geq 4}[\underline{E}_m].
\end{equation}
Since by (\ref{ConversionEis2})
\begin{equation}
\mathbb{G}_m(\tau)=\frac{4\pi i}{(m-1)!}E_m(\tau),
\end{equation}
it is easy to see that one has the following identity of generating series:
\begin{lemma}\label{LemmaF=H}
\begin{equation}
\mathcal{F}^*(X,Y;\tau)=2\mathcal{L}_{X,Y}\mathcal{H}\Big(\frac{X}{2\pi i},\frac{-Y}{2\pi i};\tau\Big).
\end{equation}
\end{lemma}
We recall from Section \ref{sectionItIntEis} that to any Eisenstein series $E_m$ we can associate a cocycle $\mathcal{C}_\gamma (a_{E_m})$ with values in $\mathbb{Q}[X,Y]_{m-2}$. Let us denote the generating function of these polynomials by
\[
\mathcal{P}_\gamma (X,Y):=\sum_{m\geq 4}\mathcal{C}_\gamma (a_{E_m}).
\]
It follows from the theory developed in \cite{MMV} that
\begin{prop}[Brown]
For any
\[
\gamma =\left( \begin{array}{cc}
a & b \\
c & d \end{array} \right)\in\mbox{SL}_2(\mathbb{Z})
\] 
we have
\begin{equation}
\mathcal{H}(X,Y;\gamma\tau)=\mathcal{H}(dX-bY,-cX+aY;\tau)-\mathcal{P}_{\gamma}(dX-bY,-cX+aY).
\end{equation}
\end{prop}
\textbf{Proof.} This is a straightforward consequence of Lemma \ref{LemmaBrown}.\\
$\square$

The last two results immediately imply the following
\begin{cor}\label{StransformF}
\begin{equation}
\mathcal{F}^*(X,Y;\gamma\tau)=\mathcal{F}^*(dX+bY,cX+aY;\tau)-2\mathcal{L}_{X,Y}\mathcal{P}_{\gamma}\Big(\frac{dX+bY}{2\pi i},\frac{-cX-aY}{2\pi i}\Big).
\end{equation}
\end{cor}
We want to use this fact to get a formula for $A_{n,r}(\gamma\tau)$. Note that we can get a refined version of Lemma \ref{E=F}:
\begin{lemma}\label{lemmaE=Fserio}
\begin{equation}
\mathcal{F}(X,Y;\tau)=\mathcal{E}(X+\tau Y,Y;\tau)
\end{equation}
\end{lemma}
\textbf{Proof.} Because of Lemma \ref{E=F}, it is enough to prove this equality for $\mathcal{E}^\infty$ and~$\mathcal{F}^\infty$. By definition, it is easy to see that
\[
\mathcal{F}^{\infty}(X,Y;\tau)=-\sum_{j,n\geq 1}\frac{(n+j-1)!\tau^j}{j!(n-1)!}\mathbb{G}^{\infty}_{n+j}X^{n-1}Y^{j-1}=\mathcal{E}^{\infty}(X,-Y;\tau).
\]
Therefore, we are left with proving the following identity:
\begin{equation}
\exp\Big(\tau Y\frac{\partial}{\partial X}\Big)\mathcal{E}^\infty (X,Y;\tau)=\mathcal{E}^{\infty}(X,-Y;\tau).
\end{equation}
This is true because
\begin{multline*}
\exp\Big(\tau Y\frac{\partial}{\partial X}\Big)\mathcal{E}^\infty (X,Y;\tau)=\\
= -\sum_{k\geq 0}\frac{(\tau Y)^k}{k!}\frac{\partial^k}{\partial X^k}\sum_{j,n\geq 1}\frac{(-1)^{j-1}(n+j-1)!}{j!(n-1)!}\mathbb{G}^{\infty}_{n+j}\tau^jX^{n-1}Y^{j-1}\\
=-\sum_{m,r\geq 1}\Big(\sum_{k=0}^{r-1}\frac{(-1)^{r-k-1}}{k!(r-k)!}\Big)\frac{(m+r-1)!}{(m-1)!}\mathbb{G}^{\infty}_{m+r}\tau^rX^{m-1}Y^{r-1}.
\end{multline*}
Since
\[
\sum_{k=0}^{r-1}\frac{(-1)^{r-k-1}}{k!(r-k)!}=-\frac{1}{r!}\big((1-1)^r-1\big),
\]
we get our claim.\\
$\square$

Let us state a refined version of Proposition \ref{PropA=E}:
\begin{prop}\label{PropA=Eseria}
\begin{equation}\label{eqA=Eseria}
\mathcal{A}(X,Y;\tau)=\mathcal{A}^{\infty}(X-\tau Y,Y)+(e^Y-1)\mathcal{E}(X,Y;\tau).
\end{equation}
\end{prop}
\textbf{Proof.} By Lemma \ref{lemmaDiffDepth1}, we have
\begin{equation}
\mathcal{A}(X,Y;\tau)=\mathcal{A}^{\infty}(X,Y)+(e^Y-1)\int_\tau^{\1}\mathcal{G}(X;z)dz+Y\frac{\partial}{\partial X}\int_\tau^{\1}\mathcal{A}(X,Y;z)dz.
\end{equation}
Iterating this identity\footnote{One needs to apply the integration procedure of Section \ref{sectionItIntEis} at each iteration step.}, we get our claim.\\
$\square$

Note that one can invert equation (\ref{eqA=Eseria}) and get
\begin{equation}\label{eqE=A}
\mathcal{E}(X,Y;\tau)=\frac{\mathcal{A}(X,Y;\tau)-\mathcal{A}^{\infty}(X-\tau Y,Y)}{e^Y-1}.
\end{equation}
We are now in the position to put all the pieces together, and get the full modular behaviour of depth one elliptic MZVs.
\begin{thm}\label{ThmFullMod}
For any $\gamma\in\mbox{SL}_2(\mathbb{Z})$, if we write
\[
(X^\prime,Y^\prime)=(dX+bY,cX+dY)
\]
we have\footnote{As we remarked in the proof of Lemma \ref{E=F}, one can write $f(x+a)=\exp\big(a\frac{\rm d}{\rm{d}x}\big)f(x)$. In the statement of this theorem  (and of the following corollary) we prefer to employ the exponential notation.}
\begin{multline}
\mathcal{A}(X,Y;\gamma\tau)=\exp\Big(-(\gamma\tau) Y\frac{\partial}{\partial X}\Big)\mathcal{A}^{\infty}(X,Y)-2(e^Y-1)\log(\eta(\gamma\tau))\\
+(e^Y-1)\exp\Big(-(\gamma\tau) Y\frac{\partial}{\partial X}\Big)\exp\Big(\tau Y^\prime\frac{\partial}{\partial X^\prime}\Big)\mathcal{E}^*(X^\prime,Y^\prime;\tau)\\
-2(e^Y-1)\exp\Big(-(\gamma\tau) Y\frac{\partial}{\partial X}\Big)\mathcal{L}_{X^\prime,Y^\prime}\mathcal{P}_{\gamma}\Big(\frac{X^\prime}{2\pi i},\frac{-Y^\prime}{2\pi i}\Big).
\end{multline}
\end{thm}
\textbf{Proof.} It is a consequence of Proposition \ref{PropA=Eseria}, Lemma \ref{lemmaE=Fserio} and Corollary \ref{StransformF}.\\
$\square$

\begin{cor}
With the same notation of the theorem, we have
\begin{multline}
\mathcal{A}(X,Y;\gamma\tau)=\exp\Big(-(\gamma\tau) Y\frac{\partial}{\partial X}\Big)\bigg(1-\frac{e^Y-1}{e^{Y^\prime}-1}\bigg)\mathcal{A}^{\infty}(X,Y)-2(e^Y-1)\big(\log\eta(\gamma\tau)-\log\eta(\tau)\big)\\
+(e^Y-1)\exp\Big(-(\gamma\tau) Y\frac{\partial}{\partial X}\Big)\Bigg(\frac{\exp\Big(\tau Y^\prime\frac{\partial}{\partial X^\prime}\Big)}{e^{Y^\prime}-1}\mathcal{A}(X^\prime,Y^\prime;\tau)\Bigg)\\
-2(e^Y-1)\exp\Big(-(\gamma\tau) Y\frac{\partial}{\partial X}\Big)\mathcal{L}_{X^\prime,Y^\prime}\mathcal{P}_{\gamma}\Big(\frac{X^\prime}{2\pi i},\frac{-Y^\prime}{2\pi i}\Big)
\end{multline}
\end{cor}
\textbf{Proof.} By equation (\ref{eqE=A}) we get
\begin{equation}
\mathcal{F}^*(X,Y;\tau)=\frac{\exp\Big(\tau Y\frac{\partial}{\partial X}\Big)\mathcal{A}(X,Y;\tau)}{e^Y-1}-\frac{\mathcal{A}^\infty (X,Y)}{e^Y-1}-\Gamma(2,\tau).
\end{equation}
This, together with the previous theorem, leads to our claim.\\
$\square$

Finally, we conclude this section with the following remark on the modular nature of depth one elliptic MZVs:
\begin{thm}\label{ThmModBehaviour}
For all $n\geq 2$ the vector spaces
\[
\big\langle A_{n,r}(\tau)|_{1-r}\,\gamma:\gamma\in\mbox{SL}_2(\mathbb{Z}) \big\rangle_\mathbb{C}
\]
are finite dimensional, as well as the vector spaces
\[
\big\langle \hat{A}_{1,r}(\tau)|_{1-r}\,\gamma:\gamma\in\mbox{SL}_2(\mathbb{Z}) \big\rangle_\mathbb{C},
\]
where $\hat{A}_{1,r}(\tau)$ was defined by (\ref{HatA}).
\end{thm}
\textbf{Proof.} We will prove this theorem for $A_{n,r}(\tau)$ with $n\geq 2$. The same proof can be repeated for $\hat{A}_{1,r}(\tau)$ (but not for $A_{1,r}(\tau)$). First of all, note that it is enough to prove that $A_{n,r}(\tau)$ can be written as a sum of functions that generate a finite dimensional vector space in weight $1-r$. From equations (\ref{A(n,0,...,0)}) and (\ref{Left=Right}) we can write $A_{n,r}(\tau)$ as a linear combination over $\mathcal{A}[(2\pi i)^{\pm 1}]$ of $1$ and $\tau^i\Gamma^{R,0}_{n+j,j-i}$, with $0\leq i\leq j-1$ and $1\leq j\leq r-1$. We know by Lemma \ref{LemmaF=H} and Lemma \ref{LemmaBrown} that
\[
\dim \Big(\big\langle \Gamma^{R}_{n,k}|_{0}\,\gamma:\gamma\in\mbox{SL}_2(\mathbb{Z}) \big\rangle_\mathbb{C}\Big)=n.
\]
Moreover, polynomials in $\tau$ of degree $\leq k$ obviously generate a finite dimensional vector space under the weight $-k$ action of SL$_2(\mathbb{Z})$, and therefore
\[
\big\langle \Gamma^{R,0}_{n,k}|_{-k}\,\gamma:\gamma\in\mbox{SL}_2(\mathbb{Z}) \big\rangle_\mathbb{C}
\]
is finite dimensional. Finally, if
\[
\big\langle f|_{k}\,\gamma:\gamma\in\mbox{SL}_2(\mathbb{Z}) \big\rangle_\mathbb{C}
\]
is finite dimensional, then it is easy to see that also
\[
\big\langle \tau^if|_{k-i-p}\,\gamma:\gamma\in\mbox{SL}_2(\mathbb{Z}) \big\rangle_\mathbb{C}
\]
must be finite dimensional for each $i,p\geq 0$. Therefore we can conclude that all summands of our expression for $A_{n,r}(\tau)$ satisfy the desired condition, and the statement of the theorem is proven.\\
$\square$

This means that each depth one elliptic MZVs $A_{n,r}(\tau)$ can be seen as the component of a weight~$1-r$ vector valued modular form, as shown concretely in the examples of Appendix \ref{AppendixVectorValued}. This is in general not true for (modular) weight $\geq 2-r$, because of Remark \ref{Slash}.

\subsection{Relations with Eichler integrals}

\begin{defn}
For all $n\geq 4$ let us consider the normalized Eisenstein series $E_{n}(\tau)$ defined by (\ref{DefEisBern}), i.e. the ones whose constant term reads $-\B_n/2n!$. An \emph{Eichler integral} of $E_n(\tau)$ is any holomorphic function $\tilde{E}_n(\tau)$ satisfying
\[
\bigg(\frac{1}{2\pi i}\frac{\partial}{\partial \tau}\bigg)^{n-1}\tilde{E}_n(\tau)=E_{n}(\tau).
\]
\end{defn}
This primitive is uniquely determined up to a degree $n-2$ polynomial. By Bol's identity (see for instance \cite{LewisZagier}), these functions satisfy
\[
\tilde{E}_n(\tau)|_{2-n}\,\gamma=\tilde{E}_n(\tau)+\tilde{P}_{n,\gamma}(\tau),
\]
where $\tilde{P}_{n,\gamma}(\tau)$ is a polynomial of degree $n-2$ depending on the choice of Eichler integral: for instance, if we chose the primitive given by the tangential base point prescription of Brown described in Section \ref{sectionItIntEis}, we would get
\[
\tilde{E}_n(\tau)=[\underline{E}_n(\tau)]\Big|_{\substack{X=\tau\\Y=1}},
\]
and therefore $\tilde{P}_{n,\gamma}(\tau)$ would coincide with the cocycle $\mathcal{C}_\gamma (a_{E_n})$ evaluated at $(X,Y)=(\tau,1)$. Here, however, we want to make a different choice, following Zagier-Gangl \cite{ZagierGangl}. For us, for each $m\geq 2$
\begin{equation}\label{EichlerIntegralZagGan}
\tilde{E}_{2m}(\tau):=\frac{1}{2}\zeta(1-2m)\frac{(2\pi i\tau)^{2m-1}}{(2m-1)!}+\frac{\zeta(2m-1)}{2}+\sum_{j\geq 1}\sigma_{1-2m}(j)q^j.
\end{equation}
This choice was dictated in \cite{ZagierGangl} by the nice arithmetic properties of the corresponding polynomial $\tilde{P}_{2m,\gamma}(\tau)$. Here, the main motivation is the following:
\begin{prop}
For all $r\geq 1$ we have
\begin{equation}
\hat{A}_{1,r}=-(2\pi i)^r\sum_{j=1}^{r-2}\frac{\B_{2+j}}{(2+j)!}\frac{\tau^{j+1}}{(r-j-1)!}-(2\pi i)^r\sum_{j=1}^{r-2}\frac{(2\pi i)^{-1-j}}{(r-j-1)!}\tilde{E}_{2+j}(\tau),
\end{equation}
where $\hat{A}_{1,r}$ is the version of $A_{1,r}$ given by (\ref{HatA}), and $\tilde{E}_{2m+1}(\tau):=0$.
\end{prop}
\textbf{Proof.} This is just a consequence of formula (\ref{FormulaAinf1r}) for the constant term $A^{\infty}_{1,r}$, formula (\ref{A(n,0,...,0)}) for A-elliptic MZVs of depth one in terms of iterated integrals of Eisenstein series and formula (\ref{ExplGammaD1}) for the explicit expansion of iterated integrals of Eisenstein series.\\
$\square$

Moreover, note that during the proof we have implicitly made use of the fact that
\[
\Gamma^L_{n,n-1}(\tau)=-\frac{2}{(n-1)!}\tilde{E}_{n}(\tau)+\frac{\zeta(n-1)}{(n-1)!}.
\]
Using this, it is straightforward to see that, for all $1\leq j\leq n-2$,
\begin{equation*}
\bigg(\frac{1}{2\pi i}\frac{\partial}{\partial \tau}\bigg)^{j}\tilde{E}_n(\tau)=-\frac{(n-1)!}{2}\Gamma^L_{n,n-1-j},
\end{equation*}
which means that all A-elliptic MZVs can easily be written in terms of the Eichler integrals (\ref{EichlerIntegralZagGan}) and their derivatives. It was shown in \cite{ZagierGangl} that repeatedly applying a certain non-holomorphic derivative on these Eichler integrals leads to interesting non-holomorphic functions, whose real part is just a real analytic Eisenstein series. This leads to an explicit link between elliptic MZVs and the simplest instances of modular graph functions, that will be explored further in \cite{Zerb17}. In the next section, we will see another way to relate real analytic Eisenstein series to A-elliptic MZVs of depth one, based on Brown's paper \cite{MMV}.

\subsection{Back to elliptic polylogarithms}

It was already noticed by Matthes \cite{MatthesMeta} that depth one elliptic MZVs are essentially the special values at $\xi=0$ of Levin's elliptic polylogarithms, i.e. special values of depth one multiple elliptic polylogarithms in the sense of Brown-Levin \cite{BrownLevin}. This correspondence was carried out by viewing depth one elliptic MZVs as the coefficient of a certain quotient of the elliptic associator. What we present here is a translation of Matthes's result to our setting.

\begin{prop}\label{eMZVellPol}
Recall the generating series $\Xi(\xi,\tau,X,Y)$ of Levin's elliptic polylogarithms defined by (\ref{DefXi}). We have that
\begin{multline}
\Big(\Xi(\xi,X,Y,\tau)-\frac{1}{2\pi i}\log(2\pi i\xi)\Big)\Big|_{\xi=0}=-\frac{\tau}{X(X-\tau Y)}-\frac{\mathcal{A}^{\infty}(X,-Y)}{2\pi i(\exp(-Y)-1)}\\
+\frac{\mathcal{A}(X-\tau Y,-Y;\tau)}{2\pi i(\exp(-Y)-1)}+\frac{1}{(e^X-1)(e^Y-1)}-\sum_{k\geq 2}\frac{\zeta(n)}{(2\pi i)^n}Y^{n-1}.
\end{multline}
\end{prop}
\textbf{Proof.} It is easy to see from Theorem \ref{TeoLEVIN} and from the definition of $\mathcal{F}(X,Y;\tau)$ that the left hand side of the statement is equal to
\[
-\frac{\tau}{X(X-\tau Y)}+\frac{1}{2\pi i}\,\mathcal{F}(X,-Y;\tau)+C^{0,0}(X,Y).
\]
Moreover, one can prove by the explicit definition of Debye polylogarithm $\underline{\Lambda}(\xi,\tau,X,Y)$ given in \cite{Levin} that
\[
C^{0,0}(X,Y)=\frac{1}{(e^X-1)}\bigg(\frac{1}{e^Y-1}-\frac{1}{Y}\bigg)-\sum_{k\geq 2}\frac{\zeta(n)}{(2\pi i)^n}Y^{n-1}.
\]
The statement then follows from Lemma \ref{lemmaE=Fserio} and Proposition \ref{PropA=Eseria}.
\\
$\square$

Now let us see what is the link with the single-valued elliptic polylogarithms considered in Section \ref{SectionSVEllPol}. This is essentially the adaptation of a result of Brown \cite{MMV} to our setting. Let us denote $\tau=\tau_1+i\tau_2$, and consider the generating function of real analytic Eisenstein series
\[
\mathcal{R}(X,Y;\tau)=\frac{\tau_2}{2\pi}\sum_{\substack{i,j\geq 1\\i+j\geq 4}}(i+j-1)!\sum_{(m,n)\neq (0,0)}\frac{(X-\tau  Y)^{i-1}(X-\overline{\tau}Y)^{j-1}}{(m\tau+n)^i(m\overline{\tau}+n)^j}.
\]
Real analytic Eisenstein series are essentially special values of the single-valued elliptic polylogarithms introduced in Section \ref{SectionSVEllPol}. We have
\begin{prop}\label{eMZVsNonHolo}
\begin{multline}
\mathcal{R}(X,Y;\tau)=\frac{1}{2}\mathcal{L}^{-1}_{X,Y}\Re\bigg(\frac{\mathcal{A}^{\infty}(2\pi iX,-2\pi iY)}{e^{-2\pi iY}-1}\bigg)-\log|\eta(\tau)|\\
-\frac{1}{2}\mathcal{L}^{-1}_{X,Y}\Re\bigg(\frac{\mathcal{A}(2\pi i(X-\tau Y),-2\pi iY;\tau)}{e^{-2\pi iY}-1}\bigg)+\sum_{k\geq 2}\frac{(2k-2)!}{2}\zeta(2k-1)Y^{2k-2}
\end{multline}
\end{prop}
\textbf{Proof.} By Lemma 9.6 of \cite{MMV}, we know that 
\[
\mathcal{R}(X,Y;\tau)=-\Re\big(\mathcal{H}(X,Y;\tau)\big)+\sum_{k\geq 2}\frac{(2k-2)!}{2}\zeta(2k-1)Y^{2k-2}.
\]
The result then follows from Lemma \ref{LemmaF=H}, Lemma \ref{lemmaE=Fserio} and Proposition \ref{PropA=Eseria}.
\\
$\square$

%% file: Chapter6.tex

\chapter{Conclusions} 

\label{ChapterConclusions} 

\section{Equivariant iterated integrals of Eisenstein series}

In order to have a clear overview of the future directions, we want to report on a new result of Brown, contained in his very recent paper \cite{BrownNewClass}.
\begin{defi}
Let $f:\mathbb{H}\longrightarrow\mathbb{C}$ be a real analytic function. We call it \emph{modular of weights} $(r,s)$ if for every $\gamma\in\mbox{SL}_2(\mathbb{Z})$ it satisfies
\begin{equation}\label{UN}
f(\gamma\tau)=(c\tau+d)^r(c\overline{\tau}+d)^sf(\tau).
\end{equation}
We denote $\mathcal{M}_{r,s}$ the space of modular functions of weights $(r,s)$ which admit an expansion of the form 
\begin{equation}\label{DOS}
f(q)\in\mathbb{C}[[q,\overline{q}]][\Im(\tau)^{\pm 1}].
\end{equation}
We also want to consider the bigraded algebra
\[
\mathcal{M}=\bigoplus_{r,s}\mathcal{M}_{r,s}.
\]
\end{defi}
Note that we have already encountered many examples of functions belonging to the algebra $\mathcal{M}$: all special values $e_{a,b}(0,\tau)$ of the single-valued elliptic polylogarithms (\ref{svEllPol}) encountered in Chapter \ref{ChapterMathBackground} are easily shown to belong to $\mathcal{M}_{-b-1,-a-1}$, but most importantly our Theorem \ref{main} implies that all modular graph functions belong to~$\mathcal{M}_{0,0}$.

Then the main result of \cite{BrownNewClass} tells us, among other things, that there exists a subalgebra $\mathcal{M}\mathcal{I}^E\subset\mathcal{M}$ generated over the algebra of single-valued MZVs $\mathcal{A}^{\rm sv}$ by certain computable linear combinations of real and imaginary parts of regularized\footnote{We are referring to the regularization introduced in Section \ref{sectionItIntEis}.} iterated integrals of Eisenstein series, such that:
\begin{itemize}
\item It carries a grading by a certain \emph{M-degree} and a filtration by the length of the iterated integrals, denoted by $\mathcal{M}\mathcal{I}^E_k\subset\mathcal{M}\mathcal{I}^E$.
\item Every element of $\mathcal{M}\mathcal{I}^E$ admits an expansion of the form
\[
f(q)\in\mathcal{A}^{\rm sv}[[q,\overline{q}]][\Im(\tau)^{\pm 1}]
\]
\item The sub-vector space of elements of fixed modular weights and M-degree $\leq m$ is finite dimensional.
\item Every element $F\in\mathcal{M}\mathcal{I}^E_k$ satisfies an inhomogeneous Laplace equation of the form
\[
(\Delta+w)F\in(E+\overline{E})[\Im(\tau)]\times\mathcal{M}\mathcal{I}^E_{k-1}+E\overline{E}[\Im(\tau)]\times \mathcal{M}\mathcal{I}^E_{k-2},
\]
where $E$ denotes the space of holomorphic Eisenstein series for SL$_2(\mathbb{Z})$.
\end{itemize}

These properties agree perfectly with all the conjectures that we have mentioned on modular graph functions (including our conjecture on single-valued multiple zeta values). Indeed, in \cite{BrownNewClass} it is conjectured that modular graph functions should belong to the class $\mathcal{M}\mathcal{I}^E$. Moreover, the linear combinations of real and imaginary parts of iterated integrals of Eisenstein series are precisely the same linear combination that one encounters for elliptic multiple zeta values.

\section{Future directions}

The conjectural picture given in this thesis for superstring amplitudes of genus zero and one can be resumed as follows:
\begin{enumerate}
\item Open string amplitudes in genus zero can be written in terms of MZVs, i.e. certain special values of (holomorphic) homotopy invariant iterated integrals on $\mathbb{P}^1_{\mathbb{C}}\setminus\{0,1,\infty\}$. This is already proven \cite{BSST}.
\item Closed string amplitudes in genus zero can be written in terms of single-valued MZVs, i.e. certain special values of real analytic single-valued versions of homotopy invariant iterated integrals on $\mathbb{P}^1_{\mathbb{C}}\setminus\{0,1,\infty\}$. This is still a conjecture \cite{ScSt}.
\item Open string amplitudes in genus one can be written in terms of elliptic MZVs, i.e. certain special values of (holomorphic) homotopy invariant iterated integrals on (configuration spaces of points in) $\mathbb{C}/(\tau\mathbb{Z}+\mathbb{Z})\setminus \{0\}$. This is already proven \cite{BMMS}.
\item Closed string amplitudes in genus one can be written in terms of certain special values of real analytic single-valued versions of homotopy invariant iterated integrals on (configuration spaces of points in) $\mathbb{C}/(\tau\mathbb{Z}+\mathbb{Z})\setminus \{0\}$. This is still a conjecture. Moreover, single-valued versions of homotopy invariant iterated integrals on (configuration spaces of points in) $\mathbb{C}/(\tau\mathbb{Z}+\mathbb{Z})\setminus \{0\}$ still need to be constructed.
\item Closed string amplitudes in genus one can be written also in terms of certain linear combinations of modular invariant real analytic versions of elliptic MZVs. This is essentially equivalent to Brown's conjecture mentioned at the end of the previous section \cite{BrownNewClass}.
\item Superstring amplitudes of $n$ points in genus one are related to superstring amplitudes of $n+2$ points in genus zero. This is a conjecture, and it still needs to be formulated in a precise way. \cite{Zerb15}.
\end{enumerate}
A first immediate goal for future investigations would consist in (making more precise, when needed, and) proving the conjectural statements in this list. Moreover, it would be extremely interesting to extend this picture to higher genera. Finally, we have left open a number of interesting questions on conical sums, modular graph functions and elliptic multiple zeta values. We hope that answering these questions could shed more light on the conjectures on superstring amplitudes listed above.

%% file: AppendixA.tex
\chapter{Proof of Theorem \ref{th1}.}

\label{AppendixCalcolo} 

In the beginning we will partially exploit the same ideas (and notations) of \cite{ZagierApp}, so we will be slightly sketchy, referring the reader to that reference for more details. 

Let us recall the definition of the function $R$:
\[
R(m_1,m_2,m_3;\alpha ,\beta ):=\sum_{(\underline{k},\underline{h},\underline{t})} \frac{\delta_0((\underline{k},\underline{h}))\delta_0((\underline{k},\underline{t}))}{|\underline{k}||\underline{h}||\underline{t}|{(\Vert\underline{k}\Vert + \Vert\underline{h}\Vert)}^\alpha{(\Vert\underline{k}\Vert + \Vert\underline{t}\Vert)}^\beta}.
\]
Note that if we have $\sum_ik_i=a$ for some $a\in\mathbb{Z}$, then we impose, using the condition in the numerator, that also $\sum_ih_i=a$ and $\sum_it_i=a$. This means that we can rewrite the series as $R(m_1,m_2,m_3;\alpha ,\beta )=R_0(m_1,m_2,m_3;\alpha ,\beta )+2R_{>0}(m_1,m_2,m_3;\alpha ,\beta )$ where
\begin{equation*}
\label{S_0}
R_0(m_1,m_2,m_3;\alpha ,\beta ):=\sum_{(\underline{k},\underline{h},\underline{t})} \frac{\delta_0(\underline{k})\delta_0(\underline{h})\delta_0(\underline{t})}{|\underline{k}||\underline{h}||\underline{t}|{(\Vert\underline{k}\Vert + \Vert\underline{h}\Vert)}^\alpha{(\Vert\underline{k}\Vert + \Vert\underline{t}\Vert)}^\beta}
\end{equation*}
and 
\begin{equation*}
R_{>0}(m_1,m_2,m_3;\alpha ,\beta ):=\sum_{a\geq 1}\sum_{(\underline{k},\underline{h},\underline{t})} \frac{\delta_a(\underline{k})\delta_a(\underline{h})\delta_a(\underline{t})}{|\underline{k}||\underline{h}||\underline{t}|{(\Vert\underline{k}\Vert + \Vert\underline{h}\Vert)}^\alpha{(\Vert\underline{k}\Vert + \Vert\underline{t}\Vert)}^\beta}.
\end{equation*}
We define, for $l\geq 1$ and for $r\geq 0$,
\[
S_{r}(l):=\sum_{\substack{k_1,\ldots,k_r\geq 1\\k_1+\cdots+k_r=l}}\frac{1}{|\underline{k}|},
\]
setting $S_{r}(l)=0$ if $r=0$ or if $r>l$.

Let us consider now $R_0(m_1,m_2,m_3;\alpha ,\beta )$: if $r_1$ is the number of positive $k_i$'s, $r_2$ and $r_3$ the same for the $h_i$'s and the $t_i$'s, and $l_1$ (resp. $l_2$ and $l_3$) is the sum of the positive $k_i$'s (resp. $h_i$'s and $t_i$'s), then
\begin{eqnarray*}
R_0(m_1,m_2,m_3;\alpha ,\beta )&=& \sum_{\substack{r_1=0,\ldots ,m_1\\r_2=0,\ldots ,m_2\\r_3=0,\ldots ,m_3}}\sum_{l_1,l_2,l_3\geq 1}\frac{\prod_{i=1}^3\binom{m_i}{r_i}S_{r_i}(l_i)S_{m_i-r_i}(l_i)}{{(2l_1+2l_2)}^\alpha {(2l_1+2l_3)}^\beta }\\
&=&\frac{1}{2^{\alpha+\beta}}\sum_{l_1,l_2,l_3\geq 1}\frac{\prod_{i=1}^3\mbox{coeff}_{x^{l_i}y^{l_i}}\big[\big(\mbox{Li}_1(x)+\mbox{Li}_1(y)\big)^{m_i}\big]}{{(l_1+l_2)}^\alpha{(l_1+l_3)}^\beta},
\end{eqnarray*}
where $\mbox{Li}_1(x)=\sum_{k\geq 1}x^k/k$.

Hence we get the generating function
\begin{multline*}
\sum_{m_1,m_2,m_3\geq 0}\frac{R_0(m_1,m_2,m_3;\alpha ,\beta )}{m_1!m_2!m_3!}X^{m_1}Y^{m_2}Z^{m_3}=\frac{1}{2^{\alpha+\beta}}\sum_{l_1,l_2,l_3\geq 1}\frac{\binom{X+l_1-1}{l_1}^2\binom{Y+l_2-1}{l_2}^2\binom{Z+l_3-1}{l_3}^2}{{(l_1+l_2)}^\alpha{(l_1+l_3)}^\beta}\\
=\frac{1}{2^{\alpha+\beta}}\sum_{l_1,l_2,l_3\geq 1}\frac{{X}^2{Y}^2{Z}^2}{{(l_1+l_2)}^\alpha{(l_1+l_3)}^\beta l_1^2l_2^2l_3^2}\prod_{n=1}^{l_1-1}\Big(1+\frac{X}{n}\Big)^2\prod_{p=1}^{l_2-1}\Big(1+\frac{Y}{p}\Big)^2\prod_{q=1}^{l_3-1}\Big(1+\frac{Z}{q}\Big)^2
\end{multline*}
\begin{multline*}
=\frac{1}{2^{\alpha+\beta}}\sum_{l_1,l_2,l_3\geq 1}\frac{{X}^2{Y}^2{Z}^2}{{(l_1+l_2)}^\alpha{(l_1+l_3)}^\beta l_1^2l_2^2l_3^2}\times \\
\times\prod_{n=1}^{l_1-1}\Big(1+4\sum_{\gamma\in\{1,2\}}\frac{(X/2)^\gamma}{n^\gamma}\Big)\prod_{p=1}^{l_2-1}\Big(1+4\sum_{\delta\in\{1,2\}}\frac{(Y/2)^\delta}{p^\delta}\Big)\prod_{q=1}^{l_3-1}\Big(1+4\sum_{\varepsilon\in\{1,2\}}\frac{(Z/2)^\varepsilon}{q^\varepsilon}\Big)\\
=\frac{1}{2^{\alpha+\beta}}\sum_{N,P,Q\geq 1}{\sum}^\prime\frac{4^{N+P+Q}{X}^{2+\underline{\gamma}}{Y}^{2+\underline{\delta}}{Z}^{2+\underline{\varepsilon}}}{2^{\underline{\gamma}+\underline{\delta}+\underline{\varepsilon}}{(l_1+l_2)}^\alpha{(l_1+l_3)}^\beta l_1^2l_2^2l_3^2n_1^{\gamma_1}\cdots n_N^{\gamma_N}p_1^{\delta_1}\cdots p_P^{\delta_P}q_1^{\varepsilon_1}\cdots q_Q^{\varepsilon_Q}},
\end{multline*}
where the sum ${\sum}^\prime$ runs over $l_1>n_1>\cdots>n_N>0$, $l_2>p_1>\cdots>p_P>0$, $l_3>q_1>\cdots>q_Q>0$ and over all the $\gamma_i$, $\delta_i$ and $\varepsilon_i$ belonging to $\{1,2\}$.

Comparing the coefficients we get
\begin{multline}\label{zero}
R_0(m_1,m_2,m_3;\alpha ,\beta )=\\
\frac{m_1!m_2!m_3!}{2^{\alpha+\beta+m_1+m_2+m_3-6}}{\sum}^{\prime\prime}\frac{2^{2(N+P+Q)}}{{(l_1+l_2)}^\alpha{(l_1+l_3)}^\beta l_1^2l_2^2l_3^2n_1^{\gamma_1}\cdots n_N^{\gamma_N}p_1^{\delta_1}\cdots p_P^{\delta_P}q_1^{\varepsilon_1}\cdots q_Q^{\varepsilon_Q}},
\end{multline}
where ${\sum}^{\prime\prime}$ runs over all the $N,P,Q$, and over all the $\gamma_i$, $\delta_i$ and $\varepsilon_i$ belonging to $\{1,2\}$ such that $\gamma_1+\cdots+\gamma_N=m_1-2$, $\delta_1+\cdots+\delta_P=m_2-2$, $\varepsilon_1+\cdots+\varepsilon_Q=m_3-2$, as well as over $l_1>n_1>\cdots>n_N>0$, $l_2>p_1>\cdots>p_P>0$, $l_3>q_1>\cdots>q_Q>0$.
Note that this sum is not zero only if all the $m_i$'s are strictly bigger than $1$. Note also that, by definition of $\mathcal{B}$, $R_0(m_1,m_2,m_3;\alpha ,\beta )\in\mathcal{B}$.

Now let us consider the more complicated sum $R_{>0}(m_1,m_2,m_3;\alpha ,\beta )$:
\begin{eqnarray*}
R_{>0}(m_1,m_2,m_3;\alpha ,\beta )&=& \sum_{a\geq 1}\sum_{\substack{r_1=0,\ldots ,m_1\\r_2=0,\ldots ,m_2\\r_3=0,\ldots ,m_3}}\sum_{l_1,l_2,l_3\geq 1}\frac{\prod_{i=1}^3\binom{m_i}{r_i}S_{r_i}(l_i+a)S_{m_i-r_i}(l_i)}{{(2l_1+2l_2+2a)}^\alpha {(2l_1+2l_3+2a)}^\beta }\\
&=&\frac{1}{2^{\alpha+\beta}}\sum_{\substack{l_1,l_2,l_3\geq 1\\a\geq 1}}\frac{\prod_{i=1}^3\mbox{coeff}_{x^{l_i+a}y^{l_i}}\big[\big(\mbox{Li}_1(x)+\mbox{Li}_1(y)\big)^{m_i}\big]}{{(l_1+l_2+a)}^\alpha{(l_1+l_3+a)}^\beta},
\end{eqnarray*}
Therefore we find the generating function
\begin{equation*}
2^{\alpha+\beta}\sum_{m_1,m_2,m_3\geq 0}\frac{R_{>0}(m_1,m_2,m_3;\alpha ,\beta )}{m_1!m_2!m_3!}X^{m_1}Y^{m_2}Z^{m_3}
\end{equation*}
\begin{eqnarray*}
&=&\sum_{\substack{l_1,l_2,l_3\geq 0\\a\geq 1}}\frac{\binom{X+l_1+a-1}{l_1+a}\binom{X+l_1-1}{l_1}\binom{Y+l_2+a-1}{l_2+a}\binom{Y+l_2-1}{l_2}\binom{Z+l_3+a-1}{l_3+a}\binom{Z+l_3-1}{l_3}}{{(l_1+l_2+a)}^\alpha{(l_1+l_3+a)}^\beta}\\
&=&\sum_{a\geq 1}\frac{\binom{X+a-1}{a}\binom{Y+a-1}{a}\binom{Z+a-1}{a}}{a^{\alpha+\beta}}\\
&+&\sum_{l_1,a\geq 1}\frac{\binom{X+l_1+a-1}{l_1+a}\binom{X+l_1-1}{l_1}\binom{Y+a-1}{a}\binom{Z+a-1}{a}}{{(l_1+a)}^{\alpha+\beta}}\\
&+&\sum_{l_2,a\geq 1}\frac{\binom{X+a-1}{a}\binom{Y+l_2+a-1}{l_2+a}\binom{Y+l_2-1}{l_2}\binom{Z+a-1}{a}}{{(l_2+a)}^{\alpha}{a}^\beta}\\
&+&\sum_{l_3,a\geq 1}\frac{\binom{X+a-1}{a}\binom{Y+a-1}{a}\binom{Z+l_3+a-1}{l_3+a}\binom{Z+l_3-1}{l_3}}{{(l_3+a)}^{\beta}{a}^\alpha}
\end{eqnarray*}
\begin{eqnarray*}
&+&\sum_{l_1,l_2,a\geq 1}\frac{\binom{X+l_1+a-1}{l_1+a}\binom{X+l_1-1}{l_1}\binom{Y+l_2+a-1}{l_2+a}\binom{Y+l_2-1}{l_2}\binom{Z+a-1}{a}}{{(l_1+l_2+a)}^\alpha{(l_1+a)}^\beta}\\
&+&\sum_{l_1,l_3,a\geq 1}\frac{\binom{X+l_1+a-1}{l_1+a}\binom{X+l_1-1}{l_1}\binom{Y+a-1}{a}\binom{Z+l_3+a-1}{l_3+a}\binom{Z+l_3-1}{l_3}}{{(l_1+a)}^\alpha{(l_1+l_3+a)}^\beta}\\
&+&\sum_{l_2,l_3,a\geq 1}\frac{\binom{X+a-1}{a}\binom{Y+l_2+a-1}{l_2+a}\binom{Y+l_2-1}{l_2}\binom{Z+l_3+a-1}{l_3+a}\binom{Z+l_3-1}{l_3}}{{(l_2+a)}^\alpha{(l_3+a)}^\beta}\\
&+&\sum_{l_1,l_2,l_3,a\geq 1}\frac{\binom{X+l_1+a-1}{l_1+a}\binom{X+l_1-1}{l_1}\binom{Y+l_2+a-1}{l_2+a}\binom{Y+l_2-1}{l_2}\binom{Z+l_3+a-1}{l_3+a}\binom{Z+l_3-1}{l_3}}{{(l_1+l_2+a)}^\alpha{(l_1+l_3+a)}^\beta}
\end{eqnarray*}
The idea is to apply to all these sums the same method shown for $R_0(m_1,m_2,m_3;\alpha ,\beta )$. Just to fix the notation, we write down explicitly what happens with the last and most complicated sum:
\begin{equation*}
\sum_{l_1,l_2,l_3,a\geq 1}\frac{\binom{X+l_1+a-1}{l_1+a}\binom{X+l_1-1}{l_1}\binom{Y+l_2+a-1}{l_2+a}\binom{Y+l_2-1}{l_2}\binom{Z+l_3+a-1}{l_3+a}\binom{Z+l_3-1}{l_3}}{{(l_1+l_2+a)}^\alpha{(l_1+l_3+a)}^\beta}
\end{equation*}
\begin{eqnarray*}
&=&\sum_{l_1,l_2,l_3,a\geq 1}\frac{{X}^2{Y}^2{Z}^2}{{(l_1+l_2+a)}^\alpha{(l_1+l_3+a)}^\beta(l_1+a)(l_2+a)(l_3+a)l_1l_2l_3}\times\\
&\times & \prod_{n=1}^{l_1+a-1}\Big(1+\frac{X}{n}\Big)\prod_{d=1}^{l_1-1}\Big(1+\frac{X}{d}\Big)\prod_{p=1}^{l_2+a-1}\Big(1+\frac{Y}{p}\Big)\prod_{e=1}^{l_2-1}\Big(1+\frac{Y}{e}\Big)\prod_{q=1}^{l_3+a-1}\Big(1+\frac{Z}{q}\Big)\prod_{f=1}^{l_3-1}\Big(1+\frac{Z}{f}\Big)\\
&=&\sum_{\substack{N,D\geq 0\\P,E\geq 0\\Q,F\geq 0}}{\sum}^{\sim}\frac{{X}^{2+N+D}{Y}^{2+P+E}{Z}^{2+Q+F}}{{(l_1+l_2+a)}^\alpha{(l_1+l_3+a)}^\beta(l_1+a)(l_2+a)(l_3+a)l_1l_2l_3n_1\cdots f_F}
\end{eqnarray*}
where ${\sum}^{\sim}$ runs over $a\geq 1$, $l_1+a>n_1>\cdots>n_N>0$, $l_1>d_1>\cdots>d_D>0$, $l_2+a> p_1>\cdots>p_P>0$, $l_2>e_1>\cdots>e_E>0$, $l_3+a>q_1>\cdots>q_Q>0$, $l_3>f_1>\cdots>f_F>0$.

Doing this for all the sums involved and comparing the coefficients, one finally obtains that $(m_1!m_2!m_3!/2^{\alpha+\beta})R_{>0}(m_1,m_2,m_3;\alpha ,\beta )$ is
\begin{eqnarray}
\label{uno}&=&{\sum}\frac{1}{a^{\alpha+\beta+3}n_1\cdots q_{m_3-1}}\\
\label{due}&+&\sum_{\substack{N,D\geq 0\\N+D=m_1-2}}{\sum}\frac{1}{{(l_1+a)}^{\alpha+\beta+1}a^2l_1n_1\cdots q_{m_3-1}}\\
\label{tre}&+&\sum_{\substack{P,E\geq 0\\P+E=m_2-2}}{\sum}\frac{1}{{(l_2+a)}^{\alpha+1}a^{\beta+2}l_2n_1\cdots q_{m_3-1}}\\
\label{quattro}&+&\sum_{\substack{Q,F\geq 0\\Q+F=m_3-2}}{\sum}\frac{1}{{(l_3+a)}^{\beta+1}a^{\alpha+2}l_3n_1\cdots f_F}\\
\label{cinque}&+&\sum_{\substack{N,D,P,E\geq 0\\N+D=m_1-2\\P+E=m_2-2}}{\sum}\frac{1}{{(l_1+l_2+a)}^{\alpha}{(l_1+a)}^{\beta+1}(l_2+a)a\;l_1l_2n_1\cdots q_{m_3-1}}
\end{eqnarray}
\begin{eqnarray}
\label{sei}&+&\sum_{\substack{N,D,Q,F\geq 0\\N+D=m_1-2\\Q+F=m_3-2}}{\sum}\frac{1}{{(l_1+a)}^{\alpha+1}{(l_1+l_3+a)}^{\beta}(l_3+a)a\;l_1l_3n_1\cdots f_F}\\
\label{sette}&+&\sum_{\substack{P,E,Q,F\geq 0\\P+E=m_2-2\\Q+F=m_3-2}}{\sum}\frac{1}{{(l_2+a)}^{\alpha+1}{(l_3+a)}^{\beta+1}a\;l_2l_3n_1\cdots f_F}\\
\label{otto}&+&\sum_{\substack{N,D,P,E\geq 0\\N+D=m_1-2\\P+E=m_2-2\\Q+F=m_3-2}}{\sum}\frac{1}{{(l_1+l_2+a)}^\alpha{(l_1+l_3+a)}^\beta(l_1+a)(l_2+a)(l_3+a)l_1l_2l_3n_1\cdots f_F}
\end{eqnarray}
The sum in (\ref{uno}) runs over $n_1>\cdots >n_{m_1-1}>0$, $p_1>\cdots >p_{m_2-1}>0$, $q_1>\cdots >q_{m_3-1}>0$, $a>\max\{n_1,p_1,q_1\}$.\\
The sum in (\ref{due}) runs over $l_1+a>n_1>\cdots >n_N>0$, $l_1>d_1>\cdots >d_D>0$, $p_1>\cdots >p_{m_2-1}>0$, $q_1>\cdots >q_{m_3-1}>0$, $a>\max\{p_1,q_1\}$, and is $0$ if $m_1=1$.\\
The sum in (\ref{tre}) runs over $n_1>\cdots >n_{m_1-1}>0$, $l_2+a>p_1>\cdots >p_P>0$, $l_2>e_1>\cdots >e_E>0$, $q_1>\cdots >q_{m_3-1}>0$, $a>\max\{n_1,q_1\}$, and is $0$ if $m_2=1$.\\
The sum in (\ref{quattro}) runs over $n_1>\cdots >n_{m_1-1}>0$, $p_1>\cdots >p_{m_2-1}>0$, $l_3+a>q_1>\cdots >q_Q>0$, $l_3>f_1>\cdots >f_F>0$, $a>\max\{n_1,p_1\}$, and is $0$ if $m_3=1$.\\
The sum in (\ref{cinque}) runs over $l_1+a>n_1>\cdots >n_N>0$, $l_1>d_1>\cdots >d_D>0$, $l_2+a>p_1>\cdots >p_P>0$, $l_2>e_1>\cdots >e_E>0$, $a>q_1>\cdots >q_{m_3-1}>0$, and is $0$ if $m_1=1$ or $m_2=1$.\\
The sum in (\ref{sei}) runs over $l_1+a>n_1>\cdots >n_N>0$, $l_1>d_1>\cdots >d_D>0$, $a>p_1>\cdots >p_{m_2-1}>0$, $l_3+a>q_1>\cdots >q_Q>0$, $l_3>f_1>\cdots >f_F>0$, and is $0$ if $m_1=1$ or $m_3=1$.\\
The sum in (\ref{sette}) runs over $a>n_1>\cdots >n_{m_1-1}>0$, $l_2+a>p_1>\cdots >p_P>0$, $l_2>e_1>\cdots >e_E>0$, $l_3+a>q_1>\cdots >q_Q>0$, $l_3>f_1>\cdots >f_F>0$, and is $0$ if $m_2=1$ or $m_3=1$.\\
The sum in (\ref{otto}) runs over $l_1+a>n_1>\cdots >n_N>0$, $l_1>d_1>\cdots >d_D>0$, $l_2+a>p_1>\cdots >p_P>0$,  $l_2>e_1>\cdots >e_E>0$, $l_3+a>q_1>\cdots >q_Q>0$, $l_3>f_1>\cdots >f_F>0$, and is $0$ if one of the $m_i$'s is $<2$.

From this formula it is not clear yet whether these numbers are in $\mathcal{B}$, so one needs to quasi-shuffle, or stuffle, some groups of variables.

In (\ref{uno}) one has to stuffle the 3 groups of ordered variables $n_i$, $p_i$, $q_i$; then setting $a>\max\{n_1,p_1,q_1\}$ we directly get MZV.\\
In (\ref{due}) one has to stuffle the 2 groups of ordered variables $p_i$, $q_i$ and the 2 groups of ordered variables $n_i$ and $l_1>d_1>\cdots >d_D$, in order to get sums of the kind, for $1\leq i\leq N$ and $N,M\geq 1$,
\[
\sum_{\substack{y_M>\cdots >y_1>0 \\ x_i+y_M>x_N>\cdots >x_1>0}}\frac{1}{x_1^{\eta_1}\cdots y_M^{\eta_{N+M}} (x_i+y_M)^\varepsilon}.
\]
Furthermore, if we stuffle the groups of ordered variables $y_M>\cdots >y_1>0$ and $x_N-x_i>\cdots >x_{i+1}-x_i>0$, then we get numbers in $\mathcal{B}$.

The same reasoning works with some obvious modification for all the other sums, and this proves our assertion.\\
$\square$

\chapter{Three vector-valued modular forms}

\label{AppendixVectorValued}

Let us write $P:=2\pi i$, $K:=P^4/720$, and let us consider the three vectors

\begin{equation*}
V_{3,2}(\tau)=
\begin{bmatrix}
P^2\tau^3 A_{3,2}(\tau)+P\tau^2 A_{2,3}(\tau)+\tau\hat{A}_{1,4}(\tau)-K\tau^4+20K\tau^2\\
P^2\tau^2 A_{3,2}(\tau)+\frac{2P\tau}{3}A_{2,3}(\tau)+\frac{1}{3}\hat{A}_{1,4}(\tau)-\frac{4K\tau^3}{3}\\
P^2\tau A_{3,2}(\tau)+\frac{P}{3}A_{2,3}(\tau)-2K\tau^2\\
P^2A_{3,2}(\tau)\\
K\tau\\
K
\end{bmatrix},
\end{equation*}
\begin{equation*}
V_{2,3}(\tau)=
\begin{bmatrix}
P\tau^2 A_{2,3}(\tau)+2\tau\hat{A}_{1,4}(\tau)+K\tau^4\\
P\tau A_{2,3}(\tau)+\hat{A}_{1,4}(\tau)+2K\tau^3\\
P A_{2,3}(\tau)\\
K\tau^2\\
K\tau\\
K
\end{bmatrix},
\end{equation*}
\begin{equation*}
V_{1,4}(\tau)=
\begin{bmatrix}
\tau\hat{A}_{1,4}(\tau)-K\tau^4\\
\hat{A}_{1,4}(\tau)\\
K\tau^3\\
K\tau^2\\
K\tau\\
K
\end{bmatrix}.
\end{equation*}

Then using the explicit formulae obtained in Section \ref{SectionDepthOne} we can see them as vector-valued modular forms for SL$_2(\mathbb{Z})$:

\begin{equation*}
V_{3,2}(\tau)\,\bigg|_{-1}T=
\begin{bmatrix}
1 & 3 & 3 & 1 & 36 & 19\\
0 & 1 & 2 & 1 & -4 & -\frac{4}{3}\\
0 & 0 & 1 & 1 & -4 & -2\\
0 & 0 & 0 & 1 & 0 & 0\\
0 & 0 & 0 & 0 & 1 & 1\\
0 & 0 & 0 & 0 & 0 & 1
\end{bmatrix}
\,\,V_{3,2}(\tau)
\end{equation*}
\begin{equation*}
V_{3,2}(\tau)\,\bigg|_{-1}S=
\begin{bmatrix}
0 & 0 & 0 & -1 & 3 & 0\\
0 & 0 & 1 & 0 & 0 & \frac{55}{3}\\
0 & -1 & 0 & 0 & -\frac{55}{3} & 0\\
1 & 0 & 0 & 0 & 0 & -3\\
0 & 0 & 0 & 0 & 0 & -1\\
0 & 0 & 0 & 0 & 1 & 0
\end{bmatrix}
\,\,V_{3,2}(\tau)
\end{equation*}
\begin{equation*}
V_{2,3}(\tau)\,\bigg|_{-2}T=
\begin{bmatrix}
1 & 2 & 1 & 6 & 2 & 1\\
0 & 1 & 1 & 6 & 6 & 2\\
0 & 0 & 1 & 0 & 0 & 0\\
0 & 0 & 0 & 1 & 2 & 1\\
0 & 0 & 0 & 0 & 1 & 1\\
0 & 0 & 0 & 0 & 0 & 1
\end{bmatrix}
\,\,V_{2,3}(\tau)
\end{equation*}
\begin{equation*}
V_{2,3}(\tau)\,\bigg|_{-2}S=
\begin{bmatrix}
0 & 0 & 1 & 1 & 0 & -5\\
0 & -1 & 0 & 0 & 0 & 0\\
1 & 0 & 0 & 5 & 0 & -1\\
0 & 0 & 0 & 0 & 0 & 1\\
0 & 0 & 0 & 0 & -1 & 0\\
0 & 0 & 0 & 1 & 0 & 0
\end{bmatrix}
\,\,V_{2,3}(\tau)
\end{equation*}
\begin{equation*}
V_{1,4}(\tau)\,\bigg|_{-3}T=
\begin{bmatrix}
1 & 1 & -4 & -6 & -4 & -1\\
0 & 1 & 0 & 0 & 0 & 0\\
0 & 0 & 1 & 3 & 3 & 1\\
0 & 0 & 0 & 1 & 2 & 1\\
0 & 0 & 0 & 0 & 1 & 1\\
0 & 0 & 0 & 0 & 0 & 1
\end{bmatrix}
\,\,V_{1,4}(\tau)
\end{equation*}
\begin{equation*}
V_{1,4}(\tau)\,\bigg|_{-3}S=
\begin{bmatrix}
0 & -1 & 1 & 0 & -5 & 0\\
1 & 0 & 0 & 5 & 0 & -1\\
0 & 0 & 0 & 0 & 0 & -1\\
0 & 0 & 0 & 0 & 1 & 0\\
0 & 0 & 0 & -1 & 0 & 0\\
0 & 0 & 1 & 0 & 0 & 0
\end{bmatrix}
\,\,V_{1,4}(\tau)
\end{equation*}

These are explicit examples of Theorem \ref{ThmModBehaviour}, which indeed predicts that $A_{3,2}(\tau)$ can be seen as a component of a weight $-1$ vector-valued modular form, $A_{2,3}(\tau)$ can be seen as a component of a weight $-2$ vector-valued modular form, and $\hat{A}_{1,4}(\tau)$ can be seen as a component of a weight $-3$ vector-valued modular form. 

%% file: main.bbl
\begin{thebibliography}{10}

\bibitem{Weinzierl14}
L.~Adams, C.~Bogner, and S.~Weinzierl.
\newblock The two-loop sunrise graph in two space-time dimensions with
  arbitrary masses in terms of elliptic dilogarithms.
\newblock {\em J. Math. Phys.}, 55(10):102301, 17, 2014.

\bibitem{AnzaiSumino}
C.~Anzai and Y.~Sumino.
\newblock Algorithms to evaluate multiple sums for loop computations.
\newblock {\em J. Math. Phys.}, 54(3):033514, 22, 2013.

\bibitem{Apery}
R.~Ap\'ery.
\newblock Irrationalit\'e de $\zeta(2)$ et $\zeta(3)$.
\newblock {\em Ast\'erisque}, 61:11--13, 1979.

\bibitem{Basu1}
A.~Basu.
\newblock Proving relations between modular graph functions.
\newblock {\em Classical Quantum Gravity}, 33(23):235011, 24, 2016.

\bibitem{BeilinsonLevin}
A.~Beilinson and A.~Levin.
\newblock The elliptic polylogarithm.
\newblock In {\em Motives ({S}eattle, {WA}, 1991)}, volume~55 of {\em Proc.
  Sympos. Pure Math.}, pages 123--190. Amer. Math. Soc., Providence, RI, 1994.

\bibitem{BelBros}
P.~Belkale and P.~Brosnan.
\newblock Matroids, motives, and a conjecture of {K}ontsevich.
\newblock {\em Duke Math. J.}, 116(1):147--188, 2003.

\bibitem{BCJ}
Z.~Bern, J.J.M. Carrasco, and H.~Johansson.
\newblock Perturbative quantum gravity as a double copy of gauge theory.
\newblock {\em Phys. Rev. Lett.}, 105(6):061602, 4, 2010.

\bibitem{BBDamgaardVanhove}
N.E.J. Bjerrum-Bohr, P.~Damgaard, and P.~Vanhove.
\newblock Minimal basis for gauge theory amplitudes.
\newblock {\em Phys.Rev.Lett.}, 103, 2009.

\bibitem{BlochRegulators}
S.~Bloch.
\newblock {\em Higher regulators, algebraic {$K$}-theory, and zeta functions of
  elliptic curves}, volume~11 of {\em CRM Monograph Series}.
\newblock American Mathematical Society, Providence, RI, 2000.

\bibitem{BlochEsnaultKreimer}
S.~Bloch, H.~Esnault, and D.~Kreimer.
\newblock On motives associated to graph polynomials.
\newblock {\em Comm. Math. Phys.}, 267(1):181--225, 2006.

\bibitem{BlochVanhove}
S.~Bloch and P.~Vanhove.
\newblock The elliptic dilogarithm for the sunset graph.
\newblock {\em J. Number Theory}, 148:328--364, 2015.

\bibitem{BradleyZhou}
D.M. Bradley and X.~Zhou.
\newblock On {M}ordell-{T}ornheim sums and multiple zeta values.
\newblock {\em Ann. Sci. Math. Qu\'ebec}, 34(1):15--23, 2010.

\bibitem{BroadKreimer}
D.J. Broadhurst and D.~Kreimer.
\newblock Knots and numbers in {$\phi^4$} theory to {$7$} loops and beyond.
\newblock {\em Internat. J. Modern Phys. C}, 6(4):519--524, 1995.

\bibitem{BMMS}
J.~Broedel, C.R. Mafra, N.~Matthes, and O.~Schlotterer.
\newblock Elliptic multiple zeta values and one-loop superstring amplitudes.
\newblock {\em J. High Energy Phys.}, (7):112, 2015.

\bibitem{BMRS}
J.~Broedel, N.~Matthes, G.~Richter, and O.~Schlotterer.
\newblock Twisted elliptic multiple zeta values and non-planar one-loop
  open-string amplitudes.
\newblock {\em arXiv:1704.03449 [hep-th]}, 2017.

\bibitem{BMS}
J.~Broedel, N.~Matthes, and O.~Schlotterer.
\newblock Relations between elliptic multiple zeta values and a special
  derivation algebra.
\newblock {\em J.Phys.}, A49:155203, 49, 2016.

\bibitem{Zerb17}
J~Broedel, N.~Matthes, O.~Schlotterer, and F.~Zerbini.
\newblock Towards single-valued elliptic polylogarithms.
\newblock {\em In preparation}.

\bibitem{BSST}
J.~Broedel, O.~Schlotterer, S.~Stieberger, and T.~Terasoma.
\newblock All order $\alpha^\prime$-expansion of superstring trees from the
  {D}rinfeld associator.
\newblock {\em Phys. Rev. D}, 89, 2014.

\bibitem{BrownSVMPL}
F.~Brown.
\newblock Single-valued multiple polylogarithms in one variable.
\newblock {\em C.R. Acad. Sci. Paris}, 338:527--532, 2004.

\bibitem{BrownFeynmanInt}
F.~Brown.
\newblock On the periods of some {F}eynman integrals.
\newblock {\em arXiv:0910.0114 [math.AG]}, 2009.

\bibitem{BrownMTM}
F.~Brown.
\newblock Mixed {T}ate motives over {$\Bbb Z$}.
\newblock {\em Ann. of Math. (2)}, 175(2):949--976, 2012.

\bibitem{MMV}
F.~Brown.
\newblock Multiple {M}odular {V}alues and the relative completion of the
  fundamental group of $\mathcal{M}_{1,1}$.
\newblock {\em arXiv:1407.5167 [math.NT]}, 2014.

\bibitem{BrownSVMZV}
F.~Brown.
\newblock Single-valued motivic periods and multiple zeta values.
\newblock {\em Forum Math. Sigma}, 2, 2014.

\bibitem{BrownNewClass}
F.~Brown.
\newblock A class of non-holomorphic modular forms i.
\newblock {\em arXiv:1707.01230 [math.NT]}, 2017.

\bibitem{BrownLevin}
F.~Brown and A.~Levin.
\newblock Multiple {E}lliptic {P}olylogarithms.
\newblock {\em arXiv:1110.6917 [math.NT]}, 2011.

\bibitem{BrownSchnetz}
F.~Brown and O.~Schnetz.
\newblock A {K}3 in {$\phi^4$}.
\newblock {\em Duke Math. J.}, 161(10):1817--1862, 2012.

\bibitem{Brown2006}
F.C.S. Brown.
\newblock P\'eriodes des espaces des modules {$\overline{M}_{0,n}$} et valeurs
  z\^etas multiples.
\newblock {\em C. R. Math. Acad. Sci. Paris}, 342(12):949--954, 2006.

\bibitem{BrownModuliSpace}
F.C.S. Brown.
\newblock Multiple zeta values and periods of moduli spaces
  {$\overline{\mathcal{M}}_{0,n}$}.
\newblock {\em Ann. Sci. \'Ec. Norm. Sup\'er. (4)}, 42(3):371--489, 2009.

\bibitem{BurgosFresan}
J.I. Burgos~Gil and J.~Fres\'an.
\newblock Multiple zeta values: from number theory to motives.
\newblock {\em Clay Mathematics Proceedings}, To appear,
  https://people.math.ethz.ch/~jfresan/mzv.pdf.

\bibitem{CalEtiEnr}
D.~Calaque, B.~Enriquez, and P.~Etingof.
\newblock Universal {KZB} equations: the elliptic case.
\newblock In {\em Algebra, arithmetic, and geometry: in honor of {Y}u. {I}.
  {M}anin. {V}ol. {I}}, volume 269 of {\em Progr. Math.}, pages 165--266.
  Birkh\"auser Boston, Inc., Boston, MA, 2009.

\bibitem{Cartier}
P.~Cartier.
\newblock Fonctions polylogarithmes, nombres polyz\^etas et groupes
  pro-unipotents.
\newblock {\em Ast\'erisque}, (282):Exp. No. 885, viii, 137--173, 2002.
\newblock S\'eminaire Bourbaki, Vol. 2000/2001.

\bibitem{Chen}
K.T. Chen.
\newblock Iterated path integrals.
\newblock {\em Bull. Amer. Math. Soc.}, 83(5):831--879, 1977.

\bibitem{DeligneGoncharov}
P.~Deligne and A.B. Goncharov.
\newblock Groupes fondamentaux motiviques de {T}ate mixte.
\newblock {\em Ann. Sci. \'Ecole Norm. Sup. (4)}, 38(1):1--56, 2005.

\bibitem{DGPR}
E.~D'Hoker, M.~B. Green, B.~Pioline, and R.~Russo.
\newblock Matching the {$D^6R^4$} interaction at two-loops.
\newblock {\em J. High Energy Phys.}, (01):031, 2015.

\bibitem{DG2014}
E.~D'Hoker and M.B. Green.
\newblock Zhang-{K}awazumi {I}nvariants and {S}uperstring {A}mplitudes.
\newblock {\em Journal of Number Theory}, 144:111, 2014.

\bibitem{DG}
E.~D'Hoker and M.B. Green.
\newblock Identities between modular graph forms.
\newblock {\em arXiv:1603.00839 [hep-th]}, 2016.

\bibitem{DGGV}
E.~D'Hoker, M.B. Green, \"O. G\"urdo\u{g}an, and P.~Vanhove.
\newblock Modular graph functions.
\newblock {\em Commun. Number Theory Phys.}, 11(1):165--218, 2017.

\bibitem{DGV2015}
E.~D'Hoker, M.B. Green, and P.~Vanhove.
\newblock On the modular structure of the genus-one type {II} superstring low
  energy expansion.
\newblock {\em JHEP}, 041(08), 2015.

\bibitem{DGV2}
E.~D'Hoker, M.B. Green, and P.~Vanhove.
\newblock Proof of a modular relation between 1-, 2- and 3-loop {F}eynman
  diagrams on a torus.
\newblock {\em arXiv:1509.00363 [hep-th]}, 2015.

\bibitem{DHokerKaidi}
E.~D'Hoker and J.~Kaidi.
\newblock Hierarchy of modular graph identities.
\newblock {\em J. High Energy Phys.}, (11):051, front matter + 49, 2016.

\bibitem{DonagiWitten}
R.~Donagi and E.~Witten.
\newblock Supermoduli space is not projected.
\newblock In {\em String-{M}ath 2012}, volume~90 of {\em Proc. Sympos. Pure
  Math.}, pages 19--71. Amer. Math. Soc., Providence, RI, 2015.

\bibitem{Drinfeld}
V.G. Drinfel'd.
\newblock On quasitriangular quasi-{H}opf algebras and on a group that is
  closely connected with ${G}al(\overline{\mathbb{q}}/\mathbb{Q})$.
\newblock {\em Algebra i Analiz}, 2(4):149--181, 1990.

\bibitem{EnriquezAss}
B.~Enriquez.
\newblock Elliptic associators.
\newblock {\em Selecta Math. (N.S.)}, 20(2):491--584, 2014.

\bibitem{Enriquez}
B.~Enriquez.
\newblock Analogues elliptiques des nombres multiz\'etas.
\newblock {\em Bull. Soc. Math. France}, 144(3):395--427, 2016.

\bibitem{GomezMafra}
H.~Gomez and C.R. Mafra.
\newblock The closed-string 3-loop amplitude and {S}-duality.
\newblock {\em J. High Energy Phys.}, (10):217, 2013.

\bibitem{GomezMafraSc}
H.~Gomez, C.R. Mafra, and O.~Schlotterer.
\newblock The two-loop superstring five-point amplitude and s-duality.
\newblock {\em Phys. Rev.}, (D 93), 2016.

\bibitem{GonchManin}
A.~B. Goncharov and Yu.~I. Manin.
\newblock Multiple {$\zeta$}-motives and moduli spaces
  {$\overline{\mathcal{M}}_{0,n}$}.
\newblock {\em Compos. Math.}, 140(1):1--14, 2004.

\bibitem{GRV}
M.~B. Green, J.~G. Russo, and P.~Vanhove.
\newblock Low energy expansion of the four-particle genus-one amplitude in type
  {II} superstring theory.
\newblock {\em JHEP}, (0802), 2008.

\bibitem{GV2000}
M.~B. Green and P.~Vanhove.
\newblock The low-energy expansion of the one loop type {II} superstring
  amplitude.
\newblock {\em Phys.Rev.}, D61, 2000.

\bibitem{GSW}
M.B. Green, J.H. Schwarz, and E.~Witten.
\newblock Superstring theory. vol. 2: Loop amplitudes, anomalies and
  phenomenology.
\newblock {\em Cambridge, Uk: Univ. Pr. (Cambridge Monographs On Mathematical
  Physics)}, 596, 1987.

\bibitem{Paycha}
L.~Guo, S.~Paycha, and B.~Zhang.
\newblock Conical zeta values and their double subdivision relations.
\newblock {\em Adv. Math.}, 252:343--381, 2014.

\bibitem{HainItInt}
R.M. Hain.
\newblock The geometry of the mixed {H}odge structure on the fundamental group.
\newblock In {\em Algebraic geometry, {B}owdoin, 1985 ({B}runswick, {M}aine,
  1985)}, volume~46 of {\em Proc. Sympos. Pure Math.}, pages 247--282. Amer.
  Math. Soc., Providence, RI, 1987.

\bibitem{IKZ}
K.~Ihara, M.~Kaneko, and D.~Zagier.
\newblock Derivation and double shuffle relations for multiple zeta values.
\newblock {\em Compos. Math.}, 142(2):307--338, 2006.

\bibitem{KLT}
H.~Kawai, D.C. Lewellen, and S.H.~H. Tye.
\newblock A relation between tree amplitudes of closed and open strings.
\newblock {\em Nuclear Phys. B}, 269(1):1--23, 1986.

\bibitem{Kitazawa}
Y.~Kitazawa.
\newblock Effective lagrangian for open superstring from five point function.
\newblock {\em Nucl.Phys.}, B289:599--608, 1987.

\bibitem{AxelValentin}
A.~Kleinschmidt and V.~Verschinin.
\newblock Tetrahedral modular graph functions.
\newblock {\em arXiv:1706.01889 [hep-th]}, 2017.

\bibitem{KontsZagier}
M.~Kontsevich and D.~Zagier.
\newblock Periods.
\newblock In {\em Mathematics unlimited---2001 and beyond}, pages 771--808.
  Springer, Berlin, 2001.

\bibitem{Lang}
S.~Lang.
\newblock {\em Complex analysis}, volume 103 of {\em Graduate Texts in
  Mathematics}.
\newblock Springer-Verlag, New York, fourth edition, 1999.

\bibitem{LeMurakami}
T.T.Q. Le and J.~Murakami.
\newblock Kontsevich's integral for the {K}auffman polynomial.
\newblock {\em Nagoya Math. J.}, 142:39--65, 1996.

\bibitem{Levin}
A.~Levin.
\newblock Elliptic polylogarithms: an analytic theory.
\newblock {\em Compositio Math.}, 106(3):267--282, 1997.

\bibitem{LewisZagier}
J.~Lewis and D.~Zagier.
\newblock Period functions for {M}aass wave forms. {I}.
\newblock {\em Ann. of Math. (2)}, 153(1):191--258, 2001.

\bibitem{LoMaSc}
P.~Lochak, N.~Matthes, and L.~Schneps.
\newblock Elliptic multiple zeta values and the elliptic double shuffle
  relations.
\newblock {\em arXiv:1703.09410 [math.NT]}, 2017.

\bibitem{MafraScSt1}
C.R. Mafra, O.~Schlotterer, and S.~Stieberger.
\newblock Complete {$N$}-point superstring disk amplitude {I}. {P}ure spinor
  computation.
\newblock {\em Nuclear Phys. B}, 873(3):419--460, 2013.

\bibitem{ManinItInt}
Y.I. Manin.
\newblock Iterated integrals of modular forms and noncommutative modular
  symbols.
\newblock In {\em Algebraic geometry and number theory}, volume 253 of {\em
  Progr. Math.}, pages 565--597. Birkh\"auser Boston, Boston, MA, 2006.

\bibitem{MatthesThesis}
N.~Matthes.
\newblock Elliptic multiple zeta values.
\newblock {\em PhD thesis}, 2016.

\bibitem{MatthesMeta}
N.~Matthes.
\newblock The meta-abelian elliptic {KZB} associator and periods of
  {E}isenstein series.
\newblock {\em arXiv:1608.00740v2 [math.NT]}, 2016.

\bibitem{MatthesDecomposition}
N.~Matthes.
\newblock Decomposition of elliptic multiple zeta values and iterated
  eisenstein integrals.
\newblock {\em arXiv:1703.09597 [math.NT]}, 2017.

\bibitem{MatthesDouble}
N.~Matthes.
\newblock Elliptic double zeta values.
\newblock {\em J. Number Theory}, 171:227--251, 2017.

\bibitem{MumfordTataI}
D.~Mumford.
\newblock {\em Tata lectures on theta. {I}}.
\newblock Modern Birkh\"auser Classics. Birkh\"auser Boston, Inc., Boston, MA,
  2007.
\newblock With the collaboration of C. Musili, M. Nori, E. Previato and M.
  Stillman, Reprint of the 1983 edition.

\bibitem{Panzer}
E.~Panzer.
\newblock Algorithms for the symbolic integration of hyperlogarithms with
  applications to {F}eynman integrals.
\newblock {\em Computer Physics Communications}, 188:148--166, 2015.

\bibitem{Rivoal}
T.~Rivoal.
\newblock La fonction z\^eta de {R}iemann prend une infinit\'e de valeurs
  irrationnelles aux entiers impairs.
\newblock {\em C. R. Acad. Sci. Paris S\'er. I Math.}, 331(4):267--270, 2000.

\bibitem{ScSt}
O.~Schlotterer and S.~Stieberger.
\newblock Motivic multiple zeta values and superstring amplitudes.
\newblock {\em J. Phys. A}, 46(47), 2013.

\bibitem{Stieb2014}
S.~Stieberger.
\newblock Closed superstring amplitudes, single-valued multiple zeta values and
  {D}eligne associator.
\newblock {\em J. Phys. A}, 47, 2014.

\bibitem{StiebTaylor}
S.~Stieberger and T.R. Taylor.
\newblock Closed string amplitudes as single-valued open string amplitudes.
\newblock {\em Nuclear Phys. B}, 881:269--287, 2014.

\bibitem{TerasomaMZV}
T.~Terasoma.
\newblock Mixed {T}ate motives and multiple zeta values.
\newblock {\em Invent. Math.}, 149(2):339--369, 2002.

\bibitem{TerasomaSelberg}
T.~Terasoma.
\newblock Selberg integrals and multiple zeta values.
\newblock {\em Compositio Math.}, 133(1):1--24, 2002.

\bibitem{TerasomaConical}
T.~Terasoma.
\newblock Rational convex cones and cyclotomic multiple zeta values.
\newblock {\em arXiv:math/0410306 [math.AG]}, 2004.

\bibitem{Veneziano}
G.~Veneziano.
\newblock Construction of a crossing-symmetric, {R}egge behaved amplitude for
  linearly rising trajectories.
\newblock {\em Nuovo Cim.}, A57:190--197, 1968.

\bibitem{Virasoro}
M.A. Virasoro.
\newblock Alternative constructions of crossing-symmetric amplitudes with
  {R}egge behaviour.
\newblock {\em Phys. Rev.}, 177:2309--2311, 1969.

\bibitem{Weil}
A.~Weil.
\newblock {\em Elliptic functions according to {E}isenstein and {K}ronecker}.
\newblock Classics in Mathematics. Springer-Verlag, Berlin, 1999.
\newblock Reprint of the 1976 original.

\bibitem{WittenSuperstringPerturbationRevisited}
E.~Witten.
\newblock Superstring perturbation theory revisited.
\newblock {\em arXiv:1209.5461 [hep-th]}, 2012.

\bibitem{Wojtk}
Z.~Wojtkowiak.
\newblock A construction of analogs of the {B}loch-{W}igner function.
\newblock {\em Math. Scand.}, 65(1):140--142, 1989.

\bibitem{ZagierApp}
D.~Zagier.
\newblock Evaluation of {$S(m,n)$}. {A}ppendix to \cite{GRV}.
\newblock pages 30--31.

\bibitem{ZagierStrings}
D.~Zagier.
\newblock Genus 0 and genus 1 string amplitudes and multiple zeta values.
\newblock {\em In preparation}.

\bibitem{ZagierDilog}
D.~Zagier.
\newblock The remarkable dilogarithm.
\newblock {\em J. Math. Phys. Sci.}, 22(1):131--145, 1988.

\bibitem{ZagierBWR}
D.~Zagier.
\newblock The {B}loch-{W}igner-{R}amakrishnan polylogarithm function.
\newblock {\em Math. Ann.}, 286(1-3):613--624, 1990.

\bibitem{ZagierPeriodsJacobi}
D.~Zagier.
\newblock Periods of modular forms and {J}acobi theta functions.
\newblock {\em Invent. Math.}, 104(3):449--465, 1991.

\bibitem{ZagierPolylog}
D.~Zagier.
\newblock Polylogarithms, {D}edekind zeta functions and the algebraic
  {$K$}-theory of fields.
\newblock In {\em Arithmetic algebraic geometry ({T}exel, 1989)}, volume~89 of
  {\em Progr. Math.}, pages 391--430. Birkh\"auser Boston, Boston, MA, 1991.

\bibitem{ZagierMZV}
D.~Zagier.
\newblock Values of zeta functions and their applications.
\newblock In {\em First {E}uropean {C}ongress of {M}athematics, {V}ol.\ {II}
  ({P}aris, 1992)}, volume 120 of {\em Progr. Math.}, pages 497--512.
  Birkh\"auser, Basel, 1994.

\bibitem{1-2-3ModForms}
D.~Zagier.
\newblock Elliptic modular forms and their applications.
\newblock In {\em The 1-2-3 of modular forms}, Universitext, pages 1--103.
  Springer, Berlin, 2008.

\bibitem{ZagierBourbaki}
D.~Zagier.
\newblock Ramanujan's mock theta functions and their applications (after
  {Z}wegers and {O}no-{B}ringmann).
\newblock {\em Ast\'erisque}, (326):Exp. No. 986, vii--viii, 143--164 (2010),
  2009.
\newblock S\'eminaire Bourbaki. Vol. 2007/2008.

\bibitem{ZagierGangl}
D.~Zagier and H.~Gangl.
\newblock Classical and elliptic polylogarithms and special values of
  {$L$}-series.
\newblock In {\em The arithmetic and geometry of algebraic cycles ({B}anff,
  {AB}, 1998)}, volume 548 of {\em NATO Sci. Ser. C Math. Phys. Sci.}, pages
  561--615. Kluwer Acad. Publ., Dordrecht, 2000.

\bibitem{Zerb15}
F.~Zerbini.
\newblock Single-valued multiple zeta values in genus 1 superstring amplitudes.
\newblock {\em Commun. Number Theory Phys.}, 10(4):703--737, 2016.

\bibitem{Zhao}
J.~Zhao.
\newblock {\em Multiple zeta functions, multiple polylogarithms and their
  special values}, volume~12 of {\em Series on Number Theory and its
  Applications}.
\newblock World Scientific Publishing Co. Pte. Ltd., Hackensack, NJ, 2016.

\bibitem{Zudilin}
V.V. Zudilin.
\newblock One of the numbers {$\zeta(5)$}, {$\zeta(7)$}, {$\zeta(9)$},
  {$\zeta(11)$} is irrational.
\newblock {\em Uspekhi Mat. Nauk}, 56(4(340)):149--150, 2001.

\end{thebibliography}
